\documentclass[
12pt,		
openright,	
twoside,  
a4paper,			
chapter=TITLE,		
french,				
spanish,			
brazil,			
english, 
 aps,
 pra,
 reprint,
 amsmath,
 amssymb
]{USPSC1}

\usepackage{verbatim}
\usepackage[export]{adjustbox}
\usepackage{subcaption}
\usepackage{amsmath}
\usepackage{amssymb}

\usepackage{dcolumn}
\usepackage{bm}
\usepackage{hyperref}
\usepackage{cleveref}
\usepackage[mathlines]{lineno}

\usepackage[T1]{fontenc}		
\usepackage[utf8]{inputenc}		
\usepackage{lmodern}			
\usepackage{lastpage}			
\usepackage{indentfirst}		
\usepackage{color}				
\usepackage{graphicx}			
\usepackage{float} 				
\usepackage{chemfig,chemmacros} 
\usepackage{microtype} 			
\usepackage{pdfpages}
\usepackage{makeidx}            
\usepackage{xspace,epsfig,amsfonts}

\usepackage{amsmath, amsthm, amssymb}
\usepackage{xcolor}
\usepackage{CJK}
\usepackage{graphicx}
\usepackage[justification=justified]{caption}
\usepackage[export]{adjustbox}
\usepackage{epsfig}
\usepackage{wasysym}
\usepackage{multirow}
\usepackage{dsfont}
\usepackage{graphicx}
\usepackage{caption}
\usepackage{subcaption}
\usepackage{float}
\usepackage{url}
\usepackage{hyperref}
\usepackage{bm}
\usepackage{verbatim}
\usepackage{booktabs}

\usepackage{cite}

\usepackage[num,overcite,abnt-emphasize=bf,abnt-thesis-year=both, abnt-repeated-author-omit=yes, abnt-last-names=abnt, abnt-etal-cite,abnt-etal-list=0, abnt-etal-text=default,abnt-and-type=e]{abntex2cite}

\renewcommand{\thefootnote}{\fnsymbol{footnote}}  

\renewcommand{\footnotesize}{\small} 

\usepackage{lipsum}				

\usepackage{multicol}	
\usepackage{multirow}	
\usepackage{longtable}	
\usepackage{threeparttablex}    
\usepackage{array}
\usepackage{dirtytalk}

\setatomsep{2em}
\setdoublesep{.6ex}
\setbondstyle{semithick}
{%
	\begin{labeling}
		{\rsnumber{R39/23/24/25}}
	}{%
\end{labeling}%
}%

\newcommand{\Tr}[1]{\mathrm{Tr}\left[#1\right]}
\newcommand{\expo}[1]{\mathrm{e}^{#1}}
\newcommand{\bra}[1]{\langle #1 \vert}
\newcommand{\ket}[1]{\vert #1 \rangle}
\newcommand{\braket}[2]{\langle #1 \vert #2 \rangle}
\newcommand{\av}[1]{\langle #1 \rangle}
\newcommand{\PP}[0]{\mathcal{P}}
\newcommand{\Q}[0]{\mathcal{Q}}
\newcommand{\eff}[0]{\mathrm{eff}}
\newcommand{\E}[0]{\mathcal{E}}

\siglaunidade{IFSC} 
\programa{DFA}
\definecolor{blue}{RGB}{41,5,195}

\makeatletter
\hypersetup{
	pdftitle={\@title}, 
	pdfauthor={\@author},
	pdfsubject={\imprimirpreambulo},
	pdfcreator={LaTeX with abnTeX2},
	pdfkeywords={abnt}{latex}{abntex}{USPSC}{trabalho acadêmico}, 
	colorlinks=true,       		
	linkcolor=blue,          	
	citecolor=blue,        		
	filecolor=magenta,      		
	urlcolor=blue,
	bookmarksdepth=4
}
\makeatother

\makeindex
\begin{document}

\selectlanguage{english}

\frenchspacing 

\renewcommand{\ABNTEXchapterfontsize}{\fontsize{12}{12}\bfseries}
\renewcommand{\ABNTEXsectionfontsize}{\fontsize{12}{12}\bfseries}
\renewcommand{\ABNTEXsubsectionfontsize}{\fontsize{12}{12}\normalfont}
\renewcommand{\ABNTEXsubsubsectionfontsize}{\fontsize{12}{12}\normalfont}
\renewcommand{\ABNTEXsubsubsubsectionfontsize}{\fontsize{12}{12}\normalfont}
\renewcommand{\thefootnote}{\Roman{footnote}}

\imprimircapa
\imprimirfolhaderosto*

\includepdf{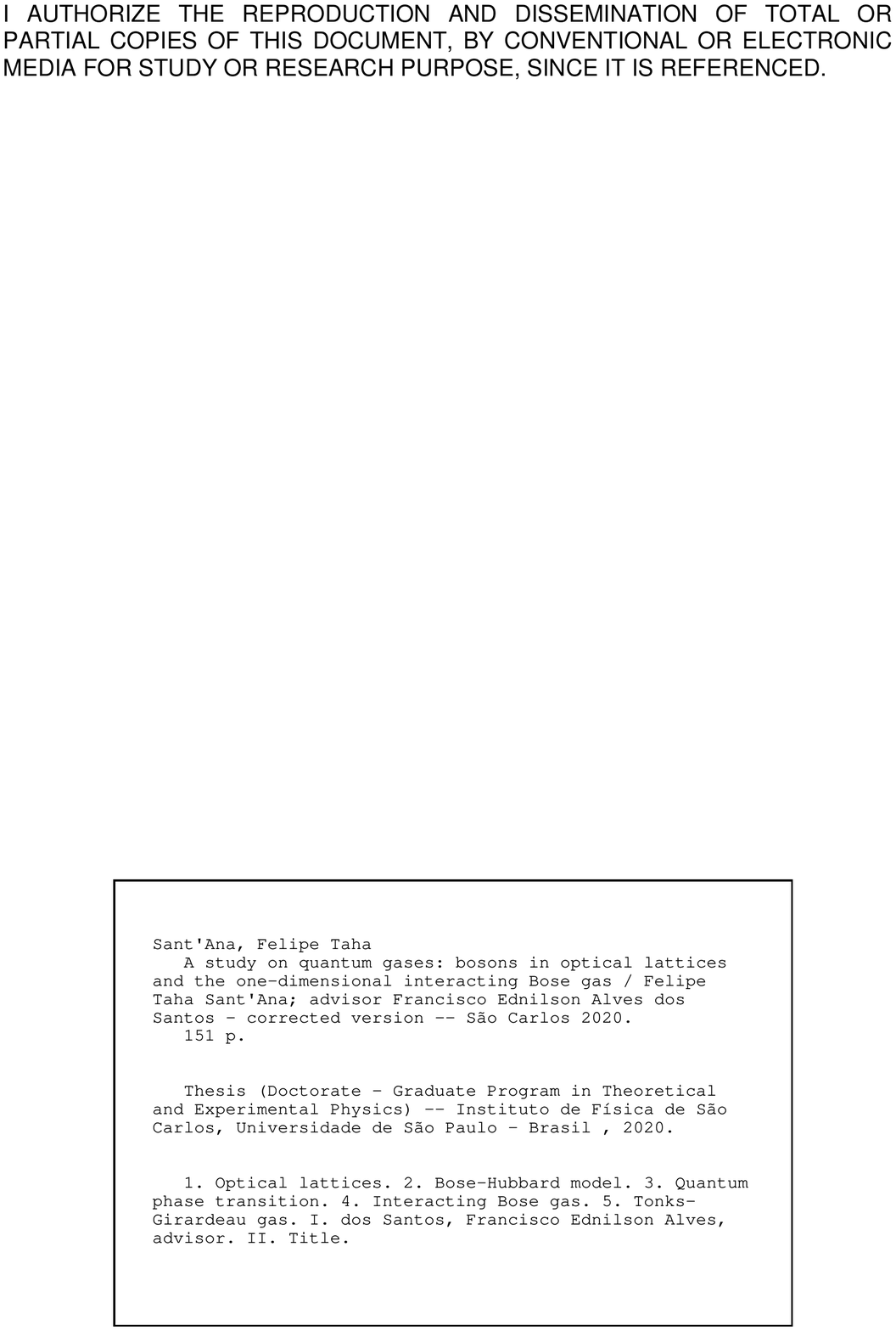}


\begin{dedicatoria}
   \vspace*{\fill}
   \centering
   \noindent
	\textit{To my parents, Leila and M\'ario, for believing in me.} \vspace*{\fill}
\end{dedicatoria}


\begin{agradecimentos}
There are many people whom I should thank for different reasons. 
Firstly, I must thank my family for the basis and for being present 
when I most needed: I am very grateful to my mother, Leila, for being so caring, 
to my father, M\'ario, for encouraging me in my studies 
during my whole life, and to my brother, Vitor, 
for sharing his enthusiasm and the rock 'n roll. 

A very special thanks goes to my wife, Fernanda, for being patient, comprehensive, 
and, above all, an awesome companion. 

I am very grateful to my friend Michael Melo for sharing his passion for physics 
and, also, for great times with beer, chess, philosophy, and countless 
epic talks.

Academically speaking, I am very grateful to my supervisor, Francisco Ednilson, 
for being present when I needed advices, for teaching and explaining 
physics when I did not understand, and for guiding me during my doctorate. 
Thanks a lot, Ednilson, for that! 
I am also very thankful to Axel Pelster, for sharing discussions, for 
supporting me when I wanted to go to Germany for a full doctorate, and, especially, 
for introducing me do Ednilson. 
In the France side of the story, I am very grateful to Mathias Albert and 
Patrizia Vignolo for receiving me, for helping me with all I needed to adapt myself 
in the new environment, and for supervising my studies during my one-year work 
at InPhyNi. Thanks a lot for that! Also, I must thank Fr\'ed\'eric H\'ebert for 
useful discussions and, especially, for providing my first 
lunch at InPhyNi. Thanks, Fred! Finally, I must thank Romain Bachelard 
for bringing me the opportunity to go to France and make it possible. 

During my doctoral studies I have made many friends. In the early days at IFSC, I 
cannot thank enough my dearest friends Marios and Sasha for sharing very good times 
and beers. Also, I must thank Juli\'an and Tiago for the "bandej\~ao" sessions and 
the barbecues. Then, in France, I really had a great time at InPhyNi. I am very
grateful to every single person I met there and who made me feel comfortable in the new 
environment. Firstly, thanks a lot to Ana and Juli\'an for the hikings, they were incredible! 
I must thank the football team members Antonin, Juli\'an, Mathias, and Vittorio, for 
making the football sessions absolutely amazing. Thanks, guys!
Also, I can not forget to thank Marius for the unstoppable humorous talks, 
those were really fun. 

This study was financed in part by the Coordenação de Aperfeiçoamento de Pessoal de Nível Superior – Brasil (CAPES) – Finance Code 001.

\end{agradecimentos}


\begin{prefacio}
I cannot remember a period of time in my life where 
I have learned about many different topics as much as I learned 
writing this thesis. Every single page came to life 
after many hours of research, reading, and 
learning, so that I could disseminate the amount of knowledge I acquired during 
this process. I wish I had more time to keep writing because 
I would appreciate very much in going deeper in 
some subjects that I found very interesting and important as a link to understand 
another one. In some parts of this work I spent 
many lines in details and algebraic manipulations that I considered to be relevant 
for the understanding of the reader, while in other parts I simply omitted the details 
because I considered them to be straightforward and not crucial 
to the understanding of the specific idea. I apologize if I failed in doing so.

This thesis is composed of two parts. The first part is concerned 
to the study of the quantum phase transition between the Mott insulator and the superfluid 
regarding bosonic atoms loaded in optical lattices. This part of my doctoral studies was conducted 
at the S\~ao Carlos Institute of Physics in Brazil under the supervision of Francisco Ednilson 
and with collaboration of Axel Pelster from the Technical University of Kaiserslautern in Germany. 
Then, I traveled to France, where I spent one year performing research on the one-dimensional interacting Bose gas at the 
Institut de Physique de Nice under the supervision of Mathias Albert and Patrizia Vignolo 
and with the collaboration of Fr\'ed\'eric H\'ebert. The research conducted in France 
corresponds to the second part of this thesis and is described in Chap. \ref{1d}.

\end{prefacio}

\begin{epigrafe}
    \vspace*{\fill}
	\begin{flushright}
		\textit{``Education is the kindling of a flame, \\
		not the filling of a vessel.''\\
		Socrates}
	\end{flushright}
\end{epigrafe}

%

\autor{Sant'Ana, F. T.}
\setlength{\absparsep}{18pt}
\begin{resumo}[Abstract]
	\begin{flushleft} 
		\setlength{\absparsep}{18pt} 
 		\SingleSpacing 
 		\imprimirautorabr~ ~\textbf{\imprimirtitulo}.	\imprimirdata.  \pageref{LastPage}p. 
		\imprimirtipotrabalho~-~\imprimirinstituicao, \imprimirlocal, 	\imprimirdata. 
 	\end{flushleft}
\OnehalfSpacing 
Bosonic atoms confined in optical lattices are described by the Bose-Hubbard model 
and can exist in two different phases, 
Mott insulator or superfluid, depending on the strength of the system parameters, 
such as the on-site interaction between particles and the hopping parameter. 
Differently from classical phase transitions, the Mott-insulator-superfluid 
transition can happen even at zero temperature, driven by quantum fluctuations, thus 
characterizing a quantum phase transition. 
For the homogeneous system, we can approximate the particle excitations 
as a mean-field over time, thus providing a local Hamiltonian, which makes possible 
the evaluation of physical properties from a single lattice site. From the Landau theory of 
second-order phase transitions, it is possible to expand the thermodynamic potential in 
a power series in terms of the order parameter, giving rise to the Mott-insulator-superfluid 
phase diagram. As the condensate density goes from a finite value to a vanishing one when the 
system transits from superfluid to a Mott insulator, it can be considered as the order 
parameter of the system.
In the vicinity of the phase boundary, it is possible to consider the hopping term 
as a perturbation, since it contains the order parameter. Thence, one can 
apply perturbation theory in order to calculate important physical quantities, 
such as the condensate density. However, due to degeneracies 
that happen to exist between every two adjacent Mott lobes, nondegenerate perturbation 
theory fails to give meaningful results for the condensate density: it predicts 
a phase transition due to the vanishing of the order parameter in a point of the phase 
diagram where no transition occurs. 
Motivated by such a misleading calculation, we develop two different degenerate perturbative 
methods to solve the degeneracy-related problems. Firstly, we develop a degenerate perturbative 
method based on Brillouin-Wigner perturbation theory to tackle the zero-temperature case. 
Afterwards, we develop another degenerate perturbative method based 
on a projection operator formalism to deal with the finite-temperature regime. Both methods 
have the common feature of separating the Hilbert subspace where the degeneracies are contained in 
from the complementary one. Therefore, such a separation of the Hilbert subspaces fixes the 
degeneracy-related problems and provides us a framework to obtain physically consistent results 
for the condensate density near the phase boundary.
Moreover, we study the one-dimensional repulsively interacting Bose gas under harmonic confinement, 
with special attention to the asymptotic behavior 
of the momentum distribution, which is a universal $k^{-4}$ decay characterized by the Tan's contact. 
The latter constitutes a direct signature of the short-range correlations in such an interacting system and 
provides valuable insights about the role of the interparticle interactions. 
From the known solutions of the system composed of two particles, we are able to acquire important knowledge about the 
manifestation of the interaction, \textit{e.g.}, the cusp condition that implies the vanishing of the 
many-body wave function whenever two particles meet. 
Then, we investigate the system constituted of $N$ interacting particles in the strongly 
interacting limit, also known as \textit{Tonks-Girardeau gas}. In such a regime, 
the strong interparticle interaction makes the bosons behave similarly to the ideal Fermi gas, 
an effect known as \textit{fermionization}. 
Because of the difficulty in analytically solving the system for $N$ particles at finite interaction, 
the Tonks-Girardeau regime provides, through the fermionization of the bosons, a favorable scenario 
to probe the Tan's contact. Therefore, within such a regime, we are able to provide an analytical formula for the Tan's contact 
in terms of the single-particle orbitals of the harmonic oscillator. 
Furthermore, we analyze the scaling properties of the Tan's contact in terms of the number of particles 
$N$ in the high-temperature regime as well as in the strongly interacting regime. Finally, 
we compare our analytical calculations of the Tan's contact to quantum Monte Carlo simulations and 
discuss some fundamental differences between the canonical and the grand-canonical ensembles. 

   \vspace{\onelineskip}
 
\end{resumo}


\setlength{\absparsep}{18pt} 
\begin{resumo}[Resumo]
	\begin{flushleft} 
			\setlength{\absparsep}{18pt} 
			\SingleSpacing 
			\imprimirautorabr~ ~\textbf{\imprimirtitleabstract}.	\imprimirdata. \pageref{LastPage}p. 
			\tipotrabalho{Tese (Doutorado em Ci\^encias)}
			\imprimirtipotrabalho~-~\imprimirinstituicao, \imprimirlocal, \imprimirdata. 
 	\end{flushleft}
\OnehalfSpacing 			
Átomos bosônicos confinados em redes ópticas são descritos pelo modelo de Bose-Hubbard
e podem existir em duas diferentes fases,
isolante de Mott ou superfluido, dependendo da força dos parâmetros do sistema,
tais como a interação local entre partículas e o parâmetro de salto.
Diferentemente das transições de fase clássicas, a transição 
entre isolante de Mott e superfluido pode ocorrer mesmo a temperatura zero, impulsionada por flutuações quânticas,
caracterizando uma transição de fase quântica. 
Para o sistema homogêneo, podemos aproximar as excitações de partículas
a um campo médio ao longo do tempo, fornecendo um Hamiltoniano local, o que torna possível
a avaliação de propriedades físicas a partir de um \'unico s\'itio da rede.
A partir da teoria de Landau de
transições de fase de segunda ordem, é possível expandir o potencial termodinâmico em
uma série de potências em termos do parâmetro de order, dando origem ao diagrama de fase. 
Como a densidade de condensado passa de um valor finito para um valor nulo quando o
sistema transita de superfluido para isolante de Mott, este pode ser considerado como sendo 
o parâmetro de ordem do sistema.
Nas proximidades da fronteira de fase, é possível considerar o termo de salto
como uma perturbação, uma vez que este contém o parâmetro de ordem. Daí, pode-se
aplicar teoria de perturbaç\~ao para calcular quantidades físicas importantes,
como a densidade de condensado. No entanto, devido a degenerescências
que existem entre dois l\'obulos de Mott adjacentes, teoria de perturbação não degenerada 
falha em fornecer resultados significativos para a densidade de condensado: esta prevê
uma transição de fase devido ao desaparecimento do parâmetro de order em um ponto do diagrama de fase 
onde nenhuma transição ocorre.
Motivados por esse cálculo enganoso, desenvolvemos dois m\'etodos perturbativos degenerados diferentes 
para resolver os problemas relacionados às degeneresc\^encias. Em primeiro lugar, desenvolvemos um m\'etodo perturbativo 
degenerado baseado em teoria de perturbação de Brillouin-Wigner para solucionar o sistema a temperatura zero.
Posteriormente, desenvolvemos outro método perturbativo degenerado
baseado em um formalismo de operadores de projeção para lidar com o regime a temperatura finita. Ambos os métodos
têm a característica comum de separar o subespaço de Hilbert onde as degenerescências estão contidas
de seu complementar. Portanto, essa separação dos subespaços de Hilbert corrige os
problemas relacionados às degenerescências e nos fornece uma estrutura para obter resultados fisicamente consistentes
para a densidade de condensado próximo \`a fronteira da fase.
Além disso, estudamos o gás de Bose unidimensional com interação repulsiva entre part\'iculas sob confinamento harmônico,
com especial atenção ao comportamento assintótico
da distribuição de momento, que \'e um decaimento universal de $ k^{-4}$ caracterizado pelo contato de Tan.
Este último constitui uma assinatura direta das correlações de curto alcance em tal sistema interagente e
fornece informações valiosas sobre o papel das interações entre partículas.
A partir das conhecidas soluç\~oes do sistema composto de duas partículas, somos capazes de 
adquirir conhecimentos importantes sobre a manifestação da interação, \textit{e.g.}, 
a condição de cúspide que implica no desaparecimento da
função de onda de muitos corpos sempre que duas partículas se encontram.
Em seguida, investigamos o sistema constituído de $N$ partículas fortemente interagentes, 
também conhecido como \textit{g\'as de Tonks-Girardeau}. Nesse regime,
a forte interação entre part\'iculas faz com que os bósons 
se comportem de maneira semelhante ao gás ideal de Fermi, 
um efeito conhecido como \textit{fermioniza\c{c}\~ao}.
Devido à dificuldade em resolver analiticamente o sistema com $N$ partículas com interação finita,
o regime de Tonks-Girardeau fornece, através da fermionização dos bósons, um cenário favorável
para o estudo do contato de Tan. Portanto, dentro de tal regime, somos capazes de fornecer uma fórmula analítica para o contato do Tan
em termos dos orbitais de uma \'unica partícula do oscilador harmônico.
Além disso, analisamos as propriedades de escalonamento do contato do Tan em termos do número de partículas
$N$ nos regimes de altas temperaturas e fortes interaç\~oes. Finalmente,
comparamos nossos cálculos analíticos do contato de Tan a simulações de Monte Carlo qu\^antico e
discutimos algumas diferenças fundamentais entre os conjuntos canônico e macrocanônico.
 
   \vspace{\onelineskip}
 
   \noindent 
   \textbf{Palavras-chave}: Redes \'opticas. Modelo de Bose-Hubbard. Transi\c{c}\~ao qu\^antica de fase. G\'as de Bose interagente. G\'as de Tonks-Girardeau.
\end{resumo}


\pdfbookmark[0]{\listfigurename}{lof}
\listoffigures*
\cleardoublepage



\begin{siglas}
    \item[BEC] Bose-Einstein Condensate/Condensation
    \item[BH] Bose-Hubbard
    \item[BWPT] Brillouin-Wigner Perturbation Theory
    \item[FTDPT] Finite-Temperature Degenerate Perturbation Theory 
    \item[MI] Mott Insulator
    \item[NDPT] Non-Degenerate Perturbation Theory
    \item[OP] Order Parameter
    \item[QMC] Quantum Monte Carlo
    \item[RSPT] Rayleigh-Schr\"odinger Perturbation Theory
    \item[SF] Superfluid
    \item[TG] Tonks-Girardeau
\end{siglas}

\begin{simbolos}
\item[$\hbar = 1.054571817 \times 10^{-34} \, \mathrm{J.s}$]  Reduced Planck's constant
\item[$k_B = 1.380649 \times 10^{-23} \, \mathrm{J}.\mathrm{K}^{-1}$]  Boltzmann's constant
\item[$ e = 1.60217662 \times 10^{-19} \, \mathrm{C}$]  Electronic charge
\item[$ \varepsilon_0 = 8.8541878128 \times 10^{-12} \, \mathrm{F}.\mathrm{m}^{-1} $]  Vacuum permittivity
\item[$ c = 299792458 \, \mathrm{m}.\mathrm{s}^{-1} $]  Speed of light
\item[$ \pi = 3.14159265359\dots $]  Archimedes' constant
\item[$ \mathrm{e}=2.71828182845\dots $]  Euler's number
\end{simbolos}
\pdfbookmark[0]{\contentsname}{toc}
\tableofcontents*
\cleardoublepage
\textual

	\chapter[Introduction]{Introduction}
	Before the quantum revolution in the early days of the twentieth century, 
	which was motivated during the 19th century by the studies of Thomas Young in the famous double-slit experiment,\cite{young} 
	the black-body radiation problem stated by Gustav Kirchhoff,\cite{kirchoff1,kirchoff2,kirchoff3} the Ludwig Boltzmann's work on 
	the statistics of possible energies of atoms and molecules in a gas,\cite{boltzmann} and 
	the works of Max Planck on the quantum hypothesis of energy,\cite{planck1,planck2,planck3,planck4,planck5} 
	the world was described by the fundamental laws of Isaac Newton on 
	gravitation and classical mechanics,\cite{newton} the unified electromagnetism theory of James Clerk Maxwell,\cite{maxwell} and the classical thermodynamics developed in the 17th century. Due to the absence of 
	a complete quantum theory in those days, 
	there was a limited access to some physical properties of matter, such as the states of matter known at that time: liquid, solid and gaseous. 
	However, such a knowledge about the forms that matter can acquire 
	was about to drastically change as a result of 
	the advent of quantum mechanics. 

	The concept of Bose-Einstein condensation (BEC) originated in 
	1925 when A. Einstein, on the basis of the work of S. N. Bose,\cite{bose} 
	which described the quantum statistical theory of light, wrote a paper 
	about the quantum theory of the ideal monoatomic gas,\cite{einstein} 
	predicting the occurrence of a phase transition at low enough 
	temperatures. In 1938, F. London argued that the phenomenon of Bose-Einstein 
	condensation was intimately related to the occurrence of superfluidity in $^{4}\mathrm{He}$.\cite{london1} Later on, he also suggested that BEC and superconductivity 
	phenomena were closely related.\cite{london2}

	An important aspect towards the understanding 
	of the Bose-Einstein condensation 
	is the thermal wavelength,\cite{huang,reif}
	that relates, by means of the de Broglie relation, 
	the wave character of a particle with mass $m$ to its temperature $T$
	through the formula
	\begin{equation}
		\lambda = \sqrt{\frac{2\pi\hbar^2}{m k_B T}}.
	\end{equation}
	As the temperature of the particles decreases, its 
	associated de Broglie wavelength increases. As a result, 
	the many wave packages related to different particles in 
	the system begin to overlap with one another, until a critical point 
	is reached and all the atoms behave as a single macroscopic wave, which is the 
	characterization of a Bose-Einstein condensate.
	The transition to the BEC phase occurs when the 
	thermal wavelength of the particles in the atomic gas becomes 
	comparable to the interatomic spacing between them
	$\lambda \sim n^{-1/3}$,\cite{ketterle} where $n$ is the density of atoms. 
	Consequently, the critical temperature in order to achieve a BEC is 
	of the order of 
	\begin{equation}
		T_c \sim n^{2/3} \frac{2\pi\hbar^2}{m k_B}.
	\end{equation}

	To create a BEC one must not just reach the 
	critical temperature, but also achieve 
	low densities, otherwise the atomic gas 
	would simply condensate into a 
	more conventional liquid or solid.\cite{ketterle} The density of a dilute gas 
	that provides an adequate environment for the emergence of 
	a BEC must be of the order of a hundred-thousandth 
	of the density of normal air, which is around 
	$10^{19} \, \mathrm{cm}^{-3}$. Typically, the 
	particle density at the center of a BEC is about
	$10^{13}-10^{15} \, \mathrm{cm}^{-3}$.\cite{pethick} Thence, 
	the temperature one must accomplish in order to 
	realize a BEC is around $10^{-5} \, \mathrm{K}$. 

	Obviously,  
	it took great amounts of scientific work throughout the twentieth century 
	for the purpose of achieving such low temperatures. A landmark 
	in the history of atomic cooling techniques are the works performed 
	during the 1970s and 1980s on laser cooling. A seminal work on laser cooling 
	techniques was performed by T. W. H\"ansch and A. L. Schawlow in 1975,\cite{hansch} where 
	they were able to achieve temperatures around $0.24 \, \mathrm{K}$. 
	Later on, in 1978, D. J. Wineland \textit{et al.}\cite{wineland} and 
	A. Ashkin\cite{ashkin} succeeded in obtaining temperatures 
	in the milikelvin, $10^{-3} \mathrm{K}$, and microkelvin scales, $10^{-6} \, \mathrm{K}$, 
	respectively. After that, many works 
	followed towards the advance of laser cooling techniques, \textit{e.g.}, 
	S.~Chu \textit{et al.}\cite{chu1} reported the cooling of neutral sodium atoms 
		in three dimensions via radiation pressure of conterpropagating laser beams, 
		attaining  temperatures around $240 \, \mu\mathrm{K}$, while 
		A.~Aspect \textit{et al.}\cite{cohen1} described a scheme 
		that allowed the cooling of $^4 \mathrm{He}$ to a temperature of $2 \, \mu\mathrm{K}$. 
		All these developments in laser cooling 
		culminated in the 1997 Nobel prize awarded to S. Chu,\cite{chu2}
		C. N. Cohen-Tannoudji,\cite{cohen2} and W. D. Phillips.\cite{bill}

		The field of laser cooling includes different techniques: 
		\textit{Sisyphus cooling};\cite{bill,dalibard,paulet}
		\textit{resolved sideband cooling};\cite{wineland,neuhauser,diedrich,monroe,hamann,eschner,schli} 
		\textit{Raman sideband cooling};\cite{kasevich,kerman} 
		\textit{velocity-selective coherent population trapping};\cite{cohen1} 
		\textit{gray molasses};\cite{weide,boiron,nath,franz,bruce} 
		\textit{cavity mediated cooling};\cite{horak} 
		\textit{Zeeman slower};\cite{cheiney,ohayon} 
		\textit{electromagnetically induced transparency}; \cite{budker,morigi,liu,safavi,haller} 
		\textit{Doppler cooling}.\cite{hansch,wineland,bill,cohen-bill} The traditional and still most used 
		is the latter one. The principle behind the \textit{Doppler cooling} technique is the following: 
		when an atom interacts with a photon \textemdash the interaction process is consisted of the absorption and 
		thereafter the emission of a photon \textemdash the velocity of the atom is changed due to momentum conservation. 
		Depending on the direction of the photon relative to the direction of the atom velocity, it can either
		increase or decrease its momentum. So, in order to eliminate the undesirable result of increasing the 
		atomic temperature, one has to tune the laser frequency just below the electronic transition frequency of the atom. This process 
		results in an overall reduction of the gas temperature, since the atoms will only interact with counterpropagating photons 
		as a result of the \textit{Doppler effect}.  

\begin{figure}[h]
			\centering
			\begin{subfigure}{0.45\columnwidth}
				\includegraphics[width=\columnwidth]{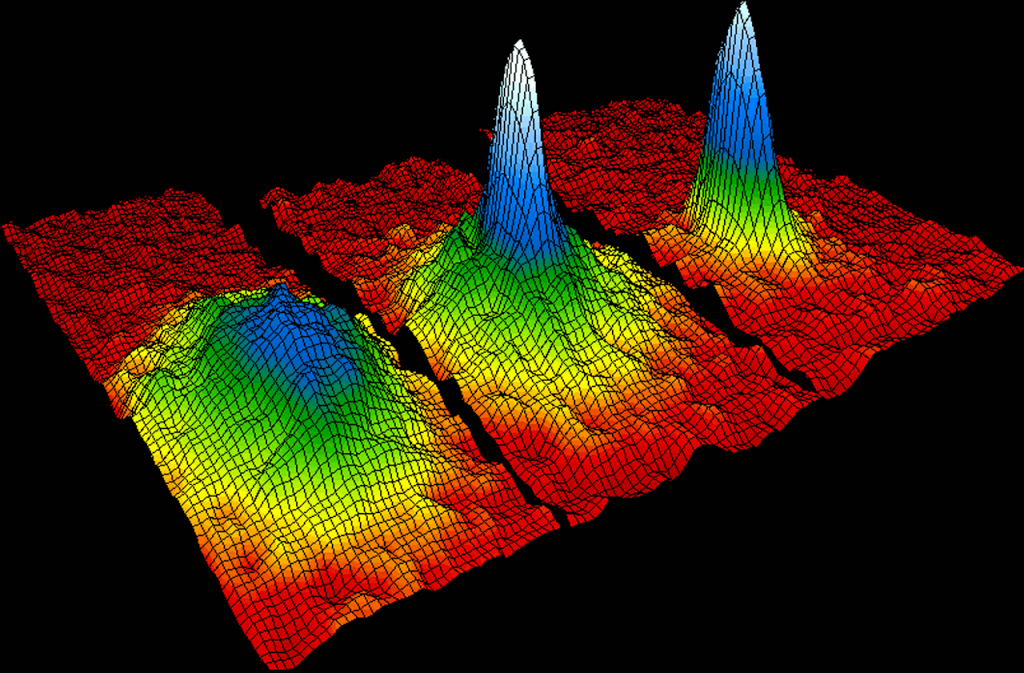}
				\caption{The respective temperatures are, from left to right, $0.4 \, \mu \mathrm{K}$, 
				$0.2 \, \mu \mathrm{K}$, and $0.05 \, \mu \mathrm{K}$.}
				\label{BEC_cornell}
			\end{subfigure}
			\qquad
				\begin{subfigure}{0.49\columnwidth}
				\includegraphics[width=\columnwidth]{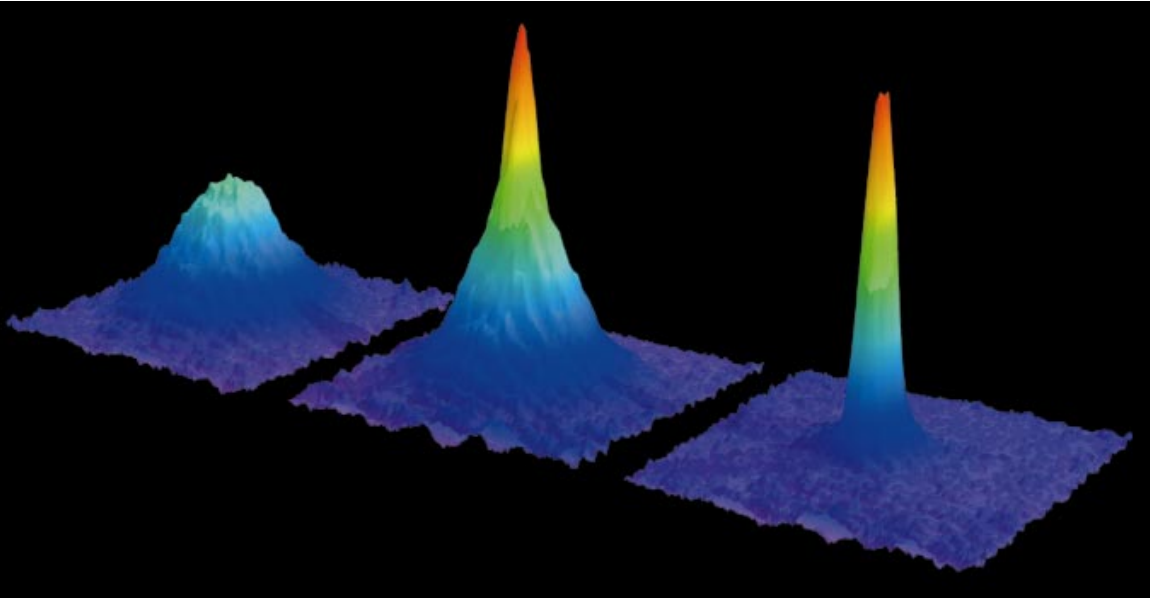}
					\caption{From left to right, the temperatures are $T>2 \, \mu\mathrm{K}$, 
					$T\sim 2\, \mu \mathrm{K}$, and $T<2\,\mu\mathrm{K}$.}
				\label{BEC_ketterle}
			\end{subfigure}
			\caption{Observation of the two first Bose-Einstein condensates by absorption imaging from the rubidium $^{87}\mathrm{Rb}$ 
			experiment\cite{anderson,cornell} in (a) and from the sodium $^{23}\mathrm{Na}$ experiment\cite{davis,ketterle} in (b).}
			Source: (a) CORNELL \textit{et al.};\cite{cornell} (b) KETTERLE.\cite{ketterle}
			\label{fig.BEC}
		\end{figure}

		Such advances in laser cooling made possible the accomplishment 
		of the \textit{Bose-Einstein condensate} in 1995 by the 
		group of Eric Cornell and Carl Wieman at the University of Colorado Boulder 
		in a gas of rubidium $^{87}\mathrm{Rb}$ atoms\cite{anderson} and 
		by the group of Wolfgang Ketterle at MIT in a gas of sodium $^{23}\mathrm{Na}$ atoms,\cite{davis} Fig. \ref{fig.BEC}. 
		In order to achieve cold enough temperatures to realize the BEC, it was necessary to combine 
		different stages: in the precooling stage, laser cooling techniques were used to make the atoms cold enough 
		to be confined in a wall-free magnetic trap; then forced evaporative cooling was applied as 
		the second stage,\cite{naoto,evket} which consists in reducing the trap depth so that the most energetic 
		atoms can escape, thus decreasing the temperature of the whole gas. For their achievements on BEC, 
		Eric Cornell, Carl Wieman and Wolfgang Ketterle were awarded the 2001 Nobel prize in physics.\cite{ketterle,cornell}

		Differently from bosons, fermionic atoms are subject 
		to the \textit{Pauli exclusion principle},\cite{pauli} which forbids two or more 
		fermions of occupying the same quantum-mechanical state. However, in view of the formation of the 
		so-called \textit{Cooper pairs},\cite{cooper} the realization of BEC in interacting Fermi gases 
		was achieved in 2003 by Greiner \textit{et al.}\cite{greiner} with potassium $^{40}\mathrm{K}$ atoms, 
		while it was also achieved thereafter with lithium $^6\mathrm{Li}$ atoms.\cite{zwi,bourdel,bart,part} 
		Such a realization in fermionic gases represented the accomplishment of a long-standing goal of ultracold 
		atoms research, the celebrated \textit{BCS-BEC crossover}, referring to the 
		Bardeen–Cooper–Schrieffer (BCS) theory of superconductivity.\cite{bcs1,bcs2}

		\section{Bosonic atoms loaded in optical lattices}
		Optical lattices are laser arrangements which enable a spatially periodic  
		trapping of atoms due to the interaction between the external 
		electric field and the induced dipole moment of the atoms.\cite{pethick,jaksch,bloch,lewenstein,pitaevskii} Such  
		artificial laser-generated periodic potentials create a propitious 
		environment to probe ultracold atoms and provide clean access 
		to physical quantities, in contrast to natural crystal lattices, where 
		disorder of many kinds, \textit{e.g.}, lattice vibrations, or the so-called \textit{phonons}, 
		contribute to undesirable features in the effective Hamiltonian that one needs to take into account 
		to describe the dynamics of the system as well as to calculate its physical properties. 
		In principle, it is always possible 
		to take only one kind of disorder into account and apply perturbation theory to evaluate physical quantities, 
		but this is simply the reduced version of the story: there happens to exist innumerable 
		kinds of noise in a crystal lattice that it is theoretically impossible to account for all their contributions 
		simultaneously. Even if one uses numerical methods, the amount of computational capacity is usually
		beyond what the computers nowadays can achieve. So, optical lattices provide a suitable setting 
		for the realization of simplified models of condensed-matter systems and the study of many-body systems. Also, optical lattices 
		allow the implementation of Richard Feynman's pioneering idea of \say{quantum simulation}:\cite{feynman1,feynman2}  
		using one quantum system to investigate another one.\cite{greiner-ol} In addition, 
		the controllability of optical lattices is much higher than most condensed-matter systems, thus providing 
		an easy control of important parameters, such as the strength of the interatomic interactions.
		
			\begin{figure}[h]
			\centering
		\includegraphics[scale=0.5]{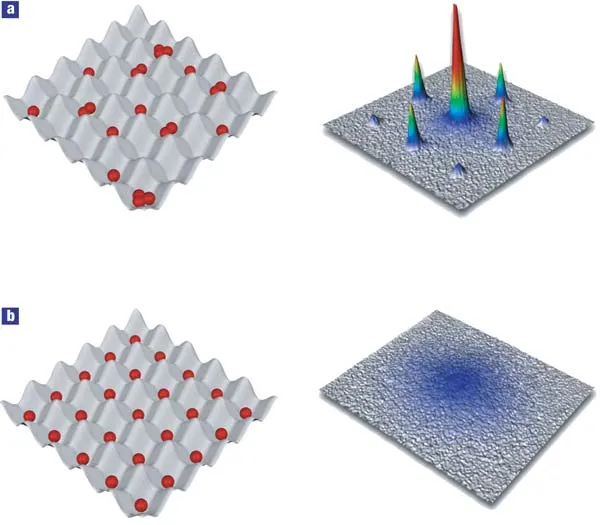}
			\caption{Schematic drawing of optical lattices in the two different phases: the superfluid in (a) 
			and the Mott insulator in (b). The superfluid phase is characterized by a high delocalization of the atoms, 
			implying well known values of the momentum, generating well defined peaks in the momentum space. On the other hand, 
			the Mott insulator phase is characterized by a high localization of the atoms, thence their momentum space image consists in a 
			blur.}
			Source: BLOCH.\cite{bloch}
			\label{fig.optical_lattice}
		\end{figure}

\begin{figure}[h]
			\centering
		\includegraphics[scale=0.6]{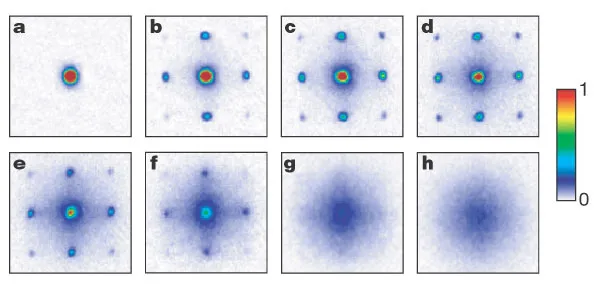}
			\caption{Time-of-flight absorption images for the potential 
			depths $V_0$ of: (a) 0, (b) $3 E_R$, (c) $7E_R$, (d) $10E_R$,
			(e) $13 E_R$, (f) $14 E_R$, (g) $16 E_R$, and (h) $20 E_R$.}
			Source: GREINER \textit{et al.}\cite{greiner1}
			\label{fig.ToF}
		\end{figure}

		A gas composed of bosonic atoms in an optical lattice can 
		be described by the Bose-Hubbard model,\cite{lewenstein,ueda} 
		which has three main parameters: the on-site interaction parameter, 
		the hopping parameter, and the chemical potential. 
		Depending on the magnitude of the parameters, the system can 
		realize two different phases, the Mott insulator or the superfluid 
		phase,\cite{fisher,sheshadri,greiner1,greiner2,greiner3,
		widera,gunter,ospelkaus,lewenstein2,gerbier} as illustrated in Fig. \ref{fig.optical_lattice}. 
		If the on-site interaction parameter is much larger than the hopping parameter, 
		the system is in the Mott insulator (MI) phase. 
		This phase is characterized by a high localization of the atoms, implying an integer 
		number of particles $n$ per lattice site and zero compressibility, $\partial n/\partial \mu =0$. 
		Also, the Mott insulator presents an energy gap for both particle and hole excitations, 
		due to the restricted mobility between neighboring sites. 
		By decreasing the amplitude of the periodic potential, so that the 
		hopping parameter becomes much larger than the atom-atom interaction parameter, 
		the system undergoes a phase transition to a superfluid (SF) phase, where a fraction of the 
		atoms become delocalized. Such a phase is characterized by zero viscosity, \textit{i.e.}, superfluidity signifies 
		the ability of carrying currents without dissipation, analogously to superconductivity. 
		Such differences in the localization of the bosons 
		make it possible to measure the phase the system is currently in through 
		time-of-flight experiments,\cite{greiner1,hoff} which are depicted in 
		Fig. \ref{fig.ToF}. The MI-SF transition can happen even at zero-temperature, 
		driven by quantum mechanical fluctuations, thus characterizing a quantum phase transition.\cite{sachdev}

		\subsection{State of the art}
		Regarding a homogeneous system, we can approximate the particle-hole excitations 
		as a mean value over time, known as mean-field approximation, thus providing a local Hamiltonian, which makes possible the evaluation of physical properties from a single lattice site. More specifically, 
		such a consideration approximates the Bose-Hubbard Hamiltonian by a sum of local Hamiltonians.\cite{fisher} 
		Following this simplification, Rayleigh-Schr\"odinger perturbation theory (RSPT) is typically 
		used for obtaining the mean-field phase diagram at zero temperature.\cite{sachdev} However, there are problems that arise from RSPT, 
		since it does not properly deal with degeneracies that occur between two consecutive Mott lobes. 
		One of such RSPT problems concerns the calculation of the condensate density, 
		which falsely vanishes between two consecutive Mott lobes.\cite{melo}

		Also, other methods have been suggested in order to improve 
		the mean-field quantum phase diagram for bosons in optical lattices, 
		\textit{e.g.}, a variational method that uses a field-theoretic concept of the effective potential.\cite{santos} 
		In addition, the MI-SF phase transition at arbitrary temperatures was investigated  using an effective-action approach.\cite{bradlyn} Furthermore, 
		a similiar method was derived for the Bose-Hubbard model within the Schwinger-Keldysh formalism in order to handle time-dependent problems at finite temperature.\cite{grass1,grass2} Likewise, Ref. \citeonline{melo} implemented a nearly degenerate perturbation theory for the zero-temperature case, which led to better results for the order parameter (OP) when compared to those from the RSPT calculations.
		The authors of Ref. \citeonline{wang} applied the Floquet theory in order to analyze the effects of a periodic modulation of the $s$-wave scattering length upon the 
		quantum phase diagram of bosons in 2D\cite{greiner3} and 3D optical lattices. 
		It turns out that nondegenerate finite-temperature perturbation theory\cite{bradlyn,ednilson} also presents degeneracy problems similar to RSPT. 
		Indeed, RSPT is equivalent to the usual finite-temperature perturbation theory in the zero-temperature limit. Therefore, degeneracy-related problems are also expected to appear at low enough temperatures.
		Moreover, beyond our considered bosonic gas in an optical lattice, 
		the degeneracy-related problems also emerge in other systems.\cite{3body,kagome,hexagonal,nietner,nietner2}

		\section{One-dimensional interacting Bose gas}
		The study of low dimensional systems is motivated by the fact that 
		many three-dimensional theories completely fail in lower dimensions, \textit{e.g.}, 
		the Landau-Fermi liquid theory\cite{landau-fermi} describing 
		interacting fermions in low temperature systems (such as metals and the liquid helium-3, $^3\mathrm{He}$) 
		simply breaks down in one dimension. The explanation for such a failure and an adequate treatment of 
		interacting fermions in 1D was firstly given by S. Tomonaga in 1950,\cite{tomonaga} and thereafter reformulated in 1963 
		by J. M. Luttinger,\cite{luttinger} with the theory receiving their names as \textit{Tomonaga-Luttinger liquid theory}. 
		The works of Tomonaga and Luttinger were complemented by D. C. Mattis and E. H. Lieb in 1964,\cite{mattis} when they noted a paradox 
		regarding the density operator commutators and used their observations to solve the model and obtain the exact solution 
		of the one-dimensional many-fermion system.

		Contrarily to higher dimensions, the role of interactions are particularly important in 1D systems. 
		This aspect can be easily elucidated by visualizing that, in higher dimensions, the particles can 
		\textit{ipso facto} avoid each other, which is completely different in 1D, where the 
		dimensional constraint assembles the inevitability of interparticle rendezvous. This implies that whenever there 
		is a single-particle excitation, a collective one will automatically emerge in the one-dimensional system.

		Another important consequence of the reduced dimensionality is the absence of BEC 
		in a uniform infinite system at nonzero temperature, $T>0$, in low dimensions $d\leq 2$.\cite{hohenberg} Even at zero 
	temperature, the one-dimensional system features the insufficient conditions for the realization 
	of BEC.\cite{pitaevskii91} Such an absence of BEC can be explained by the Hohenberg-Mermin-Wagner theorem,\cite{hohenberg,mermin} 
	that states the absence of spontaneous symmetry breaking in low dimensions $d\leq 2$.\footnote{For mathematical details on the Hohenberg-Mermin-Wagner theorem, see App. \ref{appendixB}.} 
	This theorem has, of course, 
	profound implications in the occurrence of BEC in low dimensions because the existence of BEC is associate with the 
	spontaneous breaking of the $U(1)$ gauge symmetry.
	However, realistically speaking, there is no infinite system in nature, thus 
	it is possible to realize BEC in one- and two-dimensional finite systems.\cite{widom,bagnato}

	As we have already discussed, the interparticle interactions play a special role towards the complete description 
	of the one-dimensional system. In this context, the pioneer work on the exact analysis 
	of the 1D Bose gas interacting via a delta-like potential at zero temperature was done by 
	E. H. Lieb and W. Liniger\cite{lieb-liniger,lieb} and its 
	extension to finite temperature by C. N. Yang and C. P. Yang.\cite{yang-yang} However, in real experiments the atoms 
	are harmonically trapped, thence one has to take its effect in the Hamiltonian since $V(x) = 0 \rightarrow V(x) \propto x^2$. 
	This change breaks down the integrability of the system. Nevertheless, by sufficiently increasing the interactions, it 
is possible to achieve the so-called \textit{Tonks-Girardeau gas},\cite{tonks,girardeau} where the particles behave as impenetrable hard-core spheres. 
In this regime, as first stated by M. Girardeau in 1960,\cite{girardeau} 
an effect known as \textit{fermionization} occurs, thus enabling one to reduce the original problem into a much simpler one by means of the \textit{Bose-Fermi mapping}.

In order to experimentally realize atomic gases in one dimension, 
one needs to tune the trapping frequencies in such a way that 
 $\omega_y=\omega_z=\omega_\perp \gg \omega_x$, also 
known as \textit{tight confinement} aspect. In such a  
regime, the characteristic energy scale is much greater than the 
thermal energy of the atoms $\hbar \omega_\perp \gg k_B T$. 
Furthermore, it also implies that the 
confinement characteristic length $a_\perp\equiv \sqrt{\hbar/m\omega_\perp}$ 
is much smaller than the three-dimensional scattering length of the atoms $a_{3\mathrm{D}}$ (see App. \ref{appendixA}), 
$a_\perp \ll a_{3\mathrm{D}}$.\cite{Moritz2003,Pagano2014,Salces2018} Therefore, the consequence of 
this experimental setting is that all the dynamics of the system occurs predominantly in the $x$-direction, 
characterizing a quasi one-dimensional system.

\subsection{State of the art}
An accurate description of
strongly correlated quantum systems, for an arbitrary number of particles,
is often a dare without a simple solution. Apart from the very specific
family of integrable systems,\cite{McGuire1964,Yang67,Gaudin1967,Sutherland68,LaiYang,McGuireI,McGuireII,Luther1974,Fuchs2004} 
where all observables can, in principle, be theoretically predicted,
our knowledge is, in general, limited to simplifications like the two-body case,\cite{busch,Aharony2000,Abad2005} the
thermodynamic limit,\cite{olshanii,Yao2018} the low-energy regime,\cite{Giamarchi} or the mean-field approximations.\cite{Gross1961,Pitaevskii1961} So, it is quite challenging
to extract general information from such systems, \textit{e.g.}, 
the scaling of physical observables 
with respect to the number of particles.

Considering the system composed of a quantum gas where the particles interact with each other via 
a delta-like interaction potential,
the short-range correlations are embedded in the Tan's contact $\mathcal{C}$,\cite{tan1,tan2,tan3} an experimentally relevant quantity
that determines the asymptotic behavior of the momentum distribution $n(k)$ via 
$\mathcal{C}\equiv\lim_{k\rightarrow\infty}k^4n(k)$.
This observable can be measured via
time-of-flight techniques,\cite{Stewart2010,Sagi2012,Chang2016}
via radio-frequency spectroscopy,\cite{Wild2012,Yan2019}
by Bragg spectroscopy,\cite{Hoinka2013}
by measuring the energy
variation as a function of the interaction strength,\cite{Sagi2012} or by
analyzing the three-body losses in quantum mixtures.\cite{Laurent2017}
Such a quantity depends on many physical aspects, such as the interaction energy, the density-density
correlation functions, the trapping configuration, the temperature, and 
the magnetization,\cite{Decamp2016b,Decamp2017}. Thus, it fluctuates in a 
nontrivial way with the nature and the number of particles $N$. Therefore, even in one
dimension, the behavior of $\mathcal{C}$ is not completely clarified,
especially in trapped systems, despite of many
theoretical investigations.\cite{Decamp2016b,Decamp2017,Minguzzi2002,minguzzi07,
Lewenstein-Massignan,grining,Matveeva2016,
  Patu2016,jean1,jean2018} For one-dimensional particles
trapped in a harmonic potential of frequency $\omega$,
it has been shown that, in the thermodynamic
limit, at zero temperature, the contact rescaled by $N^{5/2}$ is
an universal function of one scaling parameter, the 
adimensional interaction strength $\tilde{g}\equiv -a_0/\sqrt{N}a_{1D}$,\cite{olshanii,Matveeva2016} where $a_{\mathrm{1D}}$ is the 1D scattering length (App. \ref{appendixA}) and $a_{0}\equiv \sqrt{\hbar/m\omega}$ is the
harmonic oscillator length. Such a scaling property also holds at finite temperatures 
in the grand-canonical ensemble: the contact rescaled by $\av{N}^{5/2}$ is
an universal function of two scaling parameters, $\tilde{g}$ and $\tau\equiv T/T_F$,\cite{Yao2018,xu2015} where $T_F=N\hbar\omega/k_B$ is
the Fermi temperature.
However, for systems with a small number of particles,
the $N^{5/2}$-scaling fails. In the  zero-temperature limit,
it is possible to change the paradigm and to introduce a different scaling form
that holds for any number of particles $N\geq 2$. \cite{mateo} At finite temperature, considering the
grand-canonical ensemble, the $\av{N}^{5/2}$-scaling law holds for $N>10$.\cite{Yao2018}
However, corrections for a small number of particles have, to our knowledge,
not yet been studied in 1D, and the question of the relevance of
the statistical ensemble has also not yet been addressed. The latter is, indeed, a
crucial point, since ultracold-atom experiments are more properly described 
by the canonical ensemble or, more often,
by an average over canonical ensembles. 
This ensemble study is also 
motivated by the fact that the scaling
properties of the system are strongly affected by the statistical distribution
of the number of particles. 

\section{Thesis outline}
Now, let us provide a succint summary of each chapter.
In Chap. \ref{ol}, we present a brief description of optical lattices 
and some basic concepts towards the understanding of the atom-laser 
interacting potential. Then, we describe the solutions of the respective 
Sch\"odinger equation due to such a potential and argue, due to the periodicity of the potential, 
that the solutions can be given in terms of the Bloch functions and also 
in terms of the convenient Wannier functions. With these concepts, we have the fundamentals 
in order to interpret the Bose-Hubbard model and derive its Hamiltonian in terms of 
the important parameters. To finish, we briefly study such parameters in order to 
check how they depend on the laser potential depth as well as on the recoil energy.

In Chap. \ref{mf}, we briefly discuss the fundamentals of second-order quantum phase 
transitions, and then we introduce the Landau expansion of the thermodynamic potential 
together with the mean-field approximation in order to evaluate the 
phase boundary associated with the Mott-insulator-superfluid (MI-SF) quantum phase transition of bosonic 
particles in optical lattices. In addition, we apply nondegenerate perturbation theory (NDPT) 
at finite temperature to calculate the Landau coefficients, the condensate density, $|\Psi|^2$, 
and the density of particles, $-\partial \mathcal{F}/\partial \mu$. Consequently, we show 
that the calculations from NDPT lead to nonphysical behaviors for these two physical quantities, which 
are clear consequences of the incorrect treatment of the degeneracies that happen to occur between two 
consecutive Mott lobes. Finally, we show, as a first degenerate approach, how such problems can be fixed 
within an adequate analysis.

Chap. \ref{BWPT} is concerned with the development of the Brillouin-Wigner perturbation theory (BWPT) 
applied to the zero-temperature regime of bosons in optical lattices. We begin by introducing 
the formulation behind the BWPT, which consists in achieving a Schr\"odinger-like equation 
for an effective Hamiltonian so that it can be expanded up to the desired order in the perturbation parameter. Then we 
apply the BWPT for the case where the degenerate Hilbert subspace is consisted of one state, and check that 
the results for the condensate density are slightly improved, but still unsatisfactory, leading to the necessity of 
considering two states in the degenerate Hilbert subspace. After calculating the important physical quantities in our 
two-state approach, we realize that it produces physically consistent results for both the condensate and 
the particle densities. Afterwards, we develop a useful graphical approach for easily calculating higher-order terms in the pertubative 
expansion. Finally, we consider the effects of a harmonic trap in the system and calculate how it affects the equation of state.

In Chap. \ref{DPT}, we turn our attention to the finite-temperature scenario. 
We develop a finite-temperature degenerate perturbation theory (FTDPT) based on 
a projection operator formalism that, similarly to BWPT, separates the Hilbert space 
into the degenerate subspace and the complementary one, which is free from any degeneracy. We then 
apply our developed FTDPT to the one lattice site mean-field Bose-Hubbard Hamiltonian in order to get meaningful results for 
the condensate density as well as for the particle density in the vicinity of the MI-SF quantum 
phase transition. 

In Chap. \ref{1d}, we study the one-dimensional interacting Bose gas. We begin 
with the case with $N=2$ particles, which is an integrable system and an instructive example 
to enlighten some basic concepts that arise from considering delta-like interparticle interactions. 
We work out the details behind the calculations in order to get the relative-motion wave function and 
check the discontinuity it presents at the contact point. Then we work out the asymptotic behavior 
of the momentum distribution, that leads to a relation where one can recognize the valuable 
role of a term that depends on the second-order correlation function, the so-called \textit{Tan's contact}, $\mathcal{C}$. 
Then, we exactly calculate the two-boson contact. Subsequently, we develop an analytical expression 
for the $N$-boson contact in the Tonks-Girardeau limit. 
Following, we analyze the scaling properties of the Tan's contact within some specific temperature regimes, such as 
the zero- and the large-temperature scalings. From those, we propose a generalized scaling conjecture for all ranges of temperature. 
Furthermore, from quantum Monte Carlo (QMC) calculations, we investigate the intermediate-interaction regime, $\tilde{g} \sim 1$.
Finally, we draw a comparison between the canonical and the grand-canonical ensembles in the context of the interparticle contact.


\chapter{Bosons in optical lattices}\label{ol}
In this chapter, we discuss the fundamental concepts associated 
with the description of Bose gases loaded into optical lattices. 
We begin with the theory behind the atom-laser interaction, focusing 
on a brief discussion about the atomic energy shift due to its interaction 
with the laser-generated electric field and on the form of the periodic potential, 
which leads us to describe the solutions of the respective Sch\"odinger equation 
by applying the Bloch theorem, \textit{i.e.}, the same one used in solid state physics in order to interpret the solutions 
of a generic periodic potential in terms of a plane wave times a periodic function, which are then named as Bloch functions.
After that, we perform a description of the Bose-Hubbard model, introducing its 
general form in terms of the Hamiltonian parameters. Then, we apply a harmonic 
approximation in the laser field potential so that we can perform a first estimation 
of the Hamiltonian parameters.

\section{Atom-laser interaction}
The interaction between the laser-generated electric field and the atom electric dipole within the electric dipole approximation is given by\cite{bloch,lewenstein}
\begin{equation}
	V_{\mathrm{ext}}(\mathbf{r}) = -\mathbf{d}\cdot \boldsymbol{\E}(\mathbf{r}), 
\end{equation}
where $\mathbf{d}\equiv -e\sum_i \mathbf{r}_i$ denotes the atomic electric dipole, with $e$ being the electronic charge 
and $\mathbf{r}_i$ its distance from the nucleus, and $\boldsymbol{\E}(\mathbf{r})$ is the external electric field. 
Let us consider an atomic transition from the fundamental state $\ket{0}$ to any excited state $\ket{n}$ 
due to the external electric field. It is possible to calculate a second-order correction in the atomic ground-state energy, which is given by\cite{pethick}
\begin{equation}
	\Delta E = -\sum_n \frac{|\bra{n}\hat{V}_{\mathrm{ext}}\ket{0}|^2}{E_n-E_0} = -\frac{\alpha}{2} |\boldsymbol{\E}|^2, 
\end{equation}
where $E_0$ and $E_n$ are the energies of the fundamental and the excited state, respectively, and 
\begin{equation}
	\alpha \equiv -\frac{\partial^2 \Delta E}{\partial \E^2} = 2 \sum_n \frac{|\bra{n}\hat{\mathbf{d}}
	\cdot\hat{\boldsymbol{\epsilon}}\ket{0}|^2}{E_n-E_0},
\end{equation}
is the atomic polarizability, while $\hat{\boldsymbol{\epsilon}}$ represents the direction of the laser 
electric field. Of course that this simple static analysis already gives us some insights about the 
energy shift due to the interaction, but it does not correspond to the more realistic case. So, in order to 
approximate the discussion to the real world, let us consider a time-dependent laser-generated electric field 
$\boldsymbol{\E}(\mathbf{r},t) = \boldsymbol{\E}_0(\mathbf{r}) \expo{-i\omega t} + \mathrm{c.c.}$ 
In such a case, the energy shift due to the interaction is given by\cite{pethick}
\begin{equation}\label{deltaE}
	\Delta E = \sum_n\left(\frac{\bra{0}\hat{\mathbf{d}}\cdot\hat{\boldsymbol{\E}}\ket{n}
	\bra{n}\hat{\mathbf{d}}\cdot\hat{\boldsymbol{\E}}^\dagger\ket{0}}{E_n-E_0+\hbar\omega}
	+ \frac{\bra{0}\hat{\mathbf{d}}\cdot\hat{\boldsymbol{\E}}^\dagger\ket{n}
	\bra{n}\hat{\mathbf{d}}\cdot\hat{\boldsymbol{\E}}\ket{0}}{E_n-E_0-\hbar\omega}  \right).
\end{equation}
It is instructive to note that the first term comes from the absorption of a photon by the atom, while 
the second one comes from the emission of a photon.\cite{sakuraiadvanced}
Eq. (\ref{deltaE}) can be simplified to 
\begin{equation}\label{deltaEt}
	\begin{aligned}
	\Delta E &= \sum_n |\bra{n}\hat{\mathbf{d}}\cdot\hat{\boldsymbol{\epsilon}}\ket{0}|^2 
	|\boldsymbol{\E}_0|^2 \left[\left(E_n-E_0-\hbar\omega\right)^{-1}
	+\left(E_n-E_0+\hbar\omega\right)^{-1}\right]\\
	&=-\frac{\alpha(\omega)}{2} \langle\boldsymbol{\E}(\mathbf{r},t)^2\rangle,
	\end{aligned}
\end{equation}
where $\langle\boldsymbol{\E}(\mathbf{r},t)^2\rangle = 2 |\boldsymbol{\E}_0|^2$ represents 
the time-average of the squared electric field. Also, we have introduced the dynamical polarizability, that 
reads 
\begin{equation}
	\begin{aligned}
	a(\omega) &= \sum_n |\bra{n}\hat{\mathbf{d}}\cdot\hat{\boldsymbol{\epsilon}}\ket{0}|^2 
         \left[\left(E_n-E_0-\hbar\omega\right)^{-1}
        +\left(E_n-E_0+\hbar\omega\right)^{-1}\right]\\
		&= 2 \sum_n \frac{(E_n-E_0)|\bra{n}\hat{\mathbf{d}}\cdot\hat{\boldsymbol{\epsilon}}\ket{0}|^2}
		{(E_n-E_0)^2 - (\hbar\omega)^2}.
	\end{aligned}
\end{equation}
In many cases of interest, where the frequency of the laser field is close to 
the atomic resonance one, the polarizability can be reduced to 
\footnote{For a more profound discussion and algebraic manipulations, see 
PETHICK; SMITH\cite{pethick} and PITAEVSKII; STRINGARI.\cite{pitaevskii}}
\begin{equation}\label{eq.pola}
	\alpha(\omega) \approx \frac{|\bra{n}\hat{\mathbf{d}}\cdot\hat{\boldsymbol{\epsilon}}\ket{0}|^2}
	{E_n-E_0-\hbar\omega}.
\end{equation}

Now, let the excited state have a finite lifetime $\Gamma_n^{-1}$, where 
\begin{equation}
	\Gamma_n = \frac{4}{3} \sum_m \frac{\omega_{n,m}}{4\pi\varepsilon_0\hbar c^3}
	|\bra{n}\hat{\mathbf{d}}\cdot\hat{\boldsymbol{\epsilon}}\ket{m}|^2 
\end{equation}
is the rate of decay by spontaneous emission.\cite{bransden} In this more realistic 
scenario, it implies that the energy of the excited state must contain an additional term 
to account for the spontaneous emission, namely $E_n \to E_n -i\hbar \Gamma_n/2$. Consequently, 
the energy shift (\ref{deltaEt}) results in 
\begin{equation}
	\Delta E = \frac{\hbar}{2} \frac{\Omega_R^2}{\delta_n^2+\Gamma_n^2/4}
	\left(\delta_n-i\frac{\Gamma_n}{2}\right),
\end{equation}
where we have introduced the Rabi frequency\cite{pitaevskii,cohen,sakurai} 
$\Omega_R \equiv |\bra{n}\hat{\mathbf{d}}\cdot\hat{\boldsymbol{\E}}_0\ket{0}|/\hbar$ 
as well as the \textit{detuning}, given by the difference between the radiation field frequency and the 
frequency of the atomic transition $\delta_n \equiv \omega - (E_n-E_0)/\hbar$. In the cold atoms 
literature, $\delta_n > 0$ is known as \textit{blue detuning}, while 
$\delta_n < 0$ is known as \textit{red detuning}.

Let us turn our attention to the form of the potential $V_{\mathrm{ext}}(\mathbf{r})$ 
generated by the laser. The profile of a monochromatic Gaussian 
laser beam is given by\cite{ednilson,mandel} 
\begin{equation}
	V_{\mathrm{ext}}(\mathbf{r}) = -V_0
	\sum_{(i,j,l)} \expo{-2\left(x_i^2+x_j^2\right)/b_0^2} 
	\cos^2(k_L x_l),
\end{equation}
where $b_0$ is the laser beam waist,\cite{mandel} while 
$k_L=2\pi/\lambda$ and $\lambda$ are the laser wavevector and 
wavelength, respectively. Here, the sum is performed for the 
three possible sequence of the independent coordinate variables 
that produce different results for the argument of the sum, \textit{i.e.}, 
$(i,j,l) = (1,2,3),(1,3,2),(2,3,1)$. In the most part of cold atoms experiments, 
the atoms are concentrated around the center of the trap, thus implying $r \ll b_0$.\cite{ednilson} 
This justifies the approximation
$\expo{x} \approx 1+x+\mathcal{O}(x^2),\,|x|\ll 1$, leaving us with a simplified formula 
for the external potential
\begin{equation}
        V_{\mathrm{ext}}(\mathbf{r}) = V_0
	\left\{\frac{4 r^2}{b_0^2} -3 
	+\sum_{(i,j,l)} \sin^2(k_L x_l) 
	\left[1-\frac{2}{b_0^2}\left(x_i^2+x_j^2\right)\right]\right\}.
\end{equation}
Again, by the same reasoning as above, we perform another round of simplifications, 
that results in 
\begin{equation}\label{laserV}
	V_{\mathrm{ext}}(\mathbf{r}) 
	=V_0 \sum_i \sin^2(k_L x_i),
\end{equation}
where we have performed the shift $V_{\mathrm{ext}}(\mathbf{r}) + 3V_0 \to 
V_{\mathrm{ext}}(\mathbf{r})$, 
since $V_0$ is simply a constant.

\subsection{Solutions for the laser field potential}
Now, as the laser potential (\ref{laserV}) is a periodic one, we know from 
solid state physics\cite{ashcroft} that the solutions 
of the Schr\"odinger equation regarding noninteracting particles in such a potential,
\begin{equation}\label{schrband}
	\left[-\frac{\hbar^2}{2m}\nabla^2 + V_{\mathrm{ext}}(\mathbf{r})\right] 
	\Psi_{n,\mathbf{k}}(\mathbf{r}) = E_{n,\mathbf{k}}\Psi_{n,\mathbf{k}}(\mathbf{r}),
\end{equation}
are given by 
\begin{equation}
\Psi_{n,\mathbf{k}}(\mathbf{r}) = \expo{i \mathbf{k}\cdot\mathbf{r}}\Phi_{n,\mathbf{k}}(\mathbf{r}),
\end{equation}
where $\Phi_{n,\mathbf{k}}(\mathbf{r})$ are the so-called Bloch functions\cite{felixbloch} and they possess the same periodicity of the trapping potential. 
The wave vector is represented by $\mathbf{k}$ and $n$ is the band index.

When the potential depth is big enough and the temperature low enough so that the tunneling probability between 
neighboring sites is small, the single-particle wave functions can be approximated by a 
linear combination of localized states in each potential well. 
In order to explore this effect, it is convenient to use the more convenient Wannier functions,\cite{wannier1,wannier2} which are localized functions defined according to
\begin{equation}
W_n(\mathbf{r}-\mathbf{r}_i) \equiv \frac{1}{\sqrt{N_s}} \sum_{\mathbf{k}} 
	\expo{-i \mathbf{k}\cdot\mathbf{r}_i} \Psi_{n,\mathbf{k}}(\mathbf{r}),
\end{equation}
where $N_s$ is the number of lattice sites, $\mathbf{r}_i$ is the location of the $i$-th lattice site, 
and the sum in $\mathbf{k}$ runs over the first Brillouin zone, \textit{i.e.}, $-\pi/d \leq k_i \leq \pi/d$, 
where $d=\pi/k_L=\lambda/2$ is the lattice spacial periodicity. 
As the Bloch functions fulfill the orthonormality and the completeness properties,\cite{ashcroft,kittel} 
it is possible to derive the respective orthonormality and completeness properties for the Wannier functions:  
\begin{equation}\label{orthonormality}
	\begin{aligned}
		\int_\Re d^3 \mathbf{r} \, W_n^*(\mathbf{r}-\mathbf{r}_i)
		 W_m(\mathbf{r}-\mathbf{r}_j) &= \frac{1}{N_s}
		 \sum_{\mathbf{k},\mathbf{k}'} \expo{i(\mathbf{k}\cdot
		 \mathbf{r}_i-\mathbf{k}'\cdot\mathbf{r}_j)} \int_\Re
		 d^3 \mathbf{r} \, \Psi_{n,\mathbf{k}'}^*(\mathbf{r})
		 \Psi_{m,\mathbf{k}}(\mathbf{r}) \\
		 &=\delta_{n,m}\frac{1}{N_s} \sum_\mathbf{k}\expo{i\mathbf{k}
		 \cdot(\mathbf{r}_i-\mathbf{r}_j)} \\
		 &=\delta_{n,m}\delta_{i,j},
	\end{aligned}
\end{equation}
and 
\begin{equation}\label{completeness}
        \begin{aligned}
		\sum_{n,i} W_n^*(\mathbf{r}-\mathbf{r}_i)
                 W_n(\mathbf{r}'-\mathbf{r}_i) &= 
                 \sum_{n,\mathbf{k},\mathbf{k}'}  \Psi_{n,\mathbf{k}}^*(\mathbf{r})
		 \Psi_{n,\mathbf{k}'}(\mathbf{r}') \frac{1}{N_s} \sum_i \expo{i(\mathbf{k}-\mathbf{k}')
		 \cdot\mathbf{r}_i} \\
		 &=\sum_{n,\mathbf{k}}  \Psi_{n,\mathbf{k}}^*(\mathbf{r})
                 \Psi_{n,\mathbf{k}}(\mathbf{r}') \\
		 &=\delta(\mathbf{r}-\mathbf{r}'),
        \end{aligned}
\end{equation}
respectively.
Such conditions assure that any wave function can be written as a series 
expansion of the Wannier functions.

Now, as the laser-generated potential (\ref{laserV}) consists 
of separated contributions on each coordinate variable, the original 
three-dimensional Schr\"odinger equation can be separated into three 
identical one-dimensional Schr\"odinger equations, which read
\begin{equation}
	\left[-\frac{\hbar^2}{2m}\nabla_j^2 + V_{\mathrm{ext}}(x_j)\right] 
	\psi_{n_j,k_j}(x) = \epsilon_{n_j,k_j}\psi_{n_j,k_j}(x_j),
\end{equation}
with 
\begin{equation}
	\psi_{n,\mathbf{k}}(\mathbf{r}) = 
	\prod_j \psi_{n_{j},k_{j}}(x_j),
\end{equation}
and 
\begin{equation}
	E_{n,\mathbf{k}} = \sum_j 
	\epsilon_{n_j,k_j}.
\end{equation}
Analogously to the three-dimensional problem, each one-dimensional 
solution can be written in terms of the respective one-dimensional Bloch functions as 
\begin{equation}
	\Psi_{n_j,k_j}(x_j) = \expo{ik_j x_j} \phi_{n_j,k_j}(x_j).
\end{equation}
Consequently, the corresponding one-dimensional Wannier functions read 
\begin{equation}
	w_{n_j}\left(x_j-x^{(i)}_j\right) = \left(\frac{1}{\sqrt{N_s}}\right)^{1/3}
	\sum_{k_j} \expo{-ik_j x_j^{(i)}} \psi_{n,k_j}(x_j).
\end{equation}
Therefore, the tridimensional Bloch and Wannier functions become  
\begin{equation}
	\Phi_{\mathbf{n}}(\mathbf{r}) = \prod_j \phi_{n_j,k_j}(x_j)
\end{equation}
and 
\begin{equation}
	W_{\mathbf{n}}(\mathbf{r}-\mathbf{r}_i) = 
	\prod_j w_{n_j} \left(x_j-x^{(i)}_j\right),
\end{equation}
respectively.

\section{The Bose-Hubbard model}
The Bose-Hubbard model\cite{hubbard,bose-hubbard} provides a suitable 
description of interacting spinless bosonic atoms confined in optical lattices.
The fundamental mathematical considerations within the model are developed as follows.
Let us start with the general second-quantized Hamiltonian
\begin{equation}
\begin{aligned}\label{TBHH}
	\hat{H} =& \int d^3\mathbf{r} \hat{\Psi}^\dagger(\mathbf{r})\left[-\frac{\hbar^2}{2m}\nabla^2 + V_{\mathrm{ext}}(\mathbf{r}) - \mu \right]\hat{\Psi}(\mathbf{r}) \\
	&+ \frac{1}{2} \int d^3\mathbf{r}_1 \int d^3\mathbf{r}_2 \hat{\Psi}^\dagger(\mathbf{r}_1)
	\hat{\Psi}^\dagger(\mathbf{r}_2) V_{\mathrm{int}}(\mathbf{r}_1,\mathbf{r}_2) \hat{\Psi}(\mathbf{r}_1)\hat{\Psi}(\mathbf{r}_2),
\end{aligned}
\end{equation}
where the first term represents the single-particle Hamiltonian, with $V_{\mathrm{ext}}(\mathbf{r})$ representing the atom-laser interaction, 
and $\mu$ is the grand-canonical chemical potential; while the second term corresponds to the interparticle interaction term, where 
$V_{\mathrm{int}}(\mathbf{r}_1,\mathbf{r}_2)$ is the atomic interaction potential. The bosonic field 
operators are represented by $\hat{\Psi}^\dagger(\mathbf{r})$ and $\hat{\Psi}(\mathbf{r})$, and they obey the usual 
bosonic commutation rules 
\begin{subequations}\label{commutationpsi}
	\begin{align}
		&\left[\hat{\Psi}(\mathbf{r}), \hat{\Psi}^\dagger(\mathbf{r})\right] = \delta(\mathbf{r}-\mathbf{r'}),\\
		&\left[\hat{\Psi}(\mathbf{r}), \hat{\Psi}(\mathbf{r'})\right] = \left[\hat{\Psi}^\dagger(\mathbf{r}), \hat{\Psi}^\dagger(\mathbf{r'})\right] =0.
	\end{align}
\end{subequations}

Considering gases with low density profiles, the interaction between particles can be approximated by\cite{pethick,pitaevskii}
\begin{equation}
	V_{\mathrm{int}}(\mathbf{r}_1,\mathbf{r}_2) = \frac{4\pi \hbar^2 a_{\mathrm{3D}}}{m} \delta(\mathbf{r}_1-\mathbf{r}_2),
\end{equation}
where $a_\mathrm{3D}$ is the three-dimensional s-wave scattering length (see App. \ref{appendixA}). Thus, Eq. (\ref{TBHH}) reduces to
\begin{equation}\label{2.9}
	\hat{H} = \int d^3\mathbf{r} \hat{\Psi}^\dagger(\mathbf{r})\left[-\frac{\hbar^2}{2m}\nabla^2 + 
	V_{\mathrm{ext}}(\mathbf{r})-\mu \right]\hat{\Psi}(\mathbf{r}) + \frac{g}{2} \int d^3\mathbf{r} \hat{\Psi}^\dagger(\mathbf{r})\hat{\Psi}^\dagger(\mathbf{r})  \hat{\Psi}(\mathbf{r})\hat{\Psi}(\mathbf{r}),
\end{equation}
where $g\equiv 4\pi \hbar^2 a_{\mathrm{3D}}/m$ is the coupling constant.

Now, taking into account that ultracold atoms confined in deep periodic potentials can be regarded as occupying only the lowest Bloch band,
 we can simplify the problem by restricting ourselves to the Wannier function corresponding to $n=0$, $W_0(\mathbf{r})$.
Therefore, due to the completeness of the Wannier functions (\ref{completeness}), the field operators can then be expanded as
\begin{subequations}\label{wannier}
  \begin{align}
  	\hat{\Psi}(\mathbf{r}) &= \sum_{i} \hat{a}_i W_0(\mathbf{r}-\mathbf{r}_i),\label{10a}\\
  	\hat{\Psi}^\dagger(\mathbf{r}) &= \sum_{i} \hat{a}^\dagger_i W_0^*(\mathbf{r}-\mathbf{r}_i),\label{10b}
  \end{align}
\end{subequations}
where $\hat{a}_i$ and $\hat{a}^\dagger_i$ are, respectively, the annihilation and creation operators of particles at a given lattice site $i$.
From (\ref{commutationpsi}), it is also possible to derive the commutation rules for the lattice operators, resulting in
\begin{subequations}\label{commua}
	\begin{align}
		\left[\hat{a}_i,\hat{a}_j^\dagger \right] &= \delta_{i,j},\\
		\left[\hat{a}_i,\hat{a}_j \right] &= \left[\hat{a}_i^\dagger,\hat{a}_j^\dagger \right] = 0.
	\end{align}
\end{subequations}

By substituting (\ref{wannier}) into (\ref{2.9}), we have
\begin{equation}
\hat{H}=\frac{1}{2}\sum_{i,j,k,l}U_{ijkl}\hat{a}_{i}^{\dagger}\hat{a}_{j}^{\dagger}\hat{a}_{k}\hat{a}_{l}+\sum_{i,j}J_{ij}\hat{a}_{i}^{\dagger}\hat{a}_{j}-\sum_{i,j}\mu_{ij}\hat{a}_{i}^{\dagger}\hat{a}_{j},
\end{equation}
where the parameters read
\begin{subequations}
	\begin{align}
	U_{ijkl} &= g \int d^3\mathbf{r}\, W_0^*(\mathbf{r}-\mathbf{r}_i)W_0^*(\mathbf{r}-\mathbf{r}_j) W_0(\mathbf{r}-\mathbf{r}_k) W_0(\mathbf{r}-\mathbf{r}_l),\\
		J_{ij} &= \int d^3\mathbf{r}\, W_0^*(\mathbf{r}-\mathbf{r}_i)\left[-\frac{\hbar^2}{2m} \nabla^2+V_{\mathrm{ext}}(\mathbf{r})\right] W_0(\mathbf{r}-\mathbf{r}_j), \\
	\mu_{ij} &= \mu \int d^3\mathbf{r} \, W_0^*(\mathbf{r}-\mathbf{r}_i) W_0(\mathbf{r}-\mathbf{r}_j).        
	\end{align}
\end{subequations}

Following the discussion from UEDA,\cite{ueda} in a scenario where the confining potential is sufficiently deep, the Wannier functions are strongly localized, 
hence the overlap between the different-site-particle wave functions is small. Therefore, in this model we consider only nearest neighbors transitions and local interparticle interaction. 
Such considerations and the orthonormality of the Wannier functions lead to the Bose-Hubbard Hamiltonian
\begin{equation}\label{BH}
\hat{H}_{BH}=\frac{U}{2}\sum_{i}\hat{a}_{i}^{\dagger}\hat{a}_{i}^{\dagger}\hat{a}_{i}\hat{a}_{i}
	-J\sum_{\langle i,j\rangle}\hat{a}_{i}^{\dagger}\hat{a}_{j}-\mu\sum_{i}\hat{a}_{i}^{\dagger}\hat{a}_{i},
\end{equation}
where 
\begin{subequations}\label{parametersfinal}
	\begin{align}
		U &= g \int d^3\mathbf{r} \, |W_0(\mathbf{r})|^4, \label{Ufinal}\\
		\label{Jfinal}	J &= \int d^3\mathbf{r} \, 
		W^*_0(\mathbf{r}) \left[\frac{\hbar^2}{2m} \nabla^2-V_{\mathrm{ext}}(\mathbf{r})\right] W_0(\mathbf{r}).     
	\end{align}
\end{subequations}
Such parameters have clear interpretations: $U$ is the on-site interaction parameter between particles and 
$J$ is the hopping parameter, which describes the tunneling probability of particles between its original site to a neighboring one.

The Bose-Hubbard model predicts two different phases for the whole system depending 
on the ratio between the on-site interaction and the hopping parameters: 
if the on-site interaction between atoms is much stronger than the hopping parameter, \textit{i.e.}, $U/J\gg 1$,
the system realizes a Mott insulator phase;\cite{boer,mott1,mott2} on the other hand, 
if the hopping parameter predominates over the on-site interaction parameter, \textit{i.e.}, $U/J \ll 1$, the ground 
state of the system is a superfluid phase. Deep in the Mott insulator phase, where all atoms are 
highly localized in the potential minima, the ground state of the whole system is given by\cite{pethick,pitaevskii} 
\begin{equation}
	\ket{\Psi_{\mathrm{MI}}} = (n!)^{-N_s/2}
	\prod_{i=1}^{N_s}\left(\hat{a}_i^\dagger\right)^n \bigotimes_{N_s} \ket{0},
\end{equation}
where $N_s$ is the number of lattice sites and $n$ is the average occupation number 
per site. In the opposite scenario, \textit{i.e.}, the ground state of the system deep in the superfluid phase 
can be considered as\cite{pethick,pitaevskii} 
\begin{equation}
	\ket{\Psi_{\mathrm{SF}}} = \frac{N_s^{-N/2}}{\sqrt{N!}} 
	 \left(\sum_{i=1}^{N_s} \hat{a}_i^\dagger\right)^N \bigotimes_{N_s} \ket{0},
\end{equation}
where $N$ is the total number of particles and 
$\bigotimes_{N_s}\ket{0} = \ket{0} \otimes \ket{0} \cdots \otimes \ket{0} $ is the vacuum state.

\subsection{The Hamiltonian parameters}
For a deep periodic potential, we can consider the lowest-band Wannier function as a solution 
of the laser field potential
\begin{equation}
        \left[ -\frac{\hbar^2}{2m} \nabla_x^2
        +V_0 \sin^2(k_L x)\right] w_0(x) = E_0 w_0(x).
\end{equation}
This differential equation has approximate solutions in terms of the 
Mathieu functions\cite{handbook,table} 
\begin{equation}
	w_0(x) = 
	\mathrm{C} \left(1-\frac{\tilde{V}_0}{2},
	-\frac{\tilde{V}_0}{4}, k_L x\right) + 
	 \mathrm{S} \left(1-\frac{\tilde{V}_0}{2},
	-\frac{\tilde{V}_0}{4}, k_L x\right),
\end{equation}
where $\mathrm{C}(a,q,z)$ and $\mathrm{S}(a,q,z)$ are the 
even and odd Mathieu functions, respectively.  Here we have
defined $\tilde{V}_0 \equiv V_0/E_R$ and introduced the so-called
\textit{recoil energy} $E_R \equiv \hbar^2 k_L^2/2m$.
With such a solution, it is possible to approximately evaluate the hopping energy (\ref{Jfinal}), 
which is performed in Ref. \citeonline{zwerger} with the following result 
\begin{equation}\label{mathieu}
	J = \frac{4}{\sqrt{\pi}} E_R \tilde{V}_0^{3/4} 
	\expo{-2 \tilde{V}_0^{1/2}}.
\end{equation}

\subsubsection{Harmonic approximation}
A first approximation of the laser field potential is the harmonic approximation\cite{albus} 
$\sin^2(k_L x) \approx (k_L x)^2$. Again, let us consider the lowest-band Wannier function as a solution 
of the harmonic potential 
\begin{equation}
	\left[ -\frac{\hbar^2}{2m} \nabla_x^2 
	+V_0(k_L x)^2\right] w_0(x) = E_0 w_0(x).
\end{equation}
The solution is the known fundamental state of the harmonic oscillator 
\begin{equation}\label{w0sol}
	w_0(x) = \left(\frac{k_L^2 \tilde{V}_0^{1/2}}{\pi}\right)^{1/4}
	\exp{\left(-\frac{x^2}{2}k_L^2 \sqrt{\tilde{V}_0}\right)},
\end{equation}
with the energy given by $E_0 = E_R \sqrt{\tilde{V}_0}$. 

It follows that we can also find an expression for the the Bose-Hubbard 
parameters (\ref{parametersfinal}) from the solution $w_0(x)$ within 
the harmonic approximation. So, from (\ref{Ufinal}) we have that the 
on-site interaction energy reads 
\begin{equation}
	U = g \int_{\Re} d^3\mathbf{r} \, |W_0(\mathbf{r})|^4 
	= g \left( \int_{-\infty}^{+\infty} dx \, 
	|w_0(x)|^4 \right)^3 = \sqrt{\frac{8}{\pi}} a_{\mathrm{3D}} k_L
	E_R \tilde{V}_0^{3/4}.
\end{equation}
Similarly, from (\ref{Jfinal})
\begin{equation}
	\begin{aligned}\label{equacao2.43}
		J &= \int_\Re d^3\mathbf{r} \,
        W^*_0(\mathbf{r}) \left[\frac{\hbar^2}{2m} 
		\nabla^2-V_{\mathrm{ext}}(\mathbf{r})\right] W_0(\mathbf{r}) \\
		&= \int_{-\infty}^{+\infty} dx \, w_0^*(x-d)
		\left(\frac{\hbar^2}{2m} \nabla_x^2 - V_0 \sin^2(k_L x)\right) 
	w_0(x).
	\end{aligned}
\end{equation}
Substituting the solution (\ref{w0sol}) into (\ref{equacao2.43}) and performing the integral, 
the hopping energy results in
\begin{equation}\label{Jharmonic}
	J = \frac{E_R}{4} \left(\pi^2 \tilde{V}_0-2\tilde{V}_0^{1/2}\right)
	 \exp{\left(-\frac{\pi^2}{4}\sqrt{\tilde{V}_0}\right)}
	-\frac{E_R}{2} \tilde{V}_0
	\left( 1+\expo{\tilde{V}_0^{-1/2}} \right)
	\exp{\left( -\frac{4+\pi^2\tilde{V}_0}{4\sqrt{\tilde{V}_0}}\right)}.
\end{equation}

\begin{figure}[h]
\centering
\includegraphics[scale=1]{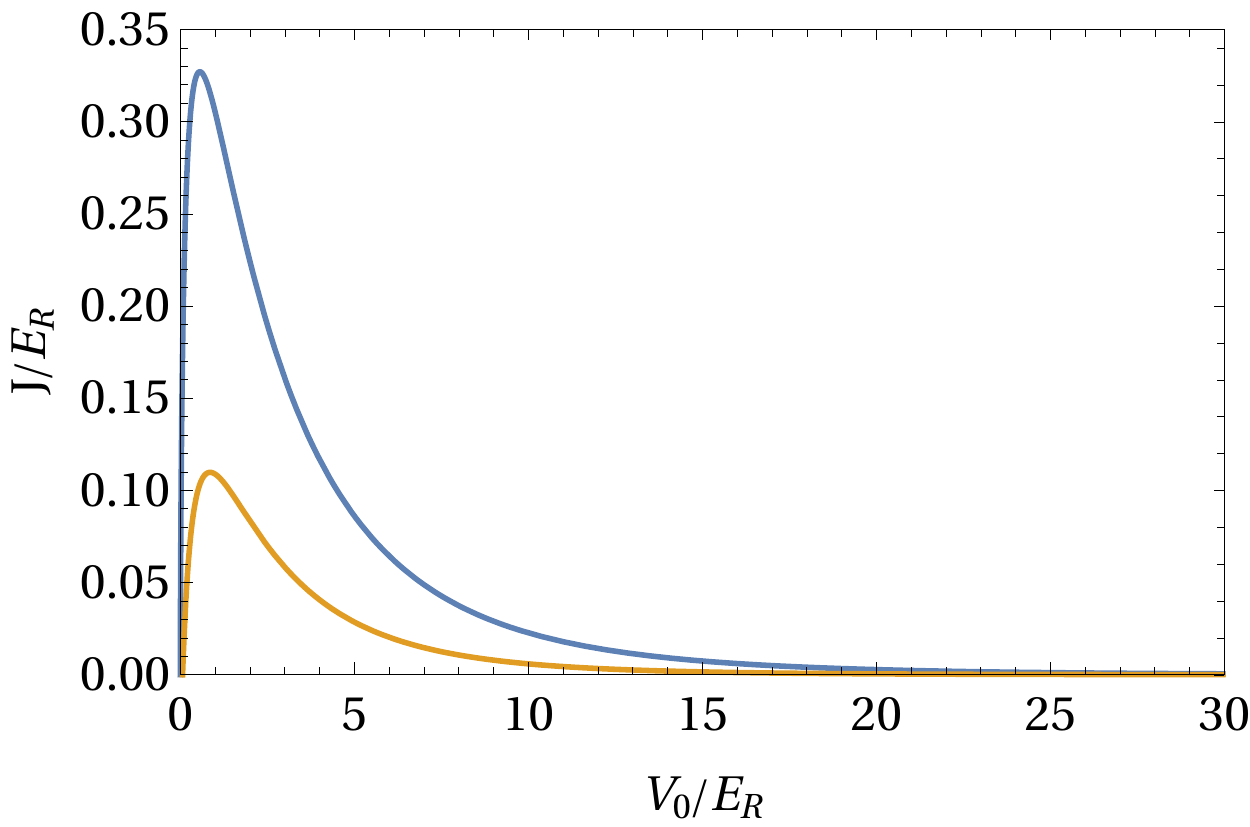}
\caption{Hopping energy from the Mathieu solution (\ref{mathieu}) (blue line) and 
	from the harmonic approximation (\ref{Jharmonic}) (yellow line).}
		Source: By the author.
\label{fig.hopping}
\end{figure}

As a comparison, we plot the hopping energy from the Mathieu solution (\ref{mathieu}) 
and from the harmonic approximation (\ref{Jharmonic}) in Fig. \ref{fig.hopping} and 
conclude, by direct observation, that the harmonic approximation does not result in a good 
enough estimation of the hopping energy for shallow potentials.

\chapter{Mott-insulator-superfluid quantum phase transition}
\label{mf}
The purpose of this chapter is the study of the main considerations taken into account 
in order to investigate the Mott-insulator-superfluid (MI-SF) quantum phase transition of bosons in optical lattices. 
We begin by introducing some basic concepts 
about second-order phase transitions. Also, we discuss the Landau assumptions for the thermodynamic potential in the vicinity of the phase transition.  
Then, we introduce the mean-field approximation, which is the main path taken in order to remove the nonlocality  
present in the Bose-Hubbard Hamiltonian, leading to a great simplification. 
Following, we perform calculations based on nondegenerate perturbation theory (NDPT), 
which results in the MI-SF phase diagram. 
Following such calculations, we show that NDPT leads to some inconsistencies due to degeneracy, 
which turns out to provide a nonphysical behavior of the order parameter.

\section{Quantum phase transitions}
In 1933, Paul Ehrenfest noted that different systems 
in thermodynamical equilibrium could present
distinct-order discontinuities in their thermodynamic potential: 
some transitions were characterized by a discontinuity in the 
first derivative of the thermodynamic potential with respect to some variable 
(which we will call \textit{order parameter} later on), which he then named 
\textit{first-order phase transitions}; others indicated 
a discontinuity in the second derivative of the thermodynamic potential, and those he called 
\textit{second-order phase transitions}.\cite{ehrenfest,jaeger}

Differently from classical phase transitions, that arise as a result of 
thermal fluctuations, quantum phase transitions can happen even at zero temperature, 
driven by quantum fluctuations.\cite{miransky,kleinert,justin,stanley}
This is the case of our considered system constituted of bosons in optical lattices: 
the transition from the Mott insulator to the superfluid phase can happen 
at $T=0$ without the effects of thermal fluctuations, thus characterizing 
a quantum phase transition. In the Mott insulating phase, the atoms 
are localized at the minima of the laser-generated potential, 
meaning that the condensate density is zero in such a regime. 
On the other hand, in the superfluid phase, the system is characterized by a 
high delocalization of the atoms, which means that it has achieved a 
nonzero condensate density. Due to this explicit change from a zero value 
to a nonzero one of the condensate density, we can regard it as being the order 
parameter of the quantum phase transition in question.

\subsection{\label{landau}Landau theory of second-order phase transitions}
Landau argued that the thermodynamic potential $\mathcal{F}$ could be written as a 
polynomial function of the order parameter in the vicinity of a phase transition.\cite{landau} 
In the case of BEC, where the order parameter is the condensate wave function $\Psi$, the Landau 
expansion could in principle be 
\begin{equation}\label{landauexpansion}
	\mathcal{F} = a_0 + a_1 |\Psi| 
	+ a_2 |\Psi|^2 + a_3 |\Psi|^3 + a_4 |\Psi|^4 + \cdots .
\end{equation}
However, in the case of the Bose-Hubbard Hamiltonian described by (\ref{BH}), 
which possesses a global $U(1)$ phase invariance, \textit{i.e.}, the Bose-Hubbard Hamiltonian 
is invariant under the transformation $\hat{a} \to \expo{i\theta} \hat{a} $, 
$a_n$ will not vanish only for even values of $n$. Therefore, 
since we are considering only small values of $|\Psi|$, 
further analysis will be held on the even Landau expansion up to fourth order, 
\begin{equation}\label{landaueven}
	\mathcal{F} \approx a_0 + a_2 |\Psi|^2 + a_4 |\Psi|^4 .
\end{equation}

\begin{figure}[h]
	\centering
\includegraphics[scale=.8]{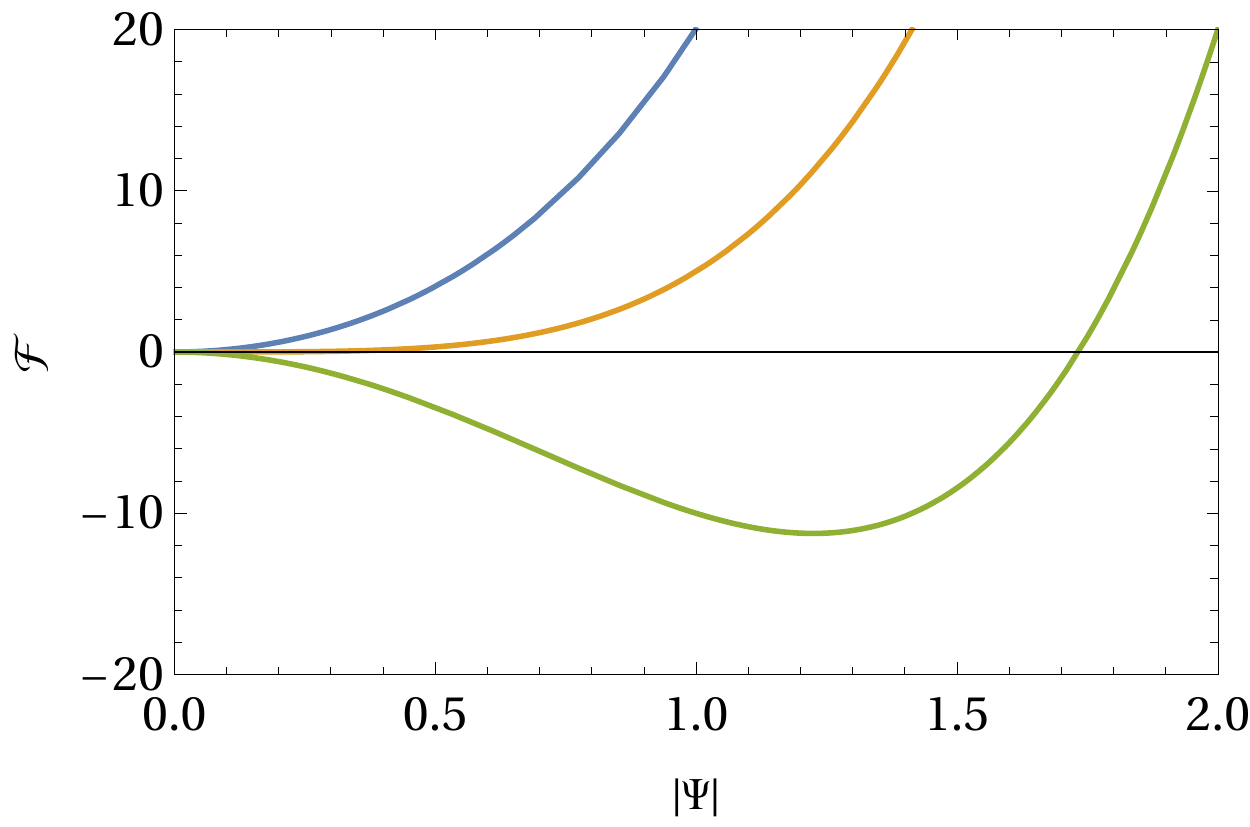}
	\caption{Landau expansion of the thermodynamic potential from (\ref{landaueven}). The blue, yellow, and green curves 
	correspond to, respectively, $a_2>0$, $a_2=0$, and $a_2<0$.}
	Source: By the author.
\label{fig.omega}
\end{figure}

From Fig. \ref{fig.omega}, for $a_4 > 0$, it is possible to realize that, for 
$a_2 > 0$, the stable state, \textit{i.e.}, the minimum of $\mathcal{F}$, 
happens at $|\Psi|=0$, which corresponds to the symmetrical phase. On the 
other hand, when $a_2 < 0$, the stable state is given by nonvanishing values 
of the order parameter, $|\Psi| \neq 0$, corresponding to the unsymmetrical phase. 
Conclusively, the condition $a_2 = 0$ defines the phase boundary between the two phases. 
Also, the solution for the unsymmetrical phase is given by 
\begin{equation}
	\frac{\partial \mathcal{F}}{\partial |\Psi|}\Bigg|_{|\Psi|\neq 0} = 0
	\Rightarrow |\Psi|^2 = -\frac{a_2}{2a_4}.
\end{equation}

\section{\label{sec2}Mean-field approximation}
Due to the non-local character of the hopping term in the Bose-Hubbard Hamiltonian (\ref{BH}), 
we perform a mean-field approximation,\cite{fisher,melo,santos,ednilson}
which consists in considering the bosonic operators as the contribution 
of its mean value summed to a fluctuation  
$\hat{a}_i=\av{\hat{a}_i}+\delta \hat{a}_i$. Thus, the hopping term 
in (\ref{BH}) reads 
\begin{equation}
		-J\sum_{\langle i,j\rangle}\hat{a}_{i}^{\dagger}\hat{a}_{j} =
	-J\sum_{\langle i,j\rangle}\left(\av{\hat{a}^\dagger_i}+\delta \hat{a}^\dagger_i\right)
	\left(\av{\hat{a}_j}+\delta \hat{a}_j\right).
\end{equation}
Neglecting quadratic terms of fluctuations we have
\begin{equation}
                -J\sum_{\langle i,j\rangle}\hat{a}_{i}^{\dagger}\hat{a}_{j} =
        -J\sum_{\langle i,j\rangle}\left(\av{\hat{a}_i^\dagger} \hat{a}_j
	+ \hat{a}_i^\dagger \av{\hat{a}_j} - \av{\hat{a}_i^\dagger}\av{\hat{a}_j}\right).
\end{equation}
Now, let us consider a homogeneous system so that the average value of the annihilation operator is 
site-independent, implying the definition $\Psi \equiv \av{\hat{a}_{i,j}}$. 
Moreover, denoting the number of nearest neighbors by $z$, we obtain the mean-field Hamiltonian,
\begin{equation}
\label{mfH}
\hat{H}_{MF}=\frac{U}{2}\sum_i \left(\hat{n}_i^{2}-\hat{n}_i\right) -\sum_i \mu\hat{n}_i -Jz\sum_i\left(\Psi^{*}\hat{a}_i+\Psi\hat{a}_i^{\dagger}-\Psi^{*}\Psi\right) ,
\end{equation}
where $\hat{n}_i=\hat{a}^{\dagger}_i\hat{a}_i$ is the number operator at the lattice site $i$. 
Note that we have used the bosonic commutation relations (\ref{commua}) 
in order to rewrite the on-site interaction term in terms of the number operator $\hat{n}$.

Since (\ref{mfH}) is a sum of local Hamiltonians, we restrict ourselves to the one lattice site Hamiltonian,
\begin{equation}
\label{H_one}
\hat{H}=\frac{U}{2}\left(\hat{n}^{2}-\hat{n}\right) -\mu\hat{n} -Jz\left(\Psi^{*}\hat{a}+\Psi\hat{a}^{\dagger}-\Psi^{*}\Psi\right).
\end{equation}

\subsection{\label{PT}Nondegenerate perturbation theory}
As mentioned before, the transition from Mott-insulator to superfluid is associated to the breakdown of 
the $U(1)$ symmetry and can then be characterized by the change in the order parameter from zero to a non-zero value. 
Since we are considering our system in the vicinity of the phase transition, where $\left| \Psi \right|$ has a small value, 
and only the hopping term depends explicitly on $\Psi$ in (\ref{H_one}), 
we can treat the hopping term as a perturbation. Thus, (\ref{H_one}) 
decomposes according to $\hat{H}=\hat{H}_0+\hat{V}$ into the unperturbed Hamiltonian
\begin{equation}\label{H0}
\hat{H}_0=\frac{U}{2}\left(\hat{n}^{2}-\hat{n}\right) -\mu\hat{n} +Jz\Psi^{*}\Psi,
\end{equation}
and the perturbation
\begin{equation}
\hat{V}=-Jz\left(\Psi^{*}\hat{a}+\Psi\hat{a}^{\dagger}\right). \label{perturbation}
\end{equation}

The unperturbed eigenenergies are
\begin{equation}
\label{En}
E_n = \frac{U}{2}\left(n^{2}-n\right)-\mu n + J z |\Psi|^2 ,
\end{equation}
where the quantum number $n=0,1,2,\dots$ indicates the number of bosons per site.

At this point, we are interested in evaluating how the perturbation changes the free energy of the system. To this purpose, we must work out the partition function, 
\begin{equation}
\label{Z1}
\mathcal{Z} = \mathrm{Tr}\left[\mathrm{e}^{-\beta \hat{H}}\right],
\end{equation}
in order to obtain the free energy of the system. 

The quantum-mechanical evolution operator within the imaginary-time formalism, \textit{i.e.}, $\hat{U}=\mathrm{e}^{-\tau \hat{H}}$, can be factorized according to
\begin{equation}
\hat{U}=\mathrm{e}^{-\tau \hat{H}_0}\hat{U}_{\mathrm{I}}(\tau),
\end{equation}
where $\hat{U}_{\mathrm{I}}(\tau)$ is the interaction picture imaginary-time evolution operator. 
Note that we are assuming $\hbar=1$.
The equation for the time evolution of such an operator is\cite{sakurai}
\begin{equation}
\label{eq21a}
\frac{d \hat{U}_{\mathrm{I}} (\tau)}{d \tau} = - \hat{V}_{\mathrm{I}} (\tau) \hat{U}_{\mathrm{I}} (\tau),
\end{equation}
with
\begin{equation}
\label{pertua}
\hat{V}_{\mathrm{I}}(\tau)=\mathrm{e}^{\tau\hat{H}_{0}}\hat{V}\mathrm{e}^{-\tau\hat{H}_{0}}.
\end{equation}

Equation (\ref{eq21a}) can be iteratively solved, 
thus allowing the construction of a perturbative expansion. 
Performing the expansion, with the initial value $\hat{U}_{\mathrm{I}}(0)=\hat{\mathds{1}}$, 
up to fourth order, we have
\begin{equation}
\begin{aligned}
	\label{UI}
	\hat{U}_{\mathrm{I}}(\beta) \approx \, & \hat{\mathds{1}}-\int_{0}^{\beta} d\tau_1 \hat{V}_{\mathrm{I}} (\tau_1) + \int_{0}^{\beta} d\tau_1 \int_{0}^{\tau_1} d\tau_2 \hat{V}_{\mathrm{I}} (\tau_1) \hat{V}_{\mathrm{I}} (\tau_2) \\
	&-\int_{0}^{\beta} d\tau_1 \int_{0}^{\tau_1} d\tau_2 \int_{0}^{\tau_2} d\tau_3 \hat{V}_{\mathrm{I}} (\tau_1) \hat{V}_{\mathrm{I}} (\tau_2) \hat{V}_{\mathrm{I}} (\tau_3) \\
	&+\int_{0}^{\beta} d\tau_1 \int_{0}^{\tau_1} d\tau_2 \int_{0}^{\tau_2} d\tau_3 \int_{0}^{\tau_3} d\tau_4 \hat{V}_{\mathrm{I}} (\tau_1) \hat{V}_{\mathrm{I}} (\tau_2) \hat{V}_{\mathrm{I}} (\tau_3) \hat{V}_{\mathrm{I}} (\tau_4).
\end{aligned}
\end{equation}

It is possible to observe, from the perturbative Hamiltonian (\ref{perturbation}), that odd-order terms in (\ref{UI}) will vanish. 
Therefore, we can restrict ourselves to the calculation of the zeroth-, second-, and fourth-order terms in (\ref{UI}).

Making use of the time-evolution operator in the interaction picture \\
$\mathcal{Z} = \Tr{\expo{-\beta \hat{H}_0}\hat{U}_{\mathrm{I}}(\beta)}$, we calculate the partition function up 
to fourth order,
\begin{equation}
	\mathcal{Z}=\sum_{n=0}^{\infty} \expo{-\beta E_n}\bra{n} \hat{U}_{\mathrm{I}}(\beta) \ket{n}
	\approx \mathcal{Z}^{(0)}+\mathcal{Z}^{(2)}+ \mathcal{Z}^{(4)},
\end{equation}
with the single-site eigenstates $\ket{n}$ corresponding to the occupation number in the Mott insulator state. 
The zeroth-order term yields
\begin{equation}
	\mathcal{Z}^{(0)} = \sum_{n=0}^\infty \mathrm{e}^{-\beta E_{n}} \bra{n}\hat{\mathds{1}}\ket{n}  = \sum_{n=0}^{\infty}\mathrm{e}^{-\beta E_{n}}.
\end{equation}

Now, let us proceed to the detailed calculation of the second- and fourth-order terms, 
$\mathcal{Z}^{(2)}$ and $\mathcal{Z}^{(4)}$, respectively. The second-order term reads
\begin{equation}\label{a1}
\mathcal{Z}^{(2)}=\sum_{n=0}^{\infty}\mathrm{e}^{-\beta E_{n}}\int_{0}^{\beta}d\tau_{1}\int_{0}^{\tau_{1}}d\tau_{2}\langle n|\hat{V}_{\mathrm{I}}(\tau_{1})\hat{V}_{\mathrm{I}}(\tau_{2})|n\rangle.
\end{equation}
Inserting (\ref{pertua}) into (\ref{a1}), we have
\begin{equation}\label{A2}
\mathcal{Z}^{(2)}=\sum_{n=0}^{\infty}\mathrm{e}^{-\beta E_{n}}\int_{0}^{\beta}d\tau_{1}\int_{0}^{\tau_{1}}d\tau_{2}\langle n|\mathrm{e}^{\tau_{1}\hat{H}_{0}}\hat{V}\mathrm{e}^{-\tau_{1}\hat{H}_{0}}\mathrm{e}^{\tau_{2}\hat{H}_{0}}\hat{V}\mathrm{e}^{-\tau_{2}\hat{H}_{0}}|n\rangle.
\end{equation}
The exponential of an Hermitian operator $\hat{\mathcal{O}}$ with 
eigenstates $\ket{\phi_\lambda}$ and respective eigenvalues $\lambda$ 
is simply given by 
\begin{equation}
	\begin{aligned}
		\expo{\hat{\mathcal{O}}} =& \, \expo{\sum_\lambda \lambda \ket{\phi_\lambda} \bra{\phi_\lambda}} \\
			       =& \sum_{n=0}^\infty \frac{1}{n!}\left(\sum_\lambda \lambda \ket{\phi_\lambda} \bra{\phi_\lambda}\right)^n \\
			       =& \sum_\lambda \left(\sum_{n=0}^\infty \frac{\lambda}{n!}\right)  \ket{\phi_\lambda} \bra{\phi_\lambda} \\
			       =& \sum_\lambda \expo{\lambda}  \ket{\phi_\lambda} \bra{\phi_\lambda}.
	\end{aligned}
\end{equation}
As $\ket{n}$ are eigenstates of $\hat{H}_0$, Eq. (\ref{A2}) reduces to
\begin{equation}
\mathcal{Z}^{(2)}=\sum_{n=0}^{\infty}\mathrm{e}^{-\beta E_{n}}\int_{0}^{\beta}d\tau_{1}\int_{0}^{\tau_{1}}d\tau_{2}\mathrm{e}^{(\tau_{1}-\tau_{2})E_{n}}\langle n|\hat{V}\mathrm{e}^{-\tau_{1}\hat{H}_{0}}\mathrm{e}^{\tau_{2}\hat{H}_{0}}\hat{V}|n\rangle.
\end{equation}
According to (\ref{perturbation}), we have
\begin{equation}
	\begin{aligned}
		\mathcal{Z}^{(2)}&=& J^{2}z^{2}\sum_{n=0}^{\infty}\mathrm{e}^{-\beta E_{n}}\int_{0}^{\beta}d\tau_{1}\int_{0}^{\tau_{1}}d\tau_{2} \,
		 \mathrm{e}^{(\tau_{1}-\tau_{2})E_{n}}\langle n|\left(\Psi^{*}\hat{a}+\Psi \hat{a}^{\dagger}\right)\mathrm{e}^{-\tau_{1}\hat{H}_{0}}\\
		&&\times\mathrm{e}^{\tau_{2}\hat{H}_{0}}\left(\Psi^{*}\hat{a}+\Psi \hat{a}^{\dagger}\right)|n\rangle,
\end{aligned}
\end{equation}
yielding
\begin{equation}
\begin{aligned}\label{A5}
	\mathcal{Z}^{(2)}&=&J^{2}z^{2}\sum_{n=0}^{\infty}\mathrm{e}^{-\beta E_{n}}\int_{0}^{\beta}d\tau_{1}\int_{0}^{\tau_{1}}d\tau_{2} \, 
	\mathrm{e}^{(\tau_{1}-\tau_{2})E_{n}}\left(\Psi\sqrt{n}\langle n-1|+\Psi^{*}\sqrt{n+1}\langle n+1|\right) \\
	&&\times\left(\Psi^{*}\sqrt{n}\mathrm{e}^{(\tau_{2}-\tau_{1})E_{n-1}}|n-1\rangle+\Psi\sqrt{n+1}\mathrm{e}^{(\tau_{2}-\tau_{1})E_{n+1}}|n+1\rangle\right).
\end{aligned}
\end{equation}
The scalar products reduce (\ref{A5}) to
\begin{equation}
\mathcal{Z}^{(2)}=J^{2}z^{2}|\Psi|^{2}\sum_{n=0}^{\infty}\mathrm{e}^{-\beta E_{n}}\int_{0}^{\beta}d\tau_{1}\int_{0}^{\tau_{1}}d\tau_{2}
	\left[n\mathrm{e}^{(\tau_{1}-\tau_{2})\Delta_{n,n-1}}+(n+1)\mathrm{e}^{(\tau_{1}-\tau_{2})\Delta_{n,n+1}}\right].
\end{equation}
Finally, the integrations yield
\begin{equation}
\begin{aligned}
	\mathcal{Z}^{(2)}&=&J^{2}z^{2}|\Psi|^{2}\sum_{n=0}^{\infty}\mathrm{e}^{-\beta E_{n}}\Bigg[n\left(
	\frac{\mathrm{e}^{\beta\Delta_{n,n-1}}-1}{\Delta_{n,n-1}^{2}}-\frac{\beta}{\Delta_{n,n-1}}\right) \\
	&&+(n+1)\left(\frac{\mathrm{e}^{\beta\Delta_{n,n+1}}-1}{\Delta_{n,n+1}^{2}}-\frac{\beta}{\Delta_{n,n+1}}\right)\Bigg],
\end{aligned}
\end{equation}
where we have introduced the abbreviation $\Delta_{i,j}\equiv E_i-E_j$ for the differences between two 
consecutive energies given by (\ref{En}).

For the fourth-order term, we have
\begin{equation}\label{A8}
\mathcal{Z}^{(4)}=\sum_{n=0}^{\infty}\mathrm{e}^{-\beta E_{n}}\int_{0}^{\beta}d\tau_{1}\int_{0}^{\tau_{1}}d\tau_{2}\int_{0}^{\tau_{2}}d\tau_{3}\int_{0}^{\tau_{3}}d\tau_{4}\langle n|\hat{V}_{\mathrm{I}}(\tau_{1})\hat{V}_{\mathrm{I}}(\tau_{2})\hat{V}_{\mathrm{I}}(\tau_{3})\hat{V}_{\mathrm{I}}(\tau_{4})|n\rangle.
\end{equation}
Inserting (\ref{H0}) and (\ref{pertua}) into (\ref{A8}) gives
\begin{equation}\label{A9}
\mathcal{Z}^{(4)}=\sum_{n=0}^{\infty}\mathrm{e}^{-\beta E_{n}}\int_{0}^{\beta}d\tau_{1}\int_{0}^{\tau_{1}}d\tau_{2}\int_{0}^{\tau_{2}}d\tau_{3}\int_{0}^{\tau_{3}}d\tau_{4}\mathrm{e}^{(\tau_{1}-\tau_{4})E_{n}}\langle n|\hat{V}\mathrm{e}^{-\tau_{1}\hat{H}_{0}}\hat{V}_{\mathrm{I}}(\tau_{2})\hat{V}_{\mathrm{I}}(\tau_{3})\mathrm{e}^{\tau_{4}\hat{H}_{0}}\hat{V}|n\rangle.
\end{equation}
According to (\ref{perturbation}), we have
\begin{equation}
\begin{aligned}\label{A10}
	\mathcal{Z}^{(4)}&=&J^{2}z^{2}\sum_{n=0}^{\infty}\mathrm{e}^{-\beta E_{n}}\int_{0}^{\beta}
	d\tau_{1}\int_{0}^{\tau_{1}}d\tau_{2}\int_{0}^{\tau_{2}}d\tau_{3}\int_{0}^{\tau_{3}}
	d\tau_{4}\mathrm{e}^{(\tau_{1}-\tau_{4})E_{n}}\left(\Psi\sqrt{n}\mathrm{e}^{(\tau_{2}-\tau_{1})E_{n-1}}\langle n-1|\right.\\
	&&\left. +\Psi^{*}\sqrt{n+1}\mathrm{e}^{(\tau_{2}-\tau_{1})E_{n+1}}\langle n+1|\right)\hat{V}
	\mathrm{e}^{-\tau_{2}\hat{H}_{0}}\mathrm{e}^{\tau_{3}\hat{H}_{0}}\hat{V}\left(\Psi^{*}\sqrt{n}
	\mathrm{e}^{(\tau_{4}-\tau_{3})E_{n-1}}|n-1\rangle \right.\\
	&&\left.+\Psi\sqrt{n+1}\mathrm{e}^{(\tau_{4}-\tau_{3})E_{n+1}}|n+1\rangle\right).
\end{aligned}
\end{equation}
Using again (\ref{H0}) and (\ref{pertua}) together with (\ref{A10}) results in
\begin{equation}
	\begin{aligned}
		\mathcal{Z}^{(4)}=&\,J^{4}z^{4}\sum_{n=0}^{\infty}\mathrm{e}^{-\beta E_{n}}\int_{0}^{\beta}
		d\tau_{1}\int_{0}^{\tau_{1}}d\tau_{2}\int_{0}^{\tau_{2}}d\tau_{3}\int_{0}^{\tau_{3}}
		d\tau_{4}\mathrm{e}^{(\tau_{1}-\tau_{4})E_{n}}\\
		&\times\left[\Psi\sqrt{n}\mathrm{e}^{(\tau_{2}-\tau_{1})E_{n-1}}\left(\Psi\sqrt{n-1}\mathrm{e}^{-\tau_{2}E_{n-2}}
		\langle n-2|+\Psi^{*}\sqrt{n}\mathrm{e}^{-\tau_{2}E_{n}}\langle n|\right)\right.\\
		&\left. +\Psi^{*}\sqrt{n+1}\mathrm{e}^{(\tau_{2}-\tau_{1})E_{n+1}}\left(\Psi\sqrt{n+1}\mathrm{e}^{-\tau_{2}E_{n}}\langle n|+\Psi^{*}\sqrt{n+2}
		\mathrm{e}^{-\tau_{2}E_{n+2}}\langle n+2|\right)\right] \\
		&\times\left[\Psi^{*}\sqrt{n}\mathrm{e}^{(\tau_{4}-\tau_{3})E_{n-1}}\left(\Psi^{*}\sqrt{n-1}\mathrm{e}^{\tau_{3}E_{n-2}}
		|n-2\rangle+\Psi\sqrt{n}\mathrm{e}^{\tau_{3}E_{n}}|n\rangle\right)\right. \\
		&\left. +\Psi\sqrt{n+1}\mathrm{e}^{(\tau_{4}-\tau_{3})E_{n+1}}\left(\Psi^{*}\sqrt{n+1}\mathrm{e}^{\tau_{3}E_{n}}|n\rangle+\Psi\sqrt{n+2}\mathrm{e}^{\tau_{3}E_{n+2}}|n+2\rangle\right)\right],
\end{aligned}
\end{equation}
which, after performing all the scalar products, reduces to
\begin{equation}
\begin{aligned}
	\mathcal{Z}^{(4)}=\,&J^{4}z^{4}|\Psi|^{4}\sum_{n=0}^{\infty}\mathrm{e}^{-\beta E_{n}}
	\int_{0}^{\beta}d\tau_{1}\int_{0}^{\tau_{1}}d\tau_{2}\int_{0}^{\tau_{2}}d\tau_{3}\int_{0}^{\tau_{3}}d\tau_{4} \\
	&\times\left[ n(n-1)\mathrm{e}^{(\tau_{1}-\tau_{4})\Delta_{n,n-1}}\mathrm{e}^{(\tau_{2}-\tau_{3})\Delta_{n-1,n-2}}
	+(n+1)(n+2)\mathrm{e}^{(\tau_{1}-\tau_{4})\Delta_{n,n+1}}\mathrm{e}^{(\tau_{2}-\tau_{3})\Delta_{n+1,n+2}}\right.\\
	&+n^{2}\mathrm{e}^{(\tau_{1}-\tau_{4})\Delta_{n,n-1}}\mathrm{e}^{(\tau_{2}-\tau_{3})\Delta_{n-1,n}}+n(n+1)
	\mathrm{e}^{(\tau_{1}-\tau_{2})\Delta_{n,n-1}}\mathrm{e}^{(\tau_{3}-\tau_{4})\Delta_{n,n+1}}\\
	&\left.+n(n+1)\mathrm{e}^{(\tau_{1}-\tau_{2})\Delta_{n,n+1}}\mathrm{e}^{(\tau_{3}-\tau_{4})\Delta_{n,n-1}}+(n+1)^{2}
	\mathrm{e}^{(\tau_{1}-\tau_{2})\Delta_{n,n+1}}\mathrm{e}^{(\tau_{3}-\tau_{4})\Delta_{n,n+1}}\right].
\end{aligned}
\end{equation}

Finally, the integrations result in
\begin{equation}
\begin{aligned}
	\mathcal{Z}^{(4)}= \, & J^{4}z^{4}|\Psi|^{4}\sum_{n=0}^{\infty}\mathrm{e}^{-\beta E_{n}}\left\{ n\left(n-1\right)
	\frac{\mathrm{e}^{\beta\Delta_{n,n-2}}-1}{\Delta_{n,n-1}\Delta_{n-1,n-2}\Delta_{n,n-2}}
	\left(\frac{1}{\Delta_{n-1,n-2}}-\frac{1}{\Delta_{n,n-2}}\right)\right.\\
	&\left. +n\left(n-1\right)\frac{\mathrm{e}^{\beta\Delta_{n,n-1}}-1}{\Delta_{n,n-1}^{2}
	\Delta_{n,n-2}}\left(\frac{1}{\Delta_{n,n-1}}+\frac{1}{\Delta_{n-1,n-2}}\right) \right. \\
	&+n\left(n-1\right)\frac{\mathrm{e}^{\beta\Delta_{n,n-1}}-1}{\Delta_{n,n-1}^{2}\Delta_{n-1,n-2}}
	\left(\frac{1}{\Delta_{n,n-1}}-\frac{1}{\Delta_{n-1,n-2}}\right) \\
	&-n\left(n-1\right)\frac{\beta}{\Delta_{n,n-1}^{2}}\left(\frac{\mathrm{e}^{\beta\Delta_{n,n-1}}}
	{\Delta_{n-1,n-2}}+\frac{1}{\Delta_{n,n-2}}\right)\\
	&+\left(n+1\right)\left(n+2\right)\frac{\mathrm{e}^{\beta\Delta_{n,n+2}}-1}{\Delta_{n,n+1}
	\Delta_{n+1,n+2}\Delta_{n,n+2}}\left(\frac{1}{\Delta_{n+1,n+2}}-\frac{1}{\Delta_{n,n+2}}\right)\\
	&+\left(n+1\right)\left(n+2\right)\frac{\mathrm{e}^{\beta\Delta_{n,n+1}}-1}{\Delta_{n,n+1}^{2}
	\Delta_{n,n+2}}\left(\frac{1}{\Delta_{n,n+1}}+\frac{1}{\Delta_{n+1,n+2}}\right)\\
	&+\left(n+1\right)\left(n+2\right)\frac{\mathrm{e}^{\beta\Delta_{n,n+1}}-1}{\Delta_{n,n+1}^{2}
	\Delta_{n+1,n+2}}\left(\frac{1}{\Delta_{n,n+1}}-\frac{1}{\Delta_{n+1,n+2}}\right) \\
	&-\left(n+1\right)\left(n+2\right)\frac{\beta}{\Delta_{n,n+1}^{2}}\left(\frac{\mathrm{e}^
	{\beta\Delta_{n,n+1}}}{\Delta_{n+1,n+2}}+\frac{1}{\Delta_{n,n+2}}\right)\\
	&+3n^{2}\frac{1-\mathrm{e}^{\beta\Delta_{n,n-1}}}{\Delta_{n,n-1}^{4}}+n^{2}\frac{\beta}
	{\Delta_{n,n-1}^{3}}\left(2+\mathrm{e}^{\beta\Delta_{n,n-1}}\right)+n^{2}\frac{\beta^{2}}{2\Delta_{n,n-1}^{2}}\\
	&+\frac{n\left(n+1\right)}{\Delta_{n,n+1}^{2}\Delta_{n-1,n+1}}\left(\frac{\mathrm{e}^{\beta\Delta_{n,n+1}}-1}
	{\Delta_{n,n+1}}+\frac{1-\mathrm{e}^{\beta\Delta_{n,n-1}}}{\Delta_{n,n-1}}\right) \\
	&+n\left(n+1\right)\frac{1-\mathrm{e}^{\beta\Delta_{n,n-1}}}{\Delta_{n,n-1}^{2}\Delta_{n,n+1}}
	\left(\frac{1}{\Delta_{n,n-1}}+\frac{1}{\Delta_{n,n+1}}\right)\\
	&+n\left(n+1\right)\frac{\beta}{\Delta_{n,n-1}\Delta_{n,n+1}}\left(\frac{1}{\Delta_{n,n-1}}
	+\frac{1}{\Delta_{n,n+1}}\right)+n\left(n+1\right)\frac{\beta^{2}}{2\Delta_{n,n-1}\Delta_{n,n+1}}\\
	&+\frac{n\left(n+1\right)}{\Delta_{n,n-1}^{2}\Delta_{n+1,n-1}}\left(\frac{\mathrm{e}^{\beta\Delta_{n,n-1}}-1}
	{\Delta_{n,n-1}}+\frac{1-\mathrm{e}^{\beta\Delta_{n,n+1}}}{\Delta_{n,n+1}}\right) \\
	&+n\left(n+1\right)\frac{1-\mathrm{e}^{\beta\Delta_{n,n+1}}}{\Delta_{n,n+1}^{2}\Delta_{n,n-1}}
	\left(\frac{1}{\Delta_{n,n+1}}+\frac{1}{\Delta_{n,n-1}}\right)\\
	&+n\left(n+1\right)\frac{\beta}{\Delta_{n,n+1}\Delta_{n,n-1}}
	\left(\frac{1}{\Delta_{n,n+1}}+\frac{1}{\Delta_{n,n-1}}\right)
	+n\left(n+1\right)\frac{\beta^{2}}{2\Delta_{n,n+1}\Delta_{n,n-1}}\\
	&\left.+3\left(n+1\right)^{2}\frac{1-\mathrm{e}^{\beta\Delta_{n,n+1}}}
	{\Delta_{n,n+1}^{4}}+\left(n+1\right)^{2}\frac{\beta}{\Delta_{n,n+1}^{3}}
	\left(2+\mathrm{e}^{\beta\Delta_{n,n+1}}\right)+\left(n+1\right)^{2}\frac{\beta^{2}}{2\Delta_{n,n+1}^{2}}\right\}.
\end{aligned}
\end{equation}

Now that we have an expression for the partition function, 
we then evaluate the free energy,
\begin{equation}\label{F}
\mathcal{F} = -\frac{1}{\beta}\ln \mathcal{Z} .
\end{equation}
By considering the natural logarithm expansion at $x=0$, $\ln(1+x)\approx x - x^2/2$, 
we get, up to fourth order in the hopping parameter, 
\begin{equation}\label{F4th}
\mathcal{F}\approx-\frac{1}{\beta} \left[\ln \mathcal{Z}^{(0)} + \frac{\mathcal{Z}^{(2)}}{\mathcal{Z}^{(0)}} + \frac{\mathcal{Z}^{(4)}}{\mathcal{Z}^{(0)}} -\frac{1}{2} \left( \frac{\mathcal{Z}^{(2)}}{\mathcal{Z}^{(0)}}\right)^2 \right].
\end{equation}

Therefore, by comparing (\ref{landauexpansion}) and (\ref{F4th}), we read off the Landau expansion coefficients:
\begin{subequations}
	\begin{align}
		a_0=&-\frac{1}{\beta}\ln \mathcal{Z}^{(0)} , \label{a01}
		\\
		a_2=&-\frac{1}{\beta} \frac{1}{|\Psi|^2} \frac{\mathcal{Z}^{(2)}}{\mathcal{Z}^{(0)}}, \label{a21}
		\\
		a_{4}=&-\frac{1}{\beta} \frac{1}{|\Psi|^4} \left[  \frac{\mathcal{Z}^{(4)}}{\mathcal{Z}^{(0)}} -\frac{1}{2} \left( \frac{\mathcal{Z}^{(2)}}{\mathcal{Z}^{(0)}}\right)^2 \right]. \label{a41}
	\end{align}
\end{subequations}

At zero temperature, we obtain results which are equivalent to RSPT. In particular, the Landau expansion coefficients reduce to:
\begin{subequations}
	\begin{align}
		\lim_{\beta\to\infty} a_0=&E_n - J z |\Psi|^2 \equiv E_n^{(0)}, \label{a0}
		\\
		\lim_{\beta\to\infty} a_2=&J z + (J z)^2 \left( \frac{n+1}{\Delta_{n,n+1}} + \frac{n}{\Delta_{n,n-1}}\right), \label{a2}
		\\
		\lim_{\beta\to\infty} a_{4}=\left(J z\right)^{4}&\left[\frac{n\left(n-1\right)}{\Delta_{n,n-1}^{2}\Delta_{n,n-2}}+\frac{\left(n+1\right)\left(n+2\right)}{\Delta_{n+1,n}^{2}\Delta_{n,n+2}}+\frac{n{}^{2}}{\Delta_{n-1,n}^{3}} \right.\nonumber \\
		&\left.+\frac{\left(n+1\right)^{2}}{\Delta_{n+1,n}^{3}}+\frac{n\left(n+1\right)}{\Delta_{n+1,n}^{2}\Delta_{n-1,n}}+\frac{n\left(n+1\right)}{\Delta_{n,n-1}^{2}\Delta_{n+1,n}}\right].\label{a4}
	\end{align}
\end{subequations}

\begin{figure}[h]
	\centering
	\includegraphics[width=.6\columnwidth]{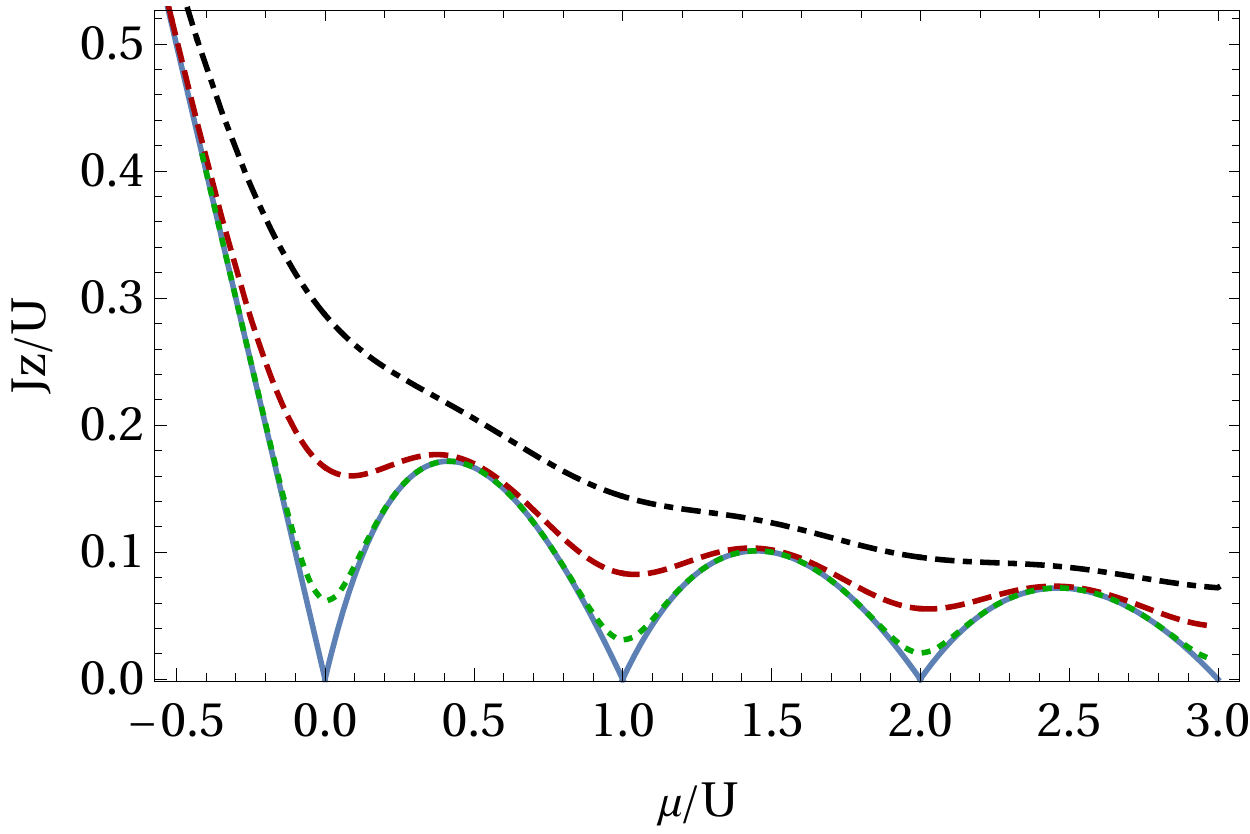}
	\caption{Phase diagrams for the inverse temperatures $\beta=5/U$ (black), $\beta=10/U$ (red), $\beta=30/U$ (green), and $\beta \rightarrow \infty$ (blue).}
	Source: SANT'ANA \textit{et al.}\cite{paper2}
	\label{pb_manyT}
\end{figure}

As previously discussed, the explicit solution of $a_2=0$ 
results in the phase boundaries between the superfluid and the Mott insulator. 
Such phase boundaries are depicted in Fig. \ref{pb_manyT} for four different temperatures.

\subsubsection{\label{inconsistencies}Nondegenerate perturbation theory inconsistencies}
As already pointed out, NDPT is expected to exhibit degeneracy-related problems. 
Indeed, by directly observing the coefficient denominators in (\ref{a2}) and (\ref{a4}), 
we clearly identify such a degeneracy problem. 
Whenever $\mu/U$ becomes an integer $n$, 
there is an equality between two consecutive energy values, for instance $E_n$ and $E_{n+1}$, thus characterizing a divergence in those expressions.

According to (\ref{landauexpansion}), we consider the Landau expansion up to fourth order for the free energy in the vicinity of a phase transition. Extremizing (\ref{landauexpansion}) with respect to the order parameter leads to
\begin{equation}\label{eq.dFdpsi}
\frac{\partial \mathcal{F}}{\partial |\Psi|^2} = a_2 + 2 a_4 |\Psi|^2 = 0,
\end{equation}
with the solution in the superfluid phase
\begin{equation}
\label{ordpara}
|\Psi|^2 = -\frac{a_2}{2 a_4}.
\end{equation}

Therefore, in order to explicitly show the degeneracy-related problems, we calculate the particle density,
\begin{equation}\label{pd}
n = -\frac{\partial \mathcal{F}}{\partial \mu},
\end{equation}
and the condensate density $|\Psi|^2$ via NDPT.

\begin{figure}[h] 
	\centering
	\begin{subfigure}{.45\columnwidth}
			\includegraphics[width=\columnwidth]{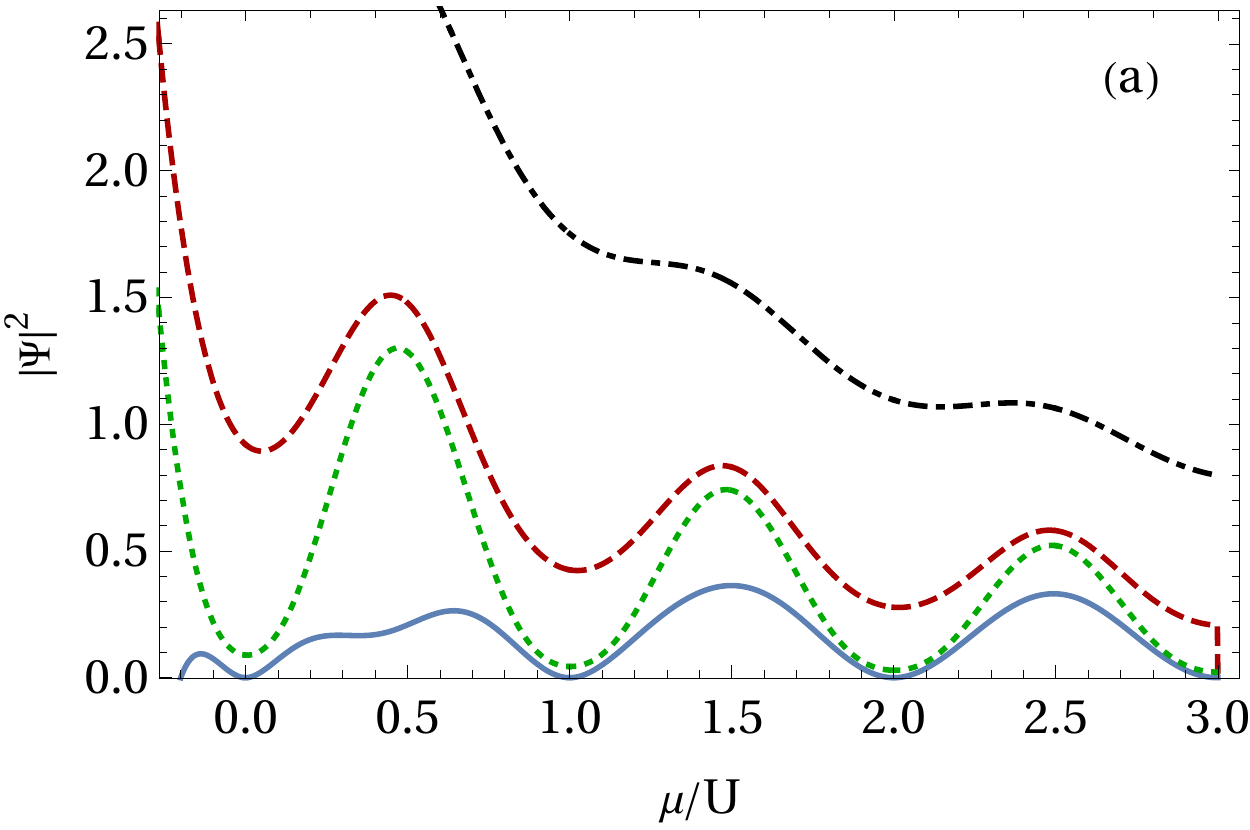}
		\label{order_parameter} \end{subfigure} \qquad
		\begin{subfigure}{.45\columnwidth}
	\includegraphics[width=\columnwidth]{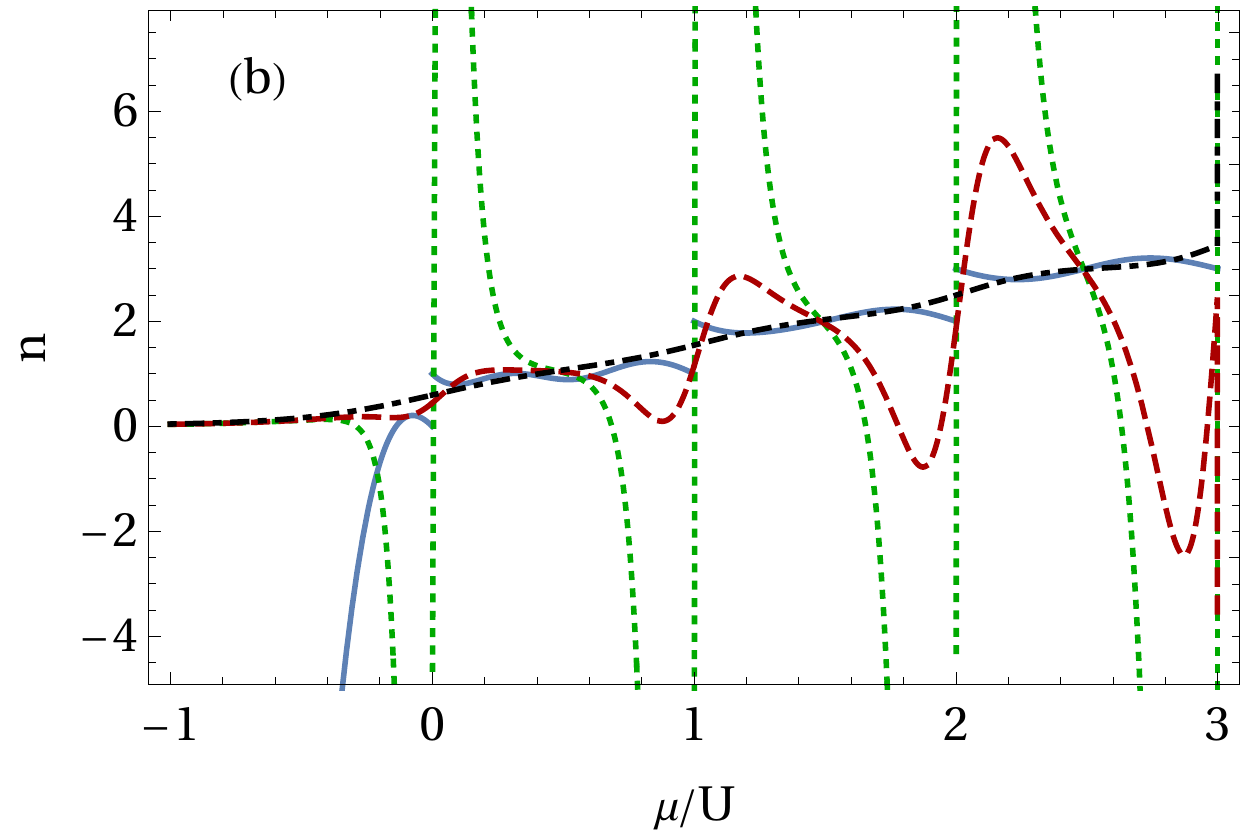}
	\label{fig.pd}
	\end{subfigure}
	\caption{Condensate density (a) from (\ref{ordpara}) and particle density (b) from (\ref{pd}) via NDPT as functions of $\mu/U$ for $Jz/U = 0.2$ as well as $\beta=5/U$ (dotted-dashed black), $\beta=10/U$ (dashed red), $\beta=30/U$ (dotted green), and $\beta \to \infty$ (continuous blue).}
	Source: SANT'ANA \textit{et al.}\cite{paper2}
	\label{nepsi}
\end{figure}

The plots of $|\Psi|^2$ and $n$ as functions of $\mu/U$ in Fig. \ref{nepsi} 
are interesting for our purposes since they reveal some nonphysical behaviors, which are consequences of NDPT: 
the order parameter approaches zero at a point where no phase transition occurs while the particle density 
shows strange behaviors, especially at the degeneracy points, presenting divergences
at $\mu/U \in \mathbb{N}$. Fig. \ref{nepsi} (a) shows equation (\ref{ordpara}) for $Jz/U=0.2$ for a varying chemical potential. 
We observe that, indeed, the OP is a well-behaved quantity in most parts of the diagram. However, it also shows an inconsistency: 
at integer values of $\mu/U$, the order parameter at the zero-temperature limit goes to zero, 
while for $T>0$ it mimics the zero-temperature behavior by decreasing its values but not vanishing.

Since, for finite temperatures, NDPT also shows the same nonphysical behavior typical of RSPT, 
in the following section we explore the method through which such problems can be fixed at finite temperature.

\subsection{Properties at zero temperature}
For large $Jz/U$, the system is in the superfluid phase, far away from the phase boundary, as the Mott insulator needs low hopping probabilities. 
Since all of our theory is based on the assumption of being close to the  phase boundary, 
we cannot obtain reliable results for values of $Jz/U$ deep in the superfluid phase. 
Nevertheless, for $Jz/U\lesssim 0.35$, we assume our model to be valid. 
While for $Jz/U=0$ we have no superfluid phase but only Mott insulator, 
it is possible to reach the superfluid phase by increasing $Jz/U$. 
Another way of changing the phase of the system from the Mott insulator to the superfluid phase is by tuning $\mu/U$ at $Jz/U>0$. 
If we start in the first Mott lobe and increase $\mu/U$, the ordered structure breaks down at some point 
and the superfluid phase is energetically more favorable and thus realized. 
For $\mu/U<0$, the system is in the superfluid phase for $Jz/U > - \mu/U$, whereas for $Jz/U < - \mu/U$ we have no particles at all, 
as depicted in Fig. \ref{pb_manyT}.

\begin{figure}[h]
\centering
      \includegraphics[width=.6\columnwidth]{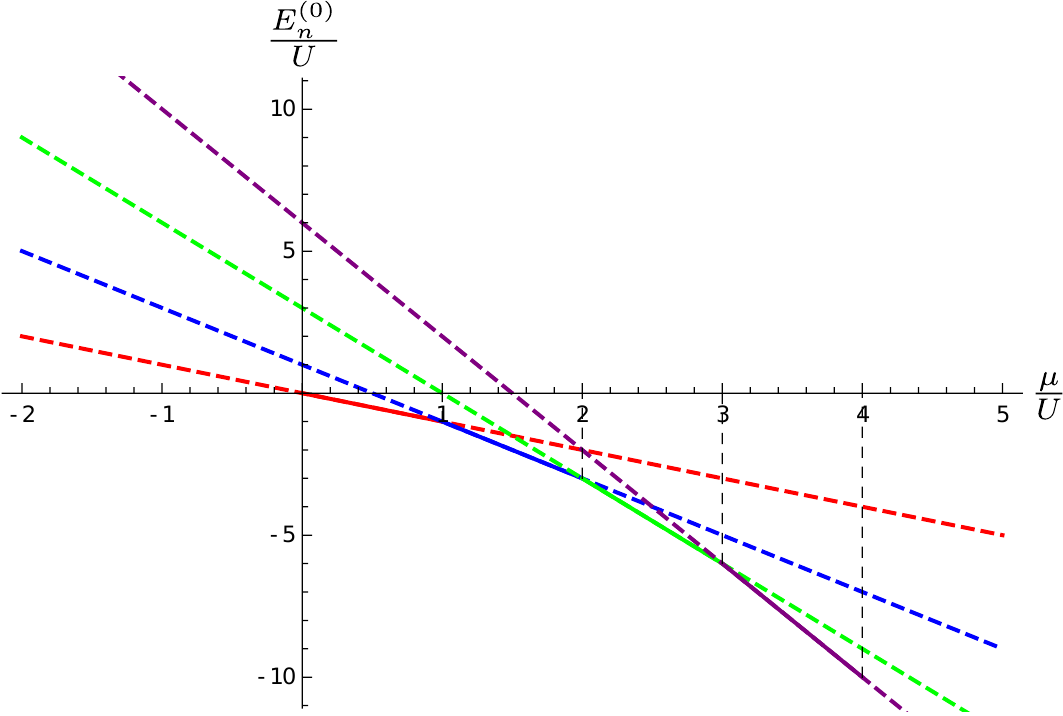}
	\caption{Unperturbed ground state energies $E_n^{(0)}=E_n-Jz|\Psi|^2$. Different lines correspond to different values for $n$ from smaller to larger slope: $n=1$ (red), $n=2$ (blue), $n=3$ (green), and $n=4$ (purple). Vertical dashed black lines correspond to the points of degeneracy. Solid colored lines represent realized lowest energies, while dashed colored lines indicate the continuation of the energy lines.}
	Source: K\"UBLER \textit{et al.}\cite{martin}
	\label{fig.Energies}
\end{figure}

After obtaining the phase boundary, 
we take a closer look at the ground state energies for increasing $n$. 
In the plot of the unperturbed energies from (\ref{a0}) 
in Fig. \ref{fig.Energies}, we see that the ground state energies have a degeneracy at integer values of $\mu/U$. 
As, for example, in between the lobes for $n=1$ (line with the smallest slope, red) 
and $n=2$ (line with the second smallest slope, blue) 
at $\mu/U=1$, we are at the degeneracy point where the energies $E_{1}^{(0)}$ and $E_{2}^{(0)}$ coincide. 
Analogous formulas are valid between 
every two neighboring lobes. 
Such degeneracies at $\mu = Un$ make any
algebraic treatment of the system quite complex.
However, since we always have only two degenerate energies to handle at a time, 
a solution to this problem can be found, as it will be shown further on.

With this degeneracy in mind, we now discuss the order parameter. 
Firstly, let us plot $\Psi^* \Psi = -a_2/2 a_4$ by using (\ref{a2}) and (\ref{a4}). 
Since $a_4$ approaches infinity at $\mu = U n$, where $E_{n}^{(0)}=E_{n+1}^{(0)}$,  
the condensate density $\Psi^* \Psi$ tends to zero 
at these points, 
which falsely indicates a phase boundary. This nonphysical behavior is depicted in Fig. \ref{fig.OPRS} through the dashed plots.

\subsection{A first degenerate correction}
One way to improve these results is to apply 
degenerate perturbation theory, 
which was done up to the first perturbative order in Ref. \citeonline{melo}.  
In the referred work, the corrected condensate density yields 
\begin{equation}\label{eq.2.92}
\Psi^* \Psi =\frac{ \left(n+1\right)}{4} - \frac{\left( \mu - Un \right)^2}{4 J^2 z^2  \left(n+1\right)}.
\end{equation}

\begin{figure}[h]
	\centering
\includegraphics[width=.6\textwidth]{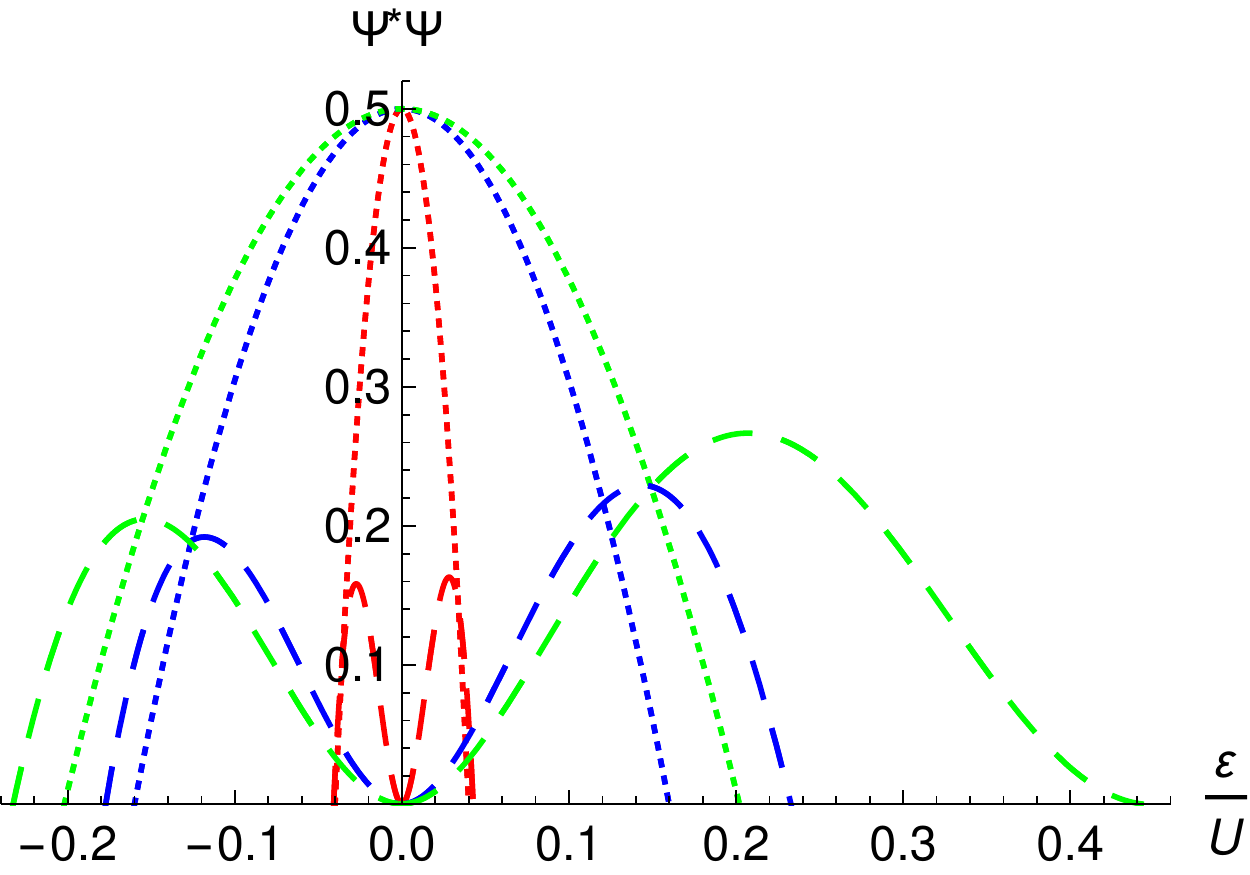}
	\caption{Condensate densities from nondegenerate perturbation theory (dashed lines) in comparison to the condensate densities 
	from degenerate perturbation theory according to (\ref{OPMelovarepsilon})\cite{melo} (dotted lines) with $\mu=Un+\varepsilon$ and $n=1$ for the left part (negative $\varepsilon/U$) 
	and $n=2$ for the right part (positive $\varepsilon/U$). The hopping strengths are, from the spacing inside to outside, 
	$Jz/U = 0.02$ (red), $Jz/U = 0.08$ (blue), and $Jz/U = 0.101$ (green). The dashed plots vanish at the mean-field phase boundary, 
	yielding a nonphysical behavior at the degeneracy. Also, they have increasing maxima for increasing $Jz/U$, and for $Jz/U = 0.101$ and $\varepsilon/U = 0.442$ 
	the lobe is just touching in one point and goes smoothly to zero. 
	The dotted plots provide a physical behavior at the degeneracies, 
	although they always present the value $\Psi^* \Psi =0.5$ for the condensate density at the degeneracies, 
	a fact that can be directly seen in (\ref{OPMelovarepsilon}). 
	For small $Jz/U$ and close to the phase boundary, the plots coincide.}
	Source:  K\"UBLER \textit{et al.}\cite{martin}
	\label{fig.OPRS}
\end{figure}

Let us now introduce the parameter $\varepsilon \equiv \mu -Un$ in order to analyze the system in the vicinity of the degeneracy, according to
\begin{equation}\label{OPMelovarepsilon}
\Psi ^* \Psi =\frac{ \left(n+1\right)}{4} - \frac{\varepsilon^2}{4 J^2 z^2  \left(n+1\right)}.
\end{equation}
The resulting condensate densities from (\ref{OPMelovarepsilon}) are depicted by the dotted curves in Fig. \ref{fig.OPRS}.

\begin{figure}[h]
	\centering
	\includegraphics[width=.6\columnwidth]{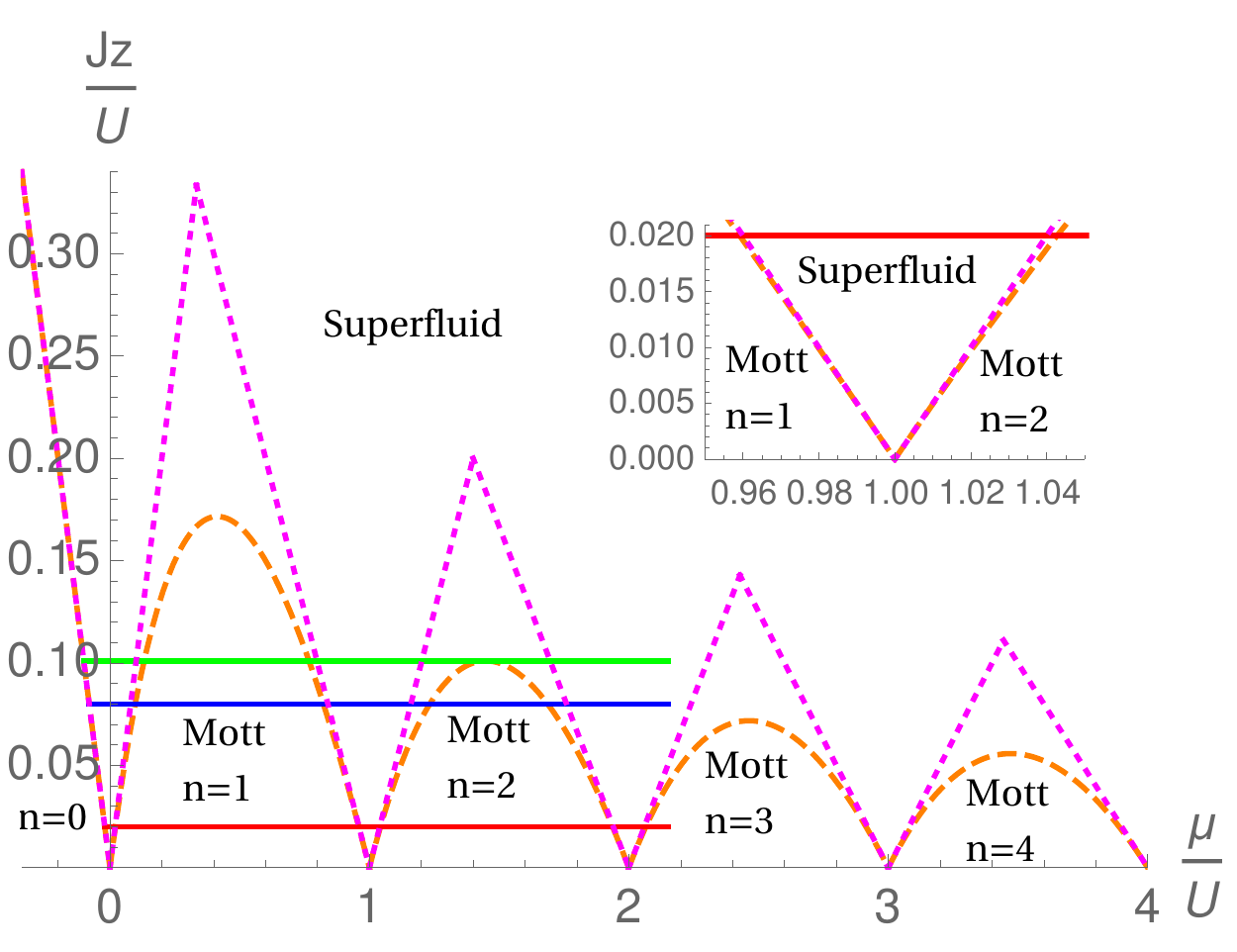}
	\caption{Zero-temperature phase boundaries for bosons in optical lattices from different treatments. 
	The nondegenerate theory \cite{thesisHoffmann} yields the dashed orange plot, while the degenerate one \cite{melo} results in the dotted magenta plot. 
	Inside the lobes the system is in the Mott insulator phase, while outside the lobes the superfluid phase takes place. 
	The number of particles per site, $n$, increases from left to right by one per lobe. 
	The three horizontal continuous lines correspond to, from bottom to top, $Jz/U=0.02$ (red), $Jz/U=0.08$ (blue), and $Jz/U=0.101$ (green). 
	They all start at the line $Jz/U=-\mu/U$, which indicates $n=0$, and end at $\mu/U=2.15$. 
	The inset shows the zoomed region between the first two Mott lobes.}
	Source: K\"UBLER \textit{et al.}\cite{martin}
	\label{fig.vieleLobes}
\end{figure}

By setting $\Psi^* \Psi =0$ in (\ref{eq.2.92}), 
we obtain the phase boundary, which is shown in Fig. \ref{fig.vieleLobes} by the dotted magenta curve. 
The phase boundary obtained out of the degenerate approach is linear in $\mu/U$, thus coinciding 
with the one from NDPT only at the vicinity of the points $\mu/U \in \mathbb{N}$. Nevertheless, for small values of $Jz/U$, 
it can be considered a good approximation (see inset in Fig. \ref{fig.vieleLobes}). 
The tips of the triangular Mott lobes (dotted magenta) correspond to $\mu/U=1/3 \approx 0.333$, 
$\mu/U=7/5=1.4$, $\mu/U=17/7\approx2.429$, and $\mu/U=31/9\approx 3.444$ for increasing $n$, 
which differ from the tips of the curved lobes (dashed orange), that correspond, respectively, to $\mu/U=\sqrt{2}-1 \approx 0.414$, 
$\mu/U=\sqrt{6}-1 \approx 1.449$, $\mu/U=2 \sqrt{3}-1 \approx 2.464$, 
and $\mu/U=2\sqrt{5}-1 \approx 3.472$. 
The horizontal continuous lines in Fig. \ref{fig.vieleLobes} correspond to, from bottom to top, $Jz/U=0.02$ (red), 
$Jz/U=0.08$ (blue), and $Jz/U=5-2\sqrt{6} \approx 0.101$ (green), 
while the latter one hits the second lobe exactly on its tip. 
These lines allow a better comparison between the dashed orange and the dotted magenta phase boundaries.

\chapter{The zero-temperature regime}
\label{BWPT}
This chapter is devoted to the development of the Brillouin-Wigner perturbation theory (BWPT) 
method to treat the degeneracy-related problems that artificially arise from NDPT for the bosonic 
lattice at zero temperature, \textit{e.g.}, 
the condensate density would vanish in a region of the quantum phase boundary where no transition 
occurs, which is a strong evidence of a nonphysical behavior generated by such an erroneous 
treatment, the NDPT. So, firstly we develop the fundamentals of the BWPT, which consists in establishing 
a Sch\"odinger-like equation with an effective Hamiltonian that can be perturbatively expanded up 
to the desired order, providing a valuable tool in order to correct the degeneracy-related problems. Then, 
we apply the BWPT in the cases where the degenerate Hilbert subspace is composed of both one and two states, 
namely one- and two-state approaches. After the evaluation for the condensate density in both treatments, we 
conclude that the one-state approach results in better results when compared to the NDPT procedure, but 
not fully overcoming the nonphysical results from the latter. Thence, we found ourselves the necessity of considering 
a degenerate Hilbert subspace composed of two states, that turns out to entirely correct the degeneracy-related 
problems from NDPT. In addition, we develop a graphical approach 
for the BWPT method that allows one to easily calculate higher-order terms 
in the perturbative expansion. Finally, we consider the effects of a harmonic trap in the equation of state of the system, 
\textit{i.e.}, how the number of particles changes with the chemical potential.

\section{Brillouin-Wigner perturbation theory}\label{sec:Appendix A: Derivation of Brillouin-Wigner}
We begin by providing a concise summary of the Brillouin-Wigner perturbation theory.\cite{bookBrillouinWigner} 
It amounts to derive an effective Hamiltonian for an arbitrarily chosen Hilbert subspace, 
which is characterized by a projection operator $\hat{\PP}$. 
To this purpose, we have to eliminate the complementary Hilbert subspace, which is spanned by the projection operator $\hat{\Q}$.

\subsection{General formalism}
Let us start by reformulating the time-independent Schr\"odinger equation, 
\begin{equation}\label{eq.2.10,5}
\hat{H}\ket{\Psi_n}=E_{n} \ket{\Psi_n},
\end{equation}
with the help of the projection operators.\cite{martin,bookBrillouinWigner} 
To this end, we insert the unity operator $\hat{\mathds{1}}=\hat{\PP}+\hat{\Q}$ 
on both sides of (\ref{eq.2.10,5}), yielding 
\begin{equation}\label{eq.2.13}
	\hat{H}\hat{\PP}\ket{\Psi_n} + \hat{H}\hat{\Q}\ket{\Psi_n} = E_{n}\hat{\PP}\ket{\Psi_n} + E_{n}\hat{\Q}\ket{\Psi_n}.
\end{equation}
Multiplying the left side of (\ref{eq.2.13}) by $\hat{\PP}$
and considering the relations $\hat{\PP}^2=\hat{\PP}$ and $\hat{\PP}\hat{\Q}=0$, we have that 
\begin{equation}\label{eq.2.14}
\hat{\PP}\hat{H}\hat{\PP}\ket{\Psi_n} + \hat{\PP}\hat{H}\hat{\Q}\ket{\Psi_n} = E_{n}\hat{\PP}\ket{\Psi_n}.
\end{equation}
Analogously, if we multiply the left side of (\ref{eq.2.13}) by $\hat{\Q}$ and make use of the 
corresponding relations $\hat{\Q}^2=\hat{\Q}$ and $\hat{\Q}\hat{\PP}=0$, we also have that 
\begin{equation}\label{eq.2.15}
\hat{\Q}\hat{H}\hat{\PP}\ket{\Psi_n} + \hat{\Q}\hat{H}\hat{\Q}\ket{\Psi_n} = E_{n}\hat{\Q}\ket{\Psi_n}.
\end{equation}

The next step consists in finding a single equation for $\hat{\PP}\ket{\Psi _n}$ 
in a similar shape to the time-independent Schr\"odinger-equation. 
So, in order to eliminate $\hat{\Q}\ket{\Psi_n}$ from (\ref{eq.2.14}), let us 
work out Eq. (\ref{eq.2.15}), 
\begin{equation}\label{eq.2.16}
\hat{\Q}\hat{H}\hat{\PP}\ket{\Psi_n} + \hat{\Q}\hat{H}\hat{\Q}^2\ket{\Psi_n} = E_{n}\hat{\Q}\ket{\Psi_n}.
\end{equation}
From rearranging and factoring out, it follows that 
\begin{equation}\label{eq.2.17}
\hat{\Q}\hat{H}\hat{\PP}\ket{\Psi_n} = \left(E_{n}-\hat{\Q}\hat{H}\hat{\Q}\right)\hat{\Q}\ket{\Psi_n}.
\end{equation}
Thus, a formal solution with respect to $\hat{\Q}\ket{\Psi_n}$ reads
\begin{equation}\label{eq.2.18}
\hat{\Q}\ket{\Psi_n} = \left(E_{n}-\hat{\Q}\hat{H}\hat{\Q}\right)^{-1}\hat{\Q}\hat{H}\hat{\PP}\ket{\Psi_n}.
\end{equation}
A further action of $\hat{\Q}$ results in
\begin{equation}\label{eq.2.19}
\hat{\Q}\ket{\Psi_n} = \hat{\Q}\left(E_{n}-\hat{\Q}\hat{H}\hat{\Q}\right)^{-1}\hat{\Q}\hat{H}\hat{\PP}\ket{\Psi_n}.
\end{equation}
Note that, at this point, we have succeeded in isolating the therm $\hat{\Q}\ket{\Psi_n}$. 
Inserting (\ref{eq.2.19}) into (\ref{eq.2.14}), we get a single equation for $\hat{\PP}\ket{\Psi_n}$:
\begin{equation}\label{eq.2.20}
\left[\hat{\PP}\hat{H}\hat{\PP}+\hat{\PP}\hat{H}
\hat{\Q}\left(E_{n}-\hat{\Q}\hat{H}\hat{\Q}\right)^{-1}
\hat{\Q}\hat{H}\hat{\PP} \right]\ket{\Psi_n} = E_{n}\hat{\PP}\ket{\Psi_n}.
\end{equation}
Splitting the Hamiltonian regarding the perturbation allows one to rewrite (\ref{eq.2.20}) as
\begin{equation}\label{eq.2.21}
\hat{\PP}\hat{H}\hat{\PP}\ket{\Psi_n}+\hat{\PP}\left(\hat{H}_0+\lambda \hat{V} \right)\hat{\Q}\left(E_{n}
	-\hat{\Q}\hat{H}\hat{\Q}\right)^{-1}\hat{\Q}\left(\hat{H}_0+\lambda \hat{V} \right)\hat{\PP}\ket{\Psi_n}
= E_{n}\hat{\PP}\ket{\Psi_n}.
\end{equation}
From the fact that $\hat{\Q}\hat{H}_0\hat{\PP}=0$, we finally obtain 
\begin{equation}\label{eq.2.23}
\hat{\PP} \left[\hat{H}
+\lambda \hat{V} \hat{\Q}\left(E_{n}-\hat{\Q}\hat{H}\hat{\Q}\right)^{-1}
	\hat{\Q}\lambda \hat{V}\right] \hat{\PP}\ket{\Psi_n}=E_{n}\hat{\PP}\ket{\Psi_n}.
\end{equation}
Equation (\ref{eq.2.23}) represents a single equation for $\hat{\PP} \ket{\Psi_{n}} $, 
which represents the basis of the Brillouin-Wigner perturbation theory.\cite{bookBrillouinWigner}

The resulting equation (\ref{eq.2.23}) has the form of a time-independent Schr\"odinger-equation
\begin{equation}\label{eq.2.24}
\hat{\PP} \hat{H}_{\eff}\hat{\PP}\ket{\Psi_n} = E_{n}\hat{\PP}\ket{\Psi_n},
\end{equation}
with the effective Hamiltonian defined as 
\begin{equation}\label{eq.2.25}
\hat{H}_{\eff} \equiv \hat{H}
+\lambda^2 \hat{V} \hat{\Q}\left(E_{n}-\hat{\Q}\hat{H}\hat{\Q}\right)^{-1}\hat{\Q} \hat{V} .
\end{equation}
Another way to represent $\hat{H}_{\eff}$ is
\begin{equation}\label{eq.2.27}
\hat{H}_{\eff}= \hat{H}_0 + \lambda\hat{V}+ \lambda^2 \hat{V} \hat{\Q}
  \left(E_{n} - \hat{\Q}\hat{H}_0\hat{\Q} - \lambda\hat{\Q}\hat{V}\hat{\Q} \right)^{-1}\hat{\Q} \hat{V}.
\end{equation}
The resolvent
\begin{equation}\label{eq.2.28}
\hat{R}(E_n)\equiv \left[E_{n} -\hat{\Q}\left(\hat{H}_0 + \lambda\hat{V}\right)\hat{\Q}\right]^{-1}
\end{equation}
can be written as a series expansion of $\lambda$ in the following way:  
\begin{equation}\label{eq.2.29}
\hat{R}(E_n)=\left(E_{n} -\hat{\Q}\hat{H}_0\hat{\Q}\right)^{-1}
	\sum_{s=0}^{\infty}\left[\lambda\hat{\Q}\hat{V}\hat{\Q}\left(E_{n} -\hat{\Q}\hat{H}_0\hat{\Q}\right)^{-1}\right]^s.
\end{equation}
Note the crucial property of (\ref{eq.2.29}): instead of the unperturbed energy eigenvalues $E_{n}^{(0)}$, 
it contains the full energy eigenvalues $E_{n}$.

Inserting (\ref{eq.2.28}) into (\ref{eq.2.27}), it follows that 
\begin{equation}\label{eq.2.30,01}
\hat{H}_{\eff}= \hat{H}_0 + \lambda\hat{V}
+ \lambda^2 \hat{V} \hat{\Q}\hat{R}(E_n)\hat{\Q} \hat{V}.
\end{equation}
As $\lambda$ approaches zero, it reproduces the unperturbed Schr\"odinger equation. 
The essential property of (\ref{eq.2.30,01}) is, however, that $E_n$ appears 
nonlinearly within the resolvent $\hat{R}(E_n)$, Eq. (\ref{eq.2.28}).

Note that the first perturbative order $\lambda\hat{V}$ in (\ref{eq.2.30,01}) corresponds to the original contribution of $\hat{H}$. 
In contrast, all higher orders in (\ref{eq.2.30,01}) originate from the resolvent term $\hat{R}(E_n)$. 
In particular, $s=0$ gives the second perturbative order, $s=1$ produces the third perturbative order, and so on. 
This fundamental difference in the origins of the perturbative orders was already evident in (\ref{eq.2.13}), 
where the term $\hat{H}\hat{\PP}$ gave rise to the zeroth and the first perturbative orders, 
while the term $\hat{H}\hat{\Q}$ gave rise to all higher orders. 
In other words, the zeroth and the first perturbative orders are contained within the Hilbert subspace $\PP$, 
while for all higher orders, the Hilbert subspace $\Q$ must be taken into account.

Now, let us calculate the correction terms of the effective Hamiltonian up to $\lambda ^4$. 
To do so, we evaluate the sum over $s$ in the resolvent formula (\ref{eq.2.29}) 
up to second order, $s=2$. This way, Eq. (\ref{eq.2.30,01}) reads 
\begin{equation}\label{eq.2.30,003}
	\begin{aligned}
\hat{H}_{\eff}=& \hat{H}_0+ \lambda\hat{V}+ \lambda^2 \hat{V} \hat{\Q}\hat{R}_0(E_n)\hat{\Q} \hat{V}
		+ \lambda^3 \hat{V} \hat{\Q}\hat{R}_0(E_n)
\hat{\Q}\hat{V}\hat{\Q}\hat{R}_0(E_n)\hat{\Q} \hat{V}
\\
&+ \lambda^4 \hat{V} \hat{\Q}\hat{R}_0(E_n)
\hat{\Q}\hat{V}\hat{\Q}\hat{R}_0(E_n)\hat{\Q}\hat{V}\hat{\Q}\hat{R}_0(E_n)\hat{\Q} \hat{V} + \cdots.
\end{aligned}
\end{equation}
Here, we have introduced the unperturbed Hamiltonian resolvent 
\begin{equation}\label{eq.Resolvent}
\hat{R}_0(E_n) \equiv \left(E_{n} -\hat{\Q}\hat{H}_0\hat{\Q}\right)^{-1}.
\end{equation}

Now, let us represent the projection operators as
$\hat{\PP}=\sum_{k\in \PP} \ket{\Psi_k^{(0)}}\bra{\Psi_k^{(0)}}$ 
and $\hat{\Q}=\sum_{l\in \Q} \ket{\Psi_l^{(0)}}\bra{\Psi_l^{(0)}}$. 
Using these relations, the matrix elements of the resolvent, Eq. (\ref{eq.Resolvent}), yield
\begin{equation}\label{eq.2.31,2}
\bra{\Psi_{l}^{(0)}}\hat{R}_0(E_n)\ket{\Psi_{l}^{(0)}} = 
	\frac{1}{E_n - E_l^{(0)}},
\end{equation}
where $l \in \Q$. 
Taking into account (\ref{eq.2.31,2}) within Eq. (\ref{eq.2.30,003}), we obtain
\begin{equation}\label{eq.2.31,003}
	\begin{aligned}
		\hat{H}_{\eff}=& \hat{H}_0 + \lambda\hat{V}+ \lambda^2 \sum_{l \in \Q}  
		\frac{\hat{V}\ket{\Psi_{l}^{(0)}}\bra{\Psi_{l}^{(0)}}\hat{V}}{E_n-E_l^{(0)}} 
		+ \lambda^3 \sum_{l,l' \in \Q}  \frac{\hat{V}\ket{\Psi_{l}^{(0)}}
		\bra{\Psi_{l}^{(0)}}\hat{V}\ket{\Psi_{l'}^{(0)}}\bra{\Psi_{l'}^{(0)}}\hat{V}}{\left(E_{n}-E_{l}^{(0)}\right)\left(E_{n}-E_{l'}^{(0)}\right)}
\\
		&+ \lambda^4 \sum_{l,l',l'' \in \Q} \frac{\hat{V}\ket{\Psi_{l}^{(0)}}\bra{\Psi_{l}^{(0)}}\hat{V}\ket{\Psi_{l'}^{(0)}}
\bra{\Psi_{l'}^{(0)}}\hat{V}\ket{\Psi_{l''}^{(0)}}
		\bra{\Psi_{l''}^{(0)}}\hat{V}}{\left(E_{n}-E_{l}^{(0)}\right)\left(E_{n}-E_{l'}^{(0)}\right)\left(E_{n}-E_{l''}^{(0)}\right)}+\cdots. 
\end{aligned}
\end{equation}
This representation of the effective Hamiltonian $\hat{H}_{\eff}$ possesses 
no operator in the denominators, 
hence it can be used as a starting point for further calculations.

Now, we want to determine an equation for the perturbed energies $E_n$. 
To this end, let us reformulate (\ref{eq.2.24}) according to 
\begin{equation}
	\sum_{k, k' \in \PP}\ket{\Psi_{k}^{(0)}}\bra{\Psi_{k}^{(0)}}
	\hat{H}_{\eff}\ket{\Psi_{k'}^{(0)}}
\braket{\Psi_{k'}^{(0)}}{\Psi_{n}}= E_n \sum_{ k' \in \PP}\ket{\Psi_{k'}^{(0)}}\braket{\Psi_{k'}^{(0)}}{\Psi_n}.
\end{equation}
Then, multiplying the left-hand side by $\bra{\Psi_{k}^{(0)}}$, 
\begin{equation}
\sum_{k,k' \in \PP}\bra{\Psi_{k}^{(0)}}\hat{H}_{\eff}\ket{\Psi_{k'}^{(0)}}
\braket{\Psi_{k'}^{(0)}}{\Psi_n}= E_n \sum_{k,k' \in \PP}
	\braket{\Psi_{k}^{(0)}}{\Psi_{k'}^{(0)}}\braket{\Psi_{k'}^{(0)}}{\Psi_n},
\end{equation}
yields 
\begin{equation}\label{eq.PsiSum}
\braket{\Psi_{k'}^{(0)}}{\Psi_n} 
	\sum_{k,k' \in \PP} 
	\left(\bra{\Psi_k^{(0)}}\hat{H}_{\eff}\ket{\Psi_{k'}^{(0)}}- E_n \delta_{k,k'} \right) = 0.
\end{equation}
In order to obtain a nontrivial solution of (\ref{eq.PsiSum}), $\braket{\Psi^{(0)}_{k'}}{\Psi_n} \neq 0$, we have to demand
\begin{equation}\label{eq.2.31,004}
\mathrm{det} \left(\bra{\Psi^{(0)}_{k}}\hat{H}_{\eff}\ket{\Psi^{(0)}_{k'}}-E_{n}\delta_{k,k'}\right)=0,
\end{equation}
where the determinant in (\ref{eq.2.31,004}) is performed with respect to $k, k' \in \PP$.

\subsection{\label{subonestate}Specific cases}
In the following, we specialize in the cases where 
the projection operator $\hat{\PP}$ consists of either one or two states.
\subsubsection{One-state approach}
Firstly, let us consider the special case where $\hat{\PP}$ contains only one state, namely
\begin{equation}\label{eq.2.30,04}
	\hat{\PP}\equiv\ket{\Psi_k^{(0)}}\bra{\Psi_k^{(0)}}.
\end{equation}
In this case, where $k=k'$, and considering that $k=n$, Eq. (\ref{eq.2.31,004}) simplifies to
\begin{equation}\label{eq.2.31,005}
E_n = \bra{\Psi^{(0)}_{n}}\hat{H}_{\eff}\ket{\Psi^{(0)}_{n}}.
\end{equation}
Inserting (\ref{eq.2.31,003}) into (\ref{eq.2.31,005}), we have that 
\begin{equation}
	\begin{aligned}
		E_n =&E^{(0)}_{n} + \lambda V_{n,n}+ \lambda^2 \sum_{l \neq n}  \frac{V_{n,l}V_{l,n}}{E_{n}-E_{l}^{(0)}} + \lambda^3 \sum_{l,l' \neq n}  \frac{V_{n,l}V_{l,l'}V_{l',n}}{\left(E_{n}-E_{l}^{(0)}\right)\left(E_{n}-E_{l'}^{(0)}\right)}
\\
&+ \lambda^4 \sum_{l,l',l'' \neq n} \frac{V_{n,l}V_{l,l'}V_{l',l''}V_{l'',n}}{\left(E_{n}-E_{l}^{(0)}\right)\left(E_{n}-E_{l'}^{(0)}\right)\left(E_{n}-E_{l''}^{(0)}\right)}+ \cdots,\label{eq.2.31,006}
\end{aligned}
\end{equation}
where we have taken into account that $\bra{\Psi^{(0)}_n} \hat{H}_0\ket{\Psi^{(0)}_n} = E^{(0)}_{n}$ 
and defined the matrix elements according to $V_{i,j} \equiv \bra{\Psi^{(0)}_{i}}\hat{V}\ket{\Psi^{(0)}_{j}}$.

Note that, due to the nonlinear appearance of $E_n$, (\ref{eq.2.31,006}) 
represents a self-consistency equation for the energies $E_n$. 
Furthermore, we observe that, up to third order, every power of $\lambda$ 
consists of one single term. 
Since $n \neq l,l',l''$, the denominator does not vanish in any situation, 
thus no divergence occurs in this perturbative representation of the perturbed energies $E_n$.

\subsubsection{Two-state approach}
Now, let us consider the case where the projection operator $\hat{\PP}$ is constituted of two states: 
\begin{equation}\label{eq.2.34,9}
	\hat{\PP}\equiv\ket{\Psi_k^{(0)}}\bra{\Psi_k^{(0)}} 
	+ \ket{\Psi_{k'}^{(0)}}\bra{\Psi_{k'}^{(0)}} .
\end{equation}
Thus, (\ref{eq.2.31,004}) yields
\begin{equation}\label{eq.2.37}
\begin{vmatrix}
H_{{\eff}, k, k} - E_n
& H_{{\eff}, k, k'}
\\
H_{{\eff}, k', k}
& H_{{\eff}, k', k'} - E_n
\end{vmatrix}
=
0,
\end{equation}
with the matrix elements defined as 
$H_{\eff,i,j} \equiv \bra{\Psi_i^{(0)}}\hat{H}_{\eff}\ket{\Psi_j^{(0)}}$.
Note that
\begin{equation}\label{eq.2.38}
\Gamma \equiv
\begin{pmatrix}
H_{{\eff}, k, k}
& H_{{\eff}, k,k'}
\\
H_{{\eff}, k', k}
&H_{{\eff}, k', k'}
\end{pmatrix}
\end{equation} 
represents a $2 \times 2$  matrix, 
since the projection operator $\hat \PP$ in (\ref{eq.2.34,9}) is composed of two states.

\section{\label{sec2}Degenerate solutions of the mean-field Bose-Hubbard Hamiltonian}
At the end of Chap. \ref{mf}, by comparing Fig. \ref{fig.vieleLobes} to Fig. \ref{fig.OPRS}, 
we have concluded that the nondegenerate approach (dashed lines) yields a reasonable quantum phase boundary, 
but an inconsistent condensate density, 
while the degenerate approach (dotted lines) yields an improved result for the order parameter, 
but a worse quantum phase boundary. 
Therefore, in order to handle both adequately, another approach is necessary. 
To this end, we stay in the perturbative picture, 
which already succeeded in reproducing the quantum phase boundary. So, in order to accurately calculate the order parameter,  
we will apply the Brillouin-Wigner perturbation theory developed in Sec. \ref{sec:Appendix A: Derivation of Brillouin-Wigner} 
in the following.

\subsection{One-state approach}\label{One-State Approach}
At first, we tackle our problem within the one-state approach 
of the Brillouin-Wigner perturbation theory as specified in Sec. \ref{subonestate}. 
To this end we consider a subspace of the Hilbert space spanned by only one eigenstate 
of the Hamiltonian (\ref{H0}), $\ket{n}$. Hence, its projection operator reads 
\begin{equation}\label{eq.6.1}
\hat{\PP}=\ket{n} \bra{n}.
\end{equation}
The ground state energy is then identified as 
$E_{n}=\bra{\Psi_{n}^{(0)}}\hat{H}_{\eff}\ket{\Psi_{n}^{(0)}}$. 
From (\ref{eq.2.31,006}), up to third order in $\lambda$, we have that 
\begin{equation}
\begin{aligned}\label{eq.8.11}
	E_{n}=&E_{n}^{(0)} + \lambda J z \Psi^{*}\Psi + \lambda ^2 J^2 z^2 \Psi^{*}\Psi \left(  \frac{n}{E_{n} - E_{n-1}^{(0)}}\right. \left. + \frac{n+1}{E_{n} - E_{n+1}^{(0)}}\right) \\
& +\lambda ^3 J^3 z^3 \left(\Psi^{*}\Psi\right)^2 \left[  \frac{n}{\left(E_{n} - E_{n-1}^{(0)}\right)^2} + \frac{n+1}{\left(E_{n} - E_{n+1}^{(0)}\right)^2}\right].
\end{aligned}
\end{equation}
As already pointed out, let us emphasize that (\ref{eq.8.11}) represents a self-consistency equation for $E_n$.

\subsubsection{Quantum phase boundary}
Now, we work out the mean-field quantum phase boundary 
within the one-state approach of the Brillouin-Wigner perturbation theory. To this end, 
we evaluate $\partial E_n\left(\Psi^*\Psi\right)/\left(\Psi \partial\Psi^*\right)$, 
with $E_n$ being the energy formula from the one-state approach up to third order in $\lambda$ according to (\ref{eq.8.11}).

We proceed by showing, in a general manner, that we can neglect all 
terms with power higher than three in $\lambda$. 
To such a purpose, we can write down a generic structure of $E_n\left(E_n,\Psi^*\Psi\right)$ from Eq. (\ref{eq.8.11}):
\begin{equation}\label{eq.(1.18)}
E_n\left(E_n,\Psi^*\Psi\right) =  
	\alpha + \Psi ^* \Psi \beta + \frac{\Psi ^* \Psi \gamma_0}{\gamma_1 + \Psi ^* \Psi \gamma_2} 
	+ \sum_{m=2}^{\infty}\frac{ \left( \Psi ^* \Psi \right)^m k_m}{P\left(E_n, \Psi ^* \Psi \right)}.
\end{equation}
The coefficients $\alpha$, $\beta$, $\gamma_0$, $\gamma_1$, $\gamma_2$, and $k_m$ are independent of $\Psi^* \Psi$, but they depend on $E_n$, 
and $P(E_n,\Psi^* \Psi)$ is a polynomial. Performing the differentiation of (\ref{eq.(1.18)}), we have that 
\begin{equation}
\begin{aligned}\label{eq.25}
\frac{1}{\Psi}\frac{\partial E_n\left(E_n,\Psi^*\Psi\right)}{\partial\Psi^*}=& \beta +  \frac{\gamma_0 \gamma_1}{\left( \gamma_1 + \Psi ^* \Psi \gamma_2 \right)^2} \\
&+\sum_{m=2}^{\infty} \left( \frac{m\left( \Psi ^* \Psi \right)^{m-1}k_m P }{P^2} \right.
\left. - \frac{ \left( \Psi ^* \Psi \right)^{m}k_m }{P^2}  \frac{1}{\Psi} \frac{\partial P}{\partial \Psi ^*} \right).
\end{aligned}
\end{equation}
Therefore, the quantum phase boundary yields 
\begin{equation}\label{eq.(1.19)}
\frac{1}{\Psi}\frac{\partial E_n\left(E_n,\Psi^*\Psi\right)}
{\partial\Psi^*}\Bigg|_{\Psi ^* \Psi=0} = \beta + \frac{\gamma_0}{\gamma_1}=0.
\end{equation}
Here, we see that all corrections from higher-order terms, $\lambda>2$, can be neglected. 
Consequently, the phase boundary does not change even if higher orders in $\lambda$ are taken into account. 

Comparing (\ref{eq.(1.19)}) to the derivative of (\ref{eq.8.11}), we identify the relevant coefficients: 
\begin{subequations}
\begin{align}
\beta=&\lambda J z,\\
\gamma_0=&\lambda ^2 J^2  z^2 \left[(2n+1)E_{n}+(n-1)E_{n-1}^{(0)} -n E_{n+1}^{(0)} \right],\\
\gamma_1=&\left(E_{n}-E_{n+1}^{(0)}\right) \left(E_{n}-E_{n-1}^{(0)}\right)\,.
\end{align}
\end{subequations}
Inserting them into (\ref{eq.(1.19)}), we obtain
\begin{equation}
	\frac{1}{\Psi}\frac{\partial E_n \left(\Psi^*\Psi\right)}{\partial\Psi^*}\Bigg|_{\Psi ^* \Psi=0}  = \lambda J z + \lambda ^2 z^2 \frac{E_{n}-E_{n-1}^{(0)} + 2 n E_{n}-n E_{n+1}^{(0)} + n E_{n-1}^{(0)}}{\left(E_{n}-E_{n+1}^{(0)}\right) \left(E_{n}-E_{n-1}^{(0)}\right)}.\label{eq29}
\end{equation}
Thus, from $(\ref{eq29})=0$, we achieve the mean-field quantum phase boundary condition 
\begin{equation}\label{form.6.7}
\frac{Jz}{U}=-\frac{1}{\lambda  U}\frac{\left(E_n-E_{n+1}^{(0)}\right) \left(E_n-E_{n-1}^{(0)}\right)}{E_n-E_{n-1}^{(0)} + 2n E_n -nE_{n+1}^{(0)}-nE_{n-1}^{(0)}},
\end{equation}
which turns out to be identical to the one obtained from RSPT in Fig. \ref{fig.vieleLobes} (dashed orange line).

\subsubsection{Energy and condensate density}
In order to calculate the energy and the condensate density 
within the one-state approach, we make use 
of $\partial E_n/\left(\Psi \partial \Psi^{*}\right)=0$ 
from (\ref{eq.8.11}), 
\begin{equation}
\begin{aligned}\label{eq.11.15}
0=&\left(E_{n} - E_{n-1}^{(0)}\right)^2 \left(E_{n} - E_{n+1}^{(0)}\right)^2 +\lambda J z \left[  n\left(E_{n} - E_{n-1}^{(0)}\right)  \right. \left(E_{n} - E_{n+1}^{(0)}\right)^2 \\
&+ \left. \left(n+1\right)\left(E_{n} - E_{n-1}^{(0)}\right)^2 \left(E_{n} - E_{n+1}^{(0)}\right)\right]
\\
& +2\lambda ^2 J^2 z^2 \Psi^{*}\Psi \left[  n\left(E_{n} - E_{n+1}^{(0)}\right)^2 \right. \left. +\left(n+1\right)\left(E_{n} - E_{n-1}^{(0)}\right)^2\right],
\end{aligned}
\end{equation}
and Eq. (\ref{eq.8.11}) itself up to second order in $\lambda$,
\begin{equation}
\begin{aligned}\label{eq.11.16}
0=&\left(E_{n} - E_{n-1}^{(0)}\right) \left(E_{n} - E_{n+1}^{(0)}\right) \left(E_{n}^{(0)}-E_{n} + \lambda J z \Psi^{*}\Psi \right)\\
& + \lambda ^2 J^2 z^2 \Psi^{*}\Psi \left[  n \left(E_{n} - E_{n+1}^{(0)}\right) \right. \left. +\left(n+1\right)\left(E_{n} - E_{n-1}^{(0)}\right) \right].
\end{aligned}
\end{equation}
Both (\ref{eq.11.15}) and (\ref{eq.11.16}) are now used to calculate the ground state energy $E_n$ and the condensate density $\Psi^* \Psi$.

The corrections on the energy are obtained by subtracting the unperturbed energy from the perturbed energy. 
From zeroth to second order, the corrections amount to $+1.08\%$. 
From second to fourth order, the corrections are $-0.05\%$. 
Furthermore, from fourth to sixth order, the corrections add up to $-0.18\%$. 
Note that for higher values of $Jz/U$ the convergence turns out to be slower.

\begin{figure}[h!]
	\centering
\includegraphics[width=.6\textwidth]{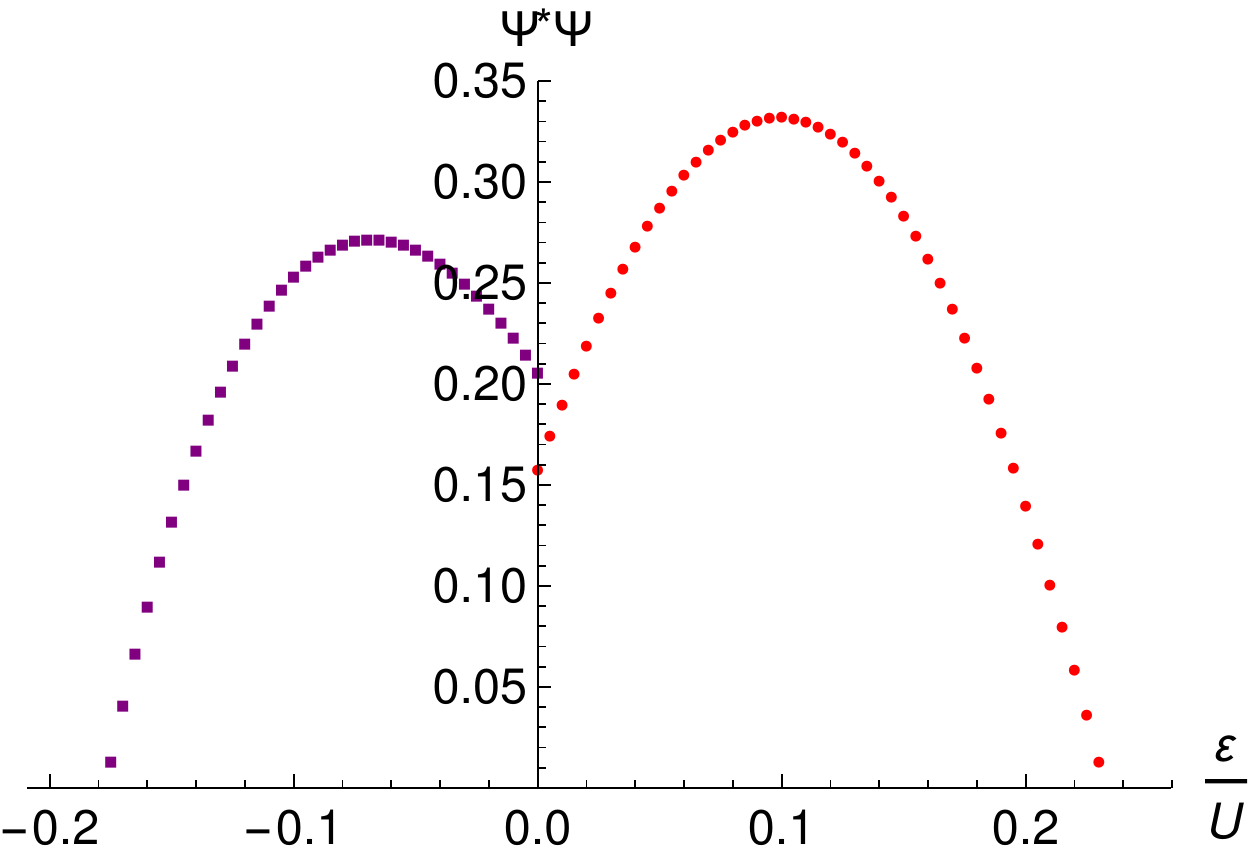}
\caption{Condensate density from the one-state approach 
	for $n=1$ (negative $\varepsilon/U$, purple squares) and $n=2$ (positive $\varepsilon/U$, red circles), with the hopping 
	strength of $Jz/U = 0.08$.}
	Source: K\"UBLER \textit{et al.}\cite{martin} 
	\label{fig.(11.2)}
\end{figure}

The condensate density $\Psi^* \Psi$ follows from iteratively solving both (\ref{eq.11.15}) and (\ref{eq.11.16}). 
The results are plotted in Fig. \ref{fig.(11.2)} for $\mu = Un + \varepsilon$, $\lambda=1$, and $Jz/U=0.08$. 
We observe that the order parameter obtained from the Brillouin-Wigner perturbation theory for the one-state approach 
according to Fig. \ref{fig.(11.2)} is better than the one obtained from Rayleigh-Schr\"odinger perturbation theory, 
where the order parameter vanishes at the degeneracy, as in Fig. \ref{fig.OPRS}. 
Nevertheless, the order parameter depicted in Fig. \ref{fig.(11.2)} is discontinuous at $\varepsilon/U=0$. 
Therefore, we conclude that it does not represent a physically acceptable result.

\subsection{Two-state approach}\label{Two-States Approach}
In the following, we will consider the degenerate Hilbert subspace composed of two states, $\ket{\Psi_{n}^{(0)}}$ and $\ket{\Psi_{n+1}^{(0)}}$. 
This choice is motivated due to the degeneracy present between two consecutive Mott lobes in the zero-temperature phase diagram of the Bose-Hubbard model. 
Any state vector is projected into the referred subspace through the projection operator
\begin{equation}
\hat{\PP}=\ket{n} \bra{n} + \ket{n+1} \bra{n+1}.
\end{equation}

\subsubsection{Quantum phase boundary}
In order to calculate the mean-field quantum phase boundary via the two-state approach, 
we start by evaluating the entries of the matrix (\ref{eq.2.38}). Up to fourth order, they read
\begin{subequations}\label{eq.11.18}
\begin{align}
\Gamma_{1,1} =& E^{(0)}_{n}+\lambda J z \Psi^* \Psi +\lambda^2 \frac{J^2 z^2 \Psi^{*} \Psi n}{E_{n} - E^{(0)}_{n-1}-\lambda J z \Psi^* \Psi}\\
	&+ \lambda^4 \frac{J^4 z^4 (\Psi^* \Psi)^2 n \left( n-1 \right)}{\left( E_{n} - E^{(0)}_{n-1}-\lambda Jz\Psi^*\Psi \right)^2 \left(  E_{n} - E^{(0)}_{n-2}-\lambda Jz\Psi^*\Psi \right)},\\
	\Gamma_{1,2}=&\,\Gamma_{2,1}^*= -\lambda J z \Psi^{*} \sqrt{n+1}, \\
\Gamma_{2,2}=& E^{(0)}_{n+1}+\lambda J z \Psi^* \Psi +\lambda^2 \frac{J^2 z^2 \Psi^{*} \Psi \left( n+2 \right)}{E_{n} - E^{(0)}_{n+2}-\lambda J z \Psi^* \Psi}\\
	&+\lambda^4 \frac{J^4 z^4 (\Psi^* \Psi)^2 \left( n+2 \right) \left( n+3 \right)}{ \left(E_{n} - E^{(0)}_{n+2}-\lambda Jz\Psi^*\Psi \right)^2 \left(  E_{n} - E^{(0)}_{n+3}-\lambda Jz\Psi^*\Psi \right) }.
\end{align}
\end{subequations}

To calculate the phase boundary, we perform 
	\begin{equation}
	\begin{aligned}\label{eq.(1.22)}
		\frac{1}{\Psi}\frac{\partial \left|\Gamma-\mathbb{I}E_n\right|}{\partial\Psi^*} \Bigg|_{\Psi^* \Psi=0}=& \lambda Jz\left[\left( E_{n}^{(0)} -E_n\right)+\left(E_{n+1}^{(0)} -E_n \right)
		-\lambda Jz\left(n+1\right)\right]\\
	&+\lambda^2 J^2 z^2\left[\frac{\left(n+2\right)\left(E_{n}^{(0)} -E_n\right)}{E_n - E_{n+2}^{(0)}}+\frac{n\left(E_{n+1}^{(0)} -E_n\right)}{E_n - E_{n-1}^{(0)}}\right] = 0,
	\end{aligned}
	\end{equation}
resulting in
	\begin{equation}\label{1form.6.7}
	\frac{Jz}{U}=\frac{- \left( 2 E_{n} - E_{n}^{(0)}- E_{n+1}^{(0)} \right) \left( E_{n} - E_{n+2}^{(0)} \right) \left( E_{n} - E_{n-1}^{(0)} \right)}
	{ \lambda n U \left( E_{n} - E_{n+1}^{(0)} \right) \left( E_{n} - E_{n+2}^{(0)} \right) + \lambda U\left[ \left(n+1\right) \left( E_{n} - E_{n+2}^{(0)} \right) 
		+ \left(n+2\right) \left( E_{n} - E_{n}^{(0)} \right)\right]},
	\end{equation}
which is the mean-field phase boundary. All higher-order corrections drop out of the formula if we set $\Psi^* \Psi =0$.
Thus, the phase boundary does not change even if higher orders in $\lambda$ are taken into account. 
To determine the perturbed energies $E_n$, we calculate the determinant of $\Gamma - \mathbb{I}E_n$ and set $\Psi^* \Psi =0$, 
which is effectively equivalent to calculate the matrix up to zeroth order. 
Hence, the roots of 
\begin{equation}
	\mathrm{det}\left(\Gamma-\mathbb{I}E_n\right)=\left( E_{n}^{(0)} - E_n \right)\left( E_{n+1}^{(0)} - E_n \right)=0
\end{equation}
are given by $E_n=E_n^{(0)}$ and $E_n=E_{n+1}^{(0)}$. 
Consequently, the mean-field phase boundary (\ref{1form.6.7}) calculated with $\lambda=1$ agrees with the previous result from Eq. (\ref{form.6.7}). 
By using the explicit forms of the unperturbed energies (\ref{a0}) together with $\mu = U +\varepsilon$, the two possible solutions  
for the perturbed energies are given by  
\begin{equation}\label{eq.(1.26)}
	\frac{E_1}{U}=E_1^{(0)} 
	=-\left(1+\frac{\varepsilon}{U}\right),
\end{equation}
and 
\begin{equation}\label{eq.(1.27)}
        \frac{E_1}{U}=E_2^{(0)}
        =-\left(1+\frac{2 \varepsilon}{U}\right).
\end{equation}
These two energies are depicted in Fig. \ref{fig.Energies} and they yield the corresponding lowest energies 
within the first and the second Mott lobes, \textit{i.e.}, for $-1<\varepsilon/U<0$, $E_1$ is the minimal energy, 
while for $0<\varepsilon/U<1$, $E_2$ turns into the lowest one. 
Therefore, in order to evaluate the phase boundary, we insert (\ref{eq.(1.26)}) and (\ref{eq.(1.27)}) into (\ref{1form.6.7}). 
According to Fig. \ref{fig.Energies}, we conclude that $E_{1}$ gives rise to the first lobe, while $E_{2}$ gives rise to the second one, 
originating the Mott-lobe structure from Fig. \ref{fig.vieleLobes}.

\subsubsection{Energy and particle density}
\begin{figure}[h!]
   \centering
   \begin{subfigure}{0.47\columnwidth}
      \includegraphics[width=\columnwidth]{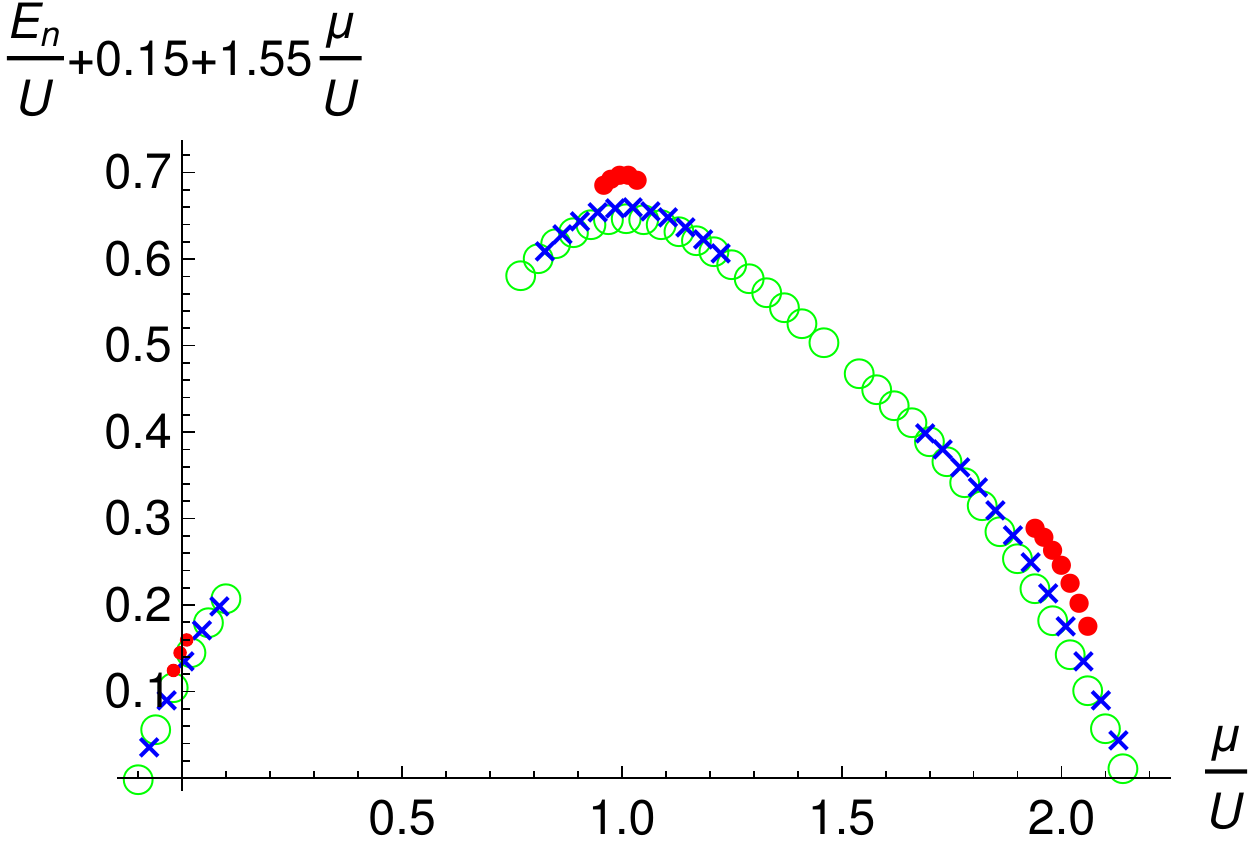}
        \caption{}
        \label{fig:1.1a}
   \end{subfigure}
   \qquad
  \begin{subfigure}{0.47\columnwidth}
       \includegraphics[width=\columnwidth]{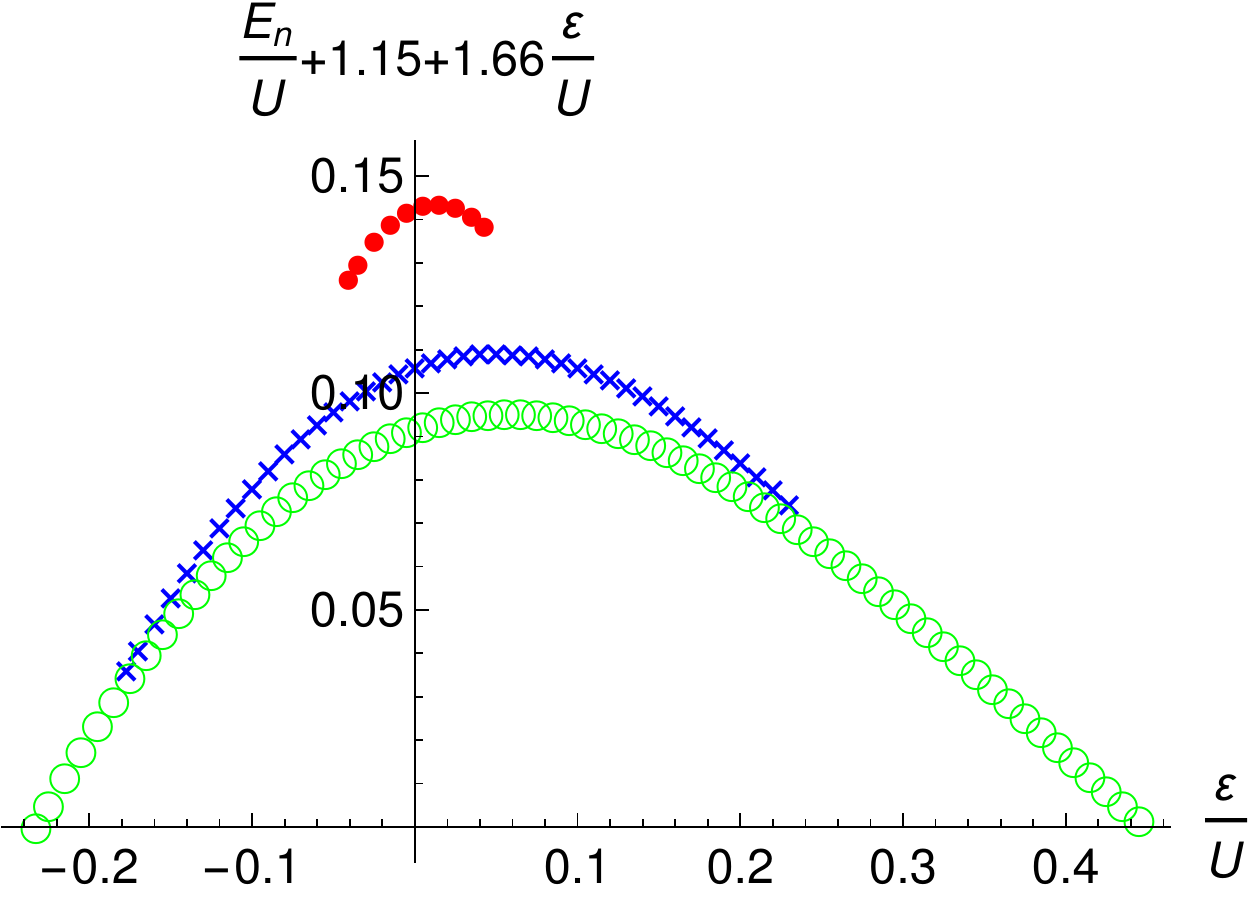}
       \caption{}
        \label{fig:1.1b}
    \end{subfigure}
	\caption{Perturbed ground state energies $E_n/U$ up to $\mathcal{O}\left(\lambda^4\right)$ between the Mott lobes, 
	inside the superfluid regions, for three different hopping values: 
	$Jz/U=0.02$ (red circles), $Jz/U=0.08$ (blue crosses), and $Jz/U=5-2\sqrt{6} \approx 0.101$ (green rings). 
	At $Jz/U=5-2\sqrt{6}$, the second lobe achieves its tip. (a) Superfluid energies between the first two Mott lobes. 
	For a better visualization, the linear equation $0.15+1.55 \mu/U$, which scales the outmost points of the green plot to zero, 
	is added to the energy. 
	(b) Zoomed region centered around the degeneracy by introducing $\mu=U+\varepsilon$. 
	For a better visualization, the linear equation $1.15+1.66 \varepsilon/U$, which scales the outmost points of the green plot to zero, is added to the energy.}
	Source: K\"UBLER \textit{et al.}\cite{martin} 
  \label{Energies}
\end{figure}

Now, we proceed to numerically calculating the perturbed ground state energies $E_{n}$ from the two conditions 
\begin{subequations}
\begin{align}
  \mathrm{det}\left(\Gamma-\mathbb{I}E_n\right)&= 0 ,\label{subeq1}\\
   \frac{1}{\Psi}\frac{\partial \left|\Gamma-\mathbb{I}E_n\right|}{\partial \Psi^*}&=0,\label{subeq2}
\end{align}
\end{subequations}
with the $\Gamma$ entries given by (\ref{eq.11.18}).
The perturbed ground state energy $E_n$ is then determined by iteratively 
solving both (\ref{subeq1}) and (\ref{subeq2}), 
resulting in Fig. \ref{Energies}, where the ground state energy $E_n$ is 
depicted as a function of the chemical potential. The calculation corresponds to $\lambda=1$. 
Note that, as we are evaluating the superfluid energy, the missing data corresponds to Mott insulating regions.

Now, we proceed to calculating the particle densities regarding both the superfluid and the Mott insulator. 
As we have already explained, a fundamental feature of the Mott insulator phase is its integer occupation number, \textit{i.e.}, 
the particle densities for the Mott insulator regions are $n=1$ within the first lobe, $n=2$ within the second one, and so on. 
Within the superfluid regions, the particle densities must be evaluated from the previously calculated energies via $-\partial E_n/\partial \mu$.
Therefore, by doing so, we achieve the results depicted in Fig. \ref{fig:002,008,0101PD} for two different hopping values. 
We can observe the effect that the increase of the hopping produces on the curves, they become smoother. 
Such an outcome has a clear interpretation: the increase of the hopping has a direct consequence on the single-particle energies, also increasing them, 
thus boosting the probabilities of the particles to hop from one site to a neighboring one. Consequently, 
the on-site characteristic occupation number changes from an integer value to a real one.

\begin{figure}[h!]
	\centering
	\begin{subfigure}{0.47\textwidth}
		\includegraphics[width=\textwidth]{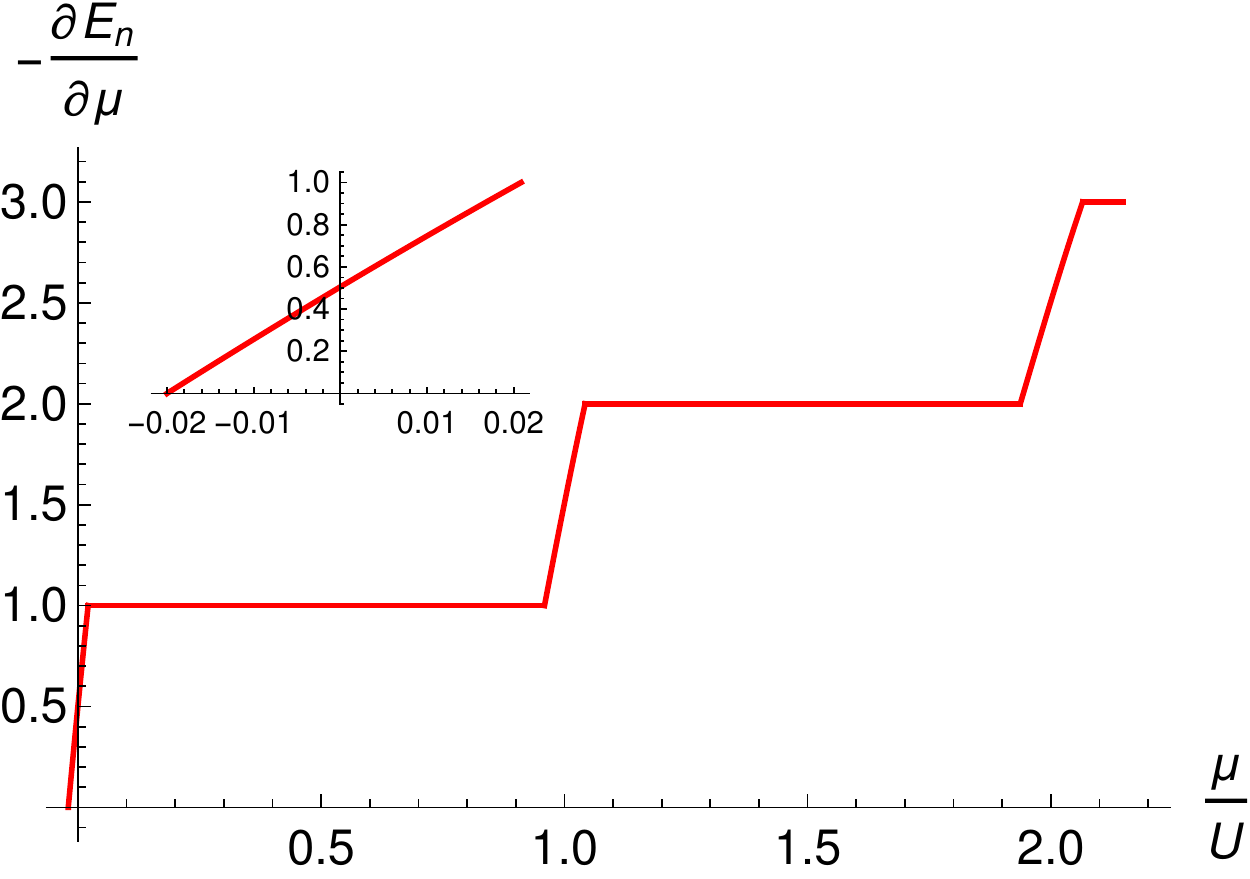}
		\caption{$Jz/U = 0.02$.}
		\label{fig:002a}
	\end{subfigure}
    \qquad 
	\begin{subfigure}{0.47\textwidth}
		\includegraphics[width=\textwidth]{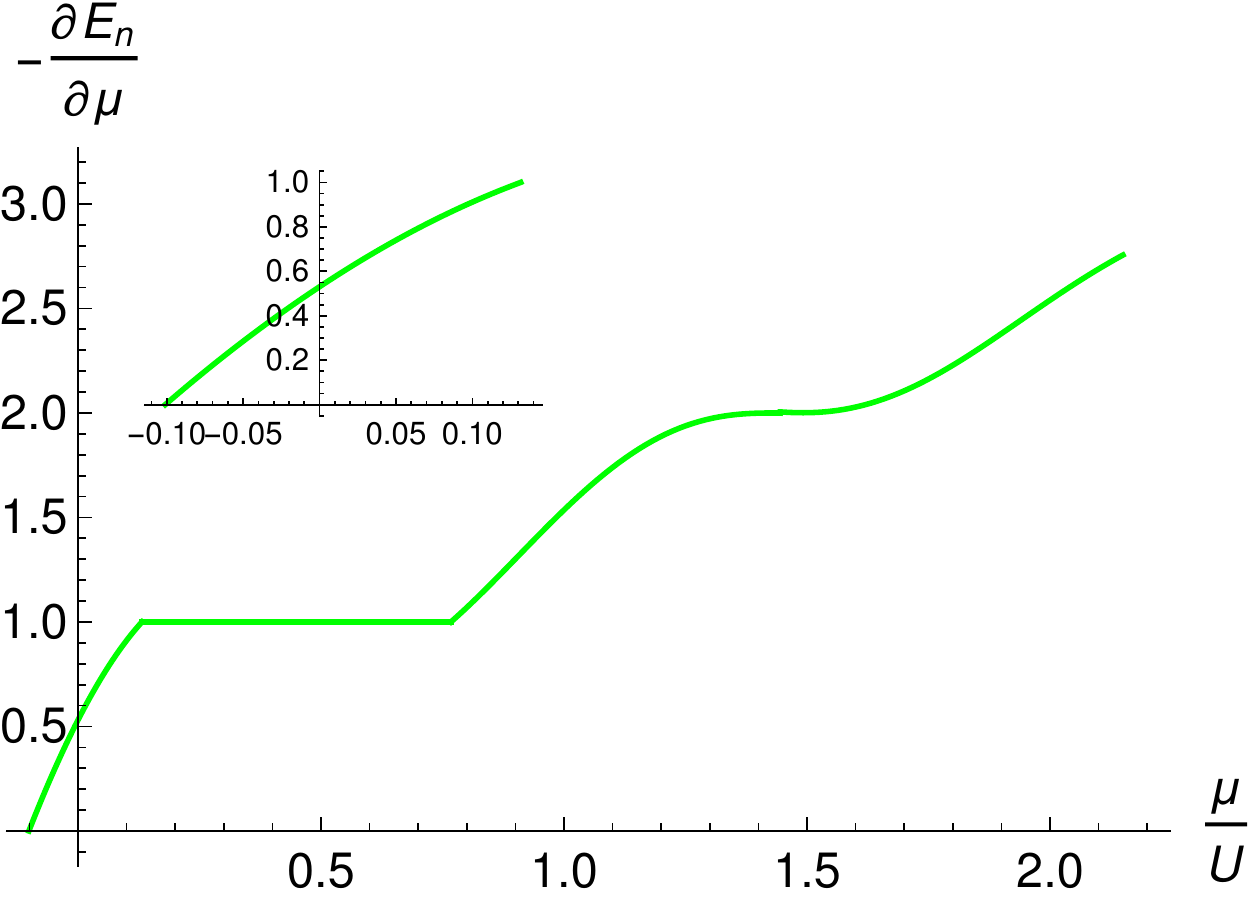}
		\caption{$Jz/U=0.101$.}
		\label{fig:0101}
	\end{subfigure}
	\caption{Particle densities $-\partial E_n/\partial \mu$ as functions of the chemical potential $\mu/U$ according to the corresponding hopping values. Horizontal lines correspond to Mott-insulating regions, while ascending curves correspond to superfluid regions. The higher the hopping, the rounder the curves become.}
	 Source: K\"UBLER \textit{et al.}\cite{martin} 
	 \label{fig:002,008,0101PD}
\end{figure}

\subsubsection{Condensate density}
The corresponding results for the condensate densities $\Psi^* \Psi$ 
are presented in Figs. \ref{fig:002,008,0101OP} and \ref{20OPS}. 
The data start at the phase boundary on the first Mott lobe, $n=1$, 
and end at the phase boundary on the second Mott lobe, $n=2$. 
Note that these different values of the occupation number $n$ are taken into account by the structure of the matrix entries (\ref{eq.11.18}).
Thus, evaluating the matrix elements with $n=1$, we get the physical results for the right half of the Mott lobe, 
while for the left half of the Mott lobe we must perform the calculation with $n=2$.
Also, we observe that every corresponding condensate density data has a maximum at $\varepsilon/U>0$.

\begin{figure}[h!]
	\centering
	\begin{subfigure}{.47\textwidth}
		\includegraphics[width=\textwidth]{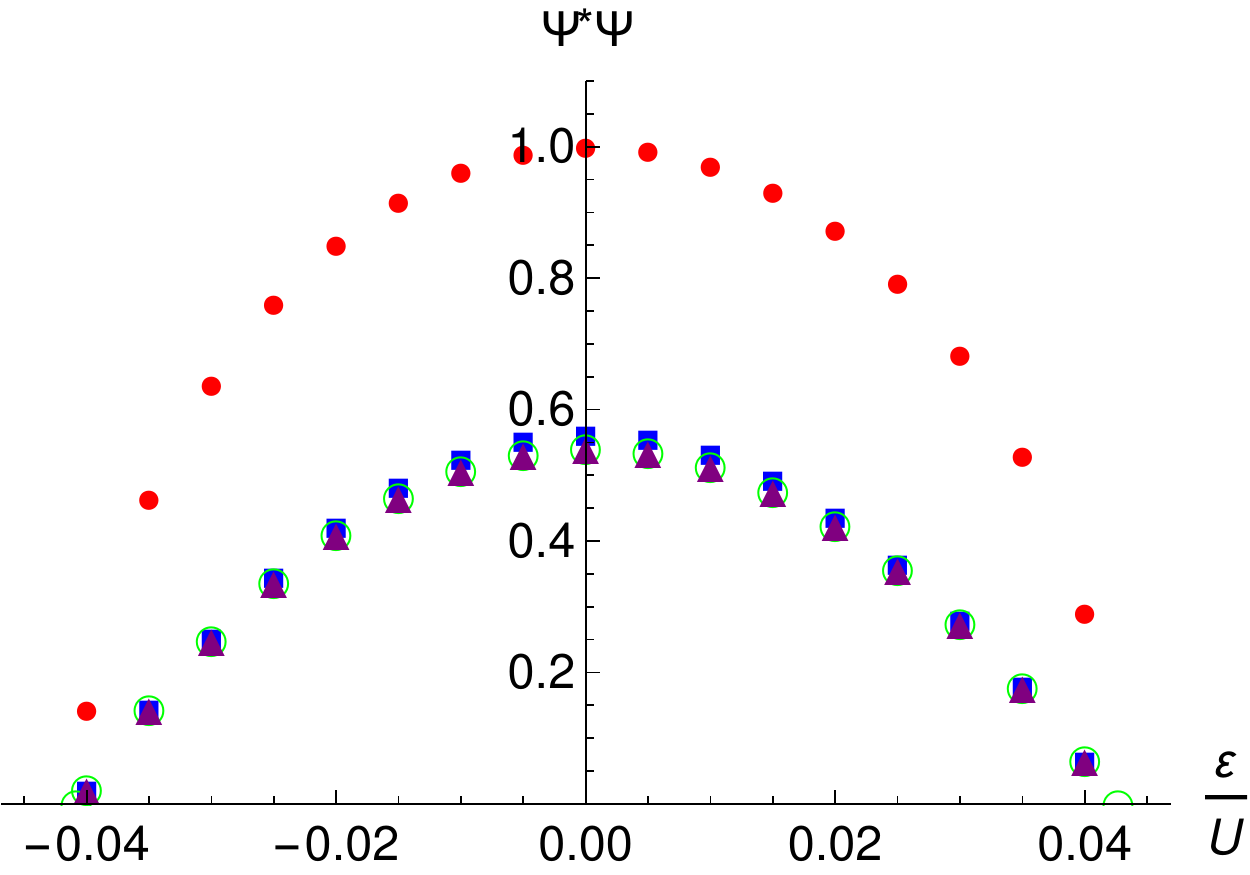}
		\caption{$Jz/U=0.02$.}
		\label{fig:002aa}
	\end{subfigure}
	\qquad
	\begin{subfigure}{.47\textwidth}
		\includegraphics[width=\textwidth]{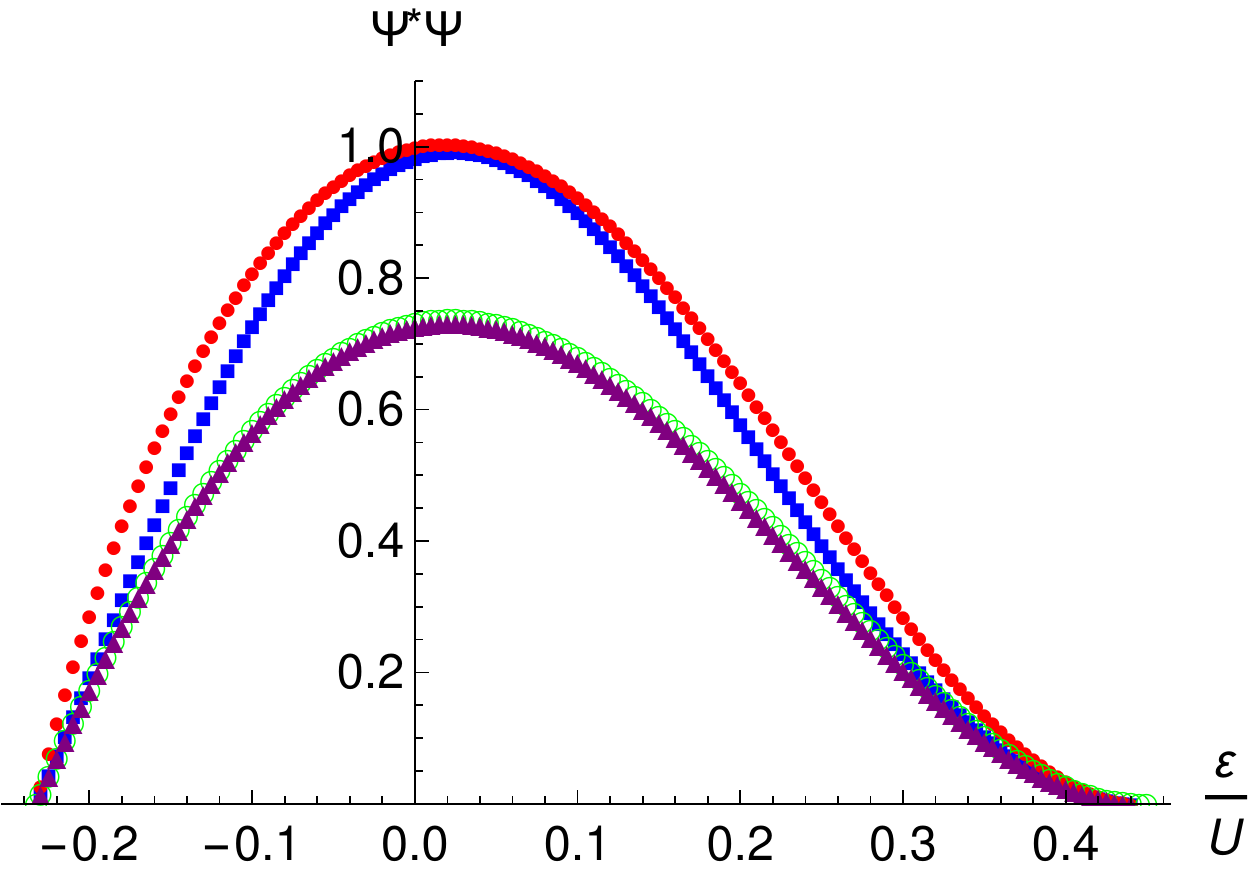}
		\caption{$Jz/U=0.101$.}
		\label{fig:0101c}
	\end{subfigure}
	\caption{Condensate densities as functions of $\varepsilon/U=\mu/U-1$ 
	according to the corresponding hopping values. 
	The curve styles correspond to, from the top to the bottom, the following corrections: $\mathcal{O}\left(\lambda\right)$ (red circles), 
	$\mathcal{O}\left(\lambda^2\right)$ (blue squares), $\mathcal{O}\left(\lambda^3\right)$ (green rings), and $\mathcal{O}\left(\lambda^4\right)$ (purple triangles). 
	For small values of $Jz/U$, and thus close to the degeneracy, the third- (green rings) and fourth-order (purple triangles) data coincide.}
	Source: K\"UBLER \textit{et al.}\cite{martin} 
	\label{fig:002,008,0101OP}
\end{figure}

\begin{figure}[h!]
	\centering
\includegraphics[width=.6\textwidth]{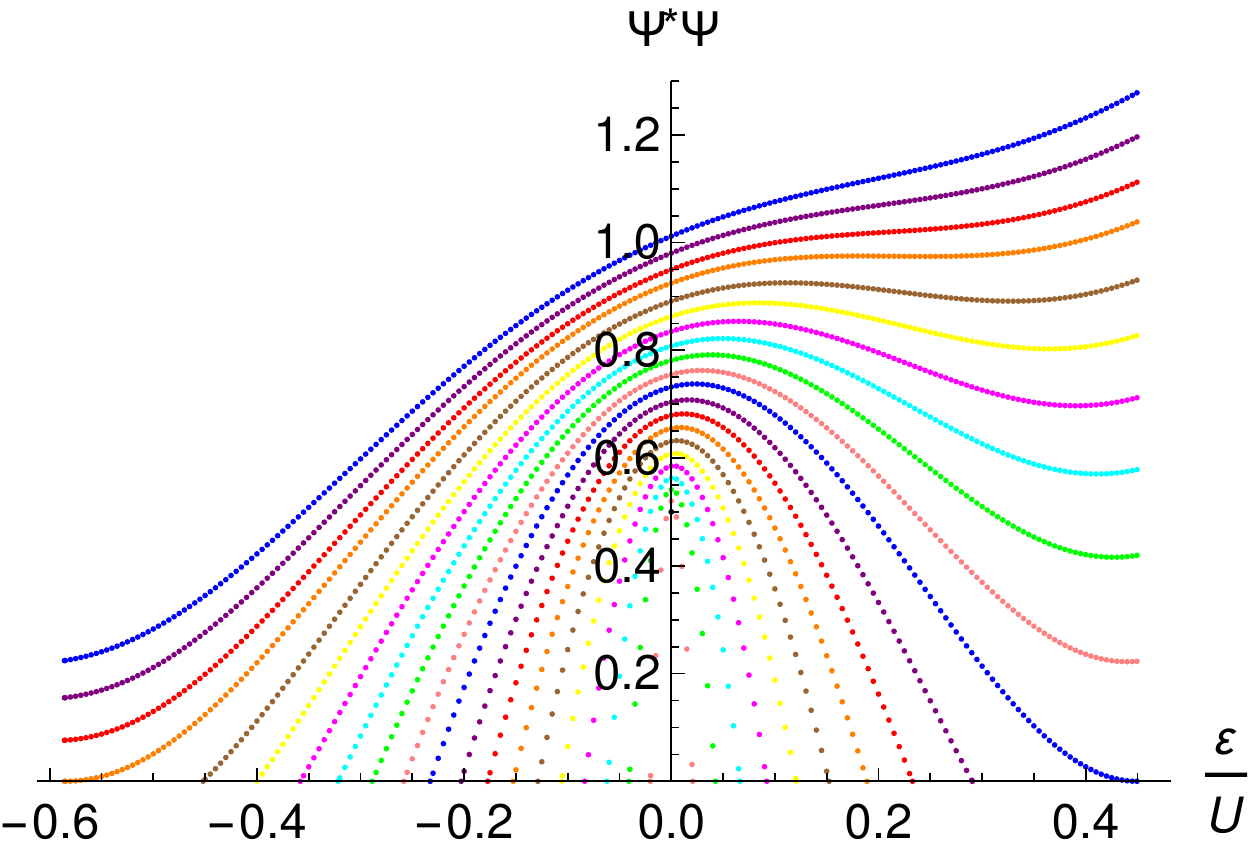}
	\caption{Condensate densities $\Psi^{*} \Psi$ as functions of 
	$\varepsilon/U=\mu/U-1$ up to $\mathcal{O}\left(\lambda^4\right)$ 
	between the first and the second Mott lobes for different hopping values: from $Jz/U=0.01$ (innermost points) until $Jz/U=0.20$ (outermost points) 
	with a step size of 0.01.}
	Source: K\"UBLER \textit{et al.}\cite{martin} 
	\label{20OPS}
\end{figure}

Fig. \ref{fig:002,008,0101OP} shows the condensate density $\Psi^* \Psi$ over $\varepsilon/U$ 
for two different hopping values and for four different orders in $\lambda$. 
As we can observe, the relative error between the condensate densities from $\mathcal{O}\left(\lambda^4\right)$ and $\mathcal{O}\left(\lambda^6\right)$ 
is about $ 0.0016 \% $, justifying the truncation of the perturbative series to fourth order in $\lambda$.

Fig. \ref{20OPS} illustrates the condensate densities $\Psi^* \Psi$ as functions of $\varepsilon/U$ for 
twenty different hopping strengths. 
Considering the hopping values from $Jz/U=0.01$ (pink dots) to $Jz/U=0.09$ (purple dots), the data behave very similarly to parabolas. 
For $Jz/U=5-2\sqrt{6} \approx 0.101$ (blue dots), 
the second Mott lobe achieves its tip, and the data touches the $\varepsilon/U$ axis at positive $\varepsilon/U$. 
From $Jz/U=0.11$ (pink dots) up to $Jz/U=0.16$ (brown dots), 
considering the region of positive $\varepsilon/U$, the data present a minimum, 
while for the negative $\varepsilon/U$ region, the data intersect the corresponding axis. 
For $Jz/U=3 - 2 \sqrt{2} \approx 0.172$ (orange dots), which corresponds to the tip of the first lobe, 
the data touches the $\varepsilon/U$ axis at negative $\varepsilon$/U. 
From $Jz/U=0.18$ (red dots) up to $Jz/U=0.20$ (blue dots), 
for which the system is found to be deeply in the superfluid phase, 
the whole graph is monotonically increasing. 
Finally, note that this is a representation of the condensate density $\Psi^* \Psi$ that provides nonzero and
continuous results at the degeneracy, an outcome that was not obtained by the Rayleigh-Schrödinger perturbation theory 
(see Fig. \ref{fig.OPRS}) nor by the Brillouin-Wigner one-state approach (see Fig. \ref{fig.(11.2)}).
Therefore, we conclude that the condensate density out of the Brillouin-Wigner two-state approach is the most 
appropriate choice and is the one that should be used for further calculations.

\subsubsection{Comparison}
\begin{figure}[h!]
	\centering
	\includegraphics[width=.6\textwidth]{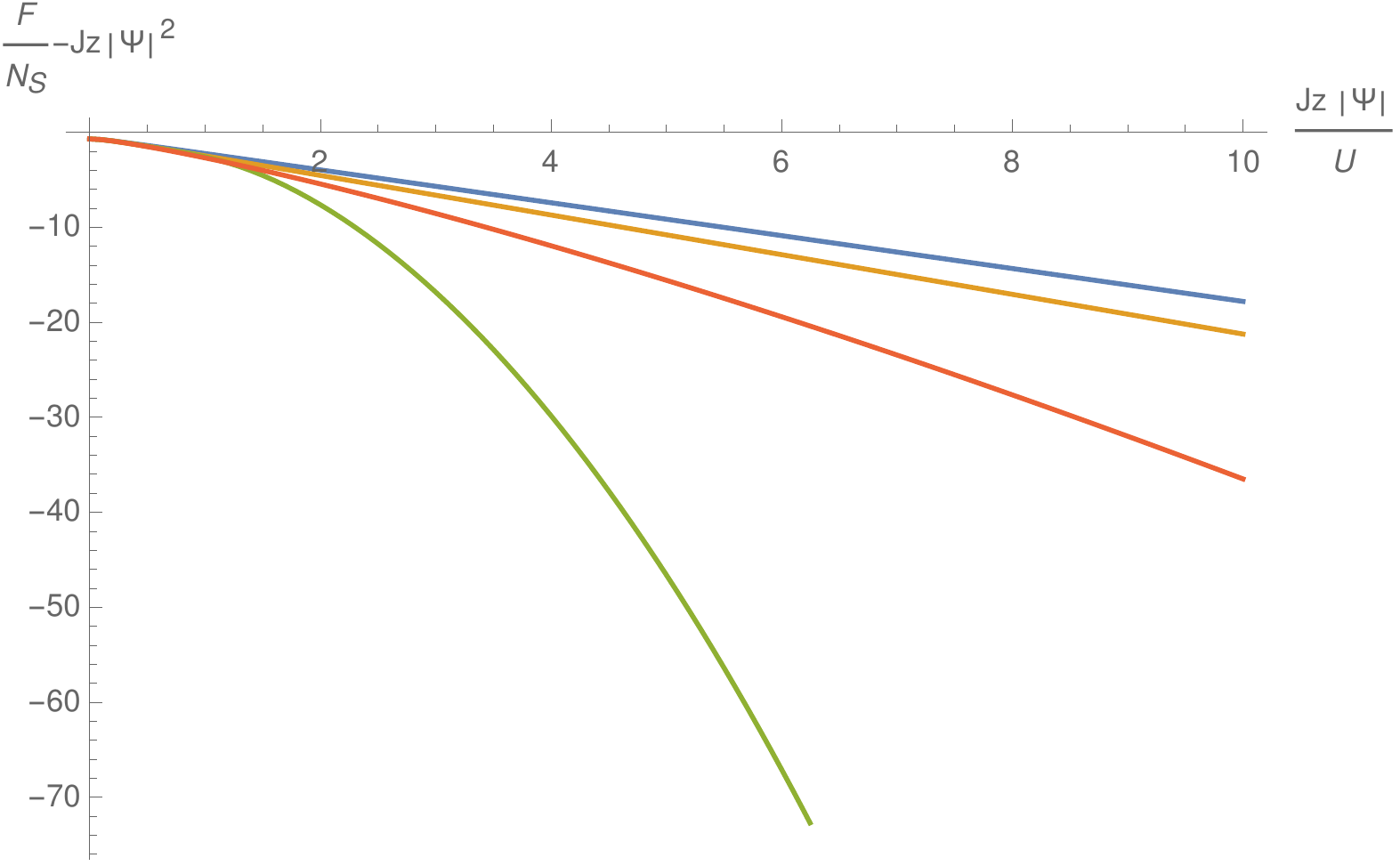}
	\caption{Ground-state energy $E_1$ out of the one-state approach for $\mu=0.7U$. 
	From top to bottom, the curves represent the numerical diagonalization 
	calculation (blue line) as well as the perturbative analytical 
	calculations up to $\mathcal{O}\left(\lambda^6\right)$ (yellow line), 
	$\mathcal{O}\left(\lambda^4\right)$ (red line), and $\mathcal{O}\left(\lambda^2\right)$ (green line). 
	Here, $N_s$ represents the number of lattice sites and $F$ 
	stands for the zero-temperature free energy.}
	Source: K\"UBLER \textit{et al.}\cite{martin} 
	\label{NumVerg1}
\end{figure}

By comparing our developed  
BWPT to the numerical diagonalization of the Bose-Hubbard Hamiltonian 
performed in M. K\"ubler \textit{et al.} (2019)\cite{martin},
we find a good convergence at small hoppings. 
In Fig. \ref{NumVerg1}, the uppermost curve (blue line) stems from the numerical calculation,  
while the remaining curves correspond to different orders from the BWPT.
As we can observe, the one-state energy is quasi-exact at small hopping values.
Moreover, as the energies from the one-state 
and the two-state approaches coincide, the two-state approach energy can also be considered as quasi-exact within the small hopping regime.

\section{Graphical approach}
In order to evaluate (\ref{eq.2.37}), it is mandatory to evaluate the matrix elements (\ref{eq.2.38}). 
It is possible to observe from the effective Hamiltonian form in Eq. (\ref{eq.2.31,003}) that, 
for higher orders in $\lambda$, there is an increase in the algebraic difficulty in calculating 
such terms. So, for the sake of simplifying the evaluation of higher-order terms of (\ref{eq.2.31,003}), we work out an efficient graphical approach.

\begin{figure}[h!]
	\centering
\includegraphics[width=\textwidth,clip]{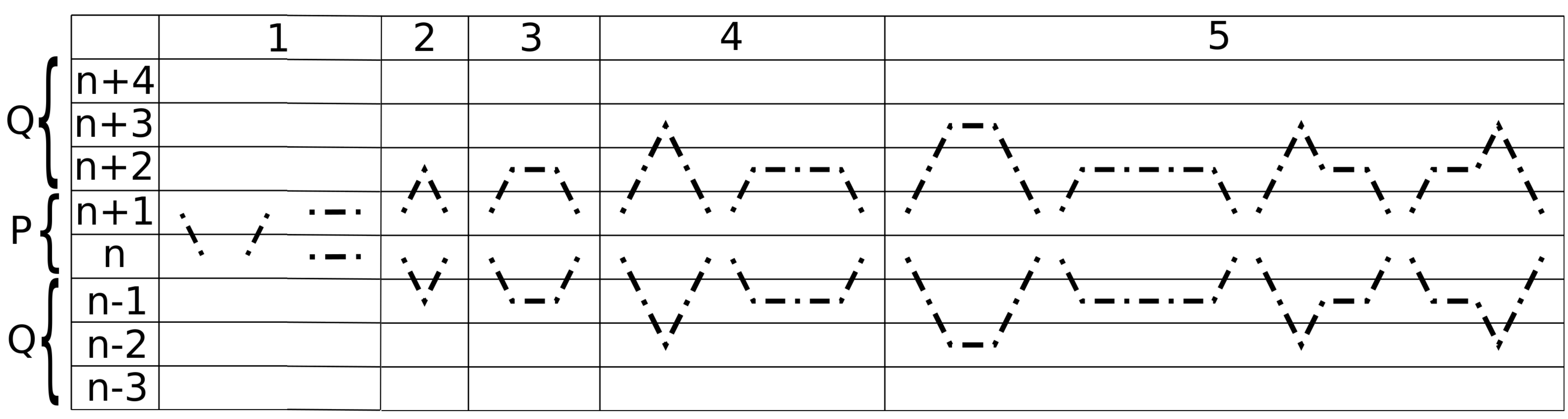}
\caption{Graphical approach for the matrix elements (\ref{eq.2.38}) from the effective Hamiltonian (\ref{eq.2.31,003}) 
	for the Bose-Hubbard mean-field Hamiltonian (\ref{H_one}) up to fifth order in the hopping term for the two-state approach.}
	Source: K\"UBLER \textit{et al.}\cite{martin} 
	\label{figIGA} 
\end{figure}

In particular, for the mean-field Hamiltonian (\ref{H_one}), we work out, within the two-state approach, 
a graphical representation for the $j$th-order terms $H_{{\eff}, k, k'}^{(j)}$ according to 
the expansion 
\begin{equation}
H_{{\eff}, k, k'}=\sum_j H_{{\eff}, k, k'}^{(j)},
\end{equation} 
which is depicted in Fig. \ref{figIGA}. 
The first row of Fig. \ref{figIGA} represents the orders in $\lambda$ 
for the respective correction terms. In the first column we have different states 
ranging from $n-3$ to $n+4$. Within the two-state matrix approach we choose $\hat{\PP}=\hat{\PP}_n + \hat{\PP}_{n+1}$, 
once there is a degeneracy between two consecutive Mott lobes in the zero-temperature phase diagram of the Bose-Hubbard model.

In order to obtain all possible graphs in Fig. \ref{figIGA}, we have to take into account the following empirical rules:
\begin{enumerate}
\item{Since $\hat{V}$ is linear in $\hat{a}$ and $\hat{a}^{\dagger}$ in Eq. (\ref{perturbation}), 
we can only go from one state to one of its nearest-neighbor states;}
\item{Because the effective Hamiltonian $\hat{H}_{\mathrm{eff}}$ in (\ref{eq.2.25}) contains only the projection operator $\hat{\Q}$, 
	but it is sandwiched by the projection operator $\hat{\PP}$ 
	according to (\ref{eq.2.24}), it follows that only the first 
		and the last states are allowed to be within $\PP$, while the intermediate states must be contained in $\Q$.}
\end{enumerate}

We interpret each graph according to the following rules:
\begin{itemize}
\item{The starting point of every graph corresponds to
\begin{equation}
S \left(m \right) = E_n - E_m^{(0)},
\end{equation}
with $m$ being the state we start the graph in.}
\item{Every line in the graph corresponds to the following terms. Ascending lines correspond to
\begin{equation}
L_A \left( m \right) = - \lambda J z \Psi \frac{\sqrt{m +1}}{E_n - E^{(0)}_{m}},
\end{equation}
with $m$ being the state the line started in.} 
\item{Descending lines correspond to
\begin{equation}
L_D \left( m \right) = - \lambda J z \Psi^* \frac{\sqrt{m}}{E_n - E^{(0)}_{m}}.
\end{equation}}
\item{Horizontal lines correspond to 
\begin{equation}
L_{H}(m)=\frac{\lambda J z \Psi^{*} \Psi}{E_{n}-E_{m}^{(0)}}.
\end{equation}}
\end{itemize}

The off-diagonal matrix elements vanish for all orders except for $\mathcal{O}(\lambda)$:
\begin{align}\label{Gleichung5.18}
&H_{{\eff}, n, n+1}^{(1)} = S(n+1) L_D(n+1)=-\lambda Jz \Psi^* \sqrt{n+1},
\\
&H_{{\eff}, n+1, n}^{(1)} =S(n) L_A(n)=-\lambda Jz \Psi \sqrt{n+1}.
\end{align}

Now we proceed to evaluating the diagonal matrix elements for ascending orders of $\lambda$. 
For $\mathcal{O}(\lambda)$ we have 
\begin{align}
&H_{{\eff}, n, n}^{(1)}=S(n+1)L_H(n+1)=\lambda J z \Psi^* \Psi,
\\
&H_{{\eff}, n+1, n+1}^{(1)}=S(n)L_H(n)=\lambda J z \Psi^* \Psi.
\end{align}
For $\mathcal{O}(\lambda^2)$ we have, correspondingly,
\begin{equation}
H_{{\eff}, n, n}^{(2)}=S(n+1) L_A(n+1)L_D(n+2)= \lambda^2 J^2 z^2 \Psi^* \Psi \frac{n+2}{E_n-E_{n+2}^{(0)}}
\end{equation}
and
\begin{equation}
H_{{\eff}, n+1, n+1}^{(2)}=S(n) L_D(n)L_A(n-1)= \lambda^2 J^2 z^2 \Psi^* \Psi \frac{ n}{E_n-E_{n-1}^{(0)} }.
\end{equation}
For $\mathcal{O}(\lambda^3)$ one obtains
\begin{equation}
H_{{\eff}, n, n}^{(3)}=S(n+1) L_A(n+1)L_H(n+2)L_D(n+2)= \lambda^3 J^3 z^3 \Psi^{*2} \Psi^2 \frac{n+2}{\left(E_n-E_{n+2}^{(0)}\right)^2 }
\end{equation}
together with
\begin{equation}
H_{{\eff}, n+1, n+1}^{(3)}=S(n) L_D(n)L_H(n-1)L_A(n-1)= \lambda^3 J^3 z^3 \Psi^{*2} \Psi^2 \frac{ n}{\left(E_n-E_{n-1}^{(0)}\right)^2 }.
\end{equation}
For $\mathcal{O}(\lambda^4)$ we find
\begin{equation}
\begin{aligned}
H_{{\eff}, n, n}^{(4)}&=S(n+1) L_A(n+1)\left[L_A(n+2)L_D(n+3) +  L_H(n+2)L_H(n+2)\right]L_D(n+2)
\\
&= \lambda^4 J^4 z^4 \Psi^{*2} \Psi^2 \frac{ \left(n+2\right) \left(n+3\right)}{\left(E_n-E_{n+2}^{(0)}\right)^2 \left(E_n-E_{n+3}^{(0)}\right)}+ \lambda^4 J^4 z^4 \Psi^{*3} \Psi^3 \frac{n+2}{\left(E_n-E_{n+2}^{(0)}\right)^3 }
\end{aligned}
\end{equation}
and
\begin{equation}
\begin{aligned}\label{Gleichung5.27}
H_{{\eff}, n+1, n+1}^{(4)}&=S(n) L_D(n)\left[L_D(n-1)L_A(n-2)+L_H(n-1)L_H(n-1)\right]L_A(n-1)
\\
&= \lambda^4 J^4 z^4 \Psi^{*2} \Psi^2 \frac{ n\left(n-1\right)}{\left(E_n-E_{n-1}^{(0)}\right)^2 \left(E_n-E_{n-2}^{(0)}\right)}+\lambda^4 J^4 z^4 \Psi^{*3} \Psi^3 \frac{n}{\left(E_n-E_{n-1}^{(0)}\right)^3 }.
\end{aligned}
\end{equation}
Finally, the fifth column, corresponding to $\mathcal{O}(\lambda^5)$, results in 
\begin{equation}
\begin{aligned}
H_{{\eff}, n, n}^{(5)}=&S(n+1) L_A(n+1)\left[
L_A(n+2)L_H(n+3)L_D(n+3)\right.\\
&+\left.L_H(n+2)L_H(n+2)L_H(n+2)\right.\\
&+\left.2L_A(n+2)L_D(n+3)L_H(n+2)
\right]L_D(n+2)
\\
=&\lambda^5 J^5 z^5 \Psi^{*3} \Psi^3 \frac{\left(n+2\right) \left(n+3\right)}{\left(E_n-E_{n+2}^{(0)}\right)^2 \left(E_n-E_{n+3}^{(0)}\right)^2} 
\\
&+2\lambda^5 J^5 z^5 \Psi^{*3} \Psi^3 \frac{\left(n+2\right) \left(n+3\right)}{\left(E_n-E_{n+2}^{(0)}\right)^3 \left(E_n-E_{n+3}^{(0)}\right)}\\
&+\lambda^5 J^5 z^5 \Psi^{*4} \Psi^4 \frac{n+2}{\left(E_n-E_{n+2}^{(0)}\right)^4 },
\end{aligned}
\end{equation}
together with
\begin{equation}
\begin{aligned}
H_{{\eff}, n+1, n+1}^{(5)}=&S(n) L_D(n)\left[
L_D(n-1)L_H(n-2)L_A(n-2)\right.
\\
&+\left.L_H(n-1)L_H(n-1)L_H(n-1)\right.\\
&+\left.2L_D(n-1)L_A(n-2)L_H(n-1)\right]L_A(n-1)
\\
=&\lambda^5 J^5 z^5 \Psi^{*3} \Psi^3 \frac{n\left(n-1\right)}{\left(E_n-E_{n-1}^{(0)}\right)^2 \left(E_n-E_{n-2}^{(0)}\right)^2}
\\
&+2\lambda^5 J^5 z^5 \Psi^{*3} \Psi^3 \frac{n\left(n-1\right)}{\left(E_n-E_{n-1}^{(0)}\right)^3 \left(E_n-E_{n-2}^{(0)}\right)}\\
&+\lambda^5 J^5 z^5 \Psi^{*4} \Psi^4 \frac{n}{\left(E_n-E_{n-1}^{(0)}\right)^4} .
\end{aligned}
\end{equation}

\section{\label{Trap}Harmonic trap}
In view of actual experiments, 
we consider now the impact of a harmonic confinement 
upon the equation of state. 
Although most traps in experiments have an ellipsoidal shape, 
for simplicity we perform calculations regarding the case of a spherical trap.\cite{martin} 
In order to add a trap to our calculations, we have to perform the so-called Thomas-Fermi approximation\cite{thomas,fermi}
\begin{equation}\label{muTRAP}
\mu = \tilde{\mu}-\frac{1}{2}m\omega^2 r^2.
\end{equation}
Here, $m$ denotes the mass of the particles and $\omega$ stands for the trap frequency. 
Thus, the chemical potential is now consisting of a trap term and the original chemical potential $\tilde{\mu}$.

This strategy effectively gives rise to the same curves as in Fig. \ref{fig:002,008,0101PD}. 
We identify $\tilde{\mu}_{max}$ as being the center of the trap, while its borders are identified by the vanishing 
points of the condensate density. In between, we have Mott insulating and superfluid regions, which produce, 
in a three-dimensional trap, a wedding-cake structure with alternating Mott insulating and superfluid shells.\cite{bloch-rev,melo,bloch2}

In order to identify the curves from Fig. \ref{fig:002,008,0101PD} 
with a real experimental setting, 
we have to determine $\tilde{\mu}$. 
This can done by integrating the curves from Fig. \ref{fig:002,008,0101PD}. 
Such a procedure results in a curve for the total number of particles, 
which allows one to determine the corresponding value of $\tilde{\mu}$. 
Thus, the number of particles reads 
\begin{equation}
N_{\mu_i, \mu_o}=-\frac{1}{a^3} \int \frac{\partial E_n}{\partial \mu} d^3\mathbf{r} 
	= -\frac{4\pi}{a^3} \int_{R_i}^{R_o}r^2\frac{\partial E_n}{\partial \mu}dr,
\end{equation}
where the radii $R_i$ and $R_o$ are the inner and the outer radius of the shell 
that we want to compute, respectively. Here, $a$ is the lattice spacing.
The following calculation is done for $Jz/U=0.101$ 
and $2 \leq n \leq 3$ (see Fig. \ref{fig:002,008,0101PD}, 
$1.69 \leq \mu/U \leq 2.15$), which is just the innermost 
superfluid shell. Hence, from the solution for the energies $E_n$, we have that 
the integration argument yields 
\begin{equation}\label{Integral1D}
N_{1.69, 2.15}=-\frac{4\pi}{a^3} \int_{R_2}^{R_3}r^2\left[13 
	- 38 \frac{\mu}{U} + 37 \left(\frac{\mu}{U}\right)^2  
	- 18 \left(\frac{\mu}{U}\right)^3 + 4 \left(\frac{\mu}{U}\right)^4 
	-  0.4 \left(\frac{\mu}{U}\right)^5 \right]dr,
\end{equation}
with
\begin{subequations}
	\begin{align}
		R_3 &= \sqrt{\frac{2(\tilde{\mu}-2.15U)}{m\omega^2}},\\
		R_2 &= \sqrt{\frac{2(\tilde{\mu}-1.69U)}{m\omega^2}}.
	\end{align}
\end{subequations}

The last step consists of inserting (\ref{muTRAP}) into (\ref{Integral1D}) and perform the integration. 
The same procedure must be repeated for all other regions 
of Fig. \ref{fig:002,008,0101PD}, namely, 
$N_{1.23, 1.69}$, $N_{0.82, 1.23}$, $N_{0.10, 0.82}$, 
and $N_{-0.08, 0.10}$, which represent the remaining superfluid and 
Mott insulating shells, respectively. 
Then, the total number of particles is obtained by summing all these contributions,  
\begin{equation}
N=N_{-0.08, 0.10}+N_{0.10, 0.82}+N_{0.82, 1.23}+N_{1.23, 1.69}+N_{1.69, 2.15}.
\end{equation}

\begin{figure}[h]
   \centering
      \includegraphics[width=.6\columnwidth]{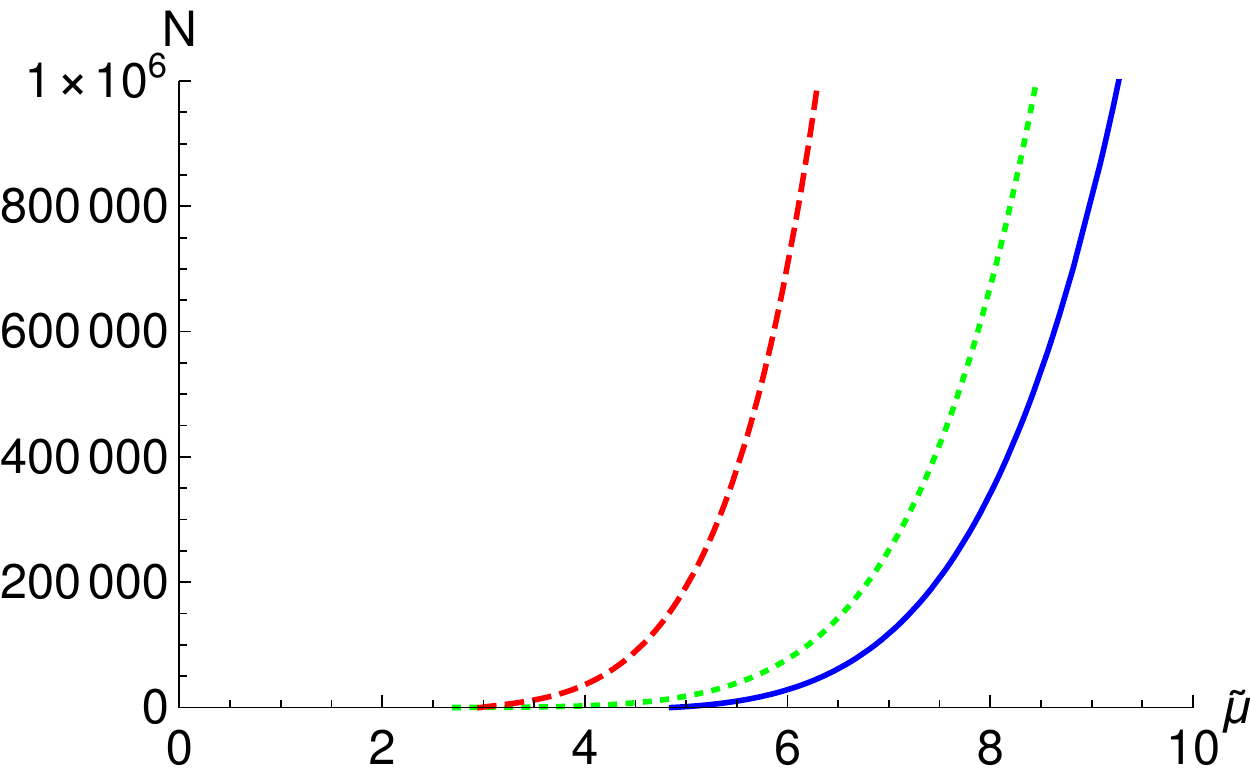}
        \caption{Equations of state $N=N(\tilde{\mu})$ for the following parameters: $m=87u$, $a=400 \mathrm{nm}$, and $\omega=48\pi \mathrm{Hz}$. From left to right, the hopping values are: $Jz/U=0.02$ (dashed red line), $Jz/U=0.101$ (dotted green line), and $Jz/U=0.08$ (continuous blue line).}
        Source: K\"UBLER \textit{et al.}\cite{martin} 
\label{fig.Imu008}
\end{figure}

The resulting equation of state $N=N(\tilde{\mu})$ is shown in Fig. \ref{fig.Imu008}. 
For small values of $\tilde{\mu}$, the particle number rapidly vanishes. 
Thus, we conclude that, for a given $\tilde{\mu}$, the minimal particle number is not achieved for $Jz/U=0$, 
where all particles are in the Mott insulator phase, nor for $Jz/U > 0.172$, 
where all particles are in the superfluid phase. 
Instead, the minimal particle number is achieved for a specific distribution of Mott insulator and superfluid, 
represented by a corresponding hopping value $Jz/U$, which can be determined by the methods introduced here.

\chapter{The system at finite temperature}
\label{DPT}
In this chapter, we introduce the finite-temperature degenerate perturbation theory (FTDPT) 
method, which consists of a degenerate perturbative calculation making use of projection operators. 
From the developed FTDPT, we evaluate the condensate densities for different temperatures and hopping values. 
Following, we turn our attention to a region between two consecutive Mott lobes, which is a 
region in the phase diagram where the superfluid clearly dominates and also a region 
where the NDPT fails at very low temperatures, 
\textit{i.e.}, it predicts a phase transition, even though there should be none. Then, we compare the results from NDPT and FTDPT in order to 
corroborate the results from our developed method. 
Finally, we calculate the particle densities for different temperatures and hopping values, 
observing the existence of the melting of the structure due to both the temperature and the hopping increase.

\section{The projection operators method}
We begin by considering two adjacent degenerate states $\ket{n}$ and $\ket{n+1}$. Firstly, we let us define the subspace of the Hilbert space in which those two state are located, $\mathcal{P}$. 
Then, the projection operator that allows us to access the respective Hilbert subspace is given by 
\begin{equation} 
	\hat{\mathcal{P}} \equiv \vert n \rangle \langle n \vert + \vert n+1 \rangle \langle n+1 \vert.
\end{equation}
Likewise, we define the complementary Hilbert subspace, $\mathcal{Q}$, in which all the remaining states are located. 
Thus, the corresponding complementary operator is defined as
\begin{equation} 
	\hat{\mathcal{Q}} \equiv \sum_{m \notin \mathcal{P}} \vert m \rangle \langle m \vert.
\end{equation}

Let us begin our analysis by considering the one-site mean-field Hamiltonian (\ref{H_one}) and regard, 
as in Sec. \ref{PT}, the hopping term (\ref{perturbation}) as a perturbation in (\ref{H0}). 
Multiplying both sides of the perturbation by the identity operator, $\hat{\mathds{1}}= \hat{\mathcal{P}}+\hat{\mathcal{Q}}$, we have that 
\begin{equation}
\begin{aligned}
	\hat{H}&=\hat{H}_{0}+\left(\hat{\mathcal{P}}+\hat{\mathcal{Q}}\right)\hat{V}\left(\hat{\mathcal{P}}+\hat{\mathcal{Q}}\right)  \\
	&=\hat{H}_{0}+\hat{\PP}\hat{V}\hat{\PP}+\hat{\PP}\hat{V}\hat{\Q}+\hat{\Q}\hat{V}\hat{\PP}+\hat{\Q}\hat{V}\hat{\Q}.
\end{aligned}
\end{equation}
This procedure allows us to define a new unperturbed Hamiltonian and a new perturbation according to
\begin{subequations}
	\begin{align}
		\hat{\mathcal{H}}_{0}&\equiv\hat{H}_{0}+\hat{\PP}\hat{V}\hat{\PP}, \label{newH0}\\
		\hat{\mathcal{V}}&\equiv\hat{\PP}\hat{V}\hat{\Q}+\hat{\Q}\hat{V}\hat{\PP}+\hat{\Q}\hat{V}\hat{\Q}. \label{newV}
	\end{align}
\end{subequations}

The Hamiltonian (\ref{newH0}), written in the basis of the unperturbed eigenstates, is a block diagonal matrix, whose only non-diagonal block is
\begin{equation}
	\hat{\mathcal{H}}_{0}^{(\mathrm{nd})}=\begin{pmatrix} E_{n} & -J z \Psi \sqrt{n+1} \\
-J z \Psi^{*}\sqrt{n+1} & E_{n+1} \end{pmatrix}.
\end{equation}
Its eigenvalues and eigenstates are respectively given by
\begin{subequations}
	\begin{align}
\mathcal{E}_{\pm} &= \frac{E_{n}+E_{n+1}}{2}
		\pm\frac{1}{2}\left[\left(E_{n}-E_{n+1}\right)^{2}+4 J^2 z^2 |\Psi|^2\left(n+1\right)\right]^{1/2}, \\
	\vert\Phi_{\pm}\rangle&=\left[1+\frac{\left|\mathcal{E}_{\pm}-E_{n}\right|^{2}}{J^2 z^2 |\Psi|^2 \left(n+1\right)}\right]^{-1/2}
		\left[\vert n\rangle+\frac{E_{n}-\mathcal{E}_{\pm}}{J z\sqrt{|\Psi|^{2}\left(n+1\right)}}\vert n+1\rangle\right]. \label{PHI+-}
	\end{align}
\end{subequations}
Note that we have dropped out the index $^{(0)}$ for the unperturbed 
eigenenergies that were used all along Chap. \ref{BWPT} for the sake of simplicity. Therefore, keep in mind that 
we use $E_n$ instead of $E_n^{(0)}$ throughout the entire current chapter.

As pointed out in Sec. \ref{PT}, we must evaluate the partition function (\ref{Z1}) in order to calculate the free energy (\ref{F}). The only difference is that now we are working with the new unperturbed Hamiltonian (\ref{newH0}) and the new perturbation (\ref{newV}). With this, the time-evolution operator now reads
\begin{equation}
\hat{U}=\mathrm{e}^{-\beta \hat{\mathcal{H}}_0}\hat{\mathcal{U}}_{\mathrm{I}}.
\end{equation}
The respective Schr\"odinger equation for the time-evolution operator in the interaction picture is given by 
\begin{equation}
\label{eq21}
\frac{d \, \hat{\mathcal{U}}_{\mathrm{I}} (\tau)}{d \tau} = - \hat{\mathcal{V}}_{\mathrm{I}} (\tau) \hat{\mathcal{U}}_{\mathrm{I}} (\tau),
\end{equation}
where 
\begin{equation}
\label{pertu}
\hat{\mathcal{V}}_{\mathrm{I}}(\tau)=\mathrm{e}^{\tau\hat{\mathcal{H}}_{0}}\left(\hat{\PP}\hat{V}\hat{\Q}+\hat{\Q}\hat{V}\hat{\PP}+\hat{\Q}\hat{V}\hat{\Q}\right)\mathrm{e}^{-\tau\hat{\mathcal{H}}_{0}}.
\end{equation}

The solution of equation (\ref{eq21}) with the initial condition $\hat{\mathcal{U}}_{\mathrm{I}}(0)=\hat{\mathds{1}}$, is given by, up to second order, 
\begin{equation}\label{U}
	\hat{\mathcal{U}}_{\mathrm{I}}(\beta) \approx \hat{\mathds{1}}-\int_{0}^{\beta} d\tau_1 \hat{\mathcal{V}}_{\mathrm{I}} (\tau_1) + \int_{0}^{\beta} d\tau_1 \int_{0}^{\tau_1} d\tau_2 \hat{\mathcal{V}}_{\mathrm{I}} (\tau_1) \hat{\mathcal{V}}_{\mathrm{I}} (\tau_2).
\end{equation}

Evaluating the partition function $\mathcal{Z}=\Tr{\expo{-\beta \hat{\mathcal{H}}_0}\hat{\mathcal{U}}_{\mathrm{I}}(\beta)}$, we have
\begin{align}
	\label{Z}
	\mathcal{Z}&= \expo{-\beta \mathcal{E}_+}\bra{\Phi_+} \hat{\mathcal{U}}_{\mathrm{I}}(\beta) \ket{\Phi_+} + \expo{-\beta \mathcal{E}_-}\bra{\Phi_-} \hat{\mathcal{U}}_{\mathrm{I}}(\beta) \ket{\Phi_-} +\sum_{m \in \Q} \expo{-\beta E_m}\bra{m} \hat{\mathcal{U}}_{\mathrm{I}}(\beta) \ket{m}.
\end{align}
Considering only the zeroth-order term from (\ref{U}) into (\ref{Z}) yields 
\begin{equation}
\label{Z0}
\mathcal{Z}^{(0)} = \expo{-\beta \mathcal{E}_+} + \expo{-\beta \mathcal{E}_-} + \sum_{m \in \Q} \expo{-\beta E_m}.
\end{equation}
Furthermore, from (\ref{perturbation}), (\ref{pertu}), and (\ref{newV}), we read off that the first-order contribution in (\ref{Z}) must vanish.

Now we proceed to performing the calculation of the second-order term,
\begin{equation}
\begin{aligned}
	\mathcal{Z}^{(2)} =&\,\expo{-\beta \mathcal{E}_+} \int_{0}^{\beta} d\tau_1 \int_{0}^{\tau_1} 
	d\tau_2 \, \bra{\Phi_+} \hat{\mathcal{V}}_{\mathrm{I}}(\tau_1) \hat{\mathcal{V}}_{\mathrm{I}}(\tau_2) \ket{\Phi_+}\\
	 &+ \expo{-\beta \mathcal{E}_-} \int_{0}^{\beta} d\tau_1 \int_{0}^{\tau_1} d\tau_2 \, \bra{\Phi_-} 
	 \hat{\mathcal{V}}_{\mathrm{I}}(\tau_1) \hat{\mathcal{V}}_{\mathrm{I}}(\tau_2) \ket{\Phi_-} \\
	&+ \sum_{m \in \Q} \expo{-\beta E_m} \int_{0}^{\beta} d\tau_1 \int_{0}^{\tau_1} d\tau_2 \, \bra{m} \hat{\mathcal{V}}_{\mathrm{I}}(\tau_1) \hat{\mathcal{V}}_{\mathrm{I}}(\tau_2) \ket{m}.
\end{aligned}
\end{equation}
We shall calculate each term separately and identify them as $\mathcal{Z}^{(2)}=\mathcal{Z}^{(2)}_+ + \mathcal{Z}^{(2)}_- + \mathcal{Z}^{(2)}_m$.
As the evaluation of $\mathcal{Z}^{(2)}_+$ and $\mathcal{Z}^{(2)}_-$ are completely equivalent, we perform a generic calculation for both contributions. 
Inserting the expression of the interaction-picture perturbation, (\ref{pertu}), into the first term, we have
\begin{equation}
\begin{aligned}
	\mathcal{Z}^{(2)}_\pm &=&\expo{-\beta \mathcal{E}_\pm} \int_{0}^{\beta} d\tau_1 \int_{0}^{\tau_1} 
	d\tau_2 \, \bra{\Phi_\pm} \expo{\tau_1 \hat{\mathcal{H}}_{0}} \left(\hat{\PP}\hat{V}\hat{\Q}
	+\hat{\Q}\hat{V}\hat{\PP}+\hat{\Q}\hat{V}\hat{\Q}\right) \expo{-\tau_1 \hat{\mathcal{H}}_{0}}\\
	&&\times \expo{\tau_2 \hat{\mathcal{H}}_{0}} \left(\hat{\PP}\hat{V}\hat{\Q}+\hat{\Q}\hat{V}\hat{\PP}+\hat{\Q}\hat{V}\hat{\Q}\right) \expo{-\tau_2 \hat{\mathcal{H}}_{0}} \ket{\Phi_\pm}.
\end{aligned}
\end{equation}
As $\ket{\Phi_\pm}$ are eigenstates of $\hat{\mathcal{H}}_0$, we get
\begin{equation}
\begin{aligned}\label{b3}
	\mathcal{Z}^{(2)}_\pm &=& \expo{-\beta \mathcal{E}_\pm} \int_{0}^{\beta} d\tau_1 \int_{0}^{\tau_1} d\tau_2 \, \expo{(\tau_1-\tau_2) \mathcal{E}_\pm} \bra{\Phi_\pm}  \left(\hat{\PP}\hat{V}\hat{\Q}+\hat{\Q}\hat{V}\hat{\PP}+\hat{\Q}\hat{V}\hat{\Q}\right) \expo{-\tau_1 \hat{\mathcal{H}}_{0}}\\
	&&\times \expo{\tau_2 \hat{\mathcal{H}}_{0}} \left(\hat{\PP}\hat{V}\hat{\Q}+\hat{\Q}\hat{V}\hat{\PP}+\hat{\Q}\hat{V}\hat{\Q}\right) \ket{\Phi_\pm}.
\end{aligned}
\end{equation}
Also, from the projection relations $\hat{\Q} \ket{\Phi_\pm} = 0$ and $\hat{\PP} \ket{\Phi_\pm} = \ket{\Phi_\pm}$, 
and having in mind that $\hat{\Q}$ and $\hat{\PP}$ are hermitian operators, (\ref{b3}) reduces to
\begin{equation}
	\mathcal{Z}^{(2)}_\pm = \expo{-\beta \mathcal{E}_\pm} \int_{0}^{\beta} d\tau_1 \int_{0}^{\tau_1} d\tau_2 \, \expo{(\tau_1-\tau_2) \mathcal{E}_\pm} \bra{\Phi_\pm}  \hat{V}\hat{\Q} \expo{-\tau_1 \hat{\mathcal{H}}_{0}} \expo{\tau_2 \hat{\mathcal{H}}_{0}} \hat{\Q}\hat{V} \ket{\Phi_\pm}.
\end{equation}
From (\ref{perturbation}), (\ref{PHI+-}), and the scalar products
\begin{subequations}
	\begin{align}
\braket{n}{\Phi_\pm} &=
			\left[1+\frac{\left|\mathcal{E}_{\pm}-E_{n}\right|^{2}}{J^2
			z^2\left|\Psi\right|^{2}\left(n+1\right)}\right]^{-1/2}, \\
			\braket{n+1}{\Phi_\pm} &=
			\left[1+\frac{\left|\mathcal{E}_{\pm}-E_{n}\right|^{2}}{J^2
			z^2\left|\Psi\right|^{2}\left(n+1\right)}\right]^{-1/2}
			\frac{E_{n}-\mathcal{E}_{\pm}}{J
			z\sqrt{\left|\Psi\right|^{2}\left(n+1\right)}},
	\end{align}
\end{subequations}
we have that 
\begin{equation}
\begin{aligned}\label{**}
	\mathcal{Z}^{(2)}_\pm &=&\expo{-\beta \mathcal{E}_\pm} \int_{0}^{\beta} d\tau_1 \int_{0}^{\tau_1} d\tau_2 \, \expo{(\tau_1-\tau_2) \mathcal{E}_\pm} J^2 z^2 \left(\Psi \braket{\Phi_\pm}{n} \sqrt{n} \bra{n-1}\right.\\
	&&\left.+ \Psi^* \braket{\Phi_\pm}{n+1}\sqrt{n+2}\bra{n+2}\right) \expo{-\tau_1 \hat{\mathcal{H}}_{0}}\expo{\tau_2 \hat{\mathcal{H}}_{0}}\\
	&&\times \left(\Psi^{*} \braket{n}{\Phi_\pm} \sqrt{n} \ket{n-1} + \Psi \braket{n+1}{\Phi_\pm}\sqrt{n+2}\ket{n+2}\right).
\end{aligned}
\end{equation}
Thus, evaluating (\ref{**}) leads to
\begin{equation}
	\begin{aligned}
		\mathcal{Z}^{(2)}_\pm &=& J^2 z^2 |\Psi|^2 \expo{-\beta \mathcal{E}_\pm} \int_{0}^{\beta} d\tau_1 \int_{0}^{\tau_1} d\tau_2 \,  \left(\expo{(\tau_1-\tau_2)\Delta_{\pm,n-1}} n \big|\braket{\Phi_\pm}{n}\big|^2 \right.\\
		&&\left.+  \expo{(\tau_1-\tau_2)\Delta_{\pm,n+2}} (n+2) \big|\braket{\Phi_\pm}{n+1}\big|^2 \right), \label{B5}
\end{aligned}
\end{equation}
where we have introduced the abbreviation $\Delta_{i,\pm} \equiv E_i - \mathcal{E}_\pm$.
Finally, performing the integrations in (\ref{B5}), the term $\mathcal{Z}^{(2)}_\pm$ results in  
\begin{equation}
\begin{aligned}\label{Z+-}
	\mathcal{Z}^{(2)}_\pm &=& J^2 z^2 |\Psi|^2 \expo{-\beta \mathcal{E}_\pm} \left[ n \big|\braket{\Phi_\pm}{n}\big|^2 \left(\frac{\expo{\beta \Delta_{\pm,n-1}}-1}{\Delta_{\pm,n-1}^2}-\frac{\beta}{\Delta_{\pm,n-1}}\right) \right.\\
	&&\left.+ (n+2) \big|\braket{\Phi_\pm}{n+1}\big|^2 \left(\frac{\expo{\beta \Delta_{\pm,n+2}}-1}{\Delta_{\pm,n+2}^2}-\frac{\beta}{\Delta_{\pm,n+2}}\right)\right].
\end{aligned}
\end{equation}

The last term to be calculated is $\mathcal{Z}_m^{(2)}$. The first steps of this calculation are similar to those from the evaluation of $\mathcal{Z}_{\pm}^{(2)}$. Therefore, we have
\begin{equation}
	\begin{aligned}
		\mathcal{Z}_m^{(2)} =& \sum_{m \in \Q} \expo{-\beta E_m} \int_{0}^{\beta} d\tau_1 \int_{0}^{\tau_1} 
		d\tau_2 \,  \bra{m} \expo{\tau_1 \hat{\mathcal{H}}_{0}} \left(\hat{\PP}\hat{V}
		\hat{\Q}+\hat{\Q}\hat{V}\hat{\PP}+\hat{\Q}\hat{V}\hat{\Q}\right) \expo{-\tau_1 \hat{\mathcal{H}}_{0}} \\
		&\times\expo{\tau_2 \hat{\mathcal{H}}_{0}} \left(\hat{\PP}\hat{V}\hat{\Q}+\hat{\Q}\hat{V}
		\hat{\PP}+\hat{\Q}\hat{V}\hat{\Q}\right) \expo{-\tau_2 \hat{\mathcal{H}}_{0}} \ket{m} \\
		=& \sum_{m \in \Q} \expo{-\beta E_m} \int_{0}^{\beta} d\tau_1 \int_{0}^{\tau_1} d\tau_2 \, 
		\expo{(\tau_1-\tau_2)E_m} \bra{m} \hat{V} \expo{-\tau_1 \hat{\mathcal{H}}_{0}} \expo{\tau_2 \hat{\mathcal{H}}_{0}} \hat{V}\ket{m} \\
		=& J^2 z^2 \sum_{m \in \Q} \expo{-\beta E_m} \int_{0}^{\beta} d\tau_1 \int_{0}^{\tau_1} d\tau_2 \, \expo{(\tau_1-\tau_2)E_m} \left(\Psi \sqrt{m} \bra{m-1} + \Psi^* \sqrt{m+1} \bra{m+1} \right)  \\
		&\times\expo{-\tau_1 \hat{\mathcal{H}}_{0}} \expo{\tau_2 \hat{\mathcal{H}}_{0}} \left(\Psi^* \sqrt{m} \ket{m-1} + \Psi \sqrt{m+1} \ket{m+1} \right).
\end{aligned}
\end{equation}
Applying the exponential operators to the eigenstates, we are left with
\begin{equation}
\begin{aligned}
	\mathcal{Z}_m^{(2)}=& \, J^2 z^2 \sum_{m \in \Q} \expo{-\beta E_m} \int_{0}^{\beta} d\tau_1 \int_{0}^{\tau_1} d\tau_2 \, \expo{(\tau_1-\tau_2)E_m} \\
	&\times \Big[\Psi \sqrt{m} \Big(\expo{-\tau_1 \mathcal{E}_+} \braket{m-1}{\Phi_+} \bra{\Phi_+} + \expo{-\tau_1 \mathcal{E}_-} \braket{m-1}{\Phi_-} \bra{\Phi_-} \\
	&+\sum_{m' \in \Q} \expo{-\tau_1 E_{m'}} \braket{m-1}{m'} \bra{m'} \Big) + \Psi^* \sqrt{m+1}  \Big(\expo{-\tau_1 \mathcal{E}_+} \braket{m+1}{\Phi_+} \bra{\Phi_+} \\
	&+ \expo{-\tau_1 \mathcal{E}_-} \braket{m+1}{\Phi_-} \bra{\Phi_-} + \sum_{m'' \in \Q} \expo{-\tau_1 E_{m''}} \braket{m+1}{m''} \bra{m''} \Big)\Big]  \\
	&\times \Big[\Psi^* \sqrt{m} \Big(\expo{\tau_2 \mathcal{E}_+} \braket{\Phi_+}{m-1} \ket{\Phi_+} + \expo{\tau_2 \mathcal{E}_-} \braket{\Phi_-}{m-1} \ket{\Phi_-} \\
	&+ \sum_{m'''\in \Q} \expo{\tau_2 E_{m'''}} \braket{m'''}{m-1} \ket{m'''} \Big) + \Psi \sqrt{m+1} \Big(\expo{\tau_2 \mathcal{E}_+} \braket{\Phi_+}{m+1} \ket{\Phi_+} \\
	&+ \expo{\tau_2 \mathcal{E}_-} \braket{\Phi_-}{m+1} \ket{\Phi_-} + \sum_{m''''\in \Q} \expo{\tau_2 E_{m''''}} \braket{m''''}{m+1} \ket{m''''} \Big) \Big].
\end{aligned}
\end{equation}
When we evaluate the multiplication among the terms between brackets, we must be aware of the fact that the cross terms, \textit{i.e.}, those that contain $\Psi^2$ or $\Psi^{*2}$ vanish
since they contain the products $\braket{m-1}{\Phi_\pm}$ and $\braket{m+1}{\Phi_\pm}$, which cannot be both non-zero because it is not possible 
that $m+1$ and $m-1$ be simultaneously equal to $n$ or $n+1$. With this, we have that 
\begin{equation}
	\begin{aligned}
		\mathcal{Z}_m^{(2)} =& \,J^2 z^2 |\Psi|^2 \sum_{m \in \Q} \expo{-\beta E_m} \int_{0}^{\beta} d\tau_1 \int_{0}^{\tau_1} d\tau_2 \, \Bigg[m \expo{(\tau_1-\tau_2) \Delta_{m,+}} \big|\braket{\Phi_+}{m-1}\big|^2 \\
		& + m \expo{(\tau_1-\tau_2) \Delta_{m,-}} \big|\braket{\Phi_-}{m-1}\big|^2 + m \sum_{m'\in \Q} \expo{(\tau_1-\tau_2) \Delta_{m,m'}} \big|\braket{m-1}{m'}\big|^2\\
		& + (m+1) \expo{(\tau_1-\tau_2) \Delta_{m,+}} \big|\braket{\Phi_+}{m+1}\big|^2 + (m+1) \expo{(\tau_1-\tau_2) \Delta_{m,-}} \big|\braket{\Phi_-}{m+1}\big|^2 \\
		& + (m+1) \sum_{m''\in \Q} \expo{(\tau_1-\tau_2) \Delta_{m,m''}} \big|\braket{m+1}{m''}\big|^2 \Bigg] .
\end{aligned}
\end{equation}
Finally, the integrations lead to
\begin{equation}
	\begin{aligned}\label{Zm}
		\mathcal{Z}_m^{(2)}=&\, J^2 z^2 |\Psi|^2 \sum_{m \in \Q} \expo{-\beta E_m} \Bigg[m \big|\braket{\Phi_+}{m-1}\big|^2 \Bigg(\frac{ \expo{\beta \Delta_{m,+}}-1}{\Delta_{m,+}^2} - \frac{\beta}{\Delta_{m,+}}\Bigg) \\
	& + m \big|\braket{\Phi_-}{m-1}\big|^2 \Bigg(\frac{ \expo{\beta \Delta_{m,-}}-1}{\Delta_{m,-}^2} - \frac{\beta}{\Delta_{m,-}}\Bigg) \\
	&+ m \sum_{m'\in \Q} \Bigg(\frac{ \expo{\beta \Delta_{m,m'}}-1}{\Delta_{m,m'}^2} - \frac{\beta}{\Delta_{m,m'}} \Bigg) \big|\braket{m-1}{m'}\big|^2 \\
	&  + (m+1) \big|\braket{\Phi_+}{m+1}\big|^2 \Bigg(\frac{ \expo{\beta \Delta_{m,+}}-1}{\Delta_{m,+}^2} - \frac{\beta}{\Delta_{m,+}}\Bigg) \\
	& + (m+1) \big|\braket{\Phi_-}{m+1}\big|^2 \Bigg(\frac{ \expo{\beta \Delta_{m,-}}-1}{\Delta_{m,-}^2} - \frac{\beta}{\Delta_{m,-}}\Bigg)  \\
	& + (m+1) \sum_{m''\in \Q} \Bigg(\frac{ \expo{\beta \Delta_{m,m''}}-1}{\Delta_{m,m''}^2} - \frac{\beta}{\Delta_{m,m''}}\Bigg) \big|\braket{m+1}{m''}\big|^2 \Bigg].
\end{aligned}
\end{equation}

Combining the contributions (\ref{Z+-}) and (\ref{Zm}), the second-order term of the partition function reads
\begin{equation}
	\begin{aligned}
		\mathcal{Z}^{(2)}=&\,J^2 z^2 |\Psi|^2 \expo{-\beta \mathcal{E}_+} \Bigg[ n \big|\braket{\Phi_+}{n}\big|^2 \Bigg(\frac{\expo{\beta \Delta_{+,n-1}}-1}{\Delta_{+,n-1}^2}-\frac{\beta}{\Delta_{+,n-1}}\Bigg) \\
	&+ (n+2) \big|\braket{\Phi_+}{n+1}\big|^2 \Bigg(\frac{\expo{\beta \Delta_{+,n+2}}-1}{\Delta_{+,n+2}^2}-\frac{\beta}{\Delta_{+,n+2}}\Bigg)\Bigg] \\
	&+J^2 z^2 |\Psi|^2 \expo{-\beta \mathcal{E}_-} \Bigg[ n \big|\braket{\Phi_-}{n}\big|^2 \Bigg(\frac{\expo{\beta \Delta_{-,n-1}}-1}{\Delta_{-,n-1}^2}-\frac{\beta}{\Delta_{-,n-1}}\Bigg) \\
	&+ (n+2) \big|\braket{\Phi_-}{n+1}\big|^2 \Bigg(\frac{\expo{\beta \Delta_{-,n+2}}-1}{\Delta_{-,n+2}^2}-\frac{\beta}{\Delta_{-,n+2}}\Bigg)\Bigg]  \\
	&+ J^2 z^2 |\Psi|^2 \sum_{m \in \Q} \expo{-\beta E_m} \Bigg[m \big|\braket{\Phi_+}{m-1}\big|^2 \Bigg(\frac{ \expo{\beta \Delta_{m,+}}-1}{\Delta_{m,+}^2} - \frac{\beta}{\Delta_{m,+}}\Bigg) \\
	&+ m \big|\braket{\Phi_-}{m-1}\big|^2 \Bigg(\frac{ \expo{\beta \Delta_{m,-}}-1}{\Delta_{m,-}^2} - \frac{\beta}{\Delta_{m,-}}\Bigg)   \\
	&  + m \sum_{m'\in \Q} \Bigg(\frac{ \expo{\beta \Delta_{m,m'}}-1}{\Delta_{m,m'}^2} - \frac{\beta}{\Delta_{m,m'}} \Bigg) \big|\braket{m-1}{m'}\big|^2 \\
	&+ (m+1) \big|\braket{\Phi_+}{m+1}\big|^2 \Bigg(\frac{ \expo{\beta \Delta_{m,+}}-1}{\Delta_{m,+}^2} - \frac{\beta}{\Delta_{m,+}}\Bigg) \\
	& + (m+1) \big|\braket{\Phi_-}{m+1}\big|^2 \Bigg(\frac{ \expo{\beta \Delta_{m,-}}-1}{\Delta_{m,-}^2} - \frac{\beta}{\Delta_{m,-}}\Bigg) \\
	&+ (m+1) \sum_{m''\in \Q} \Bigg(\frac{ \expo{\beta \Delta_{m,m''}}-1}{\Delta_{m,m''}^2} - \frac{\beta}{\Delta_{m,m''}}\Bigg) \big|\braket{m+1}{m''}\big|^2 \Bigg] .
\end{aligned}
\end{equation}
Now, taking into account that the scalar products $\braket{m-1}{m'}$ and $\braket{m+1}{m''}$ lead to one further restriction each in the summations, we finally obtain
\begin{equation}
	\begin{aligned}
	\label{Z2}
		\mathcal{Z}^{(2)}=&\,J^{2}z^{2}\vert\Psi\vert^{2}\Bigg\{ (n+2)\beta\Bigg[\big|\braket{\Phi_+}{n+1}\big|^{2}\Bigg(\frac{\mathrm{e}^{-\beta\mathcal{E}_{+}}-\mathrm{e}^{-\beta E_{n+2}}}{\Delta_{n+2,+}}\Bigg) \\
	&+\big|\braket{\Phi_-}{n+1}\big|^{2}\Bigg(\frac{\mathrm{e}^{-\beta\mathcal{E}_{-}}-\mathrm{e}^{-\beta E_{n+2}}}{\Delta_{n+2,-}}\Bigg)\Bigg] +n\beta\Bigg[\big|\braket{\Phi_+}{n}\big|^{2}\Bigg(\frac{\mathrm{e}^{-\beta\mathcal{E}_{+}}-\mathrm{e}^{-\beta E_{n-1}}}{\Delta_{n-1,+}}\Bigg) \\
	&+\big|\braket{\Phi_-}{n}\big|^{2}\Bigg(\frac{\mathrm{e}^{-\beta\mathcal{E}_{-}}-\mathrm{e}^{-\beta E_{n-1}}}{\Delta_{n-1,-}}\Bigg)\Bigg] \\
	&+\sum_{\substack{m \in \Q \\ m \neq n-1}}(m+1)\Bigg(\frac{\mathrm{e}^{-\beta E_{m+1}}-\mathrm{e}^{-\beta E_{m}}}{\Delta_{m,m+1}^{2}}-\frac{\beta\mathrm{e}^{-\beta E_{m}}}{\Delta_{m,m+1}}\Bigg) \\
	&+\sum_{\substack{m \in \Q \\ m\neq n+2}}m\Bigg(\frac{\mathrm{e}^{-\beta E_{m-1}}-\mathrm{e}^{-\beta E_{m}}}{\Delta_{m,m-1}^{2}}-\frac{\beta\mathrm{e}^{-\beta E_{m}}}{\Delta_{m,m-1}}\Bigg)\Bigg\}.
\end{aligned}
\end{equation}

From Eq. (\ref{Z2}) we observe that the differences between the degenerate energies $E_n$ and $E_{n+1}$ will no longer appear in the denominator of the free energy as it did in the NDPT treatment, thus solving the degeneracy-related problems.

\section{\label{CD}Condensate density}
Now we turn our attention to the calculation of the condensate density, which turns out to coincide to the superfluid density within the mean-field approximation.\cite{martin} 
Our degenerate approach, up to second order, results in the partition function given by $\mathcal{Z}= \mathcal{Z}^{(0)} + \mathcal{Z}^{(2)}$ 
with (\ref{Z0}) and (\ref{Z2}), which is free from any divergence despite of the degeneracies. Thus, from the partition function, the free energy of the system reads, up to second order, 
\begin{equation}
\mathcal{F} = -\frac{1}{\beta} \left[\ln \mathcal{Z}^{(0)} + \frac{\mathcal{Z}^{(2)}}{\mathcal{Z}^{(0)}} \right].
\end{equation}
Hence, we calculate the condensate density $|\Psi|^2$ by evaluating
\begin{equation}
\label{eq31}
\frac{\partial \mathcal{F}}{\partial |\Psi|^2} = 0.
\end{equation}

\begin{figure}[h] \centering
		\begin{subfigure}{.45\columnwidth}
			\includegraphics[width=\columnwidth]{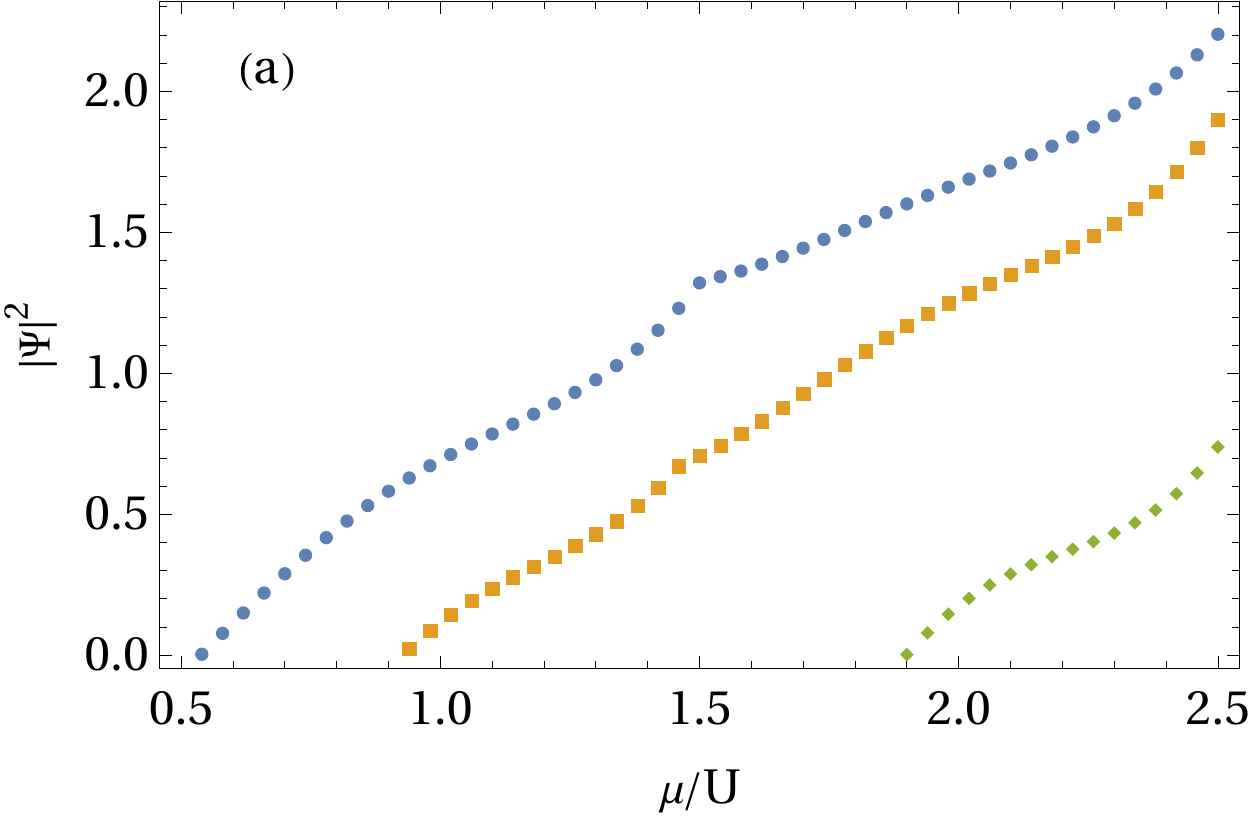}
		\label{cd_finiteT5} \end{subfigure} \qquad
		\begin{subfigure}{.45\columnwidth}
			\includegraphics[width=\columnwidth]{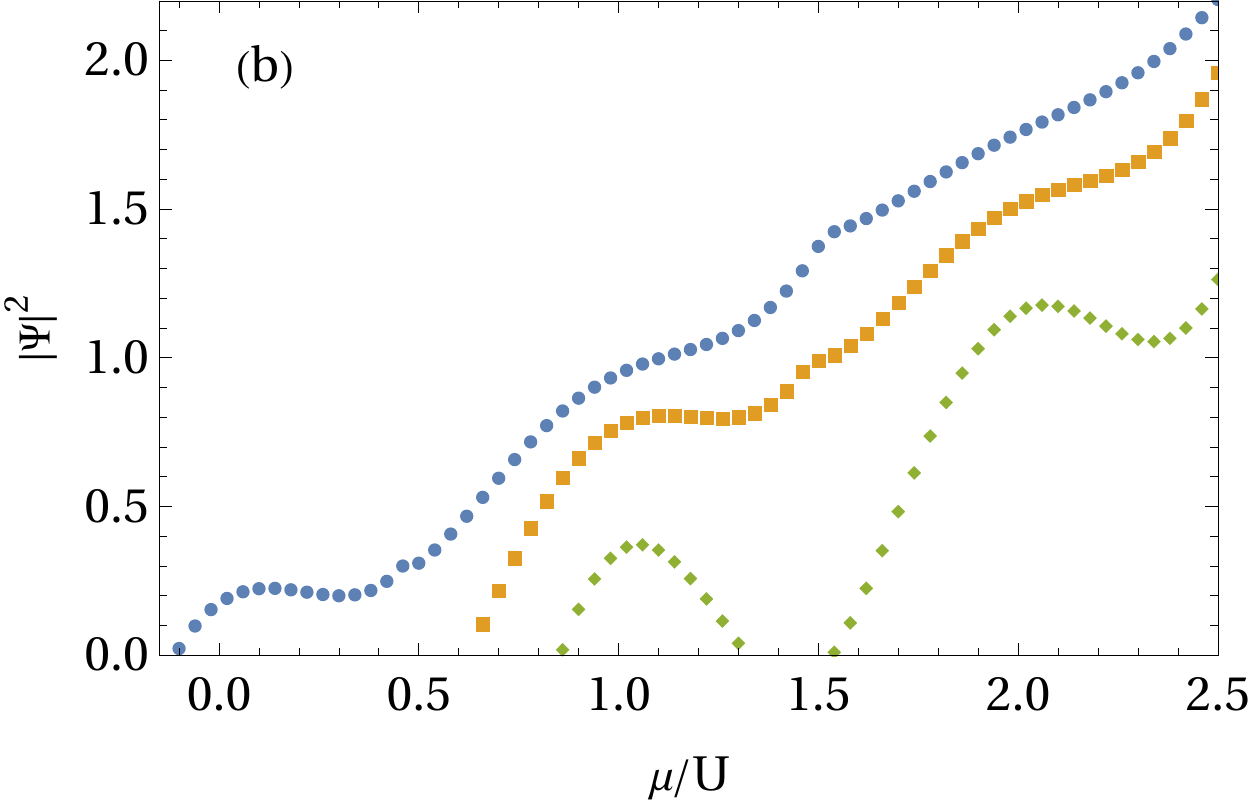}
		\label{cd_finiteT10} \end{subfigure} \qquad
		\begin{subfigure}{.45\columnwidth}
			\includegraphics[width=\columnwidth]{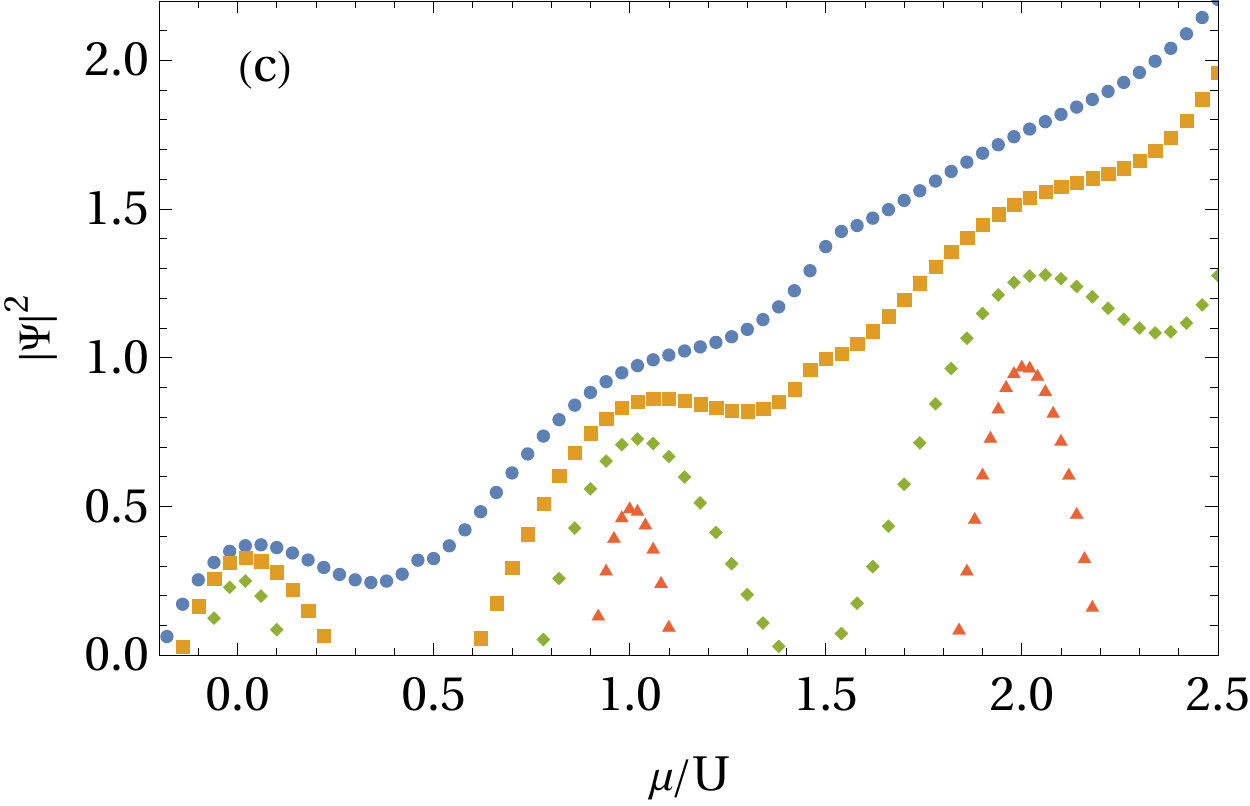}
		\label{cd_finiteT30} \end{subfigure} \qquad
		\begin{subfigure}{.45\columnwidth}
			\includegraphics[width=\columnwidth]{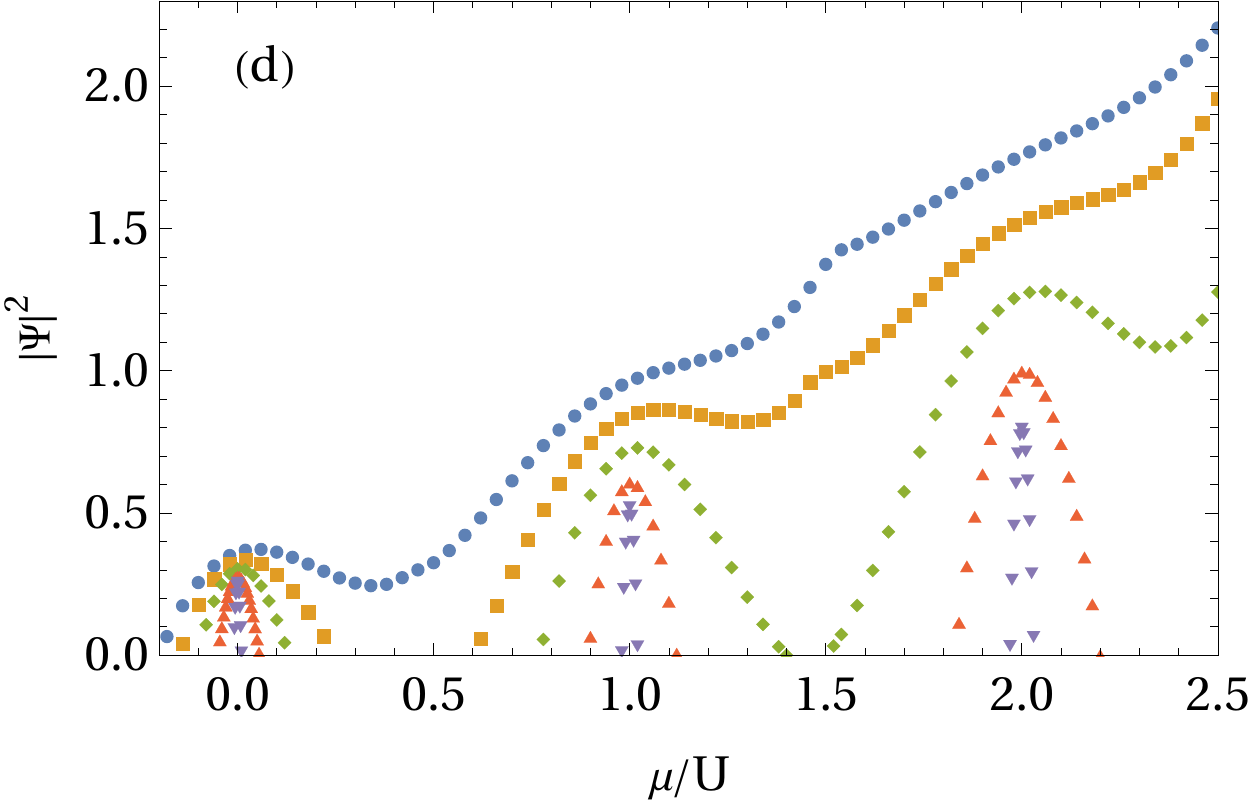}
		\label{cd_zeroT} \end{subfigure}
		\caption{Condensate densities as functions of $\mu/U$ evaluated from FTDPT via (\ref{eq31})
		for four different temperatures: (a) $\beta=5/U$, (b) $\beta=10/U$, (c) $\beta=30/U$, and (d) $T=0$. 
		Different data styles correspond to different hoppings: $Jz/U=0.2$ (blue circles), 
		$Jz/U=0.15$ (orange squares), $Jz/U=0.1$ (green rhombuses), $Jz/U=0.05$ (red triangles), and $Jz/U=0.01$ (purple inverted triangles).}
		Source: SANT'ANA \textit{et al.}\cite{paper2}
		 \label{cond_dens}
	\end{figure}

Now, applying the above-mentioned procedure for different temperatures and hopping values, the resulting condensate 
densities are depicted in Fig. \ref{cond_dens}.
In order to check the fidelity of the calculated condensate densities, we must
observe the phase boundaries evaluated by FTDPT, which emerge from 
\begin{equation} \frac{\partial \mathcal{F}}{\partial
|\Psi|^2}\Bigg|_{\Psi=0}=0.  
\end{equation} 
Such an operation leads to the same
phase diagrams evaluated by NDPT. From Fig. \ref{pb_manyT},
we read off that for small values of $Jz/U$ there are bigger
portions of values of $\mu/U$ where the condensate density can be evaluated,
since we regard the Landau expansion of the order parameter being only valid in the
vicinity of the phase transition, \textit{i.e.}, the smaller the hopping the bigger the
region of the calculated condensate density.  Therefore, we conclude that we
are able to reliably calculate $\left|\Psi\right|^2$ via FTDPT near the phase
boundary in Fig. \ref{cond_dens}. Furthermore, we observe that for $\mu/U\in\mathbb{N}$
the condensate densities no longer vanish or approach zero as they do when
calculated from NDPT.

Let us remark that the results for the zero-temperature 
condensate density from FTDPT, which is depicted in Fig.
\ref{cond_dens}(d), are similar to those obtained from BWPT in Fig. \ref{20OPS}. 
Also, it is important to note that we have restricted ourselves to the second-order term. 
This can be explained by the following: in the NDPT approach, 
the fourth-order term is necessary for the calculation of the condensate density since
it corresponds to the first nontrivial solution of the extremization equation for the free energy, 
see Eqs. (\ref{eq.dFdpsi}) and (\ref{ordpara}). However, this
is not the case for our FTDPT due to fact that the exact solution of the problem in the projected Hilbert space
automatically generates higher-order terms. Therefore, our FTDPT is an effective resummation of the power series
generated by NDPT. In the FTDPT, the second- and higher-order calculations only include extra effects due to the
non-projected Hilbert space. Indeed, it is even possible to calculate the condensate density from the zeroth-order term,
as one can observe from Eq. (\ref{Z0}), since it has an implicit dependency on the OP. Therefore,
we have restricted ourselves to the second-order correction.
 In order to check how important
the fourth-order corrections would be, we compared the 
zero-temperature results from FTDPT to the results from BWPT up to the fourth-order term in the perturbation.
The analogous results, $T=0$, are displayed in Figs \ref{20OPS} and \ref{cond_dens}(d). 
The errors between the BWPT- and FTDPT-calculated $|\Psi|^2$, for the hopping 
strengths $Jz/U=0.2$, $0.15$, $0.1$, $0.05$, and $0.01$, are $4.12\%$, $1.17\%$, $0.69\%$, 
$0.5\%$, and $0.38\%$, respectively. Consequently, we consider the errors to be small enough 
in such a way that it justifies our neglect of the fourth-order term in the FTDPT approach. 

\subsection{\label{comparison}Comparison between NDPT and FTDPT}
\begin{figure}[h] \centering \begin{subfigure}{.45\columnwidth}
		\includegraphics[width=\columnwidth]{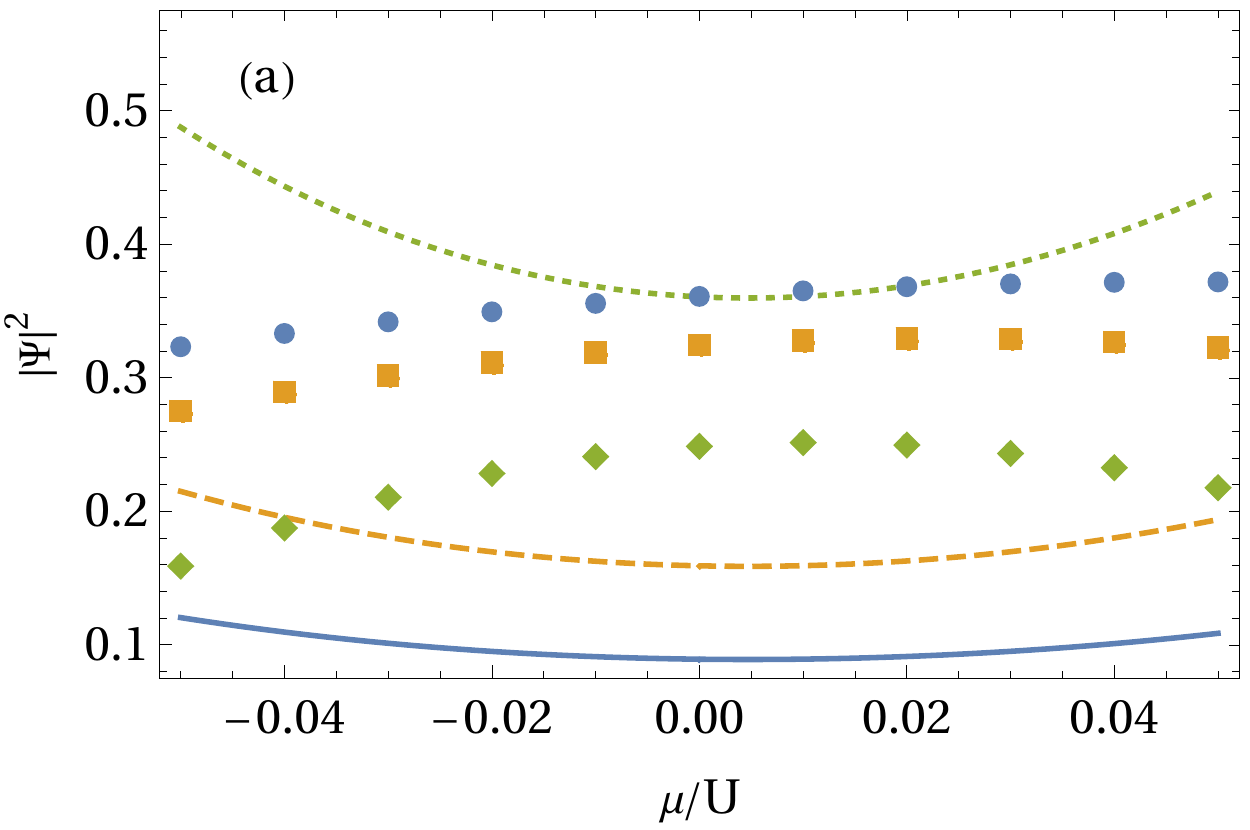} 
			\label{pl1} \end{subfigure} \qquad
			\begin{subfigure}{.45\columnwidth}
				\includegraphics[width=\columnwidth]{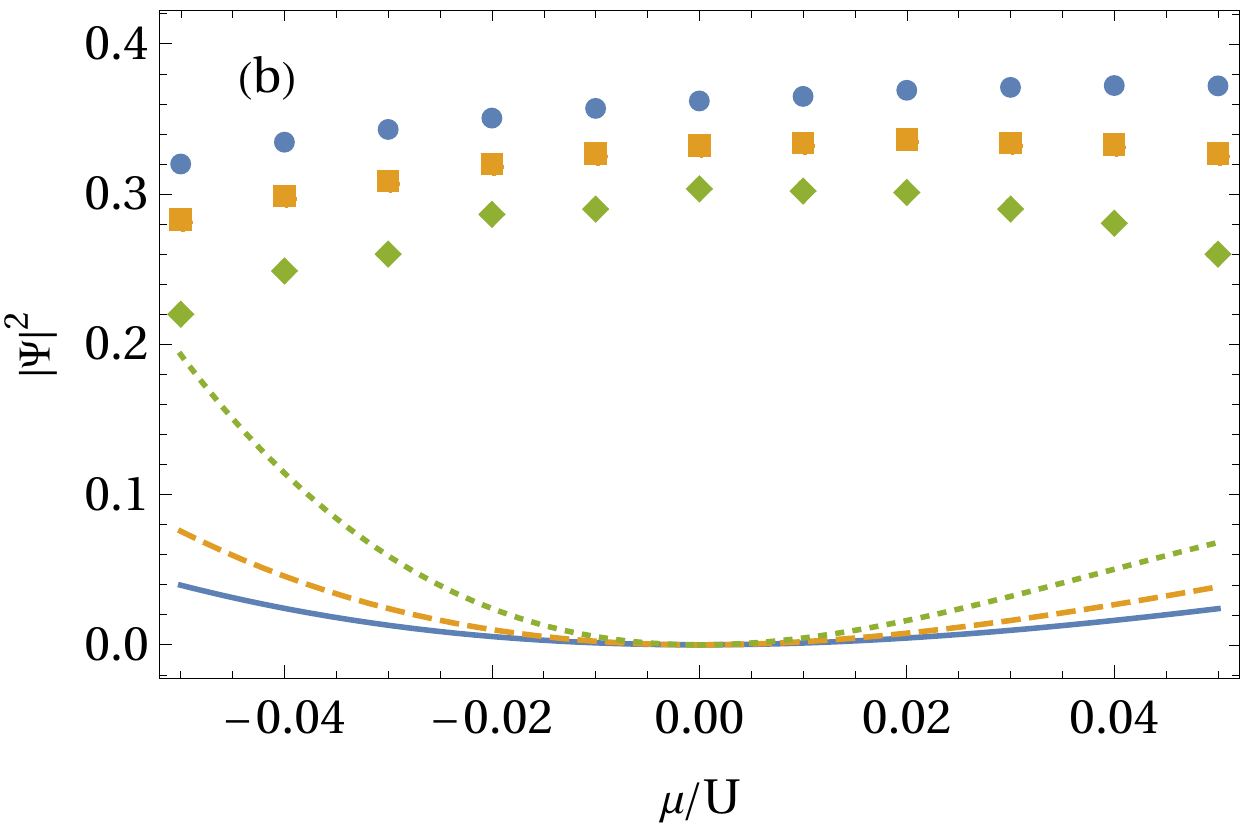}
			\label{pl2} \end{subfigure}
			\begin{subfigure}{.45\columnwidth}
		\includegraphics[width=\columnwidth]{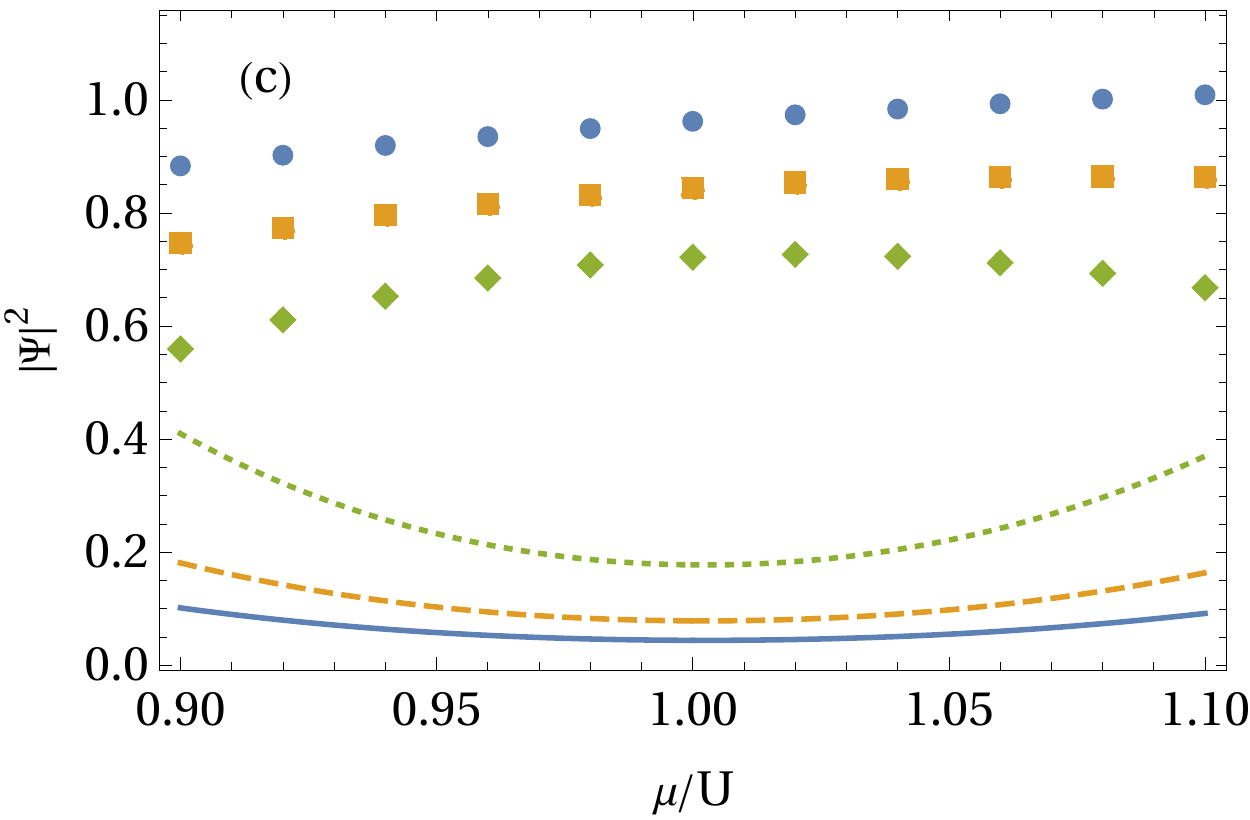} 
			\label{pl} \end{subfigure} \qquad
			\begin{subfigure}{.45\columnwidth}
				\includegraphics[width=\columnwidth]{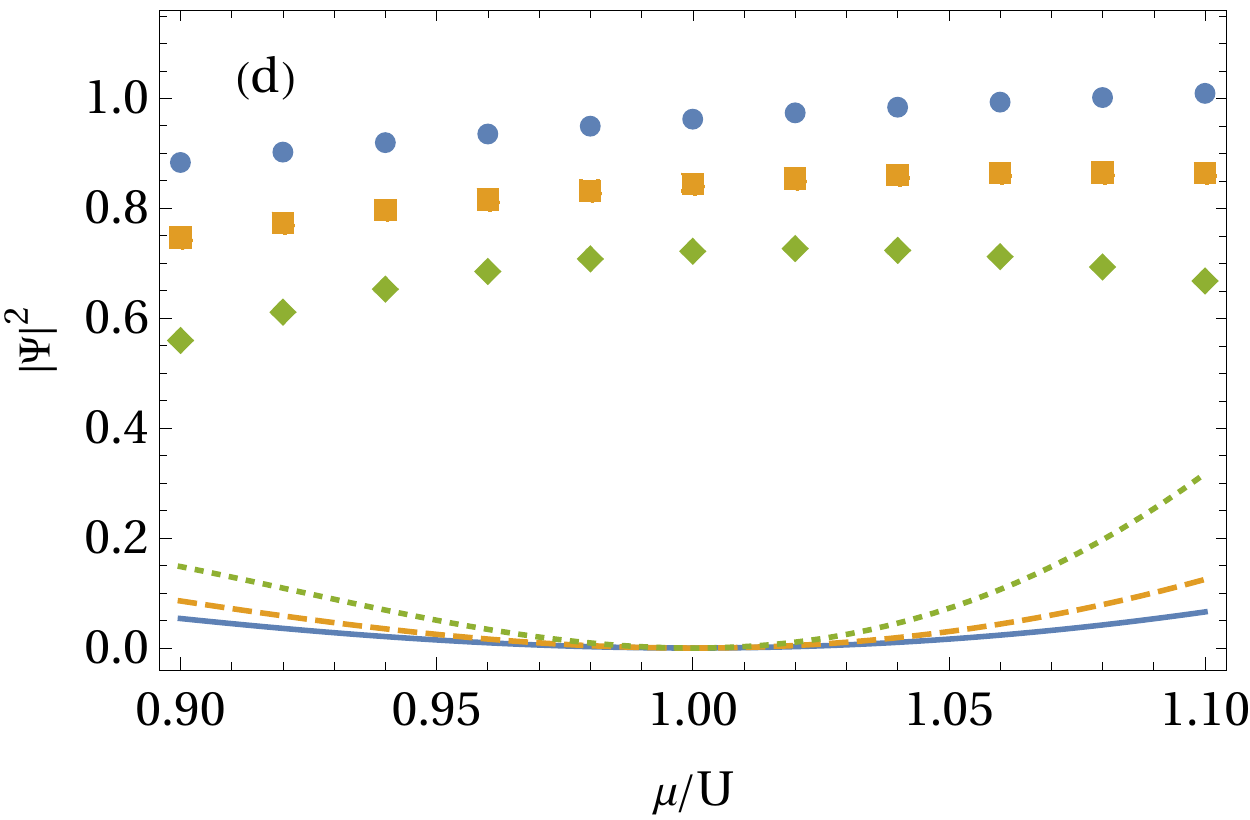}
			\label{l} \end{subfigure} \caption{Comparison between the condensate densities calculated
			via FTDPT (dots) and NDPT (lines) for the
			temperatures $\beta=30/U$ (left panel) and $T=0$ (right
			panel), and for the hoppings $Jz/U=0.2$ (blue circles and continuous blue lines), $Jz/U=0.15$
			(orange squares and dashed orange lines), and $Jz/U=0.1$ (green rhombuses and dotted green lines). (a) and 
			(b) correspond to the region between $n=0$ and 1, while (c)
			and (d) correspond to the region between the first and second lobes.}
			Source: SANT'ANA \textit{et al.}\cite{paper2}
			 \label{d}
			  \end{figure}

Now we turn our attention to the point between two consecutive Mott lobes in
order to analyze the differences between the condensate densities calculated
via NDPT and FTDPT between the Mott lobes $n=0$ and 1, and $n=1$ and 2, as shown in
Fig. \ref{d}. We observe that the NDPT gives rise to 
condensate densities that approach zero or have a decreasing behavior at the
degeneracy point, which correspond to $\mu/U=0$ for the region between $n=0$
and $n=1$ and are depicted in Figs. \ref{d}(a) and \ref{d}(b); while for the region between the
first and the second Mott lobes, \textit{i.e.}, $n=1$ and 2, the degeneracy occurs at
$\mu/U=1$ and the corresponding condensate densities are depicted in Figs. \ref{d}(c) and \ref{d}(d). Such behavior indicates an
inaccuracy of the theory, since it mimics the nonphysical vanishing of the OP
typical of RSPT, which is a direct consequence of not taking into account the
degeneracies that happen in between two consecutive Mott lobes.
While NDPT presents such a nonphysical behavior due to the incorrect treatment of
degeneracies, FTDPT gives consistent results for the condensate density between
two consecutive Mott lobes.

We observe from Fig. \ref{d} that the condensate densities
calculated via FTDPT, which are represented by the data, do not present
any decreasing behavior in the vicinity of the degeneracy, concluding that they
are consistent in all considered regions of the phase diagram. In particular,
at integer $\mu/U$ the condensate densities no longer vanish or present a decreasing behavior as they do when
calculated from NDPT. The decreasing behavior presented by the
condensate densities calculated via NDPT can clearly be observed by the curves
in Fig. \ref{d}. Such decreasing behavior is a direct
consequence of the incorrect treatment of degeneracies by NDPT, which happens to occur
between two consecutive Mott lobes.

\section{\label{density}Particle density}
\begin{figure}[h] \centering
		
		\begin{subfigure}{.45\columnwidth}
			\includegraphics[width=\columnwidth]{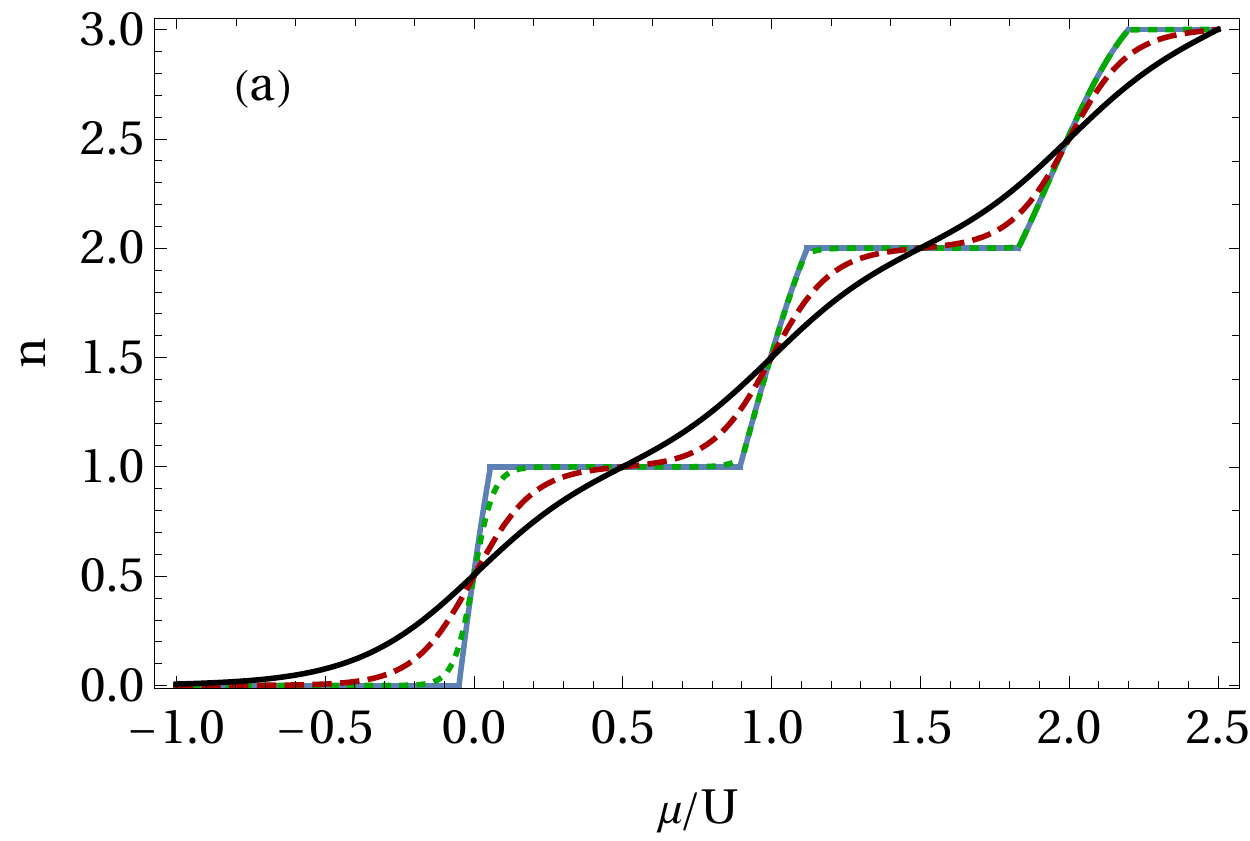} 
			\label{pd1} \end{subfigure} \qquad
			\begin{subfigure}{.45\columnwidth}
				\includegraphics[width=\columnwidth]{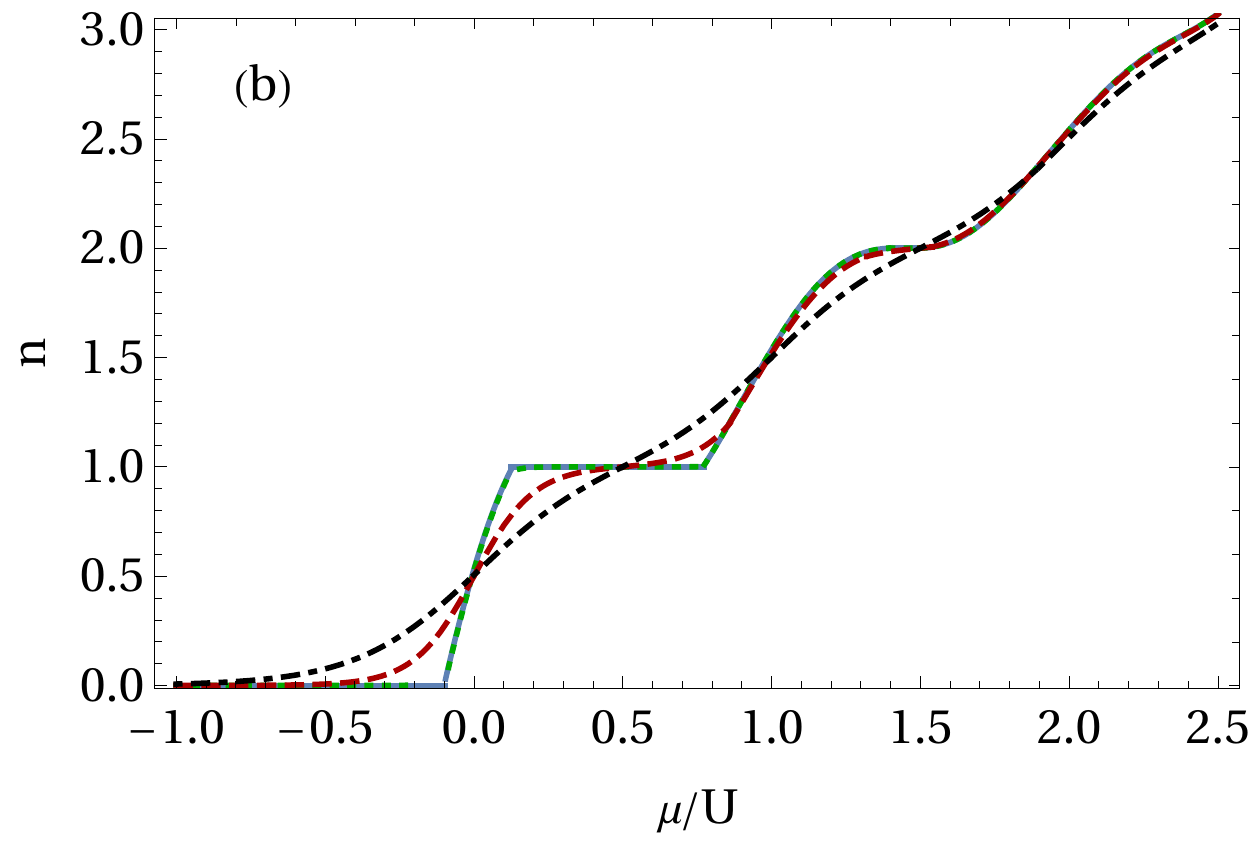} 
			\label{pd2} \end{subfigure} \caption{Equation of state for the hopping strengths (a) $Jz/U=0.05$
			and (b) $Jz/U=0.1$ and the
			temperatures $T=0$ (continuous blue), $\beta=30/U$
			(dotted green), $\beta=10/U$ (dashed red), and
			$\beta=5/U$ (dotted-dashed black).}
			Source: SANT'ANA \textit{et al.}\cite{paper2}
			 \label{partdens}
	\end{figure}

Now, let us calculate the particle density (\ref{pd}) by making use of our developed FTDPT.
We consider different temperatures and different hopping values for the purpose
of analyzing their effects on the density of particles. We plot the resulting
equation of state for two different values of the hopping parameter and four different values of the temperature,
thus observing the melting of the structure,\cite{bloch2,gerbier2} as
shown in Fig. \ref{partdens}.

We observe the effects that the change of both the temperature and the hopping have upon the
particle density in Fig. \ref{partdens}. First, we conclude that increasing
the temperature makes the particle density to vary more smoothly when compared
to those particle densities with lower temperatures. This fact is due to
thermal fluctuations, which make the system more feasible to exist in the
Mott insulator phase. Moreover, by comparing the left panel to the right one we observe
the melting of the Mott lobes due to an increased hopping, which is also very
intuitive: the particles, having more kinetic energy, are more likely to hop
from one site to another, which is a characteristic of the SF phase. 
Another factor responsible for making the curves smoother is the increase of the
chemical potential, $\mu/U$. The reason for this relies on the fact that the
bigger $\mu/U$ becomes, the smaller the Mott lobes are, as can be seen in Fig.
\ref{pb_manyT}. Thus, the system is more likely to exist in the superfluid
phase for bigger values of $\mu/U$.

Now we must turn our attention to the points of the figures where the
degeneracies happen, which correspond to $\mu/U=0,\,1$ and $2$. We observe
that our calculations lead to no nonphysical behavior happening at those regions, meaning that
our developed FTDPT method possesses no inconsistency in the calculation of the
equation of state for the mean-field approximation of bosonic atoms confined in
optical lattices. Finally, as the NDPT leads to a weird behavior of the particle densities, together 
with divergences and discontinuities in the vicinity of the degeneracies, \textit{i.e.}, $\mu/U\in\mathbb{N}$,
we also conclude that FTDPT provides reliable results for the particle density since there is no decreasing behavior 
or discontinuities in the vicinity of the degeneracies in Fig. \ref{partdens}.



\chapter{One-dimensional interacting Bose gas\label{1d}}
The purpose of this chapter is the description of 
one-dimensional repulsively interacting bosonic particles, also known as 
Lieb-Liniger gas,\cite{lieb-liniger} trapped in a harmonic confinement 
at finite temperature. 
We begin by briefly discussing the model and the solutions for the 
homogeneous gas and the role of the interactions.  
Then, we 
study the details behind the solutions for the two-particle 
case, followed by a development of the asymptotic behavior of the 
momentum distribution. In such an asymptotic context, we introduce an important 
physical quantity that gives us valuable short-range 
insights about the system, the so-called \textit{Tan's contact}. 
Moreover, we take our studies beyond the two-particle scenario up to 
the $N$-particle system. As our knowledge and abilities to solve the problem 
are reduced in such a $N>2$ scenario, \textit{i.e.}, for $N>2$ the system 
can be analytically solved only in the strongly interacting limit, 
also known as \textit{Tonks-Girardeau gas} (TG gas). 
In this limit, we are able to exactly solve the problem due to the \textit{fermionization} 
of the bosons. In the last part of this chapter, we study the scaling properties of 
the Tan's contact in all ranges of temperatures and in the intermediate- 
and strong-interaction regimes. To finalize, we compare our analytical results to quantum Monte Carlo (QMC) simulations.

\section{The model}
To begin with, let us consider the one-dimensional 
system consisted of $N$ bosons of mass $m$ repulsively interacting via a delta potential and confined in a
generic potential of the form $V(x)$. The Hamiltonian describing such a system is given by
\begin{equation}\label{H-1d-N}
	\hat{H} =  \sum_{i=1}^N \left(-\frac{\hbar^2}{2m} \frac{\partial^2}{\partial x_i^2} + V(x_i)\right) + g \sum_{i<j} \delta(x_i-x_j),
\end{equation}
where the interaction strength depends on the 1D scattering length as $g=-2\hbar^2/ma_{1\mathrm{D}}$.\footnote{For more details on the 
one-dimensional interaction strength, see App. \ref{appendixA}.}
Also, repulsive interaction means positive interaction strength, $g> 0$. 
Then, the respective stationary Schr\"odinger equation reads
\begin{equation}\label{Nwave}
	\left[\sum_{i=1}^N \left(-\frac{\hbar^2}{2m} \frac{\partial^2}{\partial x_i^2} + V(x_i)\right) 
	+ g \sum_{i<j} \delta(x_i-x_j) -E\right] \Psi\left(x_1,x_2,\dots,x_N\right) = 0.
\end{equation}
Considering any pair of particles $(i,j)$, let us integrate Eq. (\ref{Nwave}) in the vicinity of the 
interaction $x_{ij}\equiv x_i - x_j=0$, \textit{i.e.}, inside the small interval $(-\varepsilon,+\varepsilon)$:
\begin{equation}
	\begin{aligned}
		&\int_{-\varepsilon}^{+\varepsilon} \left[\left(-\frac{\hbar^2}{2m} \frac{\partial^2}{\partial x_{ij}^2} 
		+ V(x_{ij})+ g \delta(x_{ij}) -E\right) \Psi\left(x_1,x_2,\dots,x_N\right) \right] dx_{ij} = 0 \\
		&\Rightarrow -\frac{\hbar^2}{2m} \frac{\partial \Psi}{\partial x_{ij}}\Bigg|_{-\varepsilon}^{+\varepsilon}
		+ g \Psi\left(x_{ij}=0\right) = 0.
	\end{aligned}
\end{equation}
Therefore, as $\varepsilon \to 0$, the contact interaction generates a condition given by
\begin{equation}
	\label{cusp}
\left(\frac{\partial\Psi}{\partial x_{i}}-\frac{\partial\Psi}{\partial x_{j}}\right)\Bigg|_{x_{i}-x_{j}\rightarrow0^{+}}
	-\left(\frac{\partial\Psi}{\partial x_{i}}-\frac{\partial\Psi}{\partial x_{j}}\right)\Bigg|_{x_{i}-x_{j}\rightarrow0^{-}}
	=\frac{2mg}{\hbar^{2}}\Psi(x_{i}=x_{j}),
\end{equation}
which can be interpreted as the following: whenever two particles meet, there happens 
a discontinuity in the many-body wave function and it abruptly falls to zero.

\subsection{The Lieb-Liniger gas}
The case of free Bosons, $V(x)=0$, interacting via a delta-like potential 
is known as \textit{Lieb-Liniger gas} and its solutions are given by\cite{lieb-liniger}
\begin{equation}\label{liebsol}
	\Psi(x_1,\dots,x_N) = \sum_P a(P)  \exp\left(
	i\sum_{j=1}^N k_j x_j\right),
\end{equation}
where $k_j \in \Re$ with $k_1 < k_2 < \cdots < k_N$,  
and the sum in $P$ is taken over all the $N!$ 
permutations of $1,2,\dots,N$.
The coefficients read 
\begin{equation}\label{aPcoeff}
	a(P) = \prod_{1\leq i <j\leq N} \left(
	1+ \frac{i g}{k_i-k_j}\right),
\end{equation}
and the energy of the system is given by 
\begin{equation}
	E = \sum_{i=1}^N k_i^2.
\end{equation}
By substituting (\ref{aPcoeff}) into (\ref{liebsol}) and 
imposing the periodic condition on a length $L$ \\
$\Psi(x_1,\dots, x_i,\dots,x_N) = \Psi(x_1,\dots, x_i+L,\dots,x_N)$, 
one obtains a set of $N$ equations that allow the determination of the $k_i$'s:\cite{ueda} 
\begin{equation}
	k_i L = 2\pi n_i -2 \sum_{j=1}^N \tan ^{-1} \left(
	\frac{k_i-k_j}{g}\right),
\end{equation}
where $n_1 < n_2 < \cdots < n_N$ are integers if $N$ is odd and 
half-integers if $N$ is even.

\section{The two particles case}
Let us consider two particles harmonically trapped 
interacting via a delta-like potential. This case corresponds to 
considering $N=2$ and  $V(x_i)=m\omega^2x_i^2/2$ within the Hamiltonian (\ref{H-1d-N}): 
\begin{equation}
	\hat{H}=-\frac{\hbar^2}{2m} \left(\frac{\partial^2}{\partial x_1^2}+\frac{\partial^2}{\partial x_2^2}\right)
	+\frac{1}{2}m\omega^2\left(x_1^2+x_2^2\right)+g\delta(x_1-x_2).
\end{equation}
In order to solve the respective stationary Schr\"odinger equation, we perform a change of coordinates:
$x_{cm}\equiv (x_1+x_2)/2$ is the center of mass coordinate and $x_r \equiv x_1-x_2$
is the relative coordinate. Also, the effective masses are $M \equiv 2m$ and $\mu \equiv m/2$. 
With these changes the Hamiltonian reads
\begin{equation}\label{hamil}
	\hat{H}=-\frac{\hbar^2}{2M} \frac{\partial^2}{\partial x_{cm}^2} +\frac{1}{2}M\omega^{2} x_{cm}^2 
	-\frac{\hbar^2}{2 \mu} \frac{\partial^2}{\partial x_r^2} +\frac{1}{2}\mu\omega^2 x_r^2 + g\delta(x_r).
\end{equation}
The solutions for the center of mass coordinate equation are the known solutions for the harmonic oscillator,\cite{cohen,sakurai,griffiths,messiah}
\begin{equation}\label{psiCM}
	\Psi^{(cm)}_n(x_{cm}) =  \frac{\expo{-\left(x_{cm}/a_0\right)^2/2} H_n\left(x_{cm}/a_0\right)}{\pi^{1/4} \sqrt{a_0 2^n n!}},
\end{equation}
where $a_0 = \sqrt{\hbar/M \omega}$ is the harmonic oscillator length and $H_n$ are the Hermite polynomials,\cite{handbook,arfken} 
with $n\in \mathbb{N}$. The energies are given by $E^{(cm)}_n = (n+1/2)\hbar \omega$. 

\begin{figure}[h!] \centering
			\includegraphics[width=0.8\columnwidth]{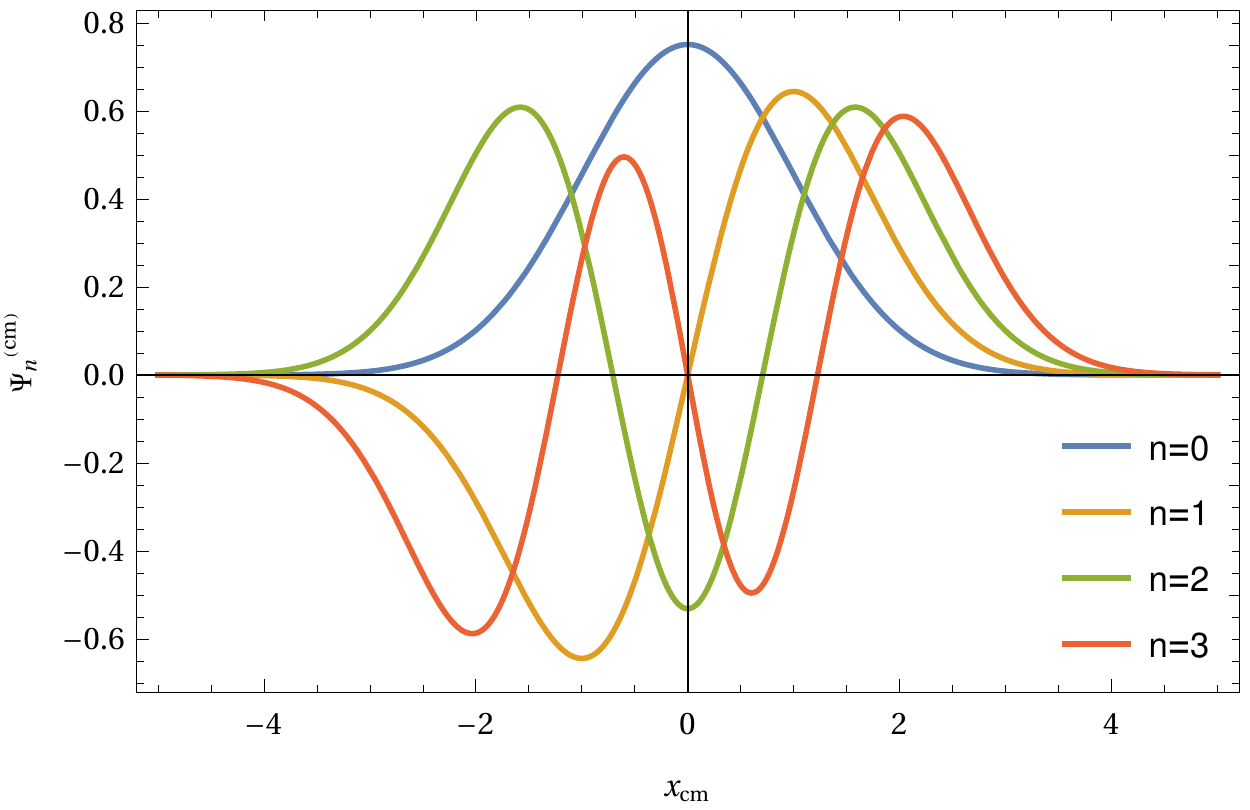}
			\caption{The fundamental and the first three excited states of the harmonic oscillator wave function (\ref{psiCM}).}
			Source: By the author.
		\label{fig.psicm}
\end{figure}

Regarding the equation for the relative coordinate, 
the solutions for the analogue three-dimensional system was given by BUSCH \textit{et al.}\cite{busch}
Here, we work out the solution for the one-dimensional system.

Firstly, let us consider the relative motion Hamiltonian given by
\begin{equation}
	\hat{H}^{(r)} =-\frac{\hbar^2}{2 \mu} \frac{\partial^2}{\partial x^2} +\frac{1}{2}\mu\omega^2 x^2 + g\delta(x).
\end{equation}
Now let us write the wave functions, which are solutions of the respective stationary 
Schr\"odinger equation 
\begin{equation}\label{Hr}
	\hat{H}^{(r)} \Psi_{\nu}^{(r)} = E_{\nu}^{(r)} \Psi_{\nu}^{(r)},
\end{equation}
as an expansion of the complete set of the known solutions of the harmonic oscillator 
$\phi_n(x)$:
\begin{equation}\label{expan}
	\Psi_{\nu}^{(r)}(x) = \sum_{n=0}^{\infty} c_n \phi_n(x).
\end{equation}
Inserting (\ref{expan}) into (\ref{Hr}) we have
\begin{equation}\label{soma1}
	\sum_{n=0}^{\infty} c_n (E_n-E_{\nu}) \phi_n(x) + g \delta(x) \sum_{m=0}^{\infty} c_m \phi_m(x) = 0.
\end{equation}
Nota that we have omitted, and will continue omitting during the following analytical demonstrations, 
the identification indexes $^{(r)}$ and $^{(cm)}$ for the sake of simplicity.

Now, multiplying (\ref{soma1}) by $\phi_j^*(x)$ and integrating it all over the real space
$\Re$, we have that 
\begin{equation}
	c_n(E_n-E_{\nu})  + g \phi_n^*(0)  \sum_m c_m \phi_m(0) = 0,
\end{equation}
where we have used the orthogonality of the $\phi$'s 
\begin{equation}
	\int_{-\infty}^{+\infty} dx \, \phi_m(x) \phi_n(x)= \delta_{m,n}.
\end{equation}

We can observe that the coefficients $c_n$ possess the following form:
\begin{equation}
	c_n = A \frac{\phi_n^*(0)}{E_n-E_{\nu}},
\end{equation}
with $A$ being a proportionality constant. Therefore, the solutions 
we are seeking for reduce to
\begin{equation}
	\Psi_{\nu}^{(r)}(x) = \expo{-x^2/2} \sum_{n=0}^{\infty}  \frac{\phi_n^*(0)}{E_n-E_{\nu}} H_n(x).
\end{equation}

As the possible energies are given by $E_{\nu} = (\nu+1/2) \hbar \omega$, we have that 
\begin{equation}
	\Psi_{\nu}^{(r)}(x) = \expo{-x^2/2} \sum_{n=0}^{\infty}  \frac{\phi_n^*(0)}{(n-\nu)\hbar \omega} H_n(x).
\end{equation}
Now we transform the Hermite polynomials into Laguerre ones through their relationships
\begin{subequations}
\begin{align}
	H_{2n}(x)&=(-1)^n 2^{2n} n! L_n^{(-1/2)}(x^2),\\
	H_{2n+1}(x)&=(-1)^n 2^{2n+1} n! x L_n^{(1/2)}(x^2),
\end{align}
\end{subequations}
and make use of the integral representation
\begin{equation}
	\frac{1}{n-\nu} = \int_0^\infty \frac{dy}{(1+y)^2} \left(\frac{y}{1+y}\right)^{n-\nu-1},
\end{equation}
so that we arrive at
\begin{equation}\label{psilaguerre}
	\Psi_{\nu}^{(r)}(x) = \expo{-x^2/2} \sum_{n=0}^{\infty} \int_0^\infty \frac{dy}{(1+y)^2} \left(\frac{y}{1+y}\right)^{n-\nu-1}
	\left[L_n^{(-1/2)}(x^2) + x L_n^{(1/2)}(x^2)\right],
\end{equation}
where we have embedded all the constants into the normalization of the wave function that we shall deal later.

From the generating function of the Laguerre polynomials
\begin{equation}
	\sum_{n=0}^{\infty} L_n^{(\alpha)} (x) t^n = \frac{\expo{-xt/(1-t)}}{(1-t)^{\alpha + 1}},
\end{equation}
(\ref{psilaguerre}) reads
\begin{equation}
        \Psi_{\nu}^{(r)}(x) = \expo{-x^2/2} \int_0^\infty \frac{dy}{(1+y)^2} \left(\frac{y}{1+y}\right)^{-\nu-1}
	\expo{-y x^2}\left[(1+y)^{1/2} + x (1+y)^{3/2}\right].
\end{equation}
At this point, we are able to recognize the integral representation of the Tricomi hypergeometric function\cite{table}
\begin{equation}
	U(a,b,z)= \frac{1}{\Gamma(a)} \int_0^\infty \expo{-z t} t^{a-1} (1+t)^{b-a-1} dt,
\end{equation}
so that
\begin{equation}\label{psiquasi}
        \Psi_{\nu}^{(r)}(x) = \expo{-x^2/2} \Gamma(-\nu) \left[U\left(-\nu,1/2,x^2\right) + x \, U\left(-\nu,3/2,x^2\right)\right],
\end{equation}
where $\Gamma(x)$ is the Euler gamma function.

Before assuming that Eq. (\ref{psiquasi}) is our final form of $\Psi^{(r)}_{\nu}$, 
we must remind ourselves that we got two solutions from the relations between the Hermite and 
the Laguerre polynomials. Hence, let us test these solutions with the help of the condition 
at $x_r=0$, Eq. (\ref{cusp}), which, for the $N=2$ case, reduces to 
\begin{equation}
\label{cuspN=2}
\frac{\partial\Psi^{(r)}_\nu}{\partial x_{r}}\Bigg|_{x_{r}\rightarrow 0^{+}}
-\frac{\partial\Psi^{(r)}_\nu}{\partial x_{r}}\Bigg|_{x_{r}\rightarrow 0^{-}}
=\frac{2\mu g}{\hbar^{2}}\Psi^{(r)}_\nu(0).
\end{equation}
The first solution unrestrictedly satisfies (\ref{cuspN=2}), while the second one 
only satisfies it for integer values of $\nu$, which is a restriction that we never imposed. 
This means that at $x_r=0$, $x_r \, U\left(-\nu,3/2,x_r^2\right) \rightarrow \infty, \, \forall \, \nu \notin \mathbb{N}$. 
Therefore, it can not be a solution because the wave function must vanish when two particles meet, 
$\Psi(x_1=x_2)=0$. In conclusion, we have that the final solution for the relative motion problem is given by
\begin{equation}
\label{psiR}
	\Psi_{\nu}^{(r)}(x_r) = \sqrt{\frac{1}{\mathcal{N}(\nu)}} \, \expo{-(x_r/a_0)^2/2} \, U\left(-\frac{\nu}{2},\frac{1}{2},\frac{x_r^2}{a_0^2}\right),
\end{equation}
where now the oscillator length reads $a_0 = \sqrt{\hbar /\mu \omega}$. Also, it is worth 
noting that we rescaled the factor $\nu$ by 2, and this can be done without loss of generality. 
We did it in order to keep the solutions more similar to the solutions of the harmonic oscillator, 
once $H_n(x) = 2^n U(-n/2,1/2,x^2)$. After all, the interacting system is simply the same as the 
harmonic oscillator except at the contact point $x_r=0$.

The normalization is given by 
\begin{equation}\label{norm}
	\mathcal{N}(\nu) = \frac{1}{2} \int dx \, \expo{-x^2/a_0^2}  \Big|U\left(-\frac{\nu}{2},\frac{1}{2},\frac{x^2}{a_0^2}\right)\Big|^2 =
		a_0 \sqrt{\pi} \frac{2^{-\nu -2}}{\Gamma (-\nu )}
	       \left[\psi\left(\frac{1-\nu }{2}\right)-\psi\left(-\frac{\nu }{2}\right)\right],  
\end{equation}
where $\psi(x)=\Gamma^\prime(x)/\Gamma(x)$ is the digamma function.\cite{handbook,arfken}

\begin{figure}[h!]
	\centering
		\begin{subfigure}{.45\columnwidth}
			\includegraphics[width=\columnwidth]{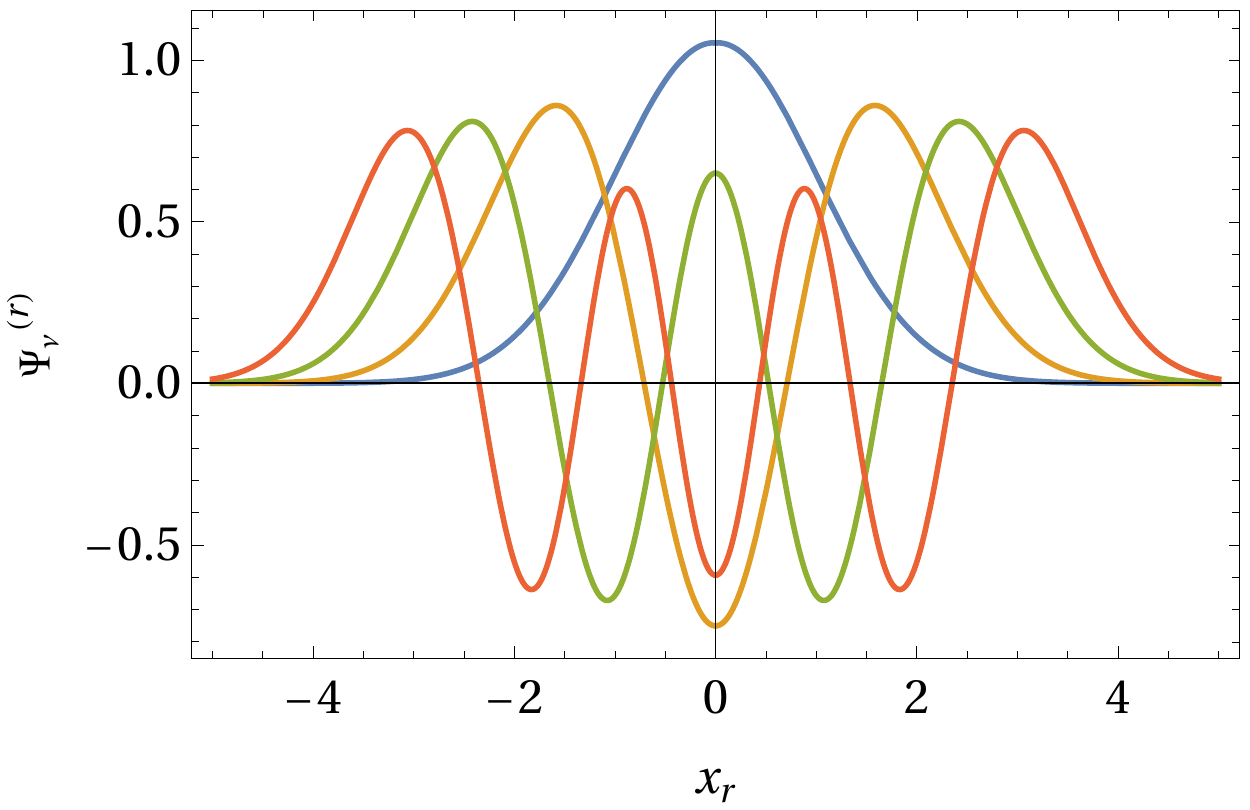}
			\caption{$\tilde{g}=0.01$.}
		\label{psir001} \end{subfigure} \qquad
		\begin{subfigure}{.45\columnwidth}
			\includegraphics[width=\columnwidth]{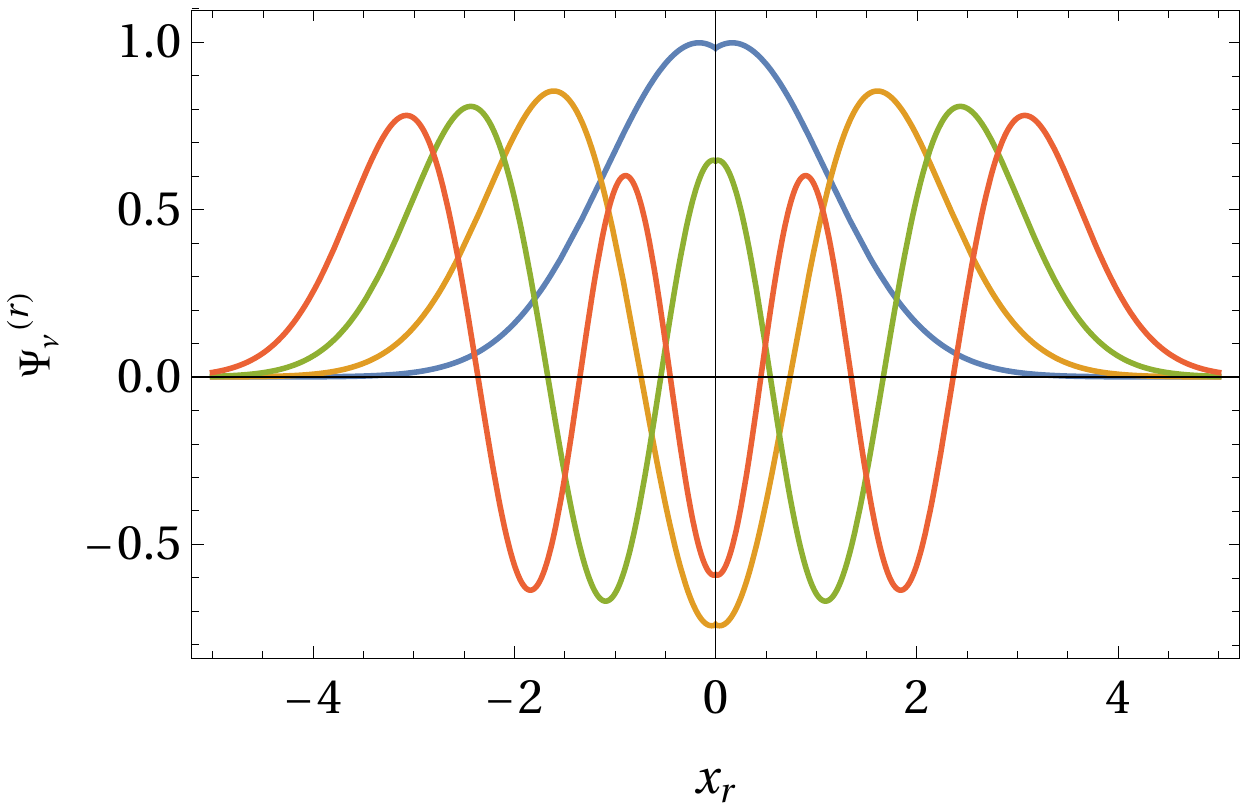}
			\caption{$\tilde{g}=0.1$.}
		\label{psir01} \end{subfigure}
		\begin{subfigure}{.45\columnwidth}
			\includegraphics[width=\columnwidth]{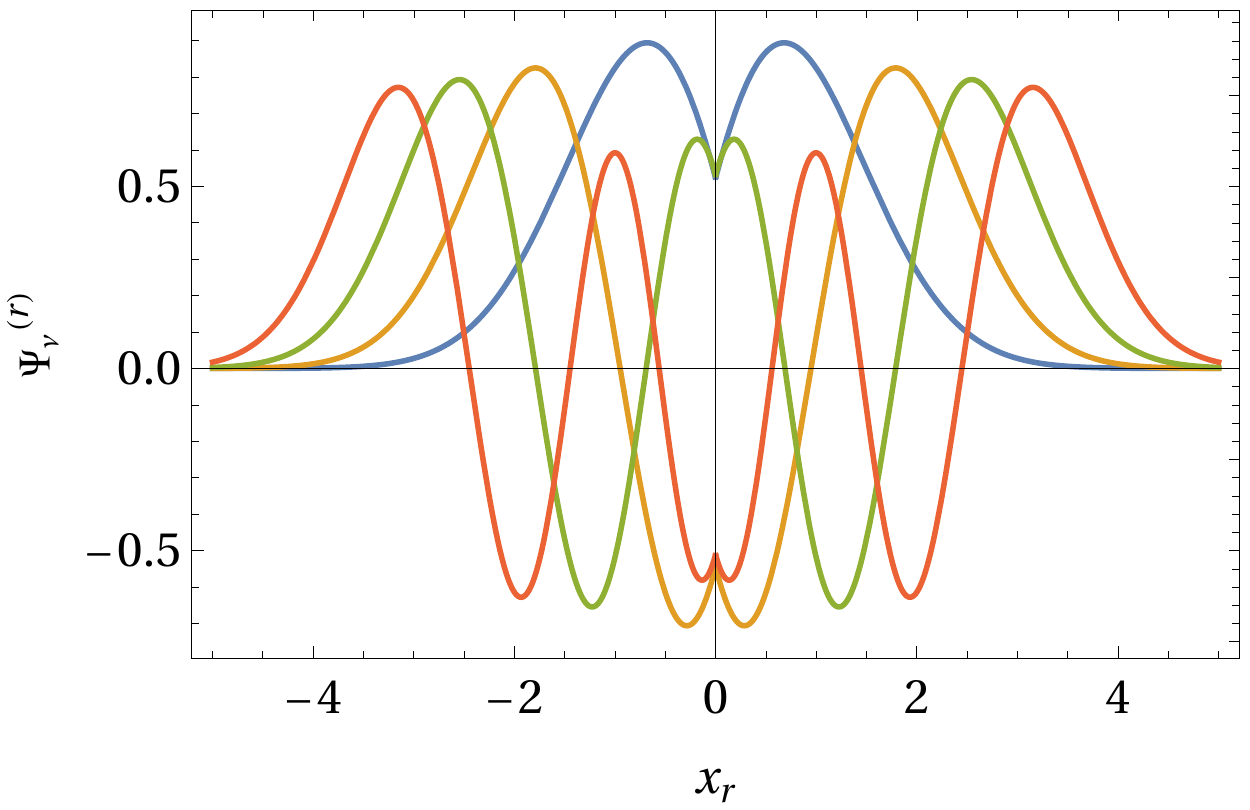}
			\caption{$\tilde{g}=1$.}
		\label{psir1} \end{subfigure}
		\begin{subfigure}{.45\columnwidth}
			\includegraphics[width=\columnwidth]{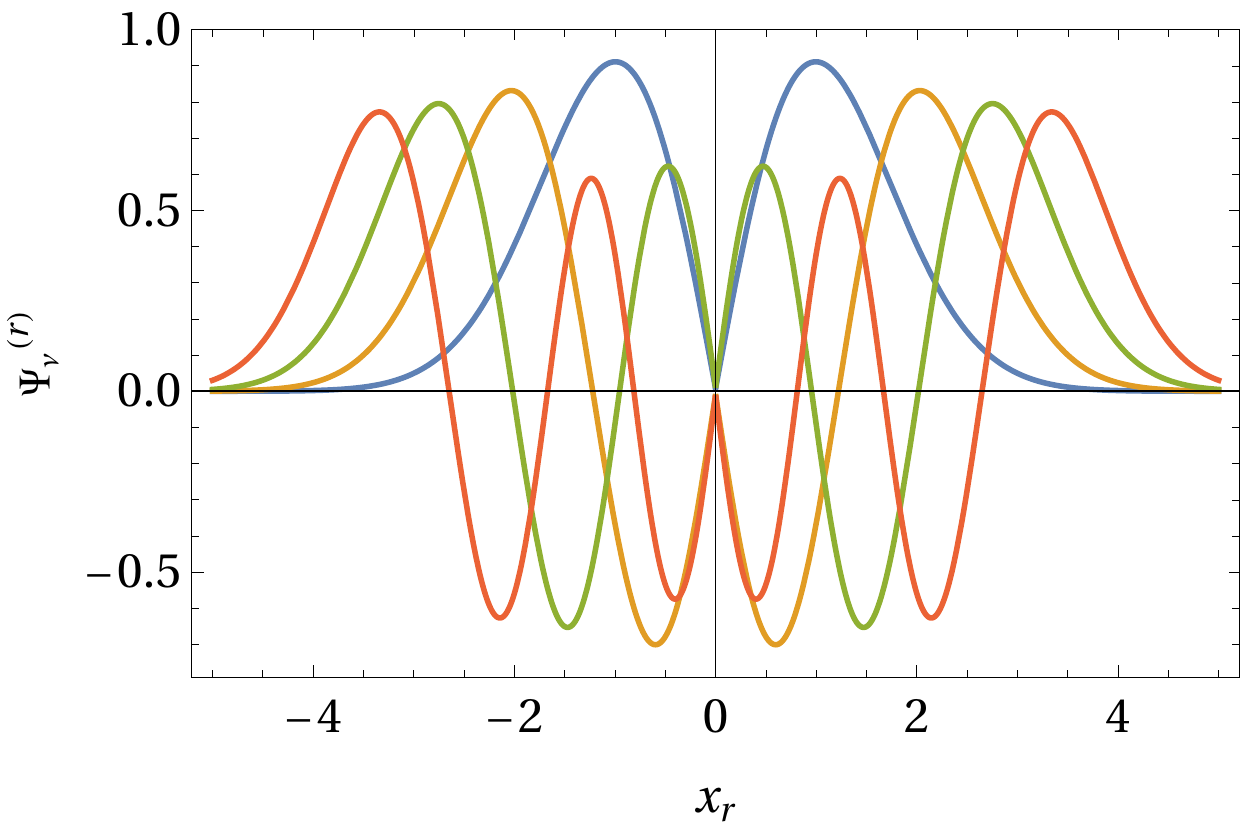}
			\caption{$\tilde{g}=100$.}
		\label{psir10} \end{subfigure}
		\caption{Relative motion wave functions for different interaction strength $\tilde{g}$. Different 
		colors correspond to different states, from the fundamental state to the third excited one. In growing order of
		 excitation they correspond to the blue, yellow, green, and red curves.}
		 Source: By the author.
		  \label{psir} 
\end{figure}

The behavior of the relative motion wave function is depicted in Fig. \ref{psir} for different values of the adimensional interaction 
parameter $\tilde{g} \equiv -a_0/\sqrt{2} a_{1\mathrm{D}}$.
We can observe its discontinuity happening at $x_1=x_2$. The higher the interaction, the more pronounced it becomes. 
Also, its expansion around $x=0$ is
\begin{equation}\label{psiexp}
	\Psi^{(r)}_\nu(x) \sim \sqrt{\frac{\pi}{\mathcal{N}(\nu)}}
	\left[\Gamma\left(\frac{1}{2}-\frac{\nu}{2}\right)^{-1}-2\,\Gamma\left(-\frac{\nu}{2}\right)^{-1}\frac{|x|}{a_0}+\mathcal{O}(x^2)\right].
\end{equation}
Conclusively, it becomes clear, not only from Fig. \ref{psir}, but also from Eq. (\ref{psiexp}), 
the $|x|$-behavior of $\Psi^{(r)}_\nu(x)$ around $x=0$.

Inserting (\ref{psiR}) into (\ref{cuspN=2}) we obtain the following relation:
\begin{equation}\label{eq.gamma}
	f(\nu) \equiv \frac{\Gamma\left(-\frac{\nu}{2}\right)}{\Gamma\left(\frac{1-\nu}{2}\right)}=-\frac{1}{\tilde{g}}.
\end{equation}
Thus, we are able to find the $\nu$'s, which are the quantum numbers with respect to the 
relative motion wave function (\ref{psiR}), by solving (\ref{eq.gamma}).

\begin{figure}[h!]
	\centering
	\includegraphics[width=\columnwidth]{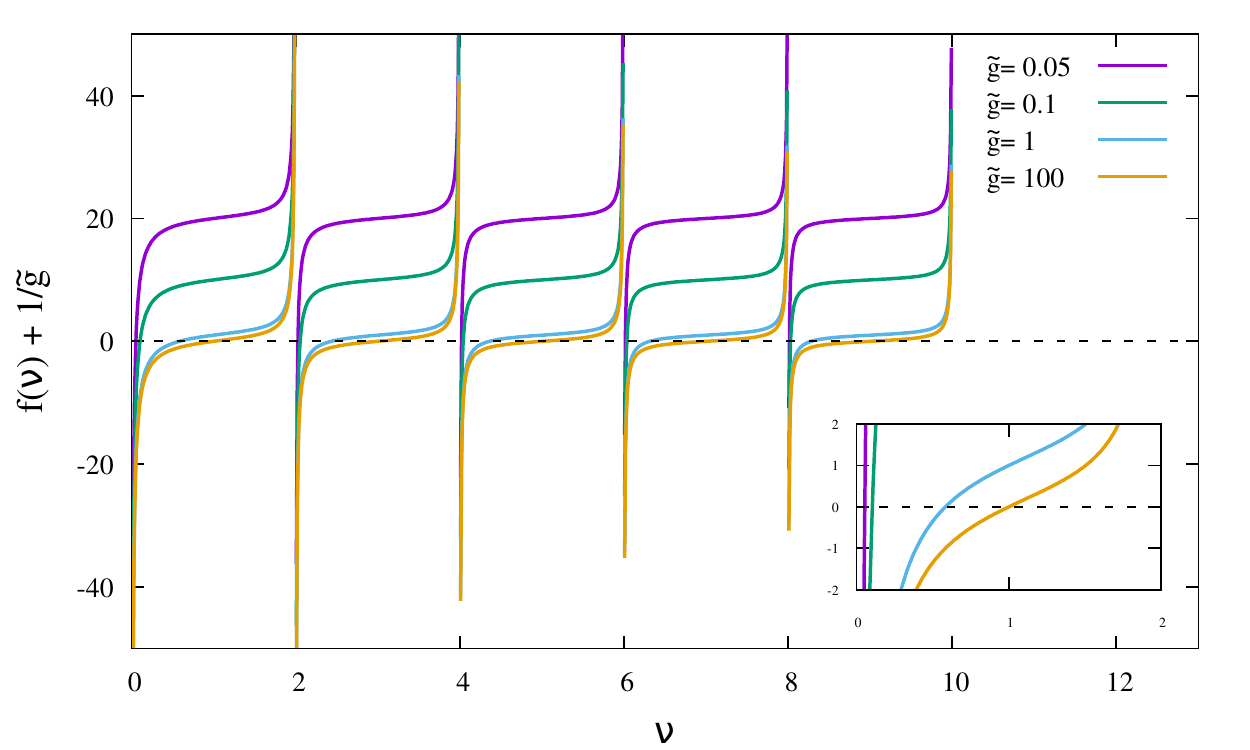}
	\caption{$f(\nu) + 1/\tilde{g}$ as a function of $\nu$ for different adimensional interaction strengths $\tilde{g}$. The zoomed inset
	helps us recognize the vanishing of the function between $\nu=0$ (weakly interacting limit) and $\nu=1$ (strongly interacting limit).}
	Source: By the author.
	\label{fig_gamma}
\end{figure}

In order to have an idea about the behavior of the function $f(\nu) + 1/\tilde{g}$, we plot it
for different values of $\tilde{g}$ in Fig. \ref{fig_gamma}. We observe that 
the solutions of (\ref{eq.gamma}) for the weakly and the strongly interacting limits are given by
\begin{equation}\label{weakstrong} \nu(n) =
\begin{cases}
	2n ,      & \text{for } \tilde{g}\ll 1 \\
	2n+1,     & \text{for } \tilde{g}\gg 1
\end{cases}, \quad \forall \, n \in \mathbb{N}.
\end{equation}

\section{The momentum distribution asymptotic behavior}
Let us begin this section with the $N$-body wave function 
solution of (\ref{Nwave}). As the $N=2$ case is solved, we can 
choose any pair of particles in the respective $N$-body problem 
so that $x_{ij}^{(cm)}=(x_i+x_j)/2$ and $x_{ij}^{(r)}=x_i-x_j$, 
so that the solution is given by
\begin{equation}
	\Psi(x_1,\dots,x_N) = \Psi(x_1,\dots,x_N)
	\Psi^{(cm)}\left(x_{ij}^{(cm)}\right) \Psi^{(r)}\left(x_{ij}^{(r)}\right). 
\end{equation}
Also, from the behavior of the relative motion wave function near $x_i=x_j$ (\ref{psiexp}), 
we have that 
\begin{equation}
	\Psi(x_1,\dots,x_N) \approx \Psi\left(x_1,\dots,x_{ij}^{(cm)},\dots,x_N\right)
	 \left(1-\sqrt{2}\frac{|x_{ij}^{(r)}|}{a_{1\mathrm{D}}} + \mathcal{O} \left({x_{ij}^{(r)}}^2\right)\right),
\end{equation}
where we have left the normalization factor and other constants out for the sake of simplicity.
Now, following the developments from Refs. \citeonline{olshanii,jeanthesis}, the Fourier transform of $\Psi$, \textit{i.e.}, its representation in the 
momentum space, is given by
\begin{equation}\label{psik}
	\begin{aligned}
		&\tilde{\Psi}(k,x_2,\dots,x_N) = \frac{1}{\sqrt{2\pi}} 
		\int dx_1 \, \expo{-i k x_1} \Psi(x_1,\dots,x_N)\\
		&\approx \frac{1}{\sqrt{2\pi}}  \int dx_1 \,
		\expo{-i k x_1} \sum_{j=2}^N \Psi\left(x_1=x_{1j}^{(cm)},\dots,x_j=x_{1j}^{(cm)},\dots,x_N\right) \left(1-\sqrt{2}\frac{|x_{1j}^{(r)}|}{a_{1\mathrm{D}}}\right).
	\end{aligned}
\end{equation}
In addition, making use of the asymptotic behavior of the Fourier transform of $f(x)=f'(x)|x-x_0|$, with 
$f'(x)$ being a smooth function,\cite{bleistein}
\begin{equation}
	\int dx \,\expo{-ikx} f(x) \underset{k\to\infty} \sim - \frac{2}{k^2} f'(x_0)\expo{-ikx_0},
\end{equation}
and that $\int dx \,\expo{-ikx} f'(x)$ falls to zero as $\mathcal{O}(k^{-2})$,\cite{bracewell} 
 (\ref{psik}) reduces to 
\begin{equation}
	\tilde{\Psi}(k,x_2,\dots,x_N) \underset{k\to\infty} \sim \frac{2 k^{-2}}{\sqrt{\pi}a_{1\mathrm{D}}}
	 \sum_{j=2}^{N} \expo{-i k x_j} \Psi\left(x_1=x_j,x_2,\dots,x_j,\dots,x_N\right).
\end{equation}
Now we proceed to the evaluation of $n(k)$ itself: 
\begin{equation}
	\begin{aligned}
		n(k) = & N \int dx_2 \dots dx_N \, |\tilde{\Psi}(k,x_2,\dots,x_N)|^2 \\
		\underset{k\to\infty} \sim &  \frac{2 N}{\pi a_{1\mathrm{D}}^2} k^{-4} \int dx_2 \dots dx_N \, \sum_{j,l=2}^N 
		\expo{-ik(x_j-x_l)} \Psi^*(x_1=x_j,x_2,\dots,x_j,\dots,x_N) \\
		&\times \Psi(x_1=x_l,x_2,\dots,x_l,\dots,x_N).
	\end{aligned}
\end{equation}
Noting that the terms $j\neq l$ all cancel out, we have
\begin{equation}
	n(k) \underset{k\to\infty} \sim \frac{2 N}{\pi a_{1\mathrm{D}}^2} k^{-4}\int dx_2 \dots dx_N \, 
	 \sum_{j=2}^N |\Psi(x_1=x_j,x_2,\dots,x_j,\dots,x_N)|^2.
\end{equation}
Moreover, because of the indistinguishability nature of quantum particles, 
it is possible to write the two-body correlation function 
as\cite{jeanthesis,glauber,patu}
\begin{equation}\label{rho2.2}
	\varrho^{(2)}(x,x') = \int dx_1 \dots dx_N \, |\Psi(x_1,\dots,x_N)|^2 
	\sum_{i\neq j} \delta(x-x_i) \delta(x'-x_j).
\end{equation}
Finally, we conclude the asymptotic behavior of the momentum distribution:  
\begin{equation}\label{nk2}
	 n(k) \underset{k\to\infty} \sim \frac{2}{\pi a_{1\mathrm{D}}^2} k^{-4} \int dx \, \varrho^{(2)}(x,x).
\end{equation}

\subsection{Tan's contact}
From the contact definition 
\begin{equation}\label{contact.def}
\mathcal{C} \equiv \lim_{k \to \infty} k^4 n(k)
\end{equation}
together with (\ref{nk2})
we have that
\begin{equation}\label{Cfinal}
        \mathcal{C} = \frac{2}{\pi a_{1\mathrm{D}}^2} \int dx \, \varrho^{(2)}(x,x).
\end{equation}
Now we want to relate the Tan's contact to the slope of the energy 
with respect to the inverse of the interaction strength $g^{-1}$. 
From the Hellmann-Feynman theorem\cite{feynman} we have that  
\begin{equation}\label{hellman}
	\begin{aligned}
		\frac{\partial E}{\partial g^{-1}} &= -g^2 \int dx_1 \dots dx_N \, 
	|\Psi(x_1,\dots,x_N)|^2 \sum_{i<j} \delta(x_i-x_j) \\
		& = -g^2  \int dx \, \varrho^{(2)}(x,x).
	\end{aligned}
\end{equation}
Therefore, from (\ref{Cfinal}) and (\ref{hellman}) we find the relation
\begin{equation}
        \mathcal{C} = -\frac{m^2}{\pi \hbar^4} \frac{\partial E}{\partial g^{-1}},
\end{equation}
which is known as \emph{Tan's sweep theorem}.\cite{tan2,1d}

\subsubsection{The two-boson contact}
Let us now proceed to the calculation of the contact for our $N=2$ system 
at finite temperature $T>0$. We begin by making use of
\begin{equation}\label{c2}
	\mathcal{C} = -\frac{m^2}{\pi \hbar^4} \frac{\partial \mathcal{F}}{\partial g^{-1}},
\end{equation}
where $\mathcal{F} = -k_B T \ln \mathcal{Z}$ is the free energy of the system, 
with $\mathcal{Z} = \sum_{n,\nu} \expo{-\beta (E_n^{(cm)} + E_{\nu}^{(r)})}$ being 
the respective partition function and $\beta = 1/k_B T$. We must note that only
the energies related to the relative motion coordinate depend on $g$ (\ref{eq.gamma}). Also,
making use of
\begin{equation}
	\frac{\partial\nu}{\partial \tilde{g}} = \frac{1}{\tilde{g}^2}\left(\frac{\partial f(\nu)}{\partial \nu}\right)^{-1},
\end{equation}
and evaluating $\partial f(\nu)/\partial \nu$,
\begin{equation}
	\frac{\partial f(\nu)}{\partial \nu} = \frac{1}{2} \frac{\Gamma\left(-\frac{\nu}{2}\right)}{\Gamma\left(-\frac{\nu}{2}+\frac{1}{2}\right)}\left[\psi\left(-\frac{\nu}{2}+\frac{1}{2}\right)-\psi \left(-\frac{\nu}{2}\right)\right],
\end{equation}
the contact, as a function of $\tilde{g}$ and $\beta$, reads
\begin{equation}
	\mathcal{C}(\tilde{g},\beta) = \frac{2^{5/2} \tilde{g}}{\pi a_{0}^{3} \mathcal{Z}}\sum_{n,\nu}\mathrm{e}^{-\beta(E_n^{(cm)}+E^{(r)}_{\nu})} 
	\left[\psi\left(-\frac{\nu}{2}\right)-\psi\left(-\frac{\nu}{2}+\frac{1}{2}\right)\right]^{-1}.
\end{equation}
It is straightforward to observe that the contact is independent of the 
center-of-mass energy $E_n^{(cm)}$, which is a direct consequence of the 
\textit{Kohn's theorem}:\cite{pitaevskii} differently from the 
relative-motion energy $E^{(r)}_{\nu}$, $E_n^{(cm)}$ is independent of 
the interatomic interactions. Thence, the contact reduces to 
\begin{equation}
	\mathcal{C}(\tilde{g},\beta) = \frac{2^{5/2} \tilde{g}}{\pi a_{0}^{3} \mathcal{Z}_r}\sum_{\nu}\mathrm{e}^{-\beta E^{(r)}_{\nu}} 
	\left[\psi\left(-\frac{\nu}{2}\right)-\psi\left(-\frac{\nu}{2}+\frac{1}{2}\right)\right]^{-1},
\end{equation}
where we have defined the relative-motion partition function $\mathcal{Z}_r\equiv \sum_{\nu}\mathrm{e}^{-\beta E^{(r)}_{\nu}}$. As $E^{(r)}_{\nu}=(\nu+1/2)\hbar\omega$, we can perform one more reduction in the contact expression, 
\begin{equation}\label{eq.contactN2}
	\mathcal{C}(\tilde{g},\beta) = \frac{2^{5/2} \tilde{g}}{\pi a_{0}^{3} \mathcal{Z}_r}\sum_{\nu}\mathrm{e}^{-\beta \hbar\omega \nu} 
	\left[\psi\left(-\frac{\nu}{2}\right)-\psi\left(-\frac{\nu}{2}+\frac{1}{2}\right)\right]^{-1}.
\end{equation}
Here, for the sake of avoiding the introduction of unnecessary terms, 
we simply transform the relative-motion partition function according to 
$\mathcal{Z}_r \to \expo{\beta \hbar \omega/2} \mathcal{Z}_r$.

\begin{figure}[h!]
	\centering
	\includegraphics[width=\columnwidth]{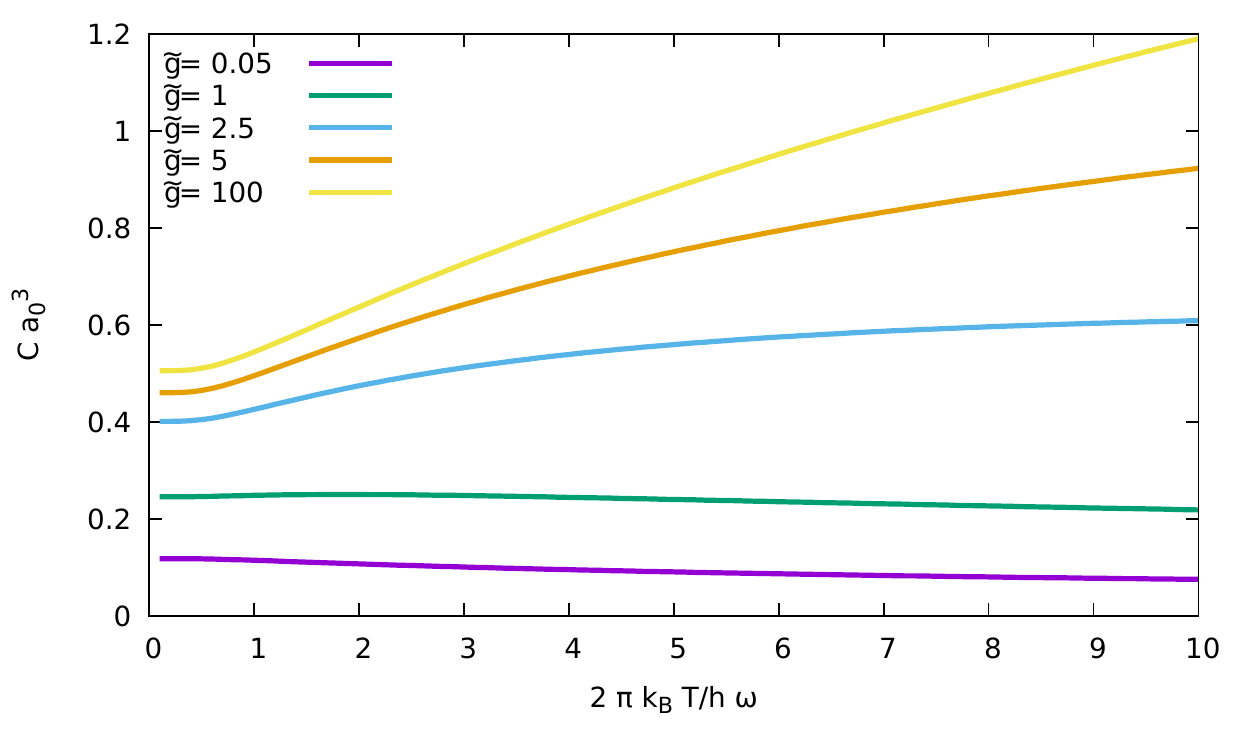}
	\caption{Tan's contact from (\ref{eq.contactN2}) 
	as a function of the adimensional temperature $k_B T/\hbar \omega$ for different values of the adimensional interaction parameter $\tilde{g}$.}
	Source: By the author.
	\label{N2vsT}
\end{figure}

The two-boson contact from (\ref{eq.contactN2}) is depicted in Fig. \ref{N2vsT} 
for different values of the adimensional interaction parameter $\tilde{g}$. From the 
referred curves, we observe that the contact increases with both the temperature and 
the interaction strength, an effect that was to be expected once both variables 
contribute to the increase of the interaction energy.

\subsubsubsection{The Tonks-Girardeau limit}
Now, let us work out a formula for the strongly interacting 
scenario, also known as \textit{Tonks-Girardeau limit} (TG limit).\footnote{Thanks to Patrizia Vignolo for working out these steps.} From the two-boson contact of Eq. (\ref{eq.contactN2}) 
and from the fact that $\nu=2j-1$ in the TG limit, we have that  
\begin{equation}\label{TG_con}
	\mathcal{C}(\tilde{g}\to\infty,\beta) = \frac{2^{5/2}}{\pi a_{0}^{3} \mathcal{Z}_r}\sum_{j>0} \mathrm{e}^{-\beta \hbar\omega (2j-1)} 
	\frac{\Gamma(1-j)}{\Gamma\left(\frac{1}{2}-j\right)}
	\left[\psi(1-j)-\psi\left(\frac{1}{2}-j\right)\right]^{-1}.
\end{equation}
Because of $\Gamma(1-j)\to\infty, \, \forall \, j\in \mathbb{N}^*$, as well as $\psi(1-j)\to \infty, \, \forall \, j \in \mathbb{N}^*$, 
we can perform the approximation $\psi(1-j)-\psi\left(\frac{1}{2}-j\right)\approx\psi(1-j)$, so that we are able to restrict ourselves to evaluating 
$\Gamma(1-j)/\psi(1-j)$. Let us begin with the definition of the gamma function $\Gamma(z)\equiv(z-1)!$, which implies 
\begin{equation}\label{z...}
\Gamma(z) = \frac{\Gamma(z+n+1)}{\prod_{i=0}^n (z+i)}, 
\end{equation}
and its derivative 
\begin{equation}
\Gamma'(z) = \frac{\Gamma'(z+n+1)}{\prod_{i=0}^n (z+i)}
 -\Gamma(z+n+1) \sum_{l=0}^n \frac{1}{(z+l)\prod_{i=0}^n (z+i)}.
\end{equation}
Hence, the digamma function reads 
\begin{equation}
\frac{\Gamma'(z)}{\Gamma(z)} = \frac{\Gamma'(z+n+1)}{\Gamma(z+n+1)}
-\sum_{\ell=0}^n \frac{1}{z+\ell}.
\end{equation}
Therefore, we have for our inverted-desired ratio the following:
\begin{equation}
\frac{\Gamma'(z)}{\Gamma^2(z)} = \frac{\Gamma'(z+n+1)}{\Gamma^2(z+n+1)}
\prod_{i=0}^n (z+i) - \sum_{\ell=0}^n \frac{1}{z+\ell} \frac{\prod_{i=0}^n (z+i)}{\Gamma(z+n+1)}.
\end{equation}
For $z=-n$, we have that 
\begin{equation}
\frac{\Gamma'(-n)}{\Gamma^2(-n)} = \frac{\Gamma'(1)}{\Gamma(-n)}
 - \sum_{\ell=0}^n \frac{\prod_{i=0}^n (i-n)}{\ell-n}.
\end{equation}
Here, all terms vanish except when $\ell=n$, 
\begin{equation}
\frac{\Gamma'(-n)}{\Gamma^2(-n)} = -\prod_{i=0}^{n-1} (i-n).
\end{equation}
Therefore, we finally arrive at our desired result: 
\begin{equation}
\frac{\Gamma'(1-j)}{\Gamma^2(1-j)} = (-1)^j (j-1)!.
\end{equation}
Consequently, Eq. (\ref{TG_con}) yields 
\begin{equation}\label{finalC2}
	\mathcal{C}(\tilde{g}\to\infty,\beta) = \frac{2^{5/2}}{\pi a_{0}^{3} \mathcal{Z}_r}\sum_{j>0} \mathrm{e}^{-\beta \hbar\omega (2j-1)} 
	\frac{(-1)^j}{\Gamma\left(\frac{1}{2}-j\right)(j-1)!}.
\end{equation}
The plot of (\ref{finalC2}) as a function of the temperature reproduces 
the yellow curve ($\tilde{g}=100$) in Fig. \ref{N2vsT}.

\begin{figure}[H]
        \centering
	\includegraphics[width=\columnwidth]{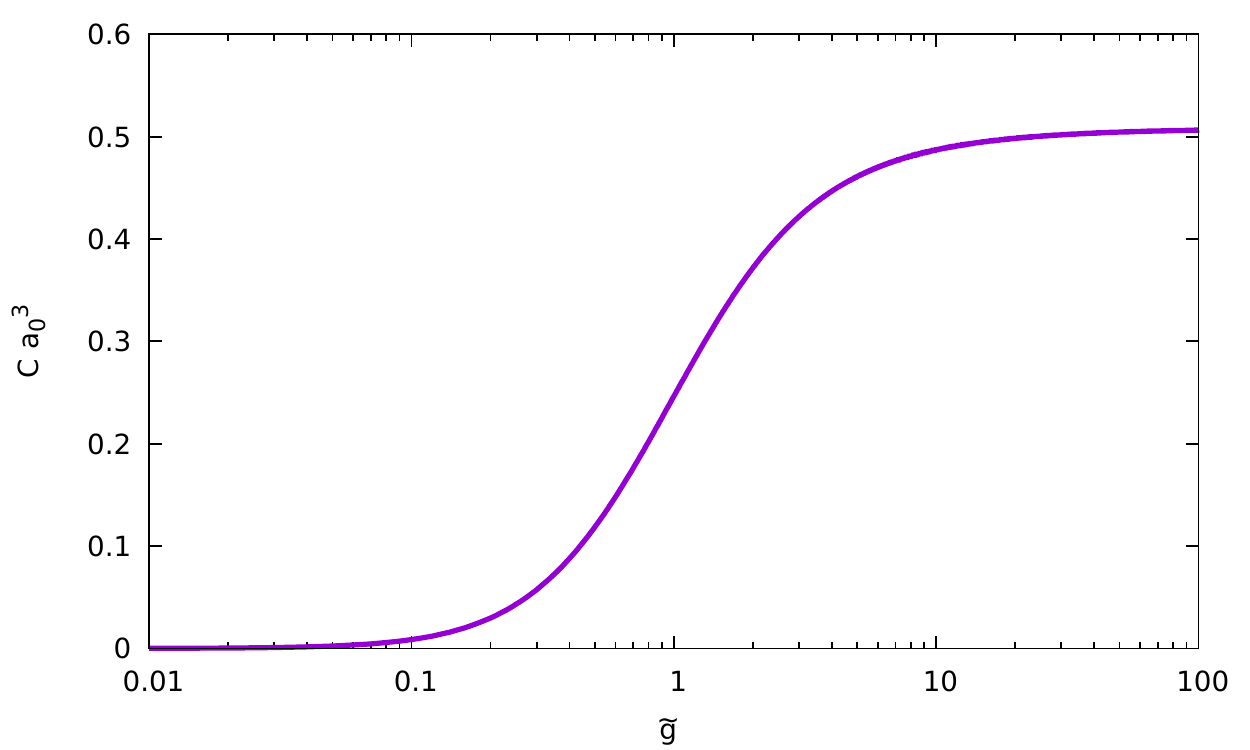}
	\caption{Zero-temperature Tan's contact from (\ref{eq.zeroT}) as a function of the adimensional interaction parameter $\tilde{g}$.}
	Source: By the author.
	\label{N2vsz}
\end{figure}

\subsubsubsection{The zero-temperature regime}
At zero temperature, the contact reduces to
\begin{equation}\label{eq.zeroT}
	\mathcal{C}(\tilde{g},\beta \to \infty) = \frac{2^{5/2} \tilde{g}}{\pi a_{0}^{3}}  
        \left[\psi\left(-\frac{\nu}{2}\right)-\psi\left(-\frac{\nu}{2}+\frac{1}{2}\right)\right]^{-1},
\end{equation}
which is depicted in Fig. \ref{N2vsz}.

\subsubsubsection{The Tonks-Girardeau limit in the zero-temperature regime}
Moreover, we can observe in Fig. \ref{N2vsz} the known zero-temperature strongly 
interacting limit as $\tilde{g}\to\infty$,
\begin{equation}\label{eq.infTg}
	\lim_{\tilde{g},\beta\to\infty} \mathcal{C}(\tilde{g},\beta)a_0^3 = \left(\frac{2}{\pi}\right)^{3/2}.
\end{equation}

\section{The Tonks-Girardeau regime for $N$ particles}
In this section we regard the strongly interacting scenario $g\rightarrow\infty$, also known as Tonks-Girardeau gas. In such a case, it is known\cite{girardeau} 
that the relationship between the many-body wave functions of such a bosonic system and a fermionic one is given by
\begin{equation}
\label{mapping}
	\Psi_{\alpha}^{(b)}(x_1,\dots,x_N) = \Theta(x_1,\dots,x_N)\Psi_{\alpha}^{(f)}(x_1,\dots,x_N),
\end{equation}
where $\Theta(x_1,\dots,x_N)\equiv\prod_{i>j}\mathrm{sgn}(x_i-x_j)$ is either $+1$ or $-1$, in order to compensate the anti-symmetrization of the fermionic wave function $\Psi^{(f)}_\alpha$, and $\alpha$ is the quantum number describing the particles in a respective set of individual quantum numbers $\{n_1,n_2,\dots,n_N\}$. 
In the strongly interacting scenario $g\rightarrow \infty$, also called Tonks-Girardeau gas, the system behaves the same way as the trapped ideal Fermi gas, whose many-body wave function is given by the Slater determinant
\begin{equation}\label{slater}
	\Psi_{\alpha}^{(f)}(x_1,\dots,x_N) = \left(N!\right)^{-1/2} \det[\phi_{n_i}(x_j)]_{n_i\in\{n_1,\dots,n_N\};\,x_j\in\{x_1,\dots,x_N\}},
\end{equation}
where $\phi_n(x)$ are the solutions of the single-particle harmonic oscillator Hamiltonian,
\begin{equation}\label{ho_sol}
\phi_n(x)= \frac{\expo{-(x/a_{0})^2/2}H_n(x/a_{0})}{\pi^{1/4}\sqrt{a_{0}2^n n!}}.
\end{equation}
This correspondence between the strongly interacting bosonic system and the noninteracting fermionic one 
is known as \emph{fermionization} of bosons.\cite{paredes, kinoshita} 
The reason behind such an effect comes from the \emph{Pauli exclusion principle}, which states that two fermions cannot occupy the same quantum state. 
In our scenario, the strong repulsion is the equivalent of such a postulate.  Therefore, 
all these considerations culminate in the fact that, 
for observables depending on $|\Psi|^2$, strongly interacting 
bosons behave as the ideal Fermi gas.

\subsection{The zero-temperature case}
The solutions for the Bose gas consisted of $N$-impenetrable particles (Tonks-Girardeau gas) 
subject to the periodic boundary condition on a length $L$ at $T=0$, 
where the solutions are given by plane waves $\phi_n(x) \propto \expo{i 2\pi nx/L}$ 
instead of (\ref{ho_sol}), were carried out in Ref. \citeonline{girardeau} and are given by
\begin{equation}
	\Psi_0^{(b)}(x_1,\dots,x_N) = \sqrt{\frac{2^{N(N-1)}}{N!L^N}} \prod_{i>j} 
	\sin\left(\frac{\pi}{L} |x_i-x_j|\right), 
\end{equation}
together with its associated ground state energy 
\begin{equation}
	E_0 =  \left(N-\frac{1}{N}\right) \frac{\hbar^2\pi^2N^2}{6mL^2}.
\end{equation}

Now, considering the effects of a harmonic trap, whose solutions are given by (\ref{ho_sol}), 
the ground state many-body wave function is given by\cite{girardeau01,forrester03} 
\begin{equation}
        \Psi_0^{(b)}(x_1,\dots,x_N) = \frac{2^{N(N-1)/4}}{a_0^{N/2}} 
	\left(N! \prod_{n=0}^{N-1}n!\sqrt{\pi}\right)^{-1/2} 
	\prod_{i=0}^N \expo{-x_i^2/2a_0^2} \prod_{1\leq j<k \leq N} |x_k-x_j|,
\end{equation}
with its respective density profile given by\cite{kolo,pv2000}
\begin{equation}
	n(x) = \frac{\expo{-x^2/a_0^2}}{\sqrt{\pi a_0}} \sum_{i=0}^{N-1} \frac{H_i^2(x/a_0)}{2^i i!} .
\end{equation}

\subsection{The finite-temperature regime}
When we are dealing with quantum systems, the notion of quantum correlation turns out to be a valuable tool in order to characterize 
a system or, more specifically, to have a measurement of how correlated the particles of a system are with each other, specially an interacting one. 
An useful quantity associated with such an idea is the so-called \emph{$j$-body density matrix},\cite{huang,pethick,pitaevskii,reichl} 
which, for the system consisted of $N$ interacting particles at temperature $T$, is given by
\begin{equation}
	\begin{aligned}\label{rhoj_oi}
		\varrho^{(j)}(x_1,\dots,x_j;x'_1,\dots,x'_j) &=& \frac{N!}{(N-j)!} \mathcal{Z}^{-1}
		\sum_{\alpha} \mathrm{e}^{-\beta E_{\alpha}} \int_\Re dx_{j+1} \dots dx_N 
		\Psi_{\alpha}^{(b)*}(x_1,\dots,x_N) \\
		&&\times\Psi_{\alpha}^{(b)} (x'_1,\dots,x'_{j},x_{j+1},\dots,x_N),
	\end{aligned}
\end{equation}
where $\mathcal{Z} = \sum_{\alpha} \mathrm{e}^{-\beta E_{\alpha}}$ is the partition function of the system and
its total energy is simply the summation of all the individual single-particle energies, \textit{i.e.},
$E_{\alpha} = \sum_{i=1}^N \epsilon_{n_i}$, with $\epsilon_{n_i} = (n_i + 1/2)\hbar \omega$.

From the \emph{Bose-Fermi mapping} relation (\ref{mapping}), Eq. (\ref{rhoj_oi}) reads
\begin{equation}
        \begin{aligned}\label{rhoj_oi2}
		&\varrho^{(j)}(x_1,\dots,x_j;x'_1,\dots,x'_j) = \frac{N!}{(N-j)!} \mathcal{Z}^{-1}
                \sum_{\alpha} \mathrm{e}^{-\beta E_{\alpha}} \int_\Re dx_{j+1} \dots dx_N
                \Theta(x_1,\dots,x_N) \\ 
		&\times \Psi_{\alpha}^{(f)}(x_1,\dots,x_N)
		\Theta(x'_1,\dots,x'_j,x_{j+1},\dots,x_N)\Psi_{\alpha}^{(f)}(x'_1,\dots,x'_j,x_{j+1},\dots,x_N).
        \end{aligned}
\end{equation}
Now we turn our focus to the integrand. It is possible to rewrite the product 
of the $\Theta$'s as\cite{lenard}
\begin{equation}
	\begin{aligned}
		\Theta(x_1,\dots,x_N)\Theta(x'_1,\dots,x'_j,x_{j+1},\dots,x_N) &=& 
	\Theta(x_1,\dots,x_j) \Theta(x'_1,\dots,x'_j) \\
		&&\times \prod_{i=j+1}^N \prod_{l=1}^{2j} \mathrm{sgn}(x_i-y_l),
	\end{aligned}
\end{equation}
with $y_1 = x_1 < y_2 = x_2 < \dots < y_j=x_j < y_{j+1} = x'_1 < \dots < y_{2j} = x'_j$.
Now let us consider the union of the disjoint intervals $(y_i,y_j)$, 
$\mathfrak{S}=\{(y_1,y_2)\cup(y_3,y_4)\cup\dots\cup(y_{2j-1},y_{2j})\}$.
It is straightforward to observe that 
\begin{equation} \prod_{i=1}^{2j} \mathrm{sgn}(x-y_i) =
\begin{cases}
	-1 ,      & x \in \mathfrak{S} \\
	+1,     & x \notin \mathfrak{S}
\end{cases}.
\end{equation}
Denoting the number of variables among $x_{j+1},\dots,x_N$ which are in $\mathfrak{S}$ by $M_{\mathfrak{S}}$, we have that
\begin{equation}
                \Theta(x_1,\dots,x_N)\Theta(x'_1,\dots,x'_j,x_{j+1},\dots,x_N)=
	\Theta(x_1,\dots,x_j) \Theta(x'_1,\dots,x'_j) (-1)^{M_\mathfrak{S}}.
\end{equation}
Consequently, Eq. (\ref{rhoj_oi2}) results in
\begin{equation}
        \begin{aligned}\label{rhoj_oi3}
		&\varrho^{(j)}(x_1,\dots,x_j;x'_1,\dots,x'_j) = \frac{N!}{(N-j)!} \mathcal{Z}^{-1}
		\Theta(x_1,\dots,x_j) \Theta(x'_1,\dots,x'_j)
                \sum_{\alpha} \mathrm{e}^{-\beta E_{\alpha}} \\
		&\times \int_\Re dx_{j+1} \dots dx_N (-1)^{M_\mathfrak{S}}
                \Psi_{\alpha}^{(f)}(x_1,\dots,x_N)
                \Psi_{\alpha}^{(f)}(x'_1,\dots,x'_j,x_{j+1},\dots,x_N).
        \end{aligned}
\end{equation}
Now, considering any integral of the form 
\begin{equation}
	I = \int_\Re dx_1 \dots \int_\Re (-1)^{M_\mathfrak{S}} f(x_1,\dots,x_j),
\end{equation}
where $M_\mathfrak{S}$ is the number of integration variables inside 
the subdomain $\mathfrak{S}$ and $f$ is a symmetric function, 
it is possible to write\cite{lenard}
\begin{equation}
	I = \sum_{m=0}^j \binom{j}{m} (-1)^m 
	\int_\mathfrak{S}dx_1 \dots dx_m
	\int_{\Re-\mathfrak{S}}dx_{m+1} \dots dx_j f(x_1,\dots,x_j).
\end{equation}
Making use of $\int_{\Re-\mathfrak{S}} dx = \int_{\Re} dx - \int_{\mathfrak{S}} dx$,
we have
\begin{equation}\label{acima}
        I = \sum_{m=0}^j \binom{j}{m} (-1)^m
	\sum_{n=0}^{j-m} \binom{j-m}{n} (-1)^n
	\int_\mathfrak{S}dx_1 \dots dx_{m+n}
        \int_{\Re}dx_{m+n+1} \dots dx_j f(x_1,\dots,x_j).
\end{equation}
Performing the summation for $m+n=i$, (\ref{acima}) reduces to
\begin{equation}
	I = \sum_{i=0}^j \binom{j}{i} (-2)^i 
	\int_\mathfrak{S} dx_1 \dots dx_i
	\int_\Re dx_{i+1}\dots dx_j f(x_1,\dots,x_j).
\end{equation}
Thence, we have that the $j$-body density matrix (\ref{rhoj_oi3}) 
can be written as
\begin{equation}
        \begin{aligned}\label{rhoj_oi4}
		\varrho^{(j)}(x_1,\dots,x_j;x'_1,\dots,x'_j) =& \frac{N!}{(N-j)!} \mathcal{Z}^{-1}
                \Theta(x_1,\dots,x_j) \Theta(x'_1,\dots,x'_j)
                \sum_{\alpha} \mathrm{e}^{-\beta E_{\alpha}} \\
		&\times \sum_{i=0}^{N-j} \binom{N-j}{i} (-2)^i \int_\mathfrak{S} 
		dx_{j+1} \dots dx_{j+i} \int_{\Re} dx_{j+i+1}\dots dx_N \\
		&\times\Psi_{\alpha}^{(f)}(x_1,\dots,x_N)
                \Psi_{\alpha}^{(f)}(x'_1,\dots,x'_j,x_{j+1},\dots,x_N).
        \end{aligned}
\end{equation}

The one-body density matrix is given by\cite{pv2013}
\begin{equation}
\begin{aligned}
	\varrho^{(1)}(x,x') =& \frac{N}{\mathcal{Z}} \sum_{\alpha} \mathrm{e}^{-\beta E_{\alpha}} 
	\sum_{j=1}^{N-1}\binom{N-1}{j} (-2)^j [\mathrm{sgn}(x-x')]^j \int_x^{x'} dx_2 \dots dx_{j+1} \\
	&\times \int_{\Re} dx_{j+2}\dots dx_N \Psi_{\alpha}^{(f)}(x,x_2,\dots,x_N) \Psi_{\alpha}^{(f)}(x',x_2,\dots,x_N).
\end{aligned}
\end{equation}
Here it is possible to recognize the $j$-body fermionic correlator as
\begin{equation}
	\varrho^{(1)}(x,x') = \sum_{j=1}^{N-1} \frac{(-2)^j}{j!} [\mathrm{sgn}(x-x')]^j \int_x^{x'} dx_2 \dots dx_{j+1} \varrho^{(j+1)}_f(x,x_2,\dots,x_{j+1};x',x_2,\dots,x_{j+1}),
\end{equation}
where
\begin{equation}
\begin{aligned}\label{rhoj}
	\varrho^{(j)}_f(x_1,\dots,x_j;x_1',\dots,x_j') &= \frac{N!}{(N-j)!}\mathcal{Z}^{-1} \sum_\alpha \mathrm{e}^{-\beta E_{\alpha}} \\
	& \times \int_{\Re} dx_{j+1} \dots dx_N \Psi_\alpha^{(f)}(x_1,\dots,x_N) \Psi_\alpha^{(f)}(x_1',\dots,x_j',x_{j+1},\dots,x_N).
\end{aligned}
\end{equation}

As we are interested in the contact, we are going to restrict ourselves to small distances, $|x'-x| \ll 1$. 
Therefore, we consider only the term $j=1$, because the terms $j>1$ produce 
negligible results in the small distance approximation:
\begin{equation}
	\begin{aligned}
		\varrho^{(1)}(x,x') \underset{x \to x'} \sim & 2 \, \mathrm{sgn}(x'-x) \int_x^{x'} dx_2 \, \varrho^{(2)}_f(x,x_2;x',x_2) \\
		\approx & 2 \, \mathrm{sgn}(x'-x) \varrho^{(2)}_f(x,R;x',R) \delta x,
	\end{aligned}
\end{equation}
where $R\equiv(x+x')/2$ and $\delta x \equiv x'-x$.

Now we proceed to the explicit evaluation of $\varrho^{(2)}$
making use of (\ref{rhoj}) together with (\ref{slater}).

\emph{N=2 particles}
\begin{equation}
\begin{aligned}
	\varrho^{(2)}_f(x,R;x',R)=&\mathcal{Z}^{-1} \sum_{n_1,n_2} \mathrm{e}^{-\beta(\epsilon_{n_1}+\epsilon_{n_2})}
\begin{vmatrix} \phi_{n_1}(x) & \phi_{n_2}(x) \\
\phi_{n_1}(R)& \phi_{n_2}(R)\\
\end{vmatrix}
\begin{vmatrix} \phi_{n_1}(x') & \phi_{n_2}(x') \\
\phi_{n_1}(x_2)& \phi_{n_2}(x_2)\\
\end{vmatrix} \\
	=& \mathcal{Z}^{-1} \sum_{n_1,n_2} \mathrm{e}^{-\beta(\epsilon_{n_1}+\epsilon_{n_2})}\left[\phi_{n_1}\left(R-\delta x/2\right)\phi_{n_2}(R)-\phi_{n_2}(R-\delta x/2)\phi_{n_1}(R)\right] \\
&\times \left[\phi_{n_1}(R+\delta x/2)\phi_{n_2}(R)-\phi_{n_2}(R+\delta x/2)\phi_{n_1}(R)\right] \\
	=& \mathcal{Z}^{-1} \sum_{n_1,n_2} \mathrm{e}^{-\beta(\epsilon_{n_1}+\epsilon_{n_2})}\left[ \left(\phi_{n_1} - \frac{\delta x}{2} \partial_R \phi_{n_1} \right)\phi_{n_2}
	- \left(\phi_{n_2} -\frac{\delta x}{2} \partial_R \phi_{n_2} \right)\phi_{n_1}\right]  \\
	&\times \left[\left(\phi_{n_1}+ \frac{\delta x}{2}\partial_R \phi_{n_1} \right) \phi_{n_2} - \left(\phi_{n_2}+ \frac{\delta x}{2}\partial_R \phi_{n_2} \right) \phi_{n_1}\right]  \\
=& \mathcal{Z}^{-1} \sum_{n_1,n_2} \mathrm{e}^{-\beta(\epsilon_{n_1}+\epsilon_{n_2})}\frac{\delta x^2}{4} \\
	&\times \left[(\phi_{n_2}\partial_R \phi_{n_1})^2+(\phi_{n_1}\partial_R \phi_{n_2})^2 - 2 \phi_{n_1} \phi_{n_2} \partial_R \phi_{n_1} \partial_R \phi_{n_2} \right].
\end{aligned}
\end{equation}

\textit{N=3 particles}
\begin{equation}
\begin{aligned}
	\varrho^{(2)}_f(x,R;x',R)=&\mathcal{Z}^{-1} \sum_{n_1,n_2,n_3} \mathrm{e}^{-\beta(\epsilon_{n_1}+\epsilon_{n_2}+\epsilon_{n_3})} \\
	& \times \int dx_3 \,
\begin{vmatrix} \phi_{n_1}(x) & \phi_{n_2}(x) & \phi_{n_3}(x)\\
\phi_{n_1}(R)& \phi_{n_2}(R)&\phi_{n_3}(R)\\
\phi_{n_1}(x_3)& \phi_{n_2}(x_3)&\phi_{n_3}(x_3)\
\end{vmatrix} \begin{vmatrix} \phi_{n_1}(x') & \phi_{n_2}(x') & \phi_{n_3}(x')\\
\phi_{n_1}(R)& \phi_{n_2}(R)&\phi_{n_3}(R)\\
\phi_{n_1}(x_3)& \phi_{n_2}(x_3)&\phi_{n_3}(x_3)\
\end{vmatrix}  \\
	=& \mathcal{Z}^{-1} \sum_{n_1,n_2,n_3} \mathrm{e}^{-\beta(\epsilon_{n_1}+\epsilon_{n_2}+\epsilon_{n_3})}
\left[\phi_{n_1}^2 \left(\frac{\delta x}{2} \partial_R \phi_{n_2}  \right)^2 +\phi_{n_1}^2 \left(\frac{\delta x}{2} \partial_R \phi_{n_3}  \right)^2 \right. \\
	&\left. +\phi_{n_2}^2 \left( \frac{\delta x}{2} \partial_R \phi_{n_1} \right)^2 +\phi_{n_2}^2 \left(\frac{\delta x}{2} \partial_R \phi_{n_3} \right)^2
+\phi_{n_3}^2 \left(\frac{\delta x}{2}\partial_R \phi_{n_1}  \right)^2 \right.  \\
	&\left. +\phi_{n_3}^2 \left(\frac{\delta x}{2} \partial_R \phi_{n_2} \right)^2  -2 \phi_{n_1} \phi_{n_2} \frac{\delta x^2}{4} \partial_R \phi_{n_1} \partial_R \phi_{n_2}  \right.  \\
	&\left.-2 \phi_{n_1} \phi_{n_3} \frac{\delta x^2}{4} \partial_R \phi_{n_1} \partial_R \phi_{n_3} -2 \phi_{n_2} \phi_{n_3} \frac{\delta x^2}{4} \partial_R \phi_{n_2} \partial_R \phi_{n_3}\right].
\end{aligned}
\end{equation}

Note that in the steps above we have used the differentiation relation
\begin{equation}
	\partial_R \phi(R) = \frac{\phi(R)-\phi(R-\delta x/2)}{\delta x/2}
\end{equation}
and the orthogonality of the $\phi$'s
\begin{equation}
	\int_{-\infty}^{+\infty} dx \, \phi_m(x) \phi_n(x)= \delta_{m,n}.
\end{equation}

Therefore, from the explicit evaluations for $N=2$ and 3 particles, we can generalize the fermionic two-body density matrix for $N$ particles as
\begin{equation}
	\begin{aligned}
		\varrho^{(2)}_f(x,R;x',R) = & \frac{(x'-x)^2}{4} \mathcal{Z}^{-1} \sum_{n_1,n_2,\dots,n_N} \mathrm{e}^{-\beta \sum_{i=1}^N \epsilon_{n_i}} \\
	& \times \sum_{j\neq k} \left\{ \left[ \phi_{n_j}(R)\partial_R \phi_{n_k}(R) \right]^2 - \phi_{n_j}(R) \phi_{n_k}(R) \partial_R \phi_{n_j}(R) \partial_R \phi_{n_k}(R)
\right\}.
	\end{aligned}
\end{equation}
Consequently, we have that  
\begin{equation}
\varrho^{(1)}(x,x') \approx \frac{|x'-x|^3}{2} F(R),
\end{equation}
with the definition
\begin{equation}
\begin{aligned}
	F(R) \equiv&  \mathcal{Z}^{-1} \sum_{n_1,n_2,\dots,n_N} \mathrm{e}^{-\beta \sum_{i=1}^N \epsilon_{n_i}} \\
	& \times \sum_{j\neq k} \left\{ \left[ \phi_{n_j}(R)\partial_R \phi_{n_k}(R) \right]^2 
	-\phi_{n_j}(R) \phi_{n_k}(R) \partial_R \phi_{n_j}(R)\partial_R \phi_{n_k}(R)
\right\}.
\end{aligned}
\end{equation}

We shall note that the limits of sums in the set ${n_1,n_2,\dots,n_N}$ were omitted, 
although they are not obvious. We must remember that in the Tonks-Girardeau gas, the bosons
\emph{fermionize} due to the strong interaction between them. This phenomena implies that 
all particles must be "found" in different states with respect to each other. For example, 
if particle 1 is in the fundamental state $n_1=0$, particle 2 must be in any excited state, 
$n_2 \in \mathbb{N} - \{0\}$. If particle 2 realizes the state $n_2=1$, then the possible 
set of states for particle 3 is $n_3 \in \mathbb{N} - \{0,1\}$. And this reasoning continues 
for all particles. Therefore the possible states particle $i$ can be found in is the set 
$\{i-1,i,\dots,\infty \}$, which are the sets for the sums that we fix here.

\subsubsection{Momentum distribution}
\begin{figure}[h!]
	\centering
		\begin{subfigure}{.45\columnwidth}
			\includegraphics[width=\columnwidth]{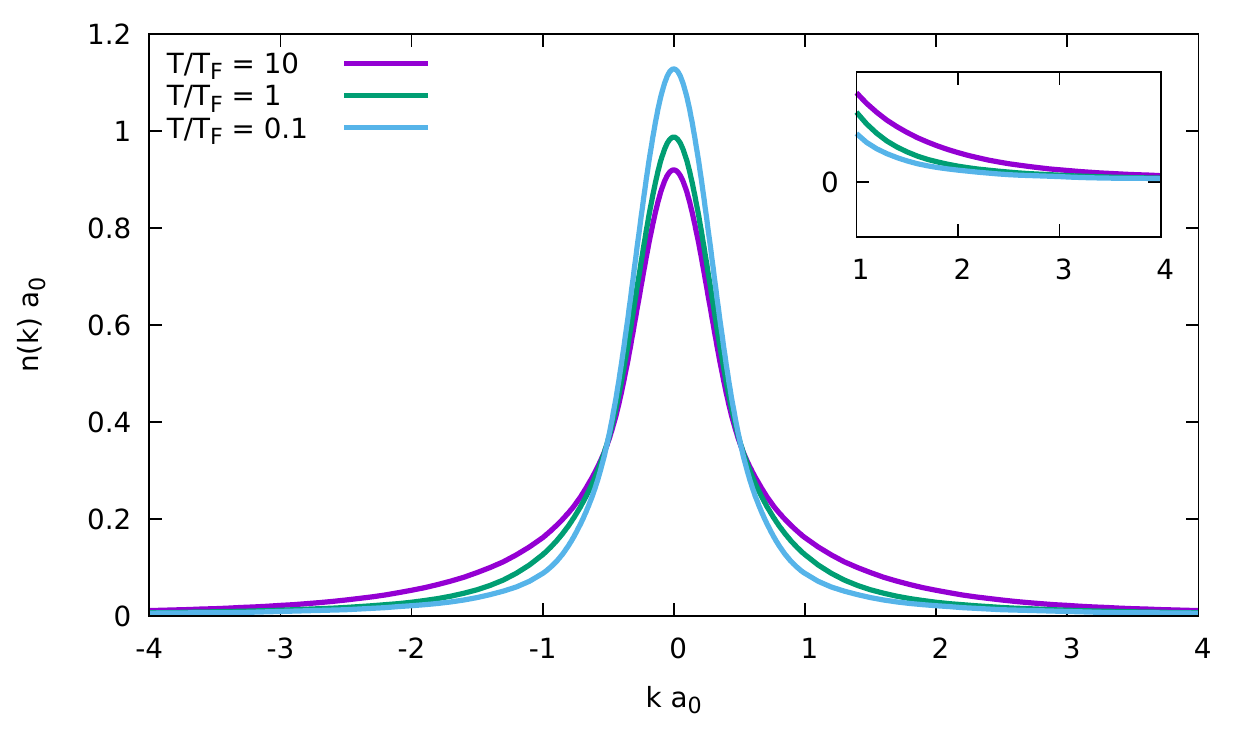}
			\caption{$N=2$.}
		\label{fig.TGN2} \end{subfigure} \qquad
		\begin{subfigure}{.45\columnwidth}
			\includegraphics[width=\columnwidth]{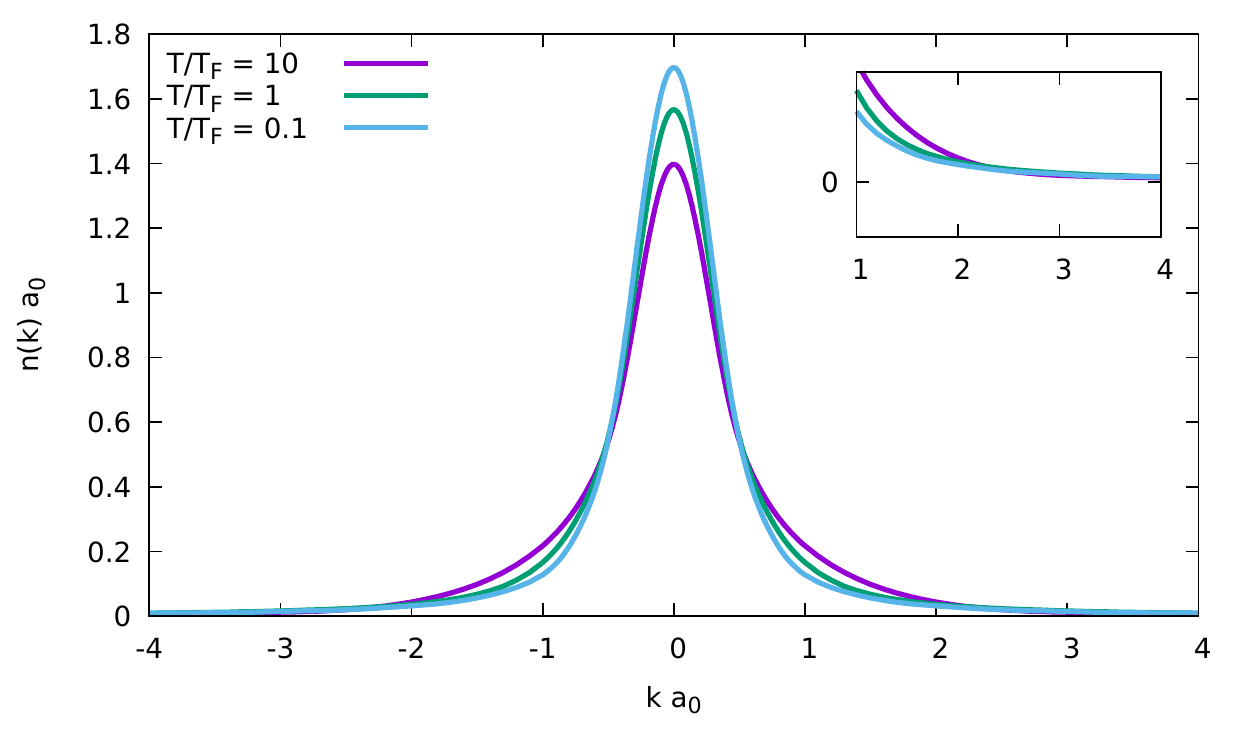}
			\caption{$N=3$.}
		\label{fig.TGN3} \end{subfigure}
		\begin{subfigure}{.45\columnwidth}
			\includegraphics[width=\columnwidth]{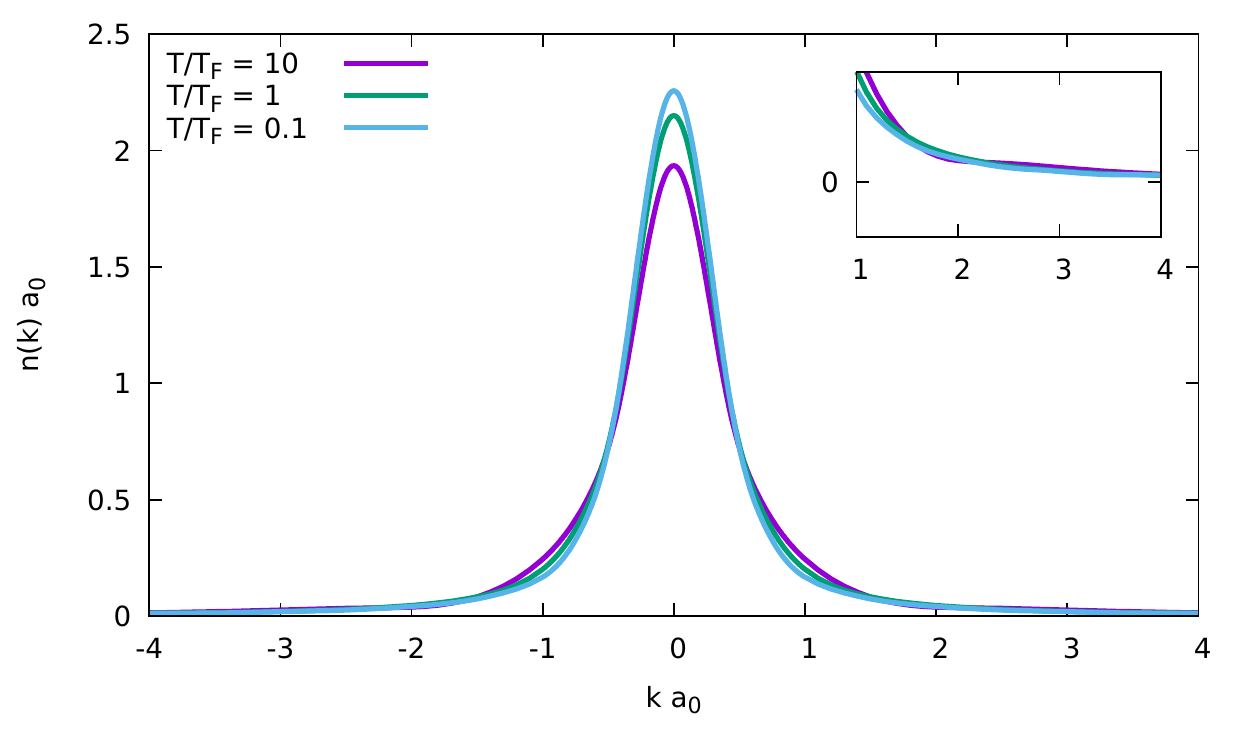}
			\caption{$N=4$.}
		\label{fig.TGN4} \end{subfigure}
		\begin{subfigure}{.45\columnwidth}
			\includegraphics[width=\columnwidth]{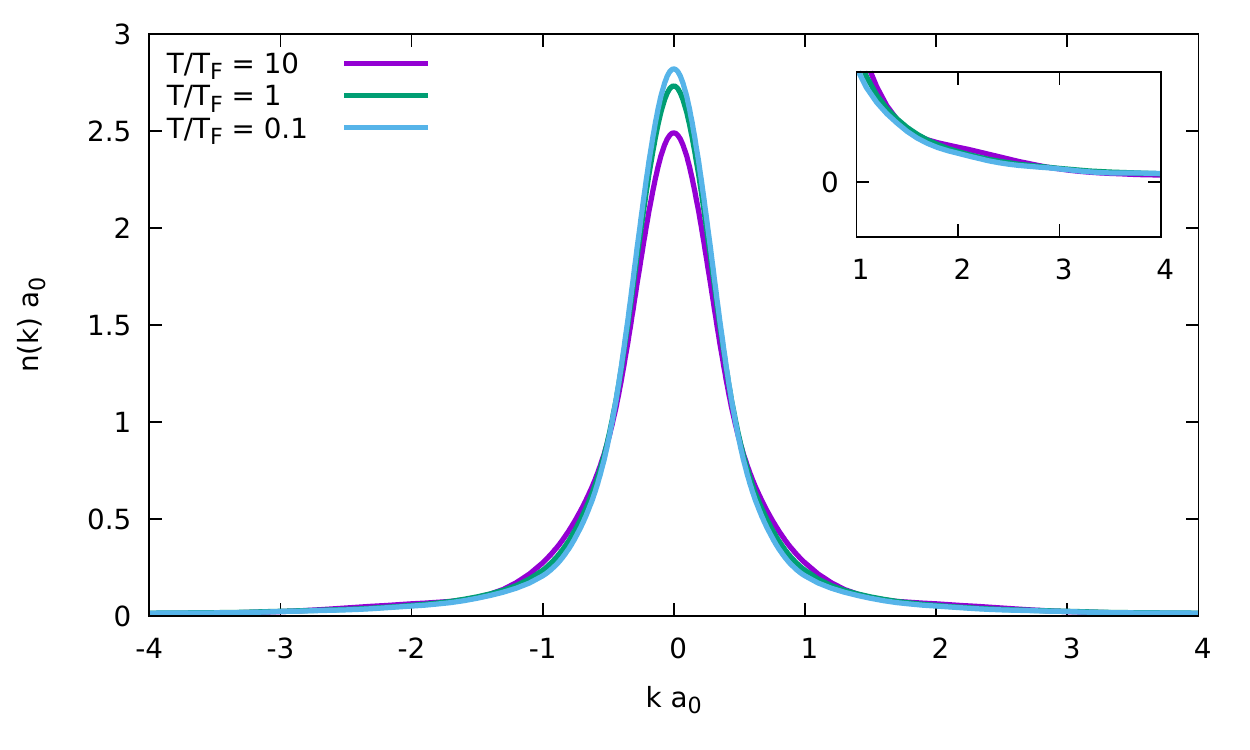}
			\caption{$N=5$.}
		\label{fig.TGN5} \end{subfigure}
		\caption{Momentum distributions for different number of particles as well as 
		different temperatures. The insets show the tails of the curves.}
		Source: By the author.
		 \label{fig.TG}
\end{figure}

As our main interest in taking the above simplification steps is the study of the Tan's contact, 
we still need to go through the momentum distribution in order to achieve our desired goal. 
Therefore, the momentum distribution, in terms of the one-body density matrix, reads
\begin{equation}
n(k) = \frac{1}{2 \pi} \int dx \int dx' \, \expo{ik(x-x')} \varrho^{(1)}(x,x'),
\end{equation}
and is depicted in Fig. \ref{fig.TG} for the number of particles from $N=2$ up to 
$N=5$ as well as for the low-, intermediate-, and high-temperature regimes.

\subsubsection{Tan's contact}
Making use of the asymptotic behavior of the Fourier transform of $|x-x_0|^{a-1} f(x)$,\cite{olshanii,pv2013,bleistein} 
\begin{equation}
	\int dx \, \expo{-ik(x-x_0)} |x-x_0|^{a-1} f(x) = \frac{2}{k^a} f(x_0) \cos (\pi a/2) \Gamma(a),
\end{equation}
and the definition of the contact $\mathcal{C} \equiv k^4 n(k)$ as $k \rightarrow \infty$ we arrive at
\begin{equation}\label{canonicalTG}
	\mathcal{C} = \frac{2}{\pi} \int_{-\infty}^{+\infty} dx \, F(x).
\end{equation}

\begin{figure}[h!]
	\centering
	\includegraphics[width=.8\columnwidth]{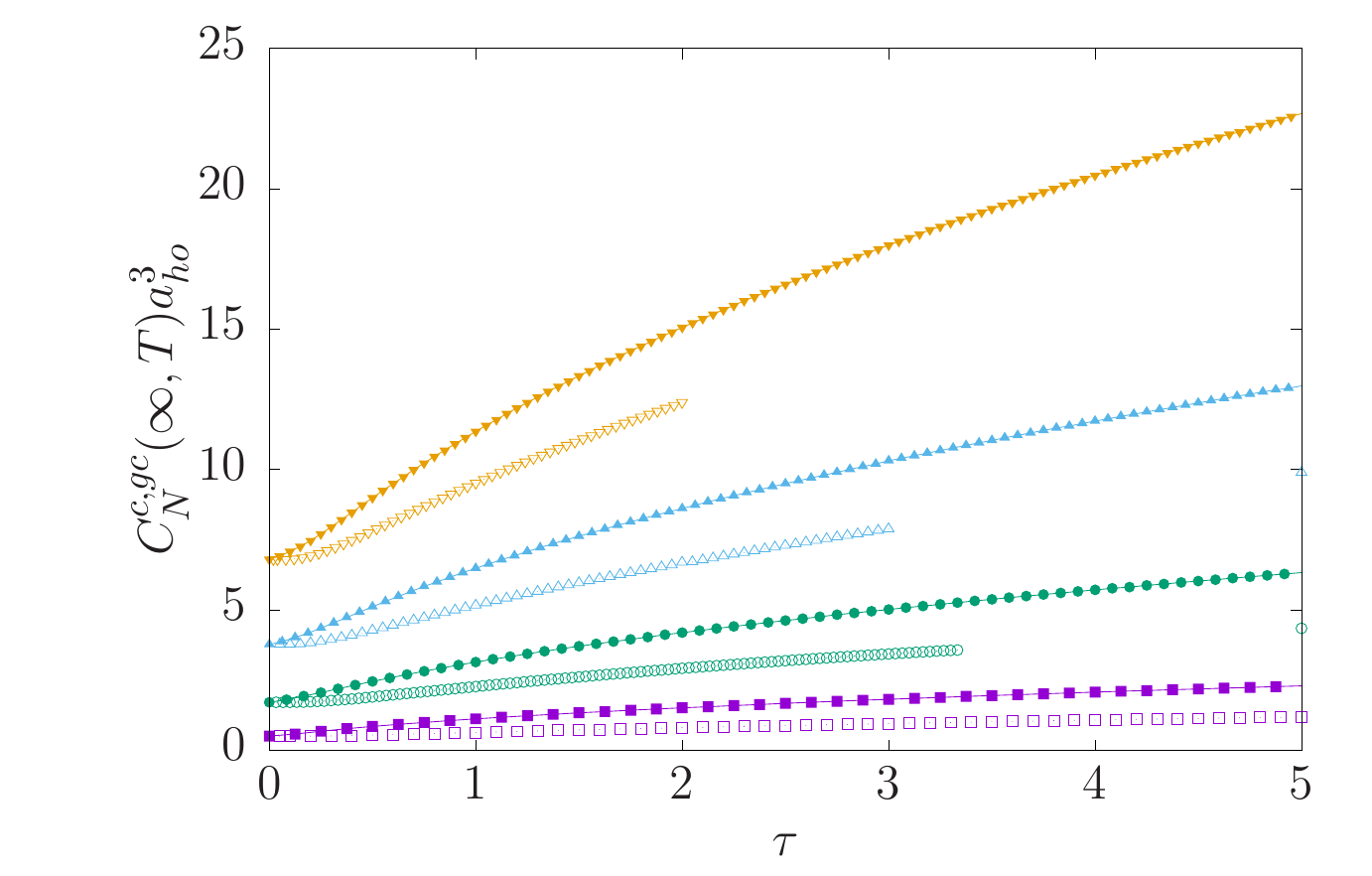}
	\caption{Tonks-Girardeau contact in the canonical ensemble (empty symbols) 
	from Eq. (\ref{canonicalTG}) and in the grand-canonical ensemble (filled symbols) 
	from Ref. \citeonline{pv2013} for $N=2$ (violet squares), $N=3$ (green circles), 
	$N=4$ (light-blue triangles), and $N=5$ (orange inverted triangles). 
	Here, $\tau\equiv T/T_F$ is the adimensional temperature and $a_{ho}\equiv \sqrt{\hbar/m\omega}$ is the harmonic oscillator length.}
	Source: SANT'ANA \textit{et al.}\cite{paper3}
	\label{fig.contactTG}
\end{figure}

The contact from (\ref{canonicalTG}) is depicted in Fig. \ref{fig.contactTG} 
for the number of particles ranging from $N=2$ to 5 in terms of the temperature, together 
with the analogous results within the grand-canonical ensemble from VIGNOLO, P.; MINGUZZI, A,\cite{pv2013} 
which we shall discuss later.

\section{The contact scaling properties}
\subsection{\label{lowT}Zero-temperature scaling}
It was shown in Ref. \citeonline{mateo} that the contact 
for $N$ bosons at $T=0$ can be expressed as a function of the two-boson 
contact, $\mathcal{C}_N=\mathcal{C}_N(\mathcal{C}_2)$. 
Also, it was verified that the scaling relation 
\begin{equation}
	f_N(\tilde{g},T=0) \equiv \frac{\mathcal{C}_N(\tilde{g},T=0)}
	{\mathcal{C}_N(\tilde{g}\to\infty,T=0)},
\end{equation}
where $\tilde{g}\equiv -a_0 a_{1\mathrm{D}}^{-1}/\sqrt{N}$ and 
$\mathcal{C}_N(\tilde{g},T=0) \propto N^{5/2} - \gamma N^\eta$,
establishes the following:   
\begin{equation}
	f_N(\tilde{g},T=0) \simeq f_2(\tilde{g},T=0).
\end{equation}
In particular, in the Tonks-Girardeau limit, $\gamma\approx 1$ and $\eta=3/4$.

For the two-particle case, from (\ref{eq.zeroT}) and (\ref{eq.infTg}), we have that 
\begin{equation}
	f_2(\tilde{g},T=0) = 2\sqrt{\pi}\tilde{g}
	\left[\psi\left(-\frac{\nu}{2}\right)-\psi\left(-\frac{\nu}{2}+\frac{1}{2}\right)\right]^{-1}.
\end{equation}

\subsection{\label{largeT}Large-temperature scaling}
When the temperature is large enough, $T\gg T_F$, where 
$T_F=N \hbar \omega/k_B$ is the Fermi temperature, 
quantum correlations become negligible in the system, so 
that the contact for $N$ particles is given by the two-particle 
contact times the number of pairs,\cite{paper3}
\begin{equation}\label{CN} 
	\mathcal{C}_N(\tilde{g},T\gg T_F) = 
	\frac{N(N-1)}{2} \mathcal{C}_2(\tilde{g},T\gg T_F).
\end{equation}
Following the development at high temperatures from Ref. \citeonline{Yao2018} and 
making use of the Euler reflection formula 
\begin{equation}
	\Gamma(x)\Gamma(1-x) = \frac{\pi}{\sin(\pi x)},
\end{equation}
it is possible to rewrite Eq. (\ref{eq.gamma}) as
\begin{equation}\label{f2}
	f(\nu) = -\cot(\pi \nu /2) 
	\frac{\Gamma\left(1/2+\nu /2\right)}{\Gamma\left(1+\nu /2\right)}.
\end{equation} 
From the asymptotic behavior of the gamma function 
\begin{equation}
	\Gamma(x) \underset{x\to\infty} \sim 
	\expo{x(\log(x)-1)+\mathcal{O}(x^{-3})} 
	\left[\sqrt{\frac{2\pi}{x}}+\mathcal{O} \left( x^{-3/2} \right)\right],
\end{equation}
we obtain the asymptotic formula of (\ref{f2}) 
\begin{equation}
	f(\nu) \underset{\nu\to\infty} \sim 
	-\sqrt{\frac{\nu}{2}} \cot(\pi \nu /2).
\end{equation}
By employing the fact that the solutions of (\ref{eq.gamma}) 
as $\tilde{g} \to \infty$ are given by $2n+1,\, n\in\mathbb{N}$, we get 
\begin{equation}
	\nu = \frac{2}{\pi} \cot^{-1}\left(
	\sqrt{\frac{2n+1}{2}} \tilde{g}^{-1} \right) + 2n.
\end{equation}

Now that we have a formula for the $\nu$'s 
in the strongly interacting limit for large temperatures, 
let us insert it in the contact expression for $N=2$. Recalling 
Eq. (\ref{c2}), we have that 
\begin{equation}\label{c2.2}
	\mathcal{C}_2(\tilde{g}\to\infty,T\gg T_F) = \frac{2^{3/2} \tilde{g}^2}{\pi a_0^3}
	\mathcal{Z}_r^{-1} \sum_n \expo{-\beta E_{\nu_n}^{(r)}} 
	\frac{\partial \nu_n}{\partial \tilde{g}},
\end{equation}
where $\mathcal{Z}_r \equiv \sum_\nu \expo{-\beta E_\nu^{(r)}}$ is the 
relative motion partition function.
By performing the derivative 
$\partial \nu_n/\partial \tilde{g}$, 
(\ref{c2.2}) yields 
\begin{equation}
	\mathcal{C}_2(\tilde{g}\to\infty,T\gg T_F) = \frac{2^{3/2} \tilde{g}^2}{\pi a_0^3}
        \mathcal{Z}_r^{-1} \sum_n \expo{-\beta E_{\nu_n}^{(r)}} 
        \frac{\sqrt{2(2n+1)}}{\pi \tilde{g}^2 \left(1+\frac{2n+1}{2\tilde{g}^2} \right)}.
\end{equation}
As we are interested in the strongly interacting limit $\tilde{g}\to\infty$, 
the term $n/\tilde{g}^2$ in the denominator of the series can be 
disregarded, leaving us with 
\begin{equation}\label{c2.3}
        \mathcal{C}_2 (\tilde{g}\to\infty,T\gg T_F) = \frac{2^{3/2}}{a_0^3}
	\mathcal{Z}_r^{-1} \expo{-\beta \hbar \omega/2} 
	\sum_n \expo{-\beta \hbar\omega (2n+1)}\sqrt{2(2n+1)}.
\end{equation}
By exploiting the fact that
\begin{equation}
	\sum_n \expo{-\alpha (2n+1)}\sqrt{2n+1}
	= \sum_n \expo{-\alpha n}\sqrt{n}
	-\sum_n \expo{-\alpha 2n}\sqrt{2n},
\end{equation}
and that the sums are given by
\begin{subequations}
\begin{align}
	\sum_n \expo{-\alpha (2n+1)} &= 
	\frac{\expo{\alpha}}{\expo{2 \alpha}-1},\\
	\sum_n \expo{-\alpha n}\sqrt{n} &=
	\mathrm{Li}_{-1/2}(\expo{-\alpha}),\\
	\sum_n \expo{-\alpha 2n}\sqrt{2n}&=
	\sqrt{2} \mathrm{Li}_{-1/2}(\expo{-2 \alpha}),
\end{align}
\end{subequations}
with $\alpha\equiv\beta\hbar\omega$, (\ref{c2.3}) yields 
\begin{equation}
        \mathcal{C}_2(\tilde{g}\to\infty,T\gg T_F) = \frac{2^{2}}{a_0^3}
	\expo{-\alpha} \left(\expo{2\alpha}-1\right)
	\left[\mathrm{Li}_{-1/2}(\expo{-\alpha})
	-\sqrt{2} \mathrm{Li}_{-1/2}(\expo{-2 \alpha})\right],
\end{equation}
where $\mathrm{Li}_n(z)$ is the polylog function.\cite{handbook}
The expasions of the polylog functions around $\alpha=0$ yield 
\begin{equation}
	\begin{aligned}
	\mathrm{Li}_{-1/2}(\expo{-\alpha})
        -\sqrt{2} \mathrm{Li}_{-1/2}(\expo{-2 \alpha}) 
	 \approx &
		\frac{\pi^{-3/2}}{4 \alpha^{3/2}} 
	+ \mathcal{O} \left(\alpha^{3/2}\right) \\
		&+\left(1-\sqrt{2}\right)\zeta(-1/2) 
		+\left(2\sqrt{2}-1\right) \alpha \zeta\left(-3/2\right)
	+\mathcal{O}(\alpha^2),
	\end{aligned}
\end{equation}
with $\zeta(x)$ being the Riemann zeta function.\cite{handbook}
Therefore, as $\alpha$ approaches zero, the contact reduces to
\begin{equation}\label{C2TGbigT}
	\mathcal{C}_2(\tilde{g}\to\infty,T\gg T_F) = \frac{2 \pi^{-3/2}}{a_0^3} 
	\sqrt{\frac{k_B T}{\hbar \omega}}
	= \left(\frac{2}{\pi}\right)^{3/2} a_0^{-3} \sqrt{\frac{T}{T_F}}.
\end{equation}

Hence, the expression for the $N$-particle contact (\ref{CN}) 
in the strongly interacting limit is given by 
\begin{equation}\label{CN2}
        \mathcal{C}_N\left(\tilde{g}\to\infty,T\gg T_F\right) =
	\frac{N(N-1)}{a_0^3 \pi^{3/2}} 	\sqrt{\frac{NT}{T_F}}.
\end{equation}
Therefore, we have the scaling function in the high-temperature regime 
for the TG limit as being  
\begin{equation}
	\begin{aligned}
	&\mathcal{C}_N\left(\tilde{g}\to\infty,T\gg T_F\right) = 
	h_N\left(\tilde{g}\to\infty,T\gg T_F\right)  \left(N^{5/2} - N^{3/2}\right) \\
	&\Rightarrow h_N\left(\tilde{g}\to\infty,T\gg T_F\right) = 
		h_2(\tilde{g}\to\infty,T\gg T_F) = \frac{\sqrt{T/T_F}}{a_0^3 \pi^{3/2}}.
	\end{aligned}
\end{equation}

\subsection{Generalized scaling conjecture}
Having studied the behavior of the contact in terms of the number 
of particles for both high ($T\gg T_F$) and 
low temperatures ($T\ll T_F$) in the strongly interacting 
limit $\tilde{g}\to\infty$, we now propose a conjecture 
for the entire range of temperatures in the aforesaid regime. 

At large temperatures, where quantum correlations play a 
minor role towards the properties of the system, the contact 
dependency on the number of particles is given by the number of 
pairs $N(N-1)$ times a $\sqrt{N}$ factor that comes from the 
Fermi temperature. As the temperature decreases, there happens 
an intensification on the $N$-dependency of the contact, from 
$N^{5/2}-N^{3/2}$ to $N^{5/2}-N^{3/4}$. Therefore, 
following such a reasoning, we have proposed the following 
scaling hypothesis:\cite{paper3} 
\begin{equation}\label{eq.conjecture}
	\mathcal{C}_N(\tilde{g}\to\infty,\tau)
	\propto N^{5/2} - N^{3/4 [1+\exp{(-2/\tau)}]},
\end{equation}
where $\tau\equiv T/T_F$.

\begin{figure}[h!]
        \centering
        \includegraphics[width=.8\columnwidth]{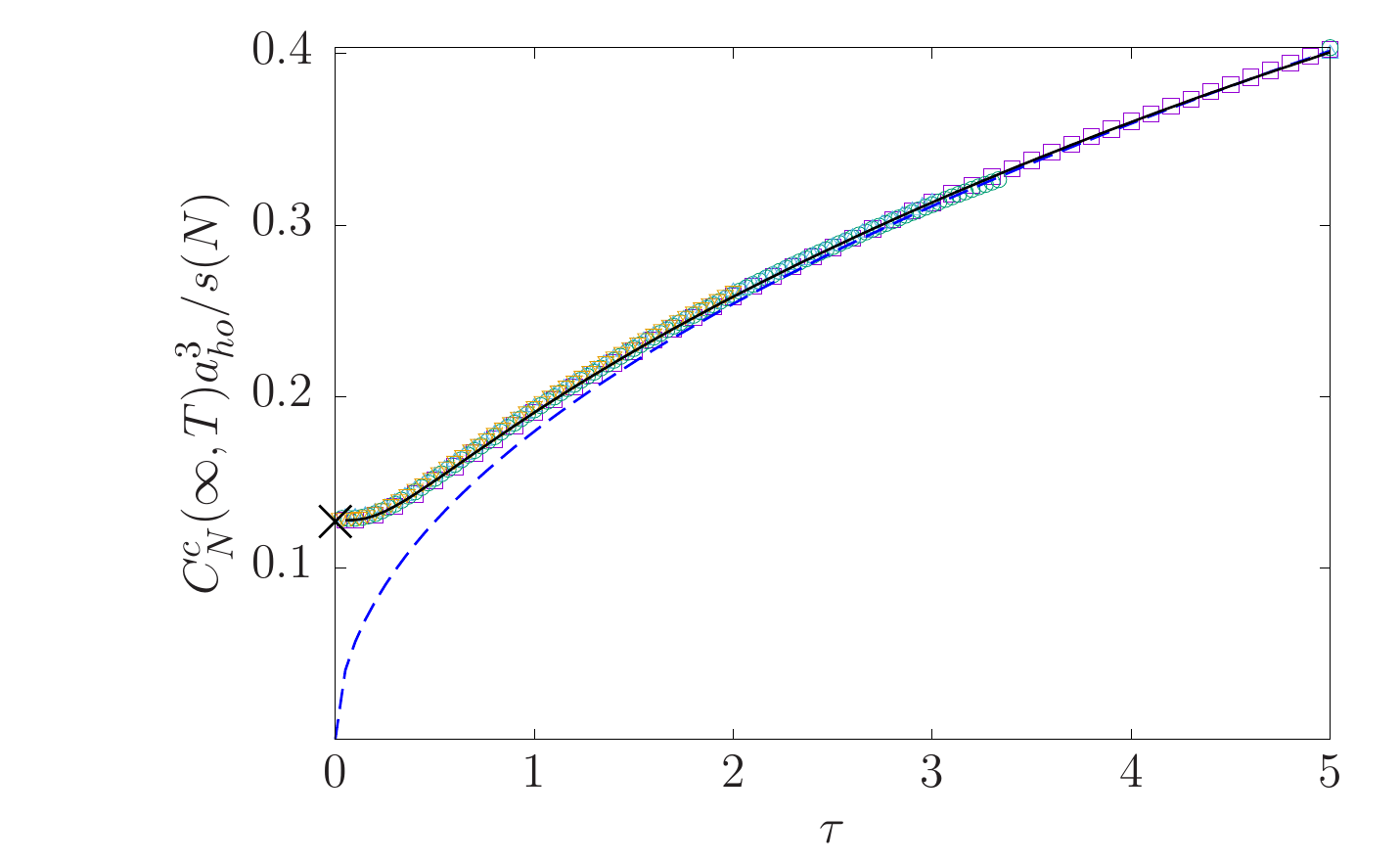}
        \caption{Tan's contacts in the Tonks-Girardeau limit from Eq. 
	(\ref{canonicalTG}) in adimensional units as functions of the 
	reduced temperature $\tau$ scaled by the generalized conjecture $s(N)\equiv N^{5/2}
	-N^{3/4[1+\exp(-2/\tau)]}$ for the respective number of particles: 
	$N=2$ (violet squares), $N=3$ (green circles), $N=4$ (blue triangles), 
	and $N=5$ (orange inverted triangles). The black cross corresponds 
	to the zero-temperature Tonks-Girardeau two-boson 
	contact from Eq. (\ref{eq.infTg}) rescaled by $s(N)$ for $\tau=0$: 
	$(2^{5/2}-2^{3/4})^{-1}\mathcal{C}_2(\tilde{g}\to\infty,T\to 0)
	=(2^{5/2}-2^{3/4})^{-1} (2/\pi)^{3/2}=0.127784$. 
	The blue dashed line corresponds to 
	$\sqrt{\tau}/\pi^{3/2}$, while the black 
	continuous line is simply the contact rescaled 
	by the generalized scaling factor, 
	\textit{i.e.}, the implicit proportionality factor in 
	Eq. (\ref{eq.conjecture}).}
	Source: SANT'ANA \textit{et al.}\cite{paper3}
	\label{fig.conjecture}
\end{figure}

Fig. \ref{fig.conjecture} displays the Tan's contacts in the strongly 
interacting limit from Eq. (\ref{canonicalTG}) scaled by 
our proposed generalized conjecture from (\ref{eq.conjecture}) for 
different number of particles ranging from 2 to 5. Hence, we 
can observe the collapse of all data on the same curve. 
Moreover, we compare it to the blue dashed line, which corresponds 
to the high-temperature behavior in the TG limit from Eq. (\ref{C2TGbigT}). 
We observe that all data, in the high-temperature regime, 
also collapse over the curve of the 
high-temperature TG-limit two-boson contact 
$\mathcal{C}_2(\tilde{g}\to\infty,T\gg T_F)$.

\subsection{Intermediate interaction strength scaling}
\begin{figure}[h!]
  \centering
    \includegraphics[width=0.49\columnwidth]{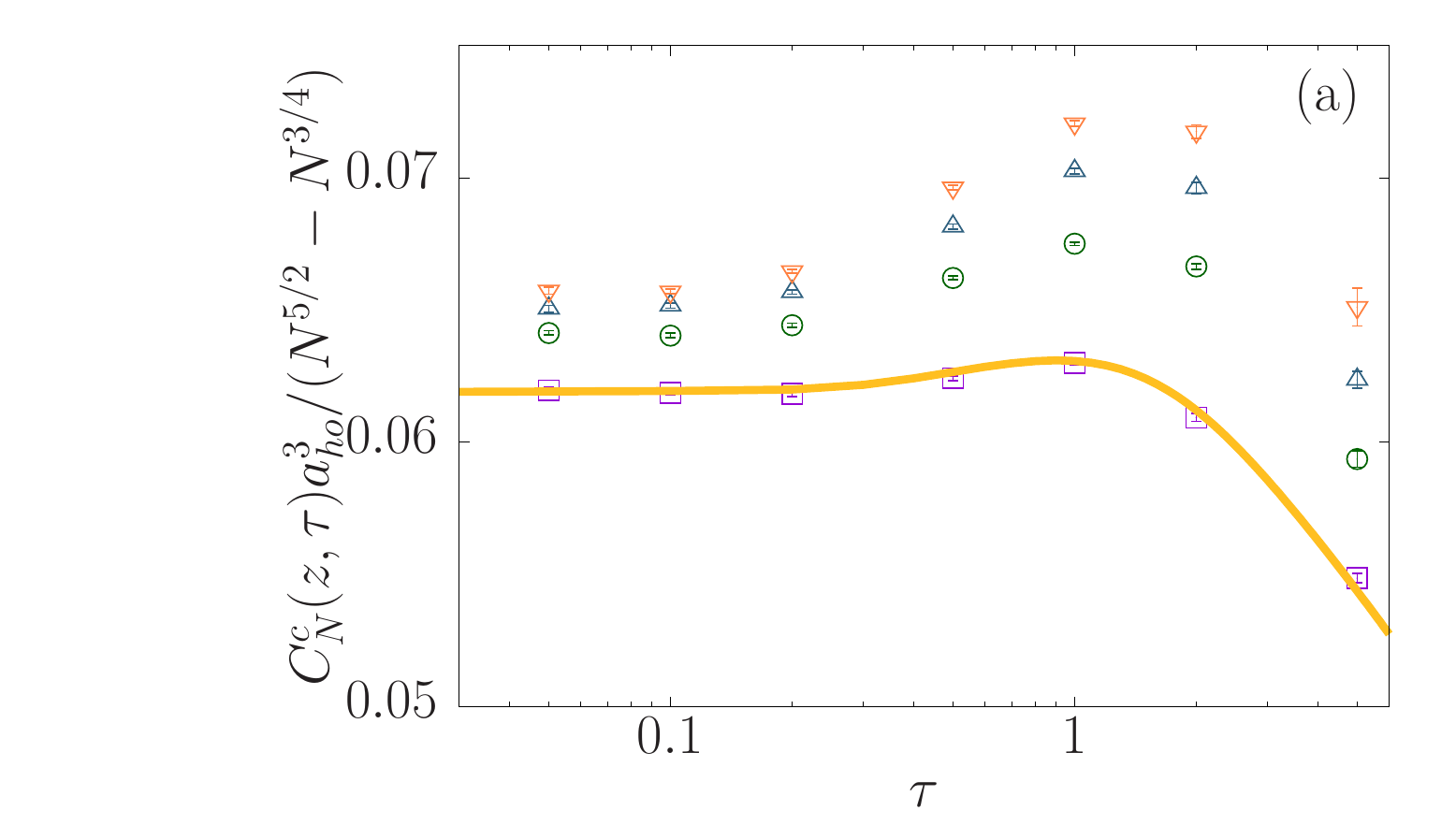}
    \includegraphics[width=0.49\columnwidth]{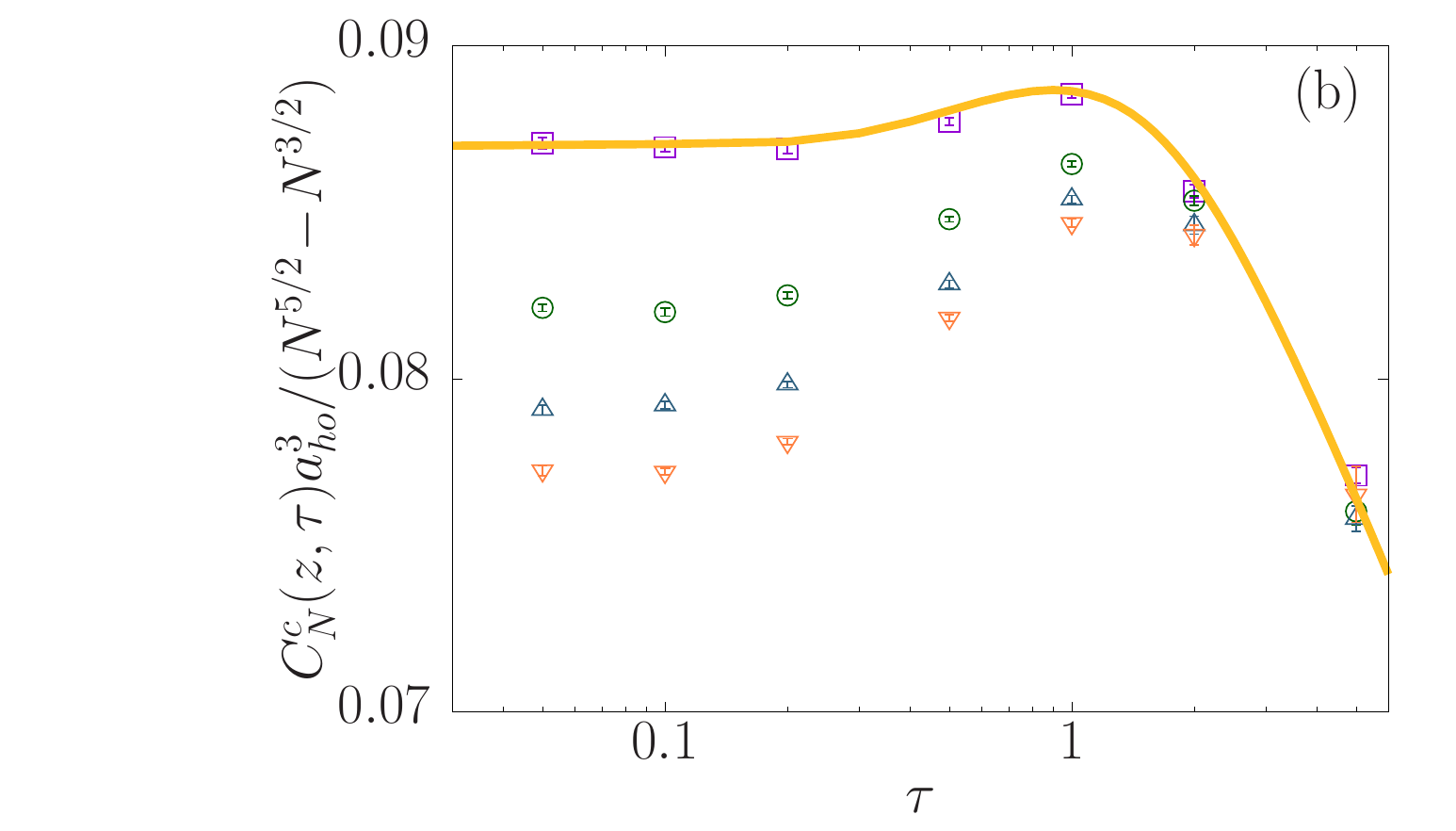}
    \includegraphics[width=0.49\columnwidth]{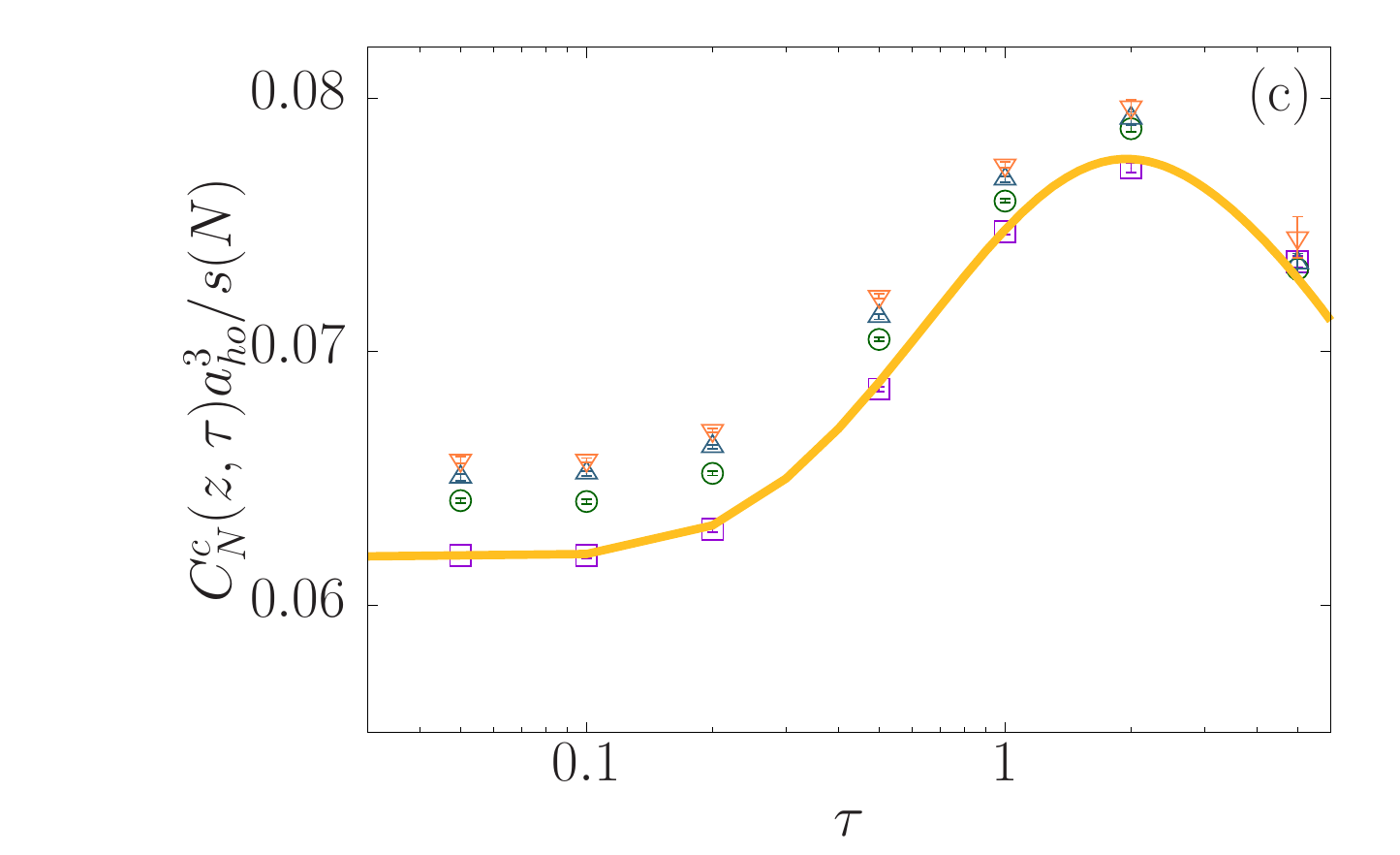}
    \includegraphics[width=0.49\columnwidth]{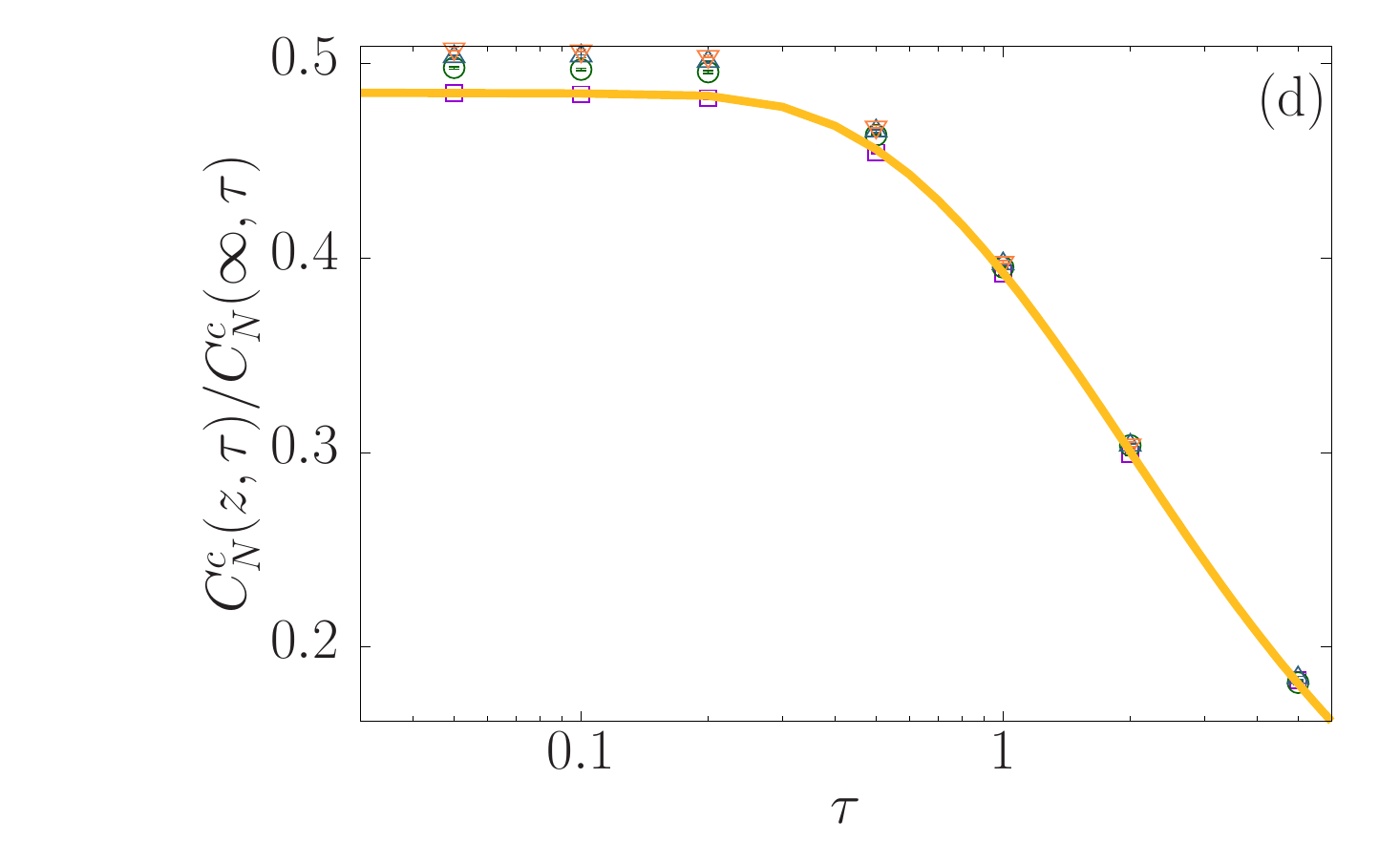}
  \caption{\label{can-G2}Tan's contacts from QMC simulations as functions 
  of the adimensional temperature $\tau$ for $z=\tilde{g}=1$. 
  The panels (a), (b), and (c) correspond to the rescaling of the contact regarding the 
    low-temperature factor $N^{5/2}-N^{3/4}$, the high-temperature factor $N^{5/2}-N^{3/2}$, and 
	the all-range-temperature factor $s(N)\equiv N^{5/2}-N^{3/4[1+\exp(-2/\tau)]}$, respectively. 
    In panel (d), the QMC data is rescaled by the TG-limit contact from Eq. (\ref{canonicalTG}). 
    The symbol styles correspond to: $N=2$ (violet squares), $N=3$ (green circles),
   $N=4$ (blue triangles), and $N=5$ (orange inverted triangles).
   The continuous yellow line corresponds to the two-boson
  contact obtained by Eq. (\ref{eq.contactN2}). 
  The QMC error bars are smaller than the symbol sizes.}
  Source: SANT'ANA \textit{et al.}\cite{paper3}
\end{figure}

\begin{figure}[h!]
   \centering
    \includegraphics[width=0.49\columnwidth]{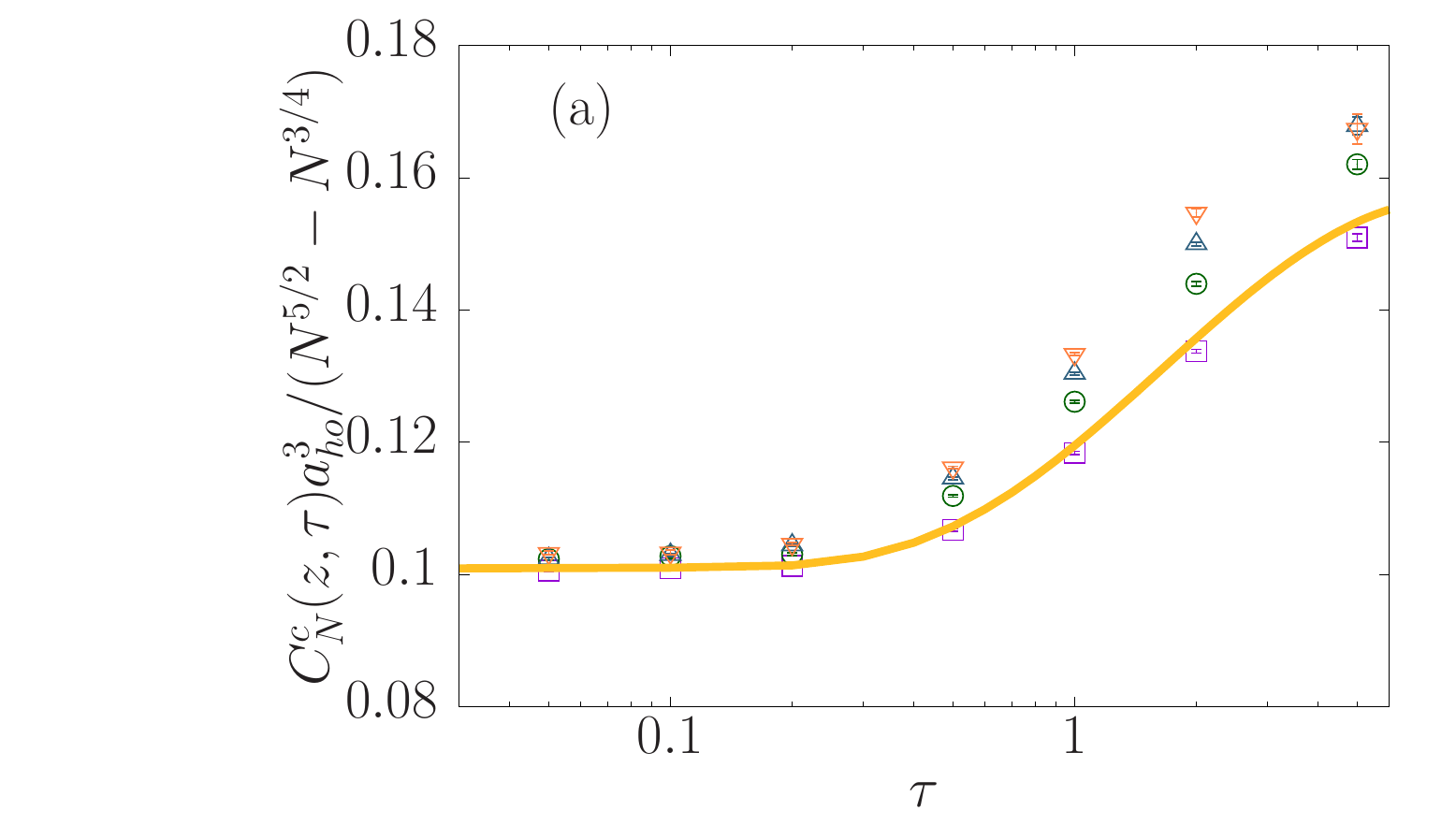}
    \includegraphics[width=0.49\columnwidth]{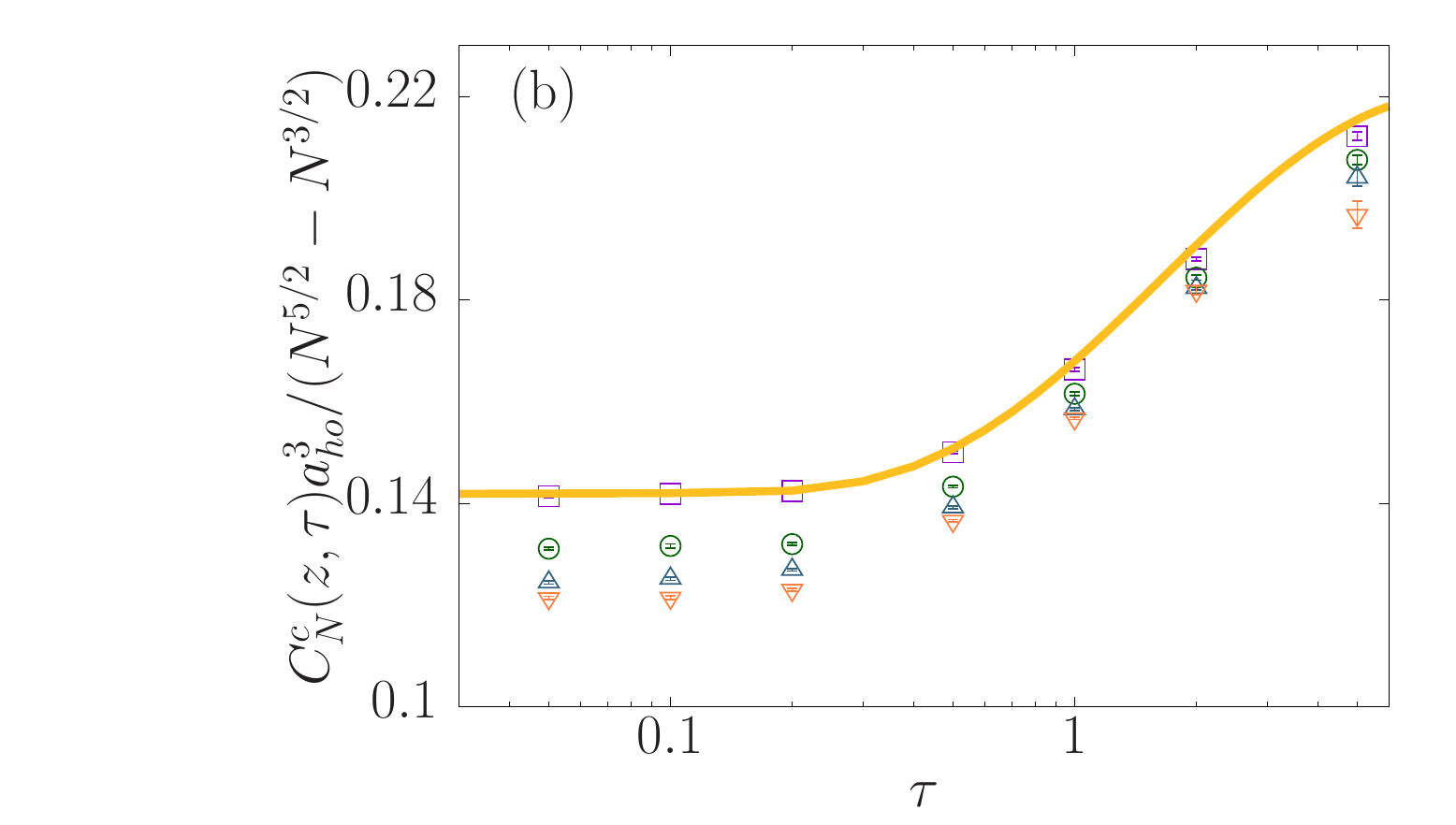}
    \includegraphics[width=0.49\columnwidth]{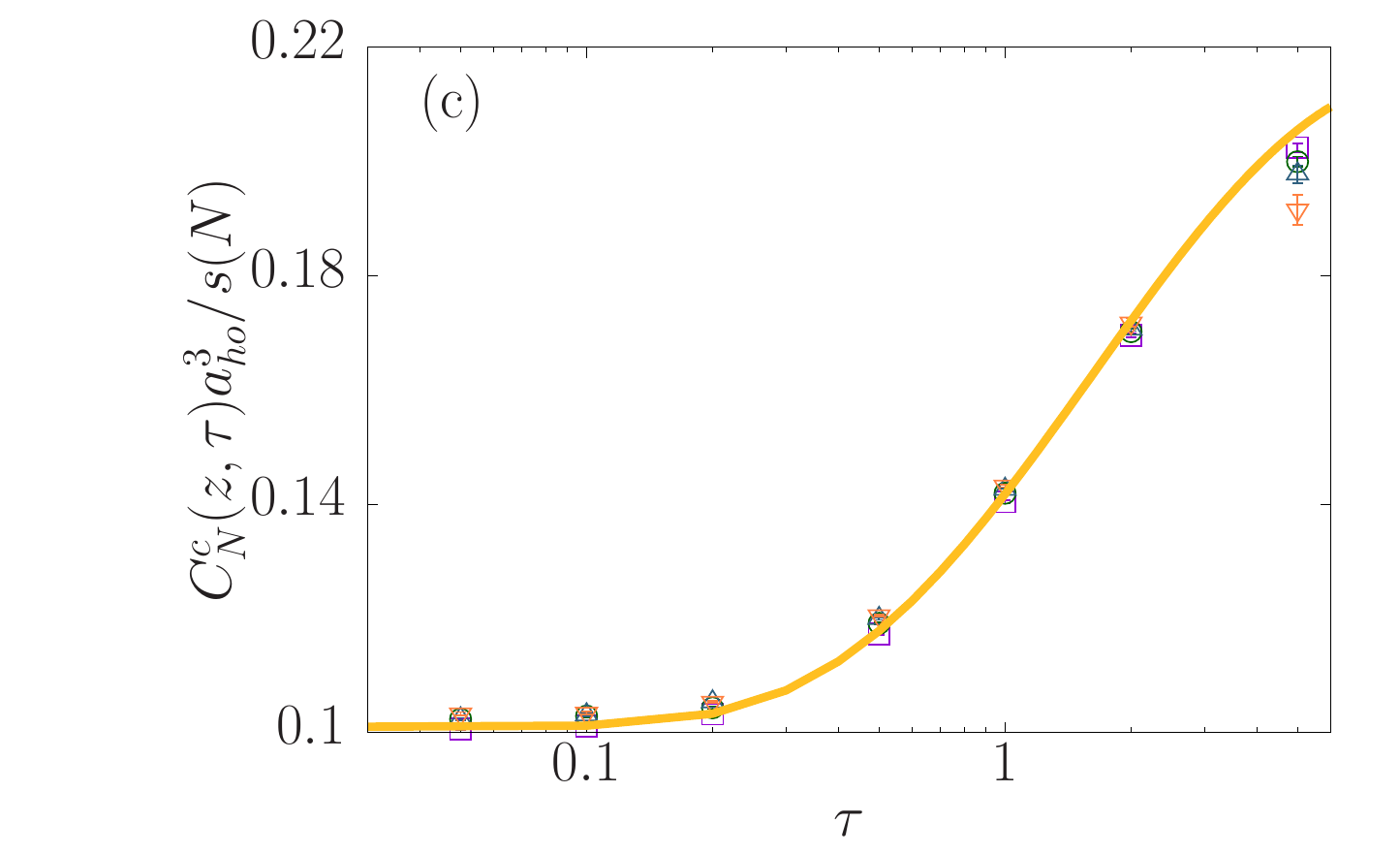}
    \includegraphics[width=0.49\columnwidth]{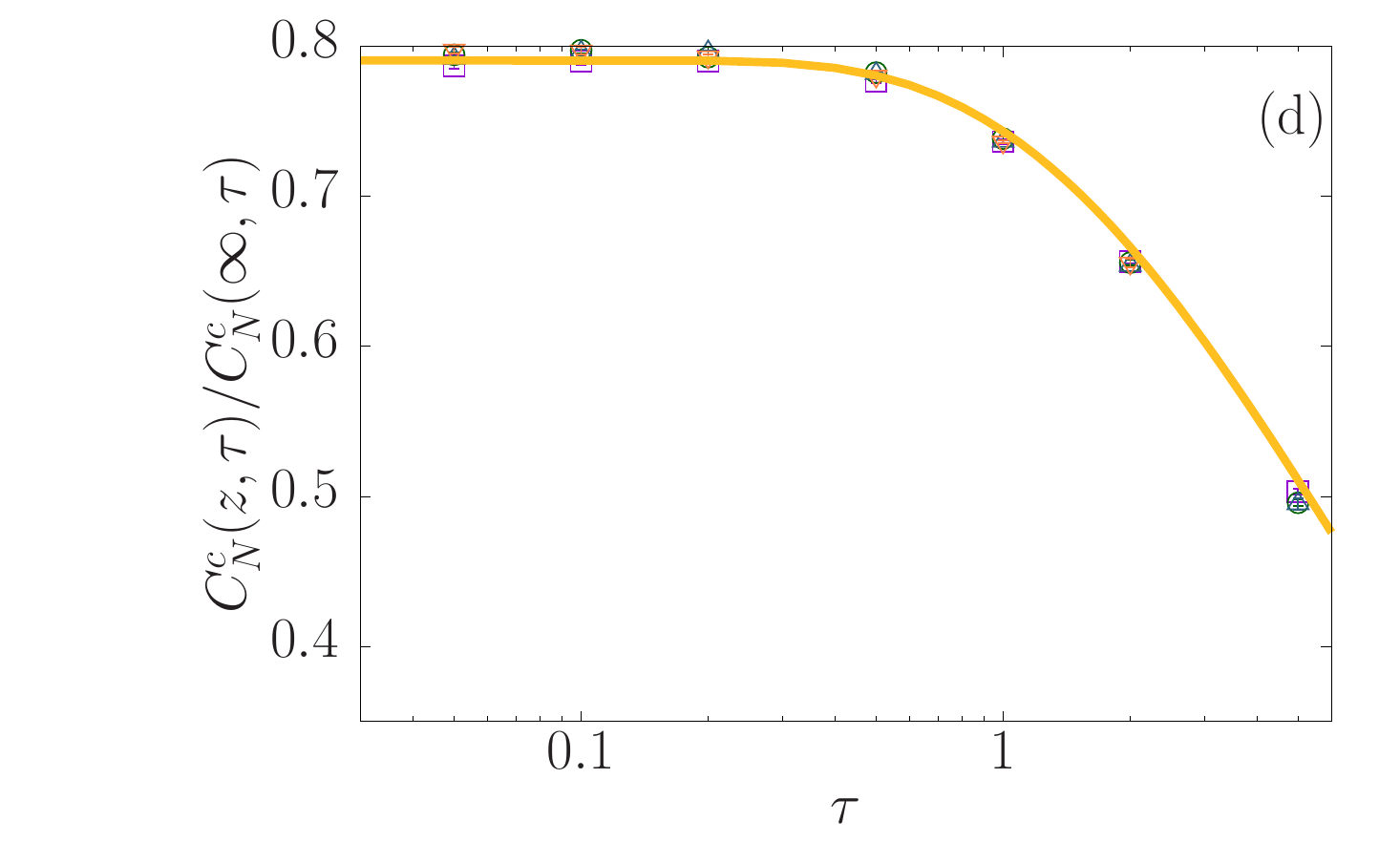}
  \caption{\label{can-G5} Tan's contacts from QMC simulations as functions 
  of the adimensional temperature $\tau$ for $z=\tilde{g}=2.5$. 
  The panels (a), (b), and (c) correspond to the rescaling of the contact regarding the 
    low-temperature factor $N^{5/2}-N^{3/4}$, the high-temperature factor $N^{5/2}-N^{3/2}$, and 
	the all-range-temperature factor $s(N)\equiv N^{5/2}-N^{3/4[1+\exp(-2/\tau)]}$, respectively. 
    In panel (d), the QMC data is rescaled by the TG-limit contact from Eq. (\ref{canonicalTG}). 
    The symbol styles correspond to: $N=2$ (violet squares), $N=3$ (green circles),
   $N=4$ (blue triangles), and $N=5$ (orange inverted triangles).
   The continuous yellow line corresponds to the two-boson
  contact obtained by Eq. (\ref{eq.contactN2}). 
  The QMC error bars are smaller than the symbol sizes.}
  Source: SANT'ANA \textit{et al.}\cite{paper3}
\end{figure}

We now turn our attention to the intermediate interaction strength scenario 
$\tilde{g} \sim 1$. In such a regime, the integrability of the system breaks down, 
so that we must rely on quantum Monte Carlo (QMC) calculations.\footnote{Thanks to Fr\'ed\'eric H\'ebert for performing the QMC simulations. 
 For the details on the QMC simulations, see SANT'ANA \textit{et al.}\cite{paper3}}
We analyze the QMC data for $\tilde{g}=1$ in Fig. \ref{can-G2} and 
for $\tilde{g}=2.5$ in Fig. \ref{can-G5}. In both figures, 
we rescale the contact by all scaling factors introduced: 
the zero-temperature scaling $N^{5/2}-N^{3/4}$ (panel (a)), 
the large-temperature scaling $N^{5/2}-N^{3/2}$ (panel (b)), 
and the generalized scaling conjecture $N^{5/2}-N^{3/4[1+\exp(-2/\tau)]}$ (panel (c)). 
In panel (d) of both Figs. \ref{can-G2} and \ref{can-G5}, we 
rescale the QMC data by the TG-limit contact from Eq. (\ref{canonicalTG}).
We observe that, the zero-temperature scaling factor in Fig. \ref{can-G2}(a) 
makes the data approach each other at small temperatures, while 
we observe a collapse of the whole data in Fig. \ref{can-G5}(a) at low temperatures. 
The same occurs at high temperatures: the data approach each other in Fig. \ref{can-G2}(b), while 
they collapse over each other in Fig. \ref{can-G5}(b). Differently, the 
generalized scaling conjecture works well within the whole temperature range, with 
an incertitude of $5\%$ for $\tilde{g}=1$ (Fig. \ref{can-G2}(c)) and 
of $1\%$ for $\tilde{g}=2.5$ (Fig. \ref{can-G5}(c)). 

Moreover, analogously to the zero-temperature analysis, we can define the ratio 
\begin{equation}\label{f_N(g,t)}
	f_N(\tilde{g},T) \equiv \frac{\mathcal{C}_N(\tilde{g},T)}
	{\mathcal{C}_N(\tilde{g}\to\infty,T)}
\end{equation}
and certify ourselves that, from Figs. \ref{can-G2}(d) and \ref{can-G5}(d), 
the relation 
\begin{equation}\label{eq.scaling}
	f_N(\tilde{g} \gtrsim 1,T\gg T_F) = f_2(\tilde{g}\gtrsim 1,T\gg T_F)
\end{equation}
holds for the whole temperature range \textemdash  
 not only for the low-temperature regime ($\tau \ll 1$) 
and for the high-temperature one ($\tau \gg 1$) as previously 
stated, but also for the intermediate-temperature regime ($\tau \sim 1$).

Now, let us summarize the results of this section. An important consequence of the scaling results
is that the contact for $N$ bosons at temperature $T$ with repulsive interaction characterized
by the adimensional interaction strength $\tilde{g}$ rescaled by the contact for $N$ strongly interacting bosons at temperature $T$,
\textit{i.e.}, $\mathcal{C}_N(\tilde{g},T)/\mathcal{C}_N({\tilde{g}\to\infty,T})$, is a universal
function of the adimensional interaction strength $\tilde{g}\equiv -a_0 a_{\mathrm{1D}}^{-1}/\sqrt{N}$
and the adimensional temperature $\tau\equiv T/T_F$. Furthermore, another important result comes from 
the generalized scaling function. Namely, the ratio between the contact for $N$ bosons at 
temperature $T$ with repulsive interaction characterized
by an interaction strength $g$ and the generalized scaling function 
$s(N)\equiv N^{5/2}-N^{3/4[1+\exp(-2/\tau)]}$, \textit{i.e.}, 
$\mathcal{C}_N(\tilde{g},T)/s(N)$, is also a universal function 
of $\tilde{g}$ and $\tau$.

\section{Comparison between ensembles}
In this section we draw a comparison between 
the contact calculated from the canonical  
ensemble evaluated in this thesis and the grand-canonical 
one from Ref. \citeonline{pv2013} at finite temperature. 
The motivation for comparing both ensembles comes from the fact 
that the scaling properties of the contact are strongly affected 
by the statistical distribution. 
Considering that in most ultracold atom experiments the number of particles 
is fixed, performing the calculations within the canonical ensemble 
is a more appropriate choice. On the other hand, the grand-canonical ensemble 
is advantageous when dealing with systems where the number of particles vary, 
such as open systems. 
In the latter, the average number of particles $\av{N}$ is then determined 
by fixing the temperature and the chemical potential, while the most remarkable feature of the former 
is that the particles can have any value for its energy and the average energy $\av{E}$ of 
the whole system is then determined by the temperature.
For the sake of clarity, in this section we introduce the index $^{(gc)}$ 
that represents the respective physical quantity in the 
grand-canonical ensemble. Moreover, we analyze the 
differences from both ensembles in the QMC simulations. 

\subsection{Analytical formula}
\begin{figure}[h!]
	\centering
	\includegraphics[width=.8\columnwidth]{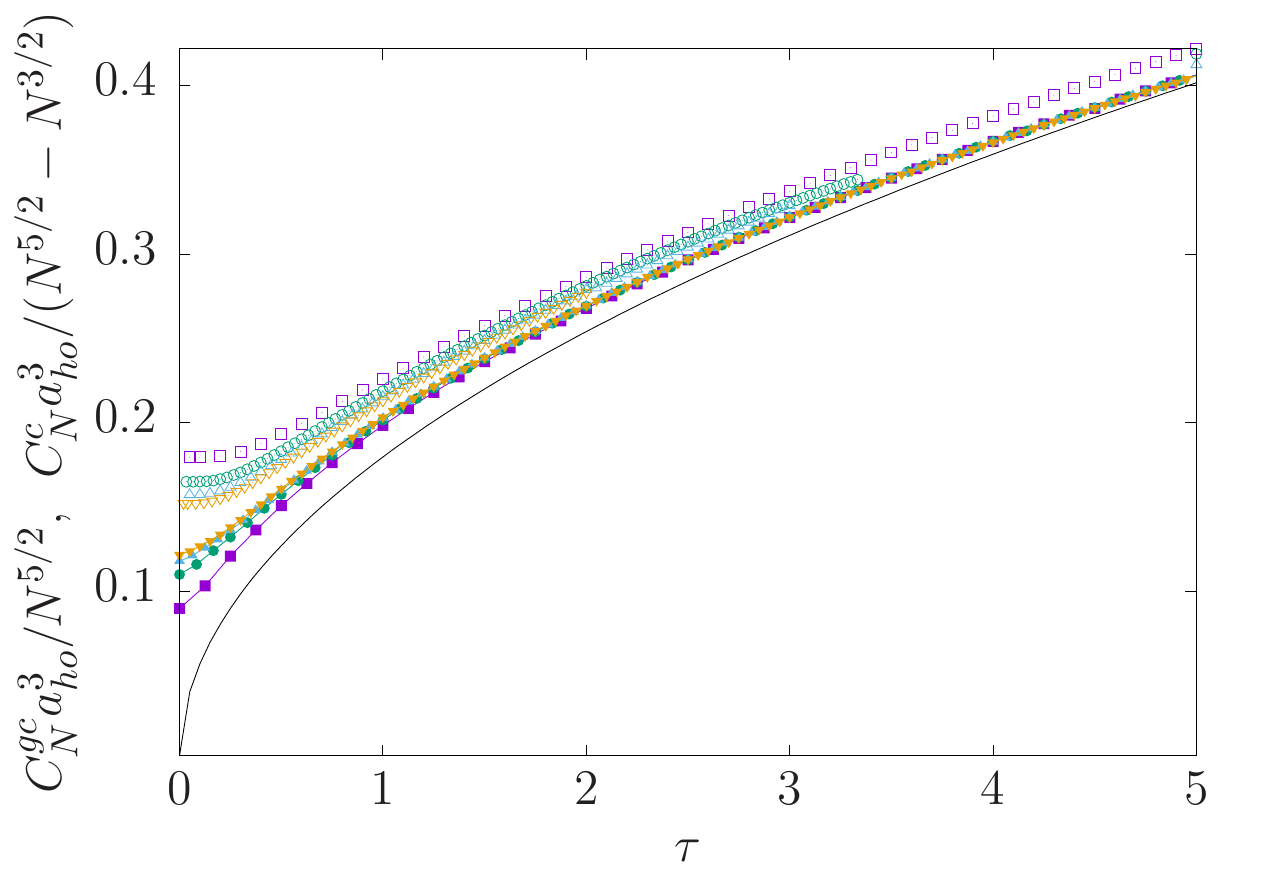}
	\caption{Canonical (empty symbols) contact in the TG limit 
	from (\ref{canonicalTG})
	rescaled by $N^{5/2}-N^{3/2}$
	and grand-canonical (filled symbols) contact from Ref. \citeonline{pv2013} 
	rescaled by $N^{5/2}$
	as functions of $\tau$ for the respective number 
	of particles: $N=2$ (violet squares), $N=3$ (green circles), 
	$N=4$ (blue triangles), and $N=5$ (orange inverted triangles). 
	The black continuous curve corresponds to $\sqrt{\tau}/\pi^{3/2}$.}
	Source: SANT'ANA \textit{et al.}\cite{paper3}
	\label{fig.comp_can-gc}
\end{figure}

In the zero-temperature limit, there is 
no physical distinction between the grand-canonical 
and the canonical ensembles. Thus, the contacts 
from both calculations scale as $N^{5/2}-N^{3/4}$. However, 
by increasing the temperature, the distinction between ensembles 
is enhanced. Such differences can be observed from Fig. \ref{fig.contactTG}, 
where the grand-canonical contact increases more rapidly when 
compared to the canonical one as the temperature rises. In fact, 
in the large-temperature scenario, the term corresponding to the number of pairs 
$N(N-1)$ in Eq. (\ref{CN}) has to be replaced by its average value 
in the grand-canonical calculations: 
\begin{equation}
\av{N(N-1)} = \av{N^2} -\av{N} = \av{N}^2.
\end{equation}
Note that this last step follows from the fact that, at large $T$, 
$\av{\Delta N^2} \simeq \av{N} $. 
So, analogously to the canonical scaling (\ref{CN}), we have for 
the grand-canonical contact that 
\begin{equation}
	\mathcal{C}^{(gc)}_N(\tilde{g},T\gg T_F) = 
	\frac{\av{N}^2}{2} \mathcal{C}_2(\tilde{g},T\gg T_F).
\end{equation}
In the TG limit, the above relation together with (\ref{CN2}) yields, 
by defining $T_F = \av{N} \hbar \omega /k_B$, 
the $\av{N}^{5/2}$-dependency, 
\begin{equation}
	\mathcal{C}^{(gc)}_N(\tilde{g}\to\infty,\tau \gg 1) = 
	\frac{\av{N}^{5/2} }{\pi^{3/2} a_0^3} \sqrt{\tau},
\end{equation}
corroborating the result from Ref. \citeonline{Yao2018}.
This result is shown in Fig. \ref{fig.comp_can-gc}.

\subsection{Quantum Monte Carlo simulations}
\begin{figure}[h!]
        \centering
        \includegraphics[width=.8\columnwidth]{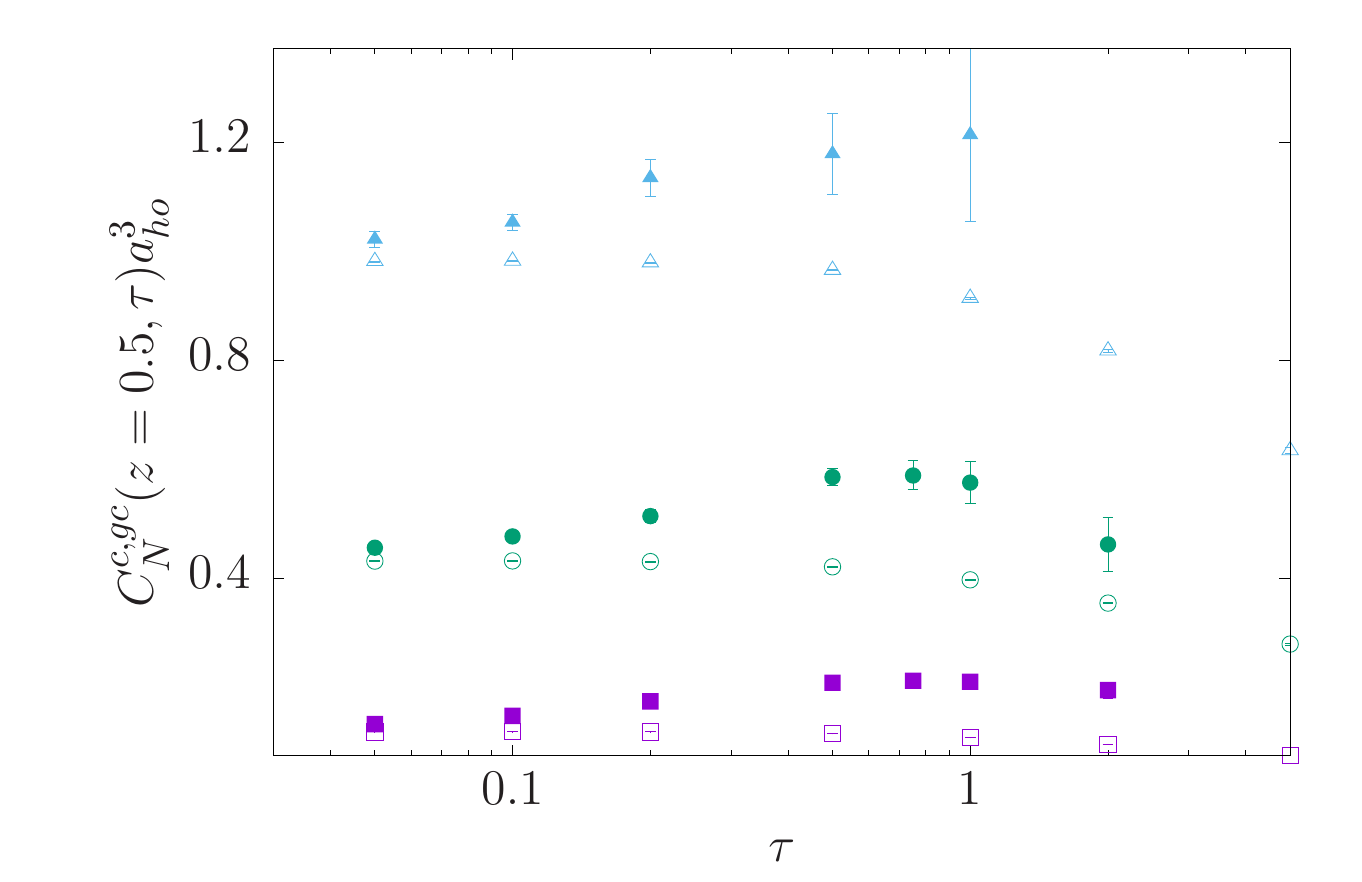}
	\caption{Tan's contacts evaluated from canonical QMC 
	simulations (empty symbols) and 
	from grand-canonical QMC simulations (filled symbols) 
	as functions of $\tau$ in the 
	weak-intermediate regime $z=\tilde{g}=0.5$ for the respective 
	number of particles: $N=2$ (violet squares), 
	$N=3$ (green circles), $N=4$ (blue triangles). 
	QMC error bars in the canonical-ensemble calculation 
	are smaller than the symbols size.}
	Source: SANT'ANA \textit{et al.}\cite{paper3}
	\label{fig.QMC-confronto}
\end{figure}

Now, let us analyze the differences between the canonical and the 
grand-canonical QMC simulations. Fig. 
\ref{fig.QMC-confronto} displays the contact from 
both ensembles calculated via 
QMC considering the weak-intermediate interaction regime, 
$\tilde{g}=0.5$. 
As expect and already discussed, the grand-canonical 
contact presents a steeper rise as the temperature 
increases when compared to the canonical one. This 
difference can be explained by the fundamentals of the grand-canonical 
ensemble: there is a probability that the system contains any number of particles; 
and such contributions, especially for $N>\av{N}$, 
produces an initial growth of the contact at low temperatures. This explanation is 
better understood by mathematical means: 
consider the fugacity term $\sum_N \expo{\beta \mu N}$, it is then 
straightforward to see that such a contribution becomes considerable 
at large values of $N$ as well as small values of $T\propto \beta^{-1}$. 
Then, at some point, the contact reaches its maximum, which was recently explained as 
the mark of the crossover between a quasicondensate and an ideal Bose gas.\cite{Yao2018} 
Afterwards, it begins to decrease. This decline is explained by the very nature of the quantum realm: 
as $T$ increases, the de Broglie wavelength decreases, then the overlap between 
individual particle waves also decreases, resulting in the overall 
drop of the total contact.

\begin{figure}[h!]
        \centering
        \includegraphics[width=.8\columnwidth]{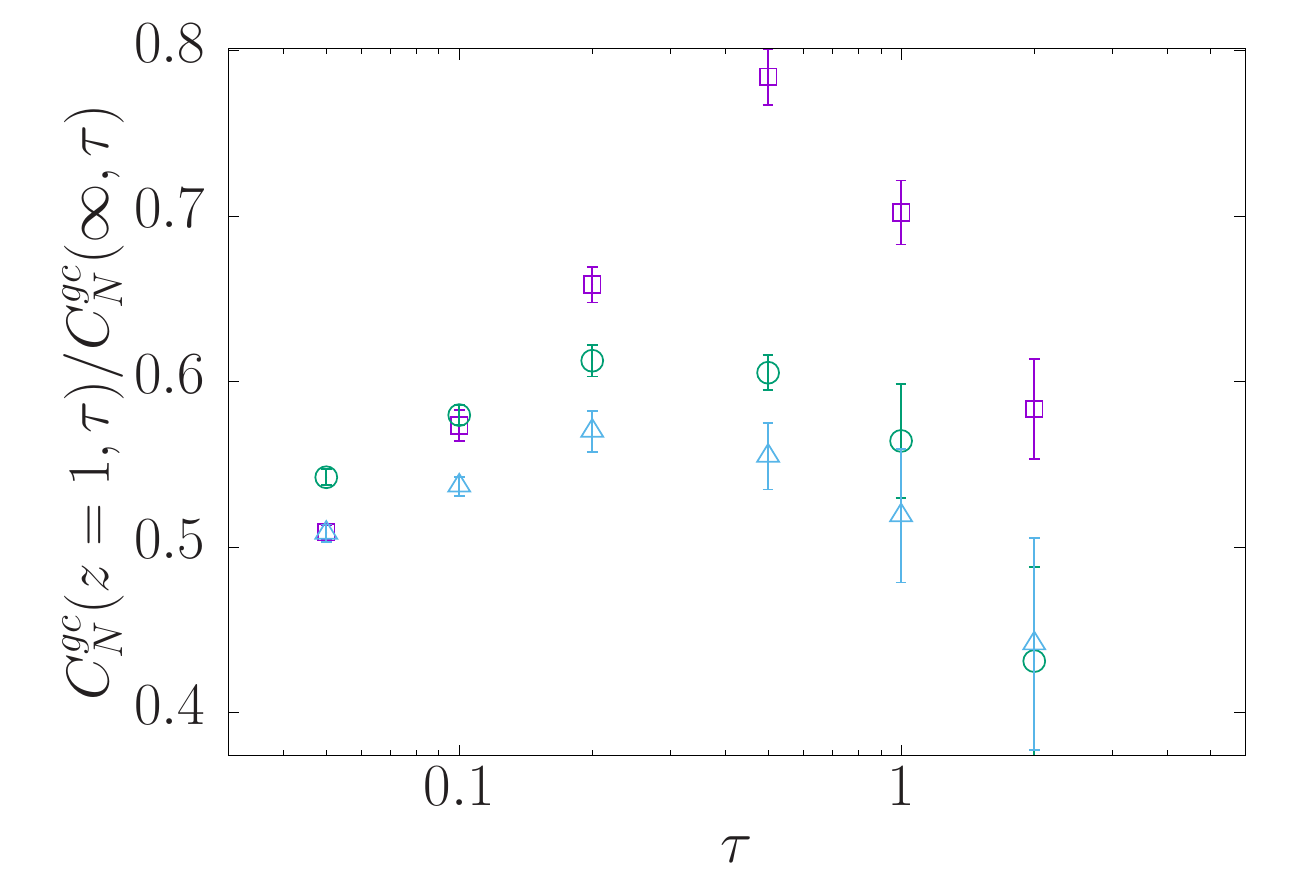}
	\caption{QMC grand-canonical contact rescaled by 
	the TG grand-canonical contact from Ref. \citeonline{pv2013} as a 
	function of $\tau$ for the respective number of particles: 
	$N=2$ (violet squares), $N=3$ (green circles), and 
	$N=4$ (blue triangles).}
	Source: SANT'ANA \textit{et al.}\cite{paper3}
	\label{fig.QMC-ratio}
\end{figure}

Now, let us test the scaling hypothesis (\ref{eq.scaling}) 
by plotting the ratio 
\begin{equation}\label{eq.6.121ultima}
	\frac{\mathcal{C}^{(gc)}_N(\tilde{g}=1,T)}
	{\mathcal{C}^{(gc)}_N(\tilde{g}\to\infty,T)},
\end{equation}
with $\mathcal{C}^{(gc)}_N(\tilde{g}=1,T)$ being the data from 
QMC calculation and $\mathcal{C}^{(gc)}_N(\tilde{g}\to\infty,T)$ 
being the analytical formula from Ref. \citeonline{pv2013}. These 
results are displayed in Fig. \ref{fig.QMC-ratio}. 
We observe that the curves do not collapse over each other. 
Thence, we conclude that the scaling hypothesis (\ref{eq.6.121ultima})
fails in the grand-canonical ensemble and that the 
grand-canonical TG contact does not embed the full $N$-dependency 
as it does in the canonical ensemble, at least in the low- and 
intermediate-temperature regimes.

It is worth remarking that the QMC simulations are limited within 
the weak- and intermediate-interaction regimes, $\tilde{g}\ll 1$ 
and $\tilde{g}\sim 1$, and also for low and intermediate 
temperatures, $\tau \ll 1$ and $\tau \sim 1$. Moreover, 
by increasing the number of particles, QMC simulations 
become very difficult, and the relative errors increase 
together with the increase of $\tau$, $\tilde{g}$, and $N$. For 
more details on the QMC calculations see SANT'ANA \textit{et al.}\cite{paper3}

\chapter{Conclusions}
In this thesis we have studied the system formed by bosonic atoms loaded 
into optical lattices and the one-dimensional repulsively interacting Bose gas under a 
harmonic confinement.
Regarding the first part, bosons in optical lattices, we focused 
on the MI-SF quantum phase transition. We introduced the general 
basis behind the Bose-Hubbard model in order to construct the BH Hamiltonian, 
which is the main mathematical quantity for the evaluation of important 
physical properties of the system. Another cornerstone of our theory is the mean-field 
approximation, which neglects some quantum fluctuations over the so-called mean-field, 
which is simply the average of the bosonic lattice operator, throughout 
the whole lattice. Within the mean-field theory, which turns out to give rise to a Hamiltonian 
which is a summation of the one-site independent Hamiltonians, we are able to consider 
such separated single lattice sites, considerably simplifying all calculations. Then, we 
introduced the general expansion of the thermodynamic potential in the vicinity of 
a second-order phase transition using the analyticity assumptions suggested by Landau. This expansion provides valuable insights 
about the phase in which the system is in terms of the series coefficients of the thermodynamic potential. 
By applying perturbation theory, it is then possible to calculate the MI-SF quantum phase diagram. 
However, it has been noted that, between every adjacent Mott lobes in the MI-SF phase boundary, 
there happens to occur degeneracies between the respective energies of these Mott lobes. 
When usual perturbation theory is applied to this system, such degeneracies lead to nonphysical results, \textit{e.g.}, 
the condensate density, which is the order parameter in this system, 
vanishes in a point of the phase diagram where actually no quantum phase transition occurs. 
This result can be considered as an inconsistency that artificially arises from such an erroneous treatment. 
Therefore, in order to correct this problem, we developed two methods based on degenerate perturbation theory 
to correctly calculate meaningful physical quantities of the system. 
Both methods are based on a projection-operator formalism, that 
enables us to separate the Hilbert subspace where the degeneracies are contained in 
from the complementary subspace, which is free from any degeneracy. 
With this we are able to solve the degeneracy problems which are
typical of nondegenerate perturbation theories. 

Firstly, we developed the Brillouin-Wigner perturbation theory 
to tackle the zero-temperature system composed of bosons in optical lattices. Such 
results were published in Ref. \citeonline{martin}. After that, 
in order to generalize for finite temperature scenarios, we developed the finite-temperature 
degenerate perturbation theory approach and published the results in Ref. \citeonline{paper2}. 
These two methods presented in this thesis are fundamentally different: 
the BWPT is developed in the context of the time-independent Schr\"odinger equation, while 
the FTDPT is developed in the context of the time-dependent Schr\"odinger equation. In the latter, 
the relation between imaginary time and temperature arises naturally by means of the correspondence between 
the time-evolution operator and the partition function, which is simply the trace of the former after performing a Wick rotation. 
By correctly treating the degeneracies between the Mott lobes, we 
were in possession of a reliable method to calculate important physical quantities, 
with special attention to the condensate density due to its role as the order parameter.
Finally, it is important to remark that both methods developed in the first part 
of the thesis, the BWPT and FTDPT, not only provide relatively simple frameworks for calculating the condensate density,
but are actually very generic approaches in the sense that they can also be applied to a wide range
of optical-lattice systems, \textit{e.g.}, out-of-equilibrium systems,\cite{grass1,grass2} 
bosonic optical lattices with three-body constraint,\cite{3body} 
and different-geometry lattices such as Kagom\'e lattice,\cite{kagome} 
triangular and hexagonal lattices,\cite{hexagonal} 
as well as the Jaynes-Cummings lattice.\cite{nietner,nietner2}

Regarding the one-dimensional repulsively 
interacting Bose gas under harmonic confinement,
we have studied the asymptotic behavior of the momentum distribution 
at both zero and finite temperatures, with special attention to the intermediate- and 
strong-interaction regimes, since the weakly interacting regime simply reduces to 
the known ideal Bose gas. 
We began our study by solving the two-particle problem, which is 
an integrable model and provides valuable insights regarding the behavior 
of the many-body wave function at the contact point: the contact 
interaction originates a cusp condition in the wave function, implying its 
vanishing whenever two particles meet. Beyond two particles, the integrability breaks down  
and we have to restrict ourselves to specific regimes, such as the strongly interacting 
limit, also known as Tonks-Girardeau limit, where there happens the so-called 
\textit{fermionization} of the bosons, \textit{i.e.}, the bosons behave as the ideal Fermi gas, 
enabling the mapping of the bosonic system into a simpler noninteracting fermionic one. 
Then, we were able to derive an analytical formula for the Tan's contact of 
$N$ particles repulsively interacting under a harmonic confinement in one dimension in terms 
of the single-particle orbitals of the harmonic oscillator in the TG limit. 
Afterwards, we investigated the scaling properties of the contact in terms of the number of particles 
$N$. Firstly, we argued that, at large temperatures, the $N$-particle contact 
can be written as the two-particle contact times the number of pairs. This comes from the fact that, 
as the temperature increases, quantum correlations can be neglected, thence we simply need to 
consider the effects of every pair of particles contacting with each other.  
More specifically, we showed that the ratio between the $N$-particle contact at finite 
interaction strength and its correspondent in the TG limit is a universal 
$N$-independent scaling function. Following such advances in the high temperature scenario, 
we proposed a generalized scaling conjecture for the entire range of temperatures in the TG limit. 
Then, we properly demonstrated that such a conjecture works well for any considered temperature.

Furthermore, we also inspected the intermediate interaction regime, $\tilde{g} \sim 1$, with the help of QMC 
simulations. In such a regime, we found analogous results to the Tonks-Girardeau limit, 
\textit{i.e.}, meaning that all our previously stated conjectures regarding the range of temperatures also 
works when $\tilde{g} \sim 1$. Consequently, we conclude that 
the contact of $N$ bosons at temperature $T$ with repulsive interaction characterized 
by the adimensional interaction strength $\tilde{g}$ rescaled by the contact of $N$ strongly interacting bosons at temperature $T$, 
\textit{i.e.}, $\mathcal{C}_N(\tilde{g},T)/\mathcal{C}_N({\tilde{g}\to\infty,T})$, is a universal 
function of the adimensional interaction strength $\tilde{g}$ 
and the reduced temperature $\tau$ for the whole range of temperatures. Likewise, 
the ratio between the contact of $N$ bosons at
temperature $T$ with finite repulsive interaction and the generalized scaling function
$s(N)$, \textit{i.e.},
$\mathcal{C}_N(\tilde{g},T)/s(N)$, is also a universal function
of $\tilde{g}$ and $\tau$ for all temperatures.

Finally, as our main calculations were performed within the canonical 
ensemble, we compared our results to the grand-canonical calculations performed in Ref. \citeonline{pv2013}. 
We have checked that, at large temperatures, both the canonical and the 
grand-canonical contacts are proportional to the two-boson contact, with 
the proportionality factor depending on the number of pairs in the canonical ensemble 
as well as the average number of pairs in the grand-canonical one. On the other hand, 
at low and intermediate temperatures, the grand-canonical contact of an average number of 
$\av{N}$ bosons cannot be written as a function of the $\av{2}$-boson contact or 
of the $\av{N}$-TG-boson contact, to the extent of the QMC simulations, which is 
the intermediate-interaction regime. 

The results obtained in the referred part of the thesis were published in Ref. \citeonline{paper3}. Such developments are properly suited for probing 
the contact as well as the role of correlations and interactions 
in experiments with a small number of
bosonic particles.\cite{Zurn2012,Wenz2013} Moreover, it is worth noting 
that the canonical ensemble calculations realized in this thesis 
correspond to most real experimental conditions, since the number of 
atoms is fixed.\cite{paper3} From a conceptual point of view, it is an important step 
forward to the understanding of the correlation and interaction effects on the 
harmonically trapped interacting Bose gas in one dimension at finite temperature,  
as well as to the enlightenment of the role of the particle-number fluctuations.

\postextual


\bibliographystyle{unsrt}
\bibliography{references}

\begin{apendicesenv}
\partapendices

\chapter{\label{appendixA}Atomic collisions in cold gases}
In this appendix we derive the general theory regarding two-body 
collisions that happens to give rise to interacting properties 
of cold atoms in three and one dimensions. 
	
\section{The three-dimensional case}
Firstly, we want to find a solution for the respective 
Schr\"odinger equation
	\begin{equation}\label{app.sch}
	\left[-\frac{\hbar^2}{2\mu} \nabla^2
		+ V(r) - E \right] \psi(\mathbf{r}) = 0
\end{equation}
in terms of the relative distance $r=r_1-r_2$ and the 
reduced mass $\mu=m/2$. Now, let us expand the 
wave function making use of the spherical harmonics:\cite{arfken,blatt}
	\begin{equation}\label{app.sol}
		\psi(\mathbf{r}) = \sum_{\ell=0}^{\infty} 
		\phi_\ell(r) Y_{\ell,0}(\theta).
	\end{equation}
Substituting the solution 
	(\ref{app.sol}) into (\ref{app.sch}) we have that 
	\begin{equation}\label{app.chi}
	\frac{d^2}{dr^2} u_\ell(r) -\frac{\ell(\ell+1)}{r^2} u_\ell(r)
	+\frac{2\mu}{\hbar^2}\left[E-V(r)\right]u_\ell(r) = 0,
\end{equation}
	where $u_\ell(r)\equiv r \phi_\ell(r)$.
In the asymptotic region $r\gg r_0$, 
where $r_0$ is the range of the interatomic potential $V(r)$, 
the solution of (\ref{app.sch}) can be regarded as the 
contributions of an incident plane wave in the $x$ direction 
and a spherical scattered wave\cite{pitaevskii} 
	\begin{equation}\label{app.asym}
		\psi(\mathbf{r}) \underset{r\to\infty}\simeq \expo{ikx} + f(\theta) \frac{\expo{ikr}}{r},
\end{equation}
	where $f(\theta)$ is the scattering amplitude 
	as a function of the angle between $\vec{r}$ and the 
	$x$-axis $\theta$, which relates the differential cross section 
$d\sigma$ to the solid angle element $d\Omega$ in elastic scattering 
processes via 
	\begin{equation}
		d\sigma = |f(\theta)|^2 d\Omega.
	\end{equation}
Therefore, for $r\gg r_0$, the term $\propto 1/r^2$ in (\ref{app.chi}) can 
be neglected, yielding the solution 
	\begin{equation}\label{app.solr}
		u_\ell(r) = A_\ell \sin \left( 
		kr -\frac{\pi \ell}{2} +\delta_\ell \right),
	\end{equation}
where $A_\ell$ is the normalization constant and $\delta_\ell$ 
is simply a phase factor. By performing the expansion 
\begin{equation}
	\expo{ikx} = \frac{1}{2ikr}
	\sum_{\ell=0}^\infty (2\ell+1) P_\ell(\cos \theta)
	\left[\expo{ikr}-\expo{-i(kr-\pi \ell)}\right]
\end{equation}
and choosing $A_\ell = (2\ell+1)i^\ell \expo{i \delta_\ell}$,\cite{pitaevskii} 
	we have that, from (\ref{app.sol}) = (\ref{app.asym}), 
	\begin{equation}
		f(\theta) = \frac{1}{2ik}
		\sum_{\ell=0}^{\infty} (2\ell+1) P_\ell(\cos \theta)
		\left(\expo{2i\delta_\ell}-1\right).
	\end{equation}

In the ultracold regime, we can consider the atomic energies low enough 
so that we must only consider the $\ell=0$ term in the expansion, also known 
as $s$-wave approximation.\cite{blatt} Within such a consideration, the scattering amplitude reduces to 
\begin{equation}
	f(\theta) \approx \frac{\expo{2i\delta_0}-1}{2ik} 
	=\frac{1}{k \cot(\delta_0) -ik}.
\end{equation}
So, we now define the so-called \textit{s-wave scattering length} as 
\begin{equation}
	a \equiv -\lim_{k\to 0} f(\theta) = -\frac{\tan(\delta_0)}{k}.
\end{equation}
It is worth noting that the scattering length $a$ is deeply related to the 
initial condition of the radial solution (\ref{app.solr}) as 
\begin{equation}
	\frac{1}{u_0(r)} \frac{d u_0(r)}{dr}\Bigg|_{r=0}
	=k \cot (\delta_0) = -\frac{1}{a}.
\end{equation}
Moreover, the definition of the scattering length implies that 
the total scattering cross section yields $\sigma=4\pi a^2$, 
which turns out to be the same as an impenetrable sphere 
of radius $a$.\cite{blatt,huang2}

It was shown in HUANG; YANG\cite{huang2} that, instead of performing all considerations 
that were done until now, the problem can be equivalently formulated 
by introducing the pseudopotential 
\begin{equation}\label{pseudoV}
	\hat{V}_{ps}(r) = \frac{4\pi\hbar^2 a}{m} 
	\delta(\mathbf{r}) \frac{\partial}{\partial r} r
\end{equation}
in the respective Schr\"odinger equation
\begin{equation}\label{app.schps}
        \left[-\frac{\hbar^2}{2\mu} \nabla^2
		+ \hat{V}_{ps}(r) -E \right] \psi(\mathbf{r}) = 0.
\end{equation}

Finally, we conclude that, via the introduction of the pseudopotential,\cite{huang2} 
the interaction parameter in the three-dimensional case is defined as 
\begin{equation}
	g\equiv\frac{4\pi\hbar^2 a_{3\mathrm{D}}}{m},
\end{equation}
with $a_{3\mathrm{D}}$ being the three-dimensional $s$-wave scattering length.

\section{The one-dimensional case}
The one-dimensional scattering amplitude within the pseudopotential approximation 
was performed by M. Olshanii in 1998.\cite{olshanii2} Thus, the considerations under such a 
treatment are: a) the atomic motion is allowed to happen freely along the $x$-axis; 
b) the harmonic potential possesses a frequency $\omega_\perp$, acting along the 
$yz$-plane; c) the interaction between particles is of the form of the pseudopotential 
introduced in Eq. (\ref{pseudoV}).
Therefore, the respective Schr\"odinder equation reads
\begin{equation}\label{app.sch2d}
	\left[-\frac{\hbar^2}{2\mu} \nabla_x^2
	+ \hat{H}_\perp + \hat{V}_{ps} - E\right]\psi(\mathbf{r}) = 0,
\end{equation}
where
\begin{equation}
	\hat{H}_\perp = -\frac{\hbar^2}{2\mu}
	\left(\nabla_y^2+\nabla_z^2\right)
	+\frac{\mu \omega_\perp^2}{2} (y^2+z^2).
\end{equation}
Assuming that the incident wave 
corresponds to a particle in the ground state of $\hat{H}_\perp$, 
$\psi = \expo{ikx} \phi_0(y,z)$, the asymptotic solution 
of (\ref{app.sch2d}) is given by 
\begin{equation}
	\psi(\mathbf{r}) \underset{x\to\infty} 
	\sim \left(\expo{ikx} + f_{even}(k)\expo{ik|x|}
	+f_{odd}(k)\mathrm{sgn}(x)\expo{ik|x|}\right)
	\phi_0(y,z),
\end{equation}
where $f_{even/odd}(k)$ is the even/odd partial wave  
scattering amplitude. For low velocities 
$k r_0\ll 1$ and the tight confinement regime $a_\perp\ll |a_{3\mathrm{D}}|$, and 
regarding the one-dimensional 
pseudopotential 
\begin{equation}\label{1dpseudo}
	V_{ps}(x)=-\frac{2\hbar^2}{ma_{1\mathrm{D}}} \delta(x),
\end{equation}
the odd scattering amplitude vanishes, 
$f_{odd}=0$, while the even one yields\footnote{For the detailed analytical calculation, see Ref. \citeonline {olshanii2}.} 
\begin{equation}
	f_{even}(k)  \approx 
	-\left(1+ika_{1\mathrm{D}}\right)^{-1},
\end{equation}
where
\begin{equation}
	a_{1\mathrm{D}} = -\frac{a_\perp^2}{2 a_{3\mathrm{D}}}
	\left(1-C\frac{a_{3\mathrm{D}}}{a_\perp}\right)
\end{equation}
is the one-dimensional $s$-wave scattering length, 
$a_\perp\equiv \sqrt{\hbar/\mu \omega_\perp}$, 
and 
\begin{equation}
	C \equiv \lim_{u\to\infty} 
	\left(\int_0^u \frac{du'}{\sqrt{u'}}
	-\sum_{u'=1}^u \frac{1}{\sqrt{u'}}
	\right) \approx 1.4603 .
\end{equation}
In conclusion, the analogous 
one-dimensional pseudopotential (\ref{1dpseudo}) 
retains the proper scattering behavior, so that 
we can define the corresponding interaction strength in 1D 
as 
\begin{equation} 
	g\equiv -\frac{2\hbar^2}{ma_{1\mathrm{D}}}.
\end{equation}

\chapter{\label{appendixB}Bose-Einstein condensation in low dimensions}
In this appendix we present a concise description of the underlying 
physics involved in low-dimensional BECs. 
Let us begin with the $D$-dimensional ideal Bose gas. The chemical potential 
$\mu$ is determined by satisfying the number of particles formula 
\begin{equation}
	\sum_\mathbf{k} \frac{1}{\expo{\beta (E_k-\mu)}-1} = N.
\end{equation}
Now, suppose the chemical potential vanishes, $\mu \rightarrow 0$, 
at a critical temperature $T_c \equiv (k_B \beta_c)^{-1}$. Then, 
as the density of states (DOS) becomes a continuous one,  $\rho(E)$, we perform the substitution 
$\sum_\mathbf{k} \rightarrow (L/2\pi)^D \int d^Dk$ and then the number of particles 
formula reads 
\begin{equation}\label{eq.dos}
	\int_0^\infty dE \frac{\rho(E)}{\expo{\beta_c E}-1} = N.
\end{equation}
However, as the DOS depends on the dimensionality as $\rho(E)\propto E^{D/2-1}$,\cite{ueda} 
the integral in (\ref{eq.dos}) diverges for $D\leq 2$. This implies that $\mu \neq 0$ in 
$D\leq 2$ and that no BEC occurs at $T>0$. 

In the 3D scenario, $\rho(E) \propto \sqrt{E}$, thence $\rho(E) \overset{E \to 0}\to 0$, 
meaning that a macroscopic occupation of the lowest-energy 
state is favorable. Differently, in 1D we have that $\rho(E) \propto E^{-1/2}$, 
implying that $\rho(E) \overset{E\to 0}\to \infty$, disfavoring the occupancy 
of the lowest-energy configuration. In 2D, the DOS is independent of $E$ and 
the analysis for the occurrence of BEC is a bit more subtle: 
the main point is that in 2D the off-diagonal long-range order breaks down, but 
a quasi long-range order associated with a topological quantum phase transition emerges.\footnote{For a detailed explanation, see UEDA.\cite{ueda}}

When the system is spatially confined, the picture changes and 
BEC can emerge in low dimensions $d\leq 2$.\cite{bagnato} This happens 
because there is a change in the DOS-dependency on the energy due to 
the dependency on the size occupied by such a nonuniform gas, \textit{e.g.}, 
consider the power law potential $V(r) \propto r^\alpha$; because of 
$L^D \propto E^{D/\alpha}$,\cite{ueda} the density of states changes according to 
$\rho(E) \propto L^D E^{D/2-1} = E^{D/\alpha + D/2-1}$. Thus, the condition 
for the integral (\ref{eq.dos}) to converge becomes 
\begin{equation}
	D > \frac{2\alpha}{\alpha+2},
\end{equation}
which results in $D>3/2$ for a linear potential $V(r) \propto r$ 
and $D>1$ for a harmonic potential $V(r) \propto r^2$. 

\section{Hohenberg-Mermin-Wagner theorem}
In this section we are concerned with the derivation 
of the \textit{Hohenberg-Mermin-Wagner theorem}, that 
states the absence of the $U(1)$-symmetry breaking at $T>0$ in $D\leq 2$ and is 
associated with the proof of Bogoliubov's inequality:\cite{mermin}
\begin{equation}\label{bogoine}
	\av{\{\hat{A},\hat{A}^\dagger\}} \av{[\hat{B}^\dagger,[\hat{H},\hat{B}]]}
	\geq 2 k_B T |\av{[\hat{A},\hat{B}]}|^2.
\end{equation}
Here $\hat{A}$ and $\hat{B}$ are arbitrary operators, $\hat{H}$ is the  
Hamiltonian of the system, $[\cdots]$ and $\{\cdots\}$ are, respectively, the commutator and the 
anticommutator, while $\av{\cdots}$ stands for the thermal average of an arbitrary 
operator $\hat{\mathcal{O}}$, 
\begin{equation}
	\av{\hat{\mathcal{O}}} \equiv \sum_n P_n 
	\bra{n}\hat{\mathcal{O}} \ket{n}, \, P_n 
	\equiv \frac{\expo{-\beta E_n}}{\Tr{\expo{-\beta \hat{H}}}},
\end{equation}
where $\hat{H}\ket{n} = E_n \ket{n}$. Let us define 
\begin{equation}
	(\hat{A},\hat{B}) \equiv \sum_{m\neq n} 
	\bra{n}\hat{A}^\dagger\ket{m} \bra{m}\hat{B}\ket{n} 
	\frac{P_m-P_n}{E_n-E_m}.
\end{equation}
Now, making use of the hyperbolic inequality $x^{-1}\tanh(x)\leq 1$, we have that 
\begin{equation}
	\begin{aligned}
		&\frac{\tanh\left[\beta(E_m-E_n)/2\right]}{\beta(E_m-E_n)/2} 
		= \frac{2(p_m-p_n)}{\beta (p_m-p_n)(E_n-E_m)}\leq 1 \\
		&\Rightarrow 
		0<\frac{p_m-p_n}{E_n-E_m} \leq \frac{\beta}{2}(p_m-p_n).
	\end{aligned}
\end{equation}
Thence, it follows that 
\begin{equation}\label{eq.(a,a)}
	(\hat{A},\hat{A}) = \sum_{m\neq n} |\bra{m}\hat{A}\ket{n}|^2
	\frac{p_m-p_n}{E_n-E_m} \leq \frac{\beta}{2} \sum_{m,n}|\bra{m}\hat{A}\ket{n}|^2
	(p_m-p_n) = \frac{\beta}{2}\av{\{\hat{A},\hat{A}^\dagger\}}.
\end{equation}
In addition, let $\hat{C}\equiv [\hat{B}^\dagger,\hat{H}]$, so we work out 
\begin{equation}\label{eq.(c,c)}
	\begin{aligned}
		(\hat{C},\hat{C}) &= \sum_{m\neq n} \bra{n}\hat{C}^\dagger\ket{m} 
		\bra{m}[\hat{B}^\dagger,\hat{H}]\ket{n} \frac{p_m-p_n}{E_n-E_m} \\
		&= \sum_{m,n} \bra{n}\hat{C}^\dagger\ket{m} \bra{m}\hat{B}^\dagger\ket{n}(p_m-p_n) \\
		&=\av{[\hat{B}^\dagger,\hat{C}^\dagger]} = \av{[\hat{B}^\dagger,[\hat{H},\hat{B}]]},
	\end{aligned}
\end{equation}
and 
\begin{equation}\label{eq.(a,c)}
        \begin{aligned}
                (\hat{A},\hat{C}) &= \sum_{m\neq n} \bra{n}\hat{A}^\dagger\ket{m}
                \bra{m}[\hat{B}^\dagger,\hat{H}]\ket{n} \frac{p_m-p_n}{E_n-E_m} \\
                &= \sum_{m,n} \bra{n}\hat{A}^\dagger\ket{m} \bra{m}\hat{B}^\dagger\ket{n}(p_m-p_n) \\
                &=\av{[\hat{B}^\dagger,\hat{A}^\dagger]}.
        \end{aligned}
\end{equation}
Now, let us introduce the Schwarz inequality\cite{handbook,arfken} 
\begin{equation}\label{schwarz}
	(\hat{A},\hat{A})(\hat{C},\hat{C})\geq |(\hat{A},\hat{C})|^2.
\end{equation}
By the direct substitution of (\ref{eq.(a,a)}), (\ref{eq.(c,c)}), and  (\ref{eq.(a,c)}) 
into (\ref{schwarz}), we prove the Bogoliubov inequality (\ref{bogoine}).

Now, let the operators be $\hat{A}=\hat{a}_{\mathbf{p}}^\dagger$ and 
$\hat{B}=\hat{\rho}_{\mathbf{p}}\equiv \sum_{\mathbf{k}}
\hat{a}_{\mathbf{k}}^\dagger \hat{a}_{\mathbf{k}+\mathbf{p}}$, then 
\begin{subequations}
	\begin{align}
		\av{\{\hat{A},\hat{A}^\dagger\}} &= 2n_\mathbf{p}+1, \, n_\mathbf{p}\equiv \av{\hat{a}_\mathbf{p}^\dagger\hat{a}_\mathbf{p}},\\
		[\hat{B}^\dagger,[\hat{H},\hat{B}]] &= [\hat{\rho}^\dagger_\mathbf{p},[\hat{H},\hat{\rho}^\dagger_\mathbf{p}]] =
		N\frac{\mathbf{p}^2}{m},\footnotemark \\
		[\hat{A},\hat{B}] &= -\hat{a}_0^\dagger.
	\end{align}
\end{subequations} 
\footnotetext{
Consider the Hamiltonian 
\begin{equation*}
	\begin{aligned}
		\hat{H} &= \sum_\mathbf{p} E_\mathbf{p}\hat{a}^\dagger_\mathbf{p} \hat{a}_\mathbf{p}
	+\frac{1}{2} \sum_{\mathbf{k},\mathbf{p},\mathbf{q}} V_\mathbf{k} \hat{a}^\dagger_{\mathbf{p}+\mathbf{k}}
	\hat{a}^\dagger_{\mathbf{q}-\mathbf{k}} \hat{a}_\mathbf{q}\hat{a}_\mathbf{p}\\
		&=  \sum_\mathbf{p} E_\mathbf{p}\hat{a}^\dagger_\mathbf{p} \hat{a}_\mathbf{p}
		+\frac{1}{2} \sum_\mathbf{k} V_\mathbf{k} \hat{\rho}^\dagger_\mathbf{k} \hat{\rho}_\mathbf{k} 
		-\frac{1}{2} N\sum_\mathbf{p} V_\mathbf{p}.
	\end{aligned}
\end{equation*}
By performing some algebraic manipulations and considering 
$E_\mathbf{k}=\mathbf{k}^2/2m$ and $\hat{N}\equiv \sum_\mathbf{k} \hat{a}^\dagger_\mathbf{k}\hat{a}_\mathbf{k}$, we have that  
\begin{subequations}
	\begin{align*}
		[\hat{H},\hat{\rho}_\mathbf{p}] &= \sum_\mathbf{k}
		(E_\mathbf{k}-E_{\mathbf{k}+\mathbf{p}}) \hat{a}_\mathbf{k}^\dagger
		\hat{a}_{\mathbf{k}+\mathbf{p}},\\
		 [\hat{\rho}^\dagger_\mathbf{p},[\hat{H},\hat{\rho}_\mathbf{p}]] &=
		 \sum_\mathbf{k}(E_{\mathbf{k}+\mathbf{p}}+E_{\mathbf{k}-\mathbf{p}}-2E_\mathbf{k})
		  \hat{a}_\mathbf{k}^\dagger\hat{a}_\mathbf{k}=\hat{N}\frac{\mathbf{p}^2}{m}.
	\end{align*}
\end{subequations}
}
By using these results into Eq. (\ref{bogoine}), the Bogoliubov inequality results in 
\begin{equation}\label{eq.np}
	n_\mathbf{p} \geq \frac{mk_B T}{p^2} \frac{|\av{\hat{a}_0}|^2}{N} -\frac{1}{2}.
\end{equation}
Therefore, it follows from (\ref{eq.np}) the absence of the $U(1)$-symmetry breaking in 
$D\leq 2$ at $T>0$ because of the divergence of the sum $\sum_\mathbf{p} n_\mathbf{p}$ 
(unless $\av{\hat{a}_0}=0$). In the zero-temperature regime, $\av{\hat{a}_0}\neq 0$ can happen in 
2D, but it cannot in 1D. In order to show this, 
let us make use of the following inequality:\cite{ueda}
\begin{equation}\label{newine}
	\av{\{\hat{A}^\dagger,\hat{A}\}}\av{\{\hat{B}^\dagger,\hat{B}\}}
	\geq |\av{[\hat{A}^\dagger,\hat{B}]}|^2.
\end{equation}
Similarly to the previous association, let the operators be 
$\hat{A}=\hat{a}_{\mathbf{p}}^\dagger$ and 
$\hat{B}=\hat{\rho}_{\mathbf{p}}$. Hence, 
\begin{subequations}
	\begin{align}
		\av{\{\hat{A},\hat{A}^\dagger\}} &= 2n_\mathbf{p}+1,\\
		\av{\{\hat{B}^\dagger,\hat{B}\}} &= 2\av{\hat{\rho}_\mathbf{p}\hat{\rho}_\mathbf{p}^\dagger}, \\
		[\hat{A},\hat{B}] &= -\hat{a}_0^\dagger.
	\end{align}
\end{subequations}
Consequently, the inequality (\ref{newine}) results in 
\begin{equation}
	n_\mathbf{p} \geq \frac{|\av{\hat{a}_0}|^2}
	{2\av{\hat{\rho}_\mathbf{p}\hat{\rho}_\mathbf{p}^\dagger}} -\frac{1}{2}.
\end{equation}
In the low-momentum limit $p\to 0$, we have that 
\begin{equation}
\av{\hat{\rho}_\mathbf{p}\hat{\rho}_\mathbf{p}^\dagger}
\leq \frac{Np}{2mv_s},
\end{equation}
where $v_s$ is the sound velocity.
\begin{proof}
Starting with
\begin{equation}
[\hat{\rho}^\dagger_\mathbf{p},[\hat{H},\hat{\rho}_\mathbf{p}]]
=\hat{N}\frac{\mathbf{p}^2}{m},
\end{equation}
multiplying both sides by $\expo{\beta \hat{H}}/\Tr{\expo{\beta \hat{H}}}$ 
and taking the trace,\cite{ueda} we obtain 
\begin{equation}\label{eq.E5}
\int_0^\infty \hbar \omega 
S(\mathbf{p},\omega) d\omega = E_\mathbf{p}N, 
\end{equation}
where
\begin{equation}
S(\mathbf{p},\omega) \equiv \frac{1}{\Tr{\expo{\beta \hat{H}}}}
\sum_{m,n} \expo{\beta E_m} \left(|\bra{m}\hat{\rho}^\dagger_\mathbf{p}\ket{n}|^2+
|\bra{m}\hat{\rho}_\mathbf{p}\ket{n}|^2\right) \delta(\hbar \omega - \hbar\omega_{nm}),
\end{equation}
is called \textit{dynamic structure factor} and $\omega_{nm}\equiv\omega_n-\omega_m$. 
From the Schwarz inequality (\ref{schwarz}), one obtains 
\begin{equation}
S(\mathbf{p}) = \int_{-\infty}^{+\infty} S(\mathbf{p},\omega) d\omega 
\leq \sqrt{\int_{-\infty}^{+\infty} \hbar \omega S(\mathbf{p},\omega) d\omega 
\int_{-\infty}^{+\infty} (\hbar \omega)^{-1} S(\mathbf{p},\omega) d\omega},
\end{equation}
with $S(\mathbf{p})$ being called \textit{static structure factor}. From 
(\ref{eq.E5}), the inequality reduces to 
\begin{equation}
S(\mathbf{p}) \leq \sqrt{2 E_\mathbf{p} N
 \int_{-\infty}^{+\infty} (\hbar \omega)^{-1} S(\mathbf{p},\omega) d\omega}.
\end{equation}
In the low-momentum limit, it can be shown that\cite{ueda}  
\begin{equation}
\lim_{p\to 0}  \int_{-\infty}^{+\infty} (\hbar \omega)^{-1} S(\mathbf{p},\omega) d\omega
= \frac{N\hbar}{m v_s^2}, 
\end{equation}
which results in  
\begin{equation}
S(\mathbf{p})\Big|_{p\to 0}
\leq \frac{Np}{mv_s}.
\end{equation}
\end{proof}

Consequently, we have that 
\begin{equation}\label{ineq.np}
n_\mathbf{p} \geq \frac{mv_s |\hat{a}_0|^2}{2Np} - \frac{1}{2}.
\end{equation}
The aftermath of Eq. (\ref{ineq.np}) is the following: 
because $\sum_\mathbf{p} n_\mathbf{p}$ diverges in 1D, we conclude 
that the $U(1)$-symmetry breaking does not occur, even at $T=0$, unless $\av{\hat{a}_0}$ 
vanishes. On the other hand, for $D=2$, the sum $\sum_\mathbf{p} n_\mathbf{p}$ is a convergent one, thence 
it is possible the appearance of the $U(1)$-symmetry breaking.

To conclude this appendix, we showed that the Hohenberg-Mermin-Wagner theorem 
forbids the breaking of the $U(1)$ symmetry in low dimensions. However, BEC 
can exist even without the presence of the $U(1)$-symmetry breaking.\cite{ueda}

\chapter{\label{appendixC}Bose-Einstein condensation and superfluidity}
The occurrence of BEC is associated with an important quantity that, depending on 
the textbook, is known as \textit{one-body density matrix} or \textit{one-body correlator}, 
\begin{equation}
	\varrho^{(1)}(\mathbf{r},\mathbf{r}') \equiv 
	\Tr{\hat{\varrho}\hat{\Psi}^\dagger(\mathbf{r})\hat{\Psi}(\mathbf{r}')}
	=\av{\hat{\Psi}^\dagger(\mathbf{r})\hat{\Psi}(\mathbf{r}')},
\end{equation}
where $\hat{\varrho}$ is the density operator, while 
$\hat{\Psi}^\dagger(\mathbf{r})$ and $\hat{\Psi}(\mathbf{r})$ 
are the field operators that create and annihilate a particle at the spatial 
position $\mathbf{r}$, respectively.
This physical quantity has a clear interpretation: it is the probability 
amplitude of the annihilation of a particle at position $\mathbf{r}'$ followed 
by the creation of a particle at position $\mathbf{r}$. As the system undergoes BEC, 
the waves of the individual particles overlap over each other, enhancing their 
indistinguishability and, consequently, producing the effect 
of increasing the correlation between long-distance particles. 
Mathematically, this fact corresponds to\cite{yang62}
\begin{equation}
\lim_{|\mathbf{r}-\mathbf{r}'|\to\infty} \varrho^{(1)}(\mathbf{r},\mathbf{r}') \neq 0,
\end{equation}
and is known as \textit{off-diagonal long-range order} (ODLRO).

As we shall discuss, the concepts of BEC and superfluidity are strongly related, 
although there is a subtle relation between them. For example, the special case of 
an ideal gas that undergoes BEC presents no superfluidity and a 2D superfluid presents 
no BEC. Despite of these two referred cases, there are many systems in which BEC and 
superfluidity coexist. 

Let us consider a time-dependent system and its associated one-body density matrix 
\begin{equation}
	\varrho^{(1)}(\mathbf{r},\mathbf{r}';t) 
	=\av{\hat{\psi}^\dagger(\mathbf{r},t)\hat{\psi}(\mathbf{r}',t)}
	=\sum_\nu n_\nu(t) \psi_\nu^*(\mathbf{r},t) \psi_\nu(\mathbf{r}',t),
\end{equation}
with $n_\nu (t)$ satisfying 
\begin{equation}
\sum_\nu \int_0^\infty n_\nu(t) dt = N.
\end{equation}
Considering that $\nu=0$ represents the BEC mode, it follows that 
\begin{equation}
	\lim_{|\mathbf{r}-\mathbf{r}'|\to\infty}\varrho^{(1)}(\mathbf{r},\mathbf{r}';t) 
	=n_0 \psi_0^*(\mathbf{r},t) \psi_0(\mathbf{r}',t).
\end{equation}
We can interpret 
\begin{equation}
\Psi(\mathbf{r},t) \equiv \sqrt{n_0}\psi_0(\mathbf{r},t)
\end{equation}
as being the condensate wave function as well as 
\begin{equation}
\varrho(\mathbf{r},t) \equiv \Psi^*(\mathbf{r},t) \Psi(\mathbf{r},t)
\end{equation}
as being the particle density in the BEC state. 
From the continuity equation 
\begin{equation}
\frac{\partial}{\partial t} \varrho(\mathbf{r},t) + \nabla \cdot \mathbf{j}(\mathbf{r},t)=0,
\end{equation}
the current of particles yields 
\begin{equation}
\mathbf{j}(\mathbf{r},t)= -i\frac{\hbar}{2m}\left[
\Psi^*(\mathbf{r},t)\nabla \Psi(\mathbf{r},t)- \Psi(\mathbf{r},t)\nabla \Psi^*(\mathbf{r},t) \right].
\end{equation}
Decomposing the BEC wave function into an amplitude times a phase, 
\begin{equation}
\Psi(\mathbf{r},t) = A(\mathbf{r},t)\expo{i\phi(\mathbf{r},t)},
\end{equation}
the density and the current possess the form 
\begin{subequations}
\begin{align}
\varrho(\mathbf{r},t) &= A^2(\mathbf{r},t),\\
\mathbf{j}(\mathbf{r},t) &= \frac{\hbar}{m}A^2(\mathbf{r},t)\nabla \phi(\mathbf{r},t).
\end{align}
\end{subequations}
Now, from the definition of the superfluid velocity, 
\begin{equation}
\mathbf{v}_s(\mathbf{r},t)\equiv\frac{\mathbf{j}(\mathbf{r},t)}{\varrho(\mathbf{r},t)} 
= \frac{\hbar}{m} \nabla \phi(\mathbf{r},t),
\end{equation}
it is possible to realize the connection between BEC and superfluidity:  
the phase of the condensate wave function $\phi$ plays the role of the velocity potential 
in the superfluid.

\end{apendicesenv}

\begin{anexosenv}

\partanexos

\chapter{Publications}
M.~Kübler, F.~T. Sant'Ana, F.~E.~A. dos Santos, and A.~Pelster,
\newblock Improving mean-field theory for bosons in optical lattices via
  degenerate perturbation theory,
\newblock {\em Phys. Rev. A} \textbf{99}, 063603 (2019).

Felipe Taha Sant'Ana, Axel Pelster, and Francisco Ednilson Alves dos Santos,
\newblock Finite-temperature degenerate perturbation theory for bosons in optical lattices,
\newblock {\em  Phys. Rev. A} \textbf{100}, 043609 (2019).

F.~T.~Sant'Ana, F.~H\'ebert, V.~Rousseau, M.~Albert, and P.~Vignolo,
\newblock Scaling properties of Tan's contact: Embedding pairs and correlation effect in the Tonks-Girardeau limit,
\newblock {\em  Phys. Rev. A} \textbf{100}, 063608 (2019).

\end{anexosenv}

%
%





\end{document}